\documentclass[aps,prd,floatfix,showpacs,amsmath,amsfonts,amssymb,superscriptaddress]{revtex4}

\usepackage{tabularx}
\usepackage{subfig}
\usepackage[dvipsnames]{xcolor}
\usepackage{comment}
\usepackage{enumitem}
\usepackage{graphicx}
\usepackage{float}
\usepackage{dcolumn}
\usepackage{bm}
\usepackage{mathrsfs}
\usepackage{amsmath}
\usepackage{amssymb}
\usepackage{amsfonts}
\usepackage{dsfont}
\usepackage{url}

\begin{document}

\title{Quasinormal Modes of Gauss--Bonnet Black Holes via the Spectral Method: Scalar, Vector, and Tensor Perturbations}

\author{Davide Batic}
\email{davide.batic@ku.ac.ae}
\affiliation{Mathematics Department, Khalifa University of Science and Technology, PO Box 127788, Abu Dhabi, United Arab Emirates}
\author{Denys Dutykh}
\email{denys.dutykh@ku.ac.ae}
\affiliation{Mathematics Department, Khalifa University of Science and Technology, PO Box 127788, Abu Dhabi, United Arab Emirates}
\date{\today}

\begin{abstract}
We present a unified study of scalar, vector, and tensor quasinormal modes (QNMs) of Schwarzschild black holes corrected by a Gauss--Bonnet (GB) term in higher dimensions. Using a high-precision Chebyshev spectral method, we map the QNM spectra across $D\in\{5,6,7,8,10,11,12,26\}$ well beyond the regime where sixth-order WKB and characteristic-integration techniques remain reliable. Across the three spin sectors, we find several robust signatures of higher-curvature dynamics: the appearance of overdamped purely imaginary modes, non-monotonic behaviour in the real parts of higher overtones, and a strong amplification of the dimensionless QNM frequencies in string-motivated dimensions. In the scalar and vector sectors, we uncover an exact isospectrality between the scalar monopole ($\ell=0$) and vector dipole ($\ell=1$) at vanishing GB coupling, and we provide an analytic proof based on a Darboux factorisation of the corresponding Hamiltonians. In the tensor sector, we obtain the first numerical confirmation of the long-predicted instability in six dimensions; its onset is sharply captured by the Cohn--Calogero bound and leads to the mass threshold $GM\leq 158.1\,\alpha^{3/2}$. No analogous instability is found for $D\geq 7$, and no tensor isospectrality occurs. Converting the dimensionless frequencies to physical units suggests that the amplified modes in higher dimensions may enter the sensitivity window of future space-based detectors such as DECIGO. The merged analysis provides a comprehensive benchmark for QNMs in Einstein--Gauss--Bonnet gravity and highlights the limitations of standard approximation schemes in the strong-coupling and high-overtone regimes.
\end{abstract}

\maketitle

\section{Introduction}

The study of black hole perturbations and their associated QNMs has become a central topic in gravitational physics. These damped resonances encode information about the underlying spacetime geometry, the nature of the compact object, and, potentially, deviations from general relativity. With the advent of gravitational-wave astronomy, ringdown spectroscopy has transformed QNMs from a purely theoretical subject into an observational probe of strong-field gravity \cite{Kokkotas1999LR, Berti2009CQG, Abbott2016PRL, Dreyer2004CQG, Giesler2019PRX}.

Among the most extensively studied higher-curvature extensions of general relativity is Einstein--Gauss--Bonnet gravity. The GB term is singled out by string theory as a special quadratic-curvature correction that appears in the low-energy effective action \cite{Zwiebach1985PLB, Gross1987NPB}. While it is topological in four dimensions \cite{Lanczos1938AM, Lovelock1971JMP}, it becomes dynamically relevant in higher-dimensional spacetimes \cite{Boulware1985PRL}. This makes higher-dimensional Schwarzschild--Gauss--Bonnet black holes a natural arena in which to investigate how string-inspired corrections modify black hole spectra, stability, and potential observational signatures.

Earlier studies of QNMs in Gauss--Bonnet black holes established several important features of the scalar, vector, and tensor sectors, but they also exposed substantial limitations of standard approximation schemes. In particular, WKB-based analyses are often restricted to low overtones and simple single-barrier potentials, whereas the effective potentials arising in the GB problem can develop negative pockets, multiple peaks, or strong-coupling deformations that are not captured reliably by those methods \cite{Dotti2005CQG, Dotti2005PRD, Konoplya2005PRDa, Abdalla2005PRD, Chakrabarti2007GRG, Daghigh2012PRD, Moura2007CQG}. As a consequence, strong-coupling phenomena, high overtones, overdamped modes, and instability thresholds have remained only partially understood.

In this work we merge the scalar, vector, and tensor analyses into a single comprehensive study. We consider perturbations in dimensions $D\in\{5,6,7,8,10,11,12,26\}$ and employ a high-precision Chebyshev spectral method that has already proved effective in a variety of QNM problems \cite{Batic2024EPJC, Batic2024PRD, Batic2025CQG, Batic2024CQG, Batic2025EPJC, Batic2025PRSA}. This unified treatment makes it possible to compare the three spin sectors on the same footing and to disentangle genuinely universal features from sector-specific behaviour.

Our analysis reveals a rich and coherent picture. In all sectors, higher-curvature effects reorganise the spectra through overdamped modes, overtone non-monotonicity, and characteristic morphology changes in the complex-frequency plane. In the scalar and vector sectors we uncover an exact scalar--vector isospectrality at vanishing GB coupling, relating the scalar monopole to the vector dipole in every spacetime dimension $D\geq 5$; Appendix~\ref{AppendixA} shows that this property follows from a Darboux factorisation of the corresponding Schr\"odinger operators. In the tensor sector we obtain the first direct numerical confirmation of the six-dimensional instability predicted long ago on analytical grounds \cite{Dotti2005CQG, Dotti2005PRD}, and we show that its onset is sharply tracked by the Cohn--Calogero bound. We also find that the dimensionless QNM frequencies are strongly amplified in string-motivated dimensions, suggesting that future space-based detectors such as DECIGO may provide an observational handle on higher-curvature corrections \cite{Kawamura2008JPCS, Kawamura2019IJMPD, Kawamura2021PTEP}.

The paper is organised as follows. Section II reviews the background geometry and the scalar sector. Section III summarises the numerical method. Sections IV, VI, and VIII present the scalar, vector, and tensor spectra, respectively, while Sections V and VII derive the vector and tensor master equations in the form required by the spectral method. Section IX collects the main conclusions, and Appendix~\ref{AppendixA} contains the analytic proof of scalar--vector isospectrality.

\section{Background geometry and scalar perturbations}

Let the spacetime dimension be denoted by $D = n+2$ where $n \geqslant 3$. In natural units, where $c = G_N = 1$, the line element is given by \cite{Iyer1989CQG}
\begin{equation}\label{phantomWH}
  ds^2=-f(r)dt^2+\frac{dr^2}{f(r)}+r^2 d\Omega^2_n
\end{equation}
where 
\begin{equation}\label{f}
f(r)=1+\frac{r^2}{2\widetilde{\alpha}k}\left[1-\sqrt{1+\frac{8M\widetilde{\alpha}k}{r^{n+1}}}\right],\quad
\widetilde{\alpha}=(n-1)(n-2)\alpha,\quad
k=\int d\Omega_n   
\end{equation}
with
\begin{equation}
d\Omega_n^2=d\vartheta_n^2+\sin^2{\vartheta_n}\left[d\vartheta^2_{n-1}+\sin^2{\vartheta_{n-1}}\left(
d\vartheta^2_{n-2}+\cdots+\sin^2{\vartheta_3}\left(d\vartheta^2_2+\sin^2{\vartheta_2}d\vartheta^2_1\right)\cdots\right)\right].
\end{equation}
Here, $\vartheta_1\in[0,2\pi)$ and $\vartheta_i\in [0,\pi)$ for all $i\in\{2,3,\cdots,n\}$. According to \cite{Iyer1989CQG}, the horizon equation for the Schwarzschild metric with GB correction is 
\begin{equation}
    r_h^{D-3}+\widetilde{\alpha}k r_h^{D-5}=2M,
\end{equation}
where $r_h$ denotes the horizon position and $M$ represents the total mass of the black hole. In order to study the roots of the above equation, we recall that $M$ has dimensions of $L^{n-1}$. Hence, we can introduce the rescalings \cite{Iyer1989CQG}
\begin{equation}\label{rescalings}
\rho=\frac{r}{M^\frac{1}{n-1}},\quad
\widehat{\alpha}=\frac{\widetilde{\alpha}k}{M^\frac{2}{n-1}},
\end{equation}
by means of which we rewrite the function $f(r)$ as follows
\begin{equation}\label{frho}
     f(\rho)=1+\frac{\rho^2}{2\widehat{\alpha}}\left(1-\sqrt{1+\frac{8\widehat{\alpha}}{\rho^{n+1}}}\right)=1-\frac{4\rho^2}{\rho^{n+1}+\sqrt{\rho^{2n+2}+8\widehat{\alpha}\rho^{n+1}}}.
\end{equation}
Moreover, the equation governing the event horizon becomes
\begin{equation}\label{event}
    \rho_h^{n-1}+\widehat{\alpha}\rho_h^{n-3}=2.
\end{equation}
\begin{figure}[ht!] 
    \includegraphics[width=0.3\textwidth]{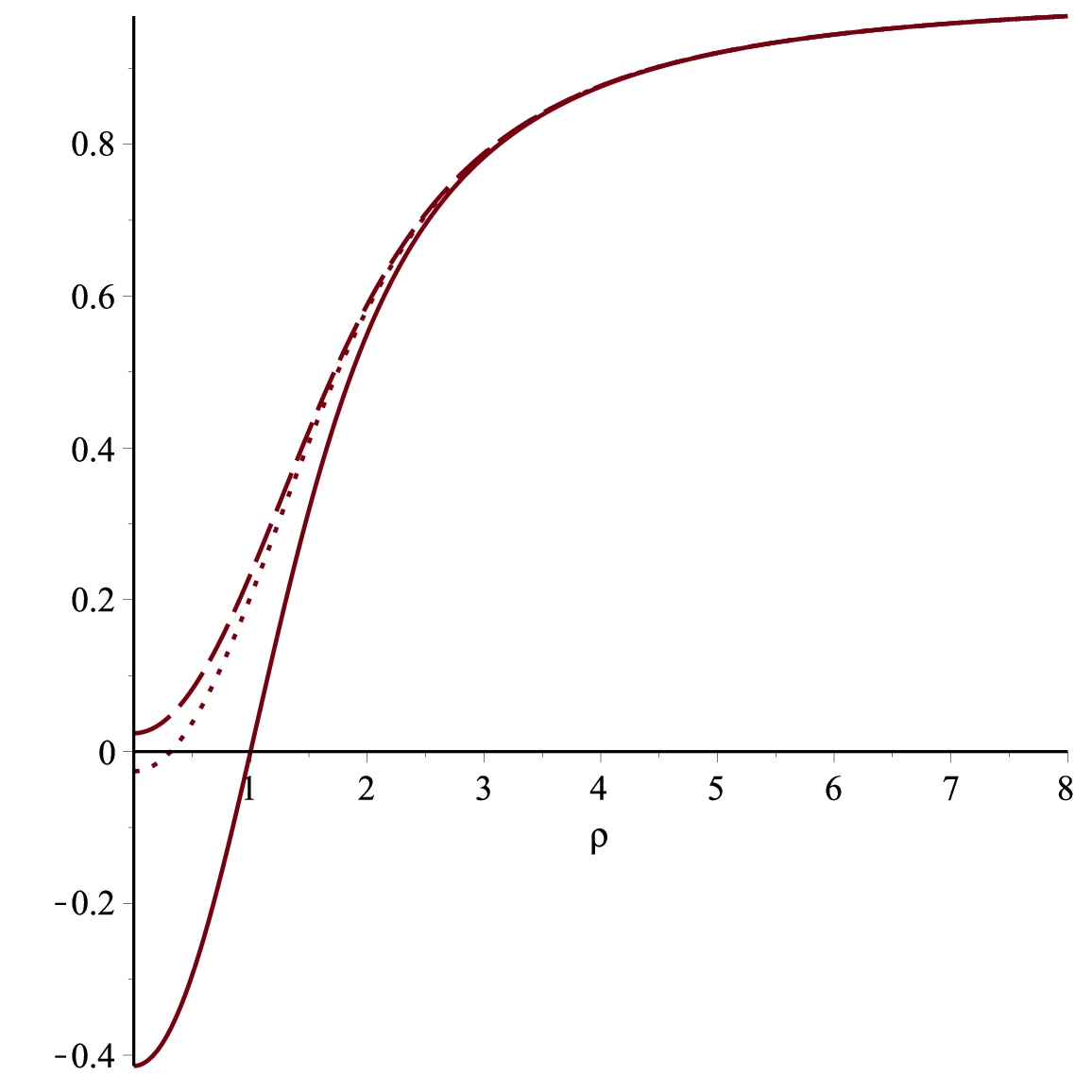}
    \includegraphics[width=0.3\textwidth]{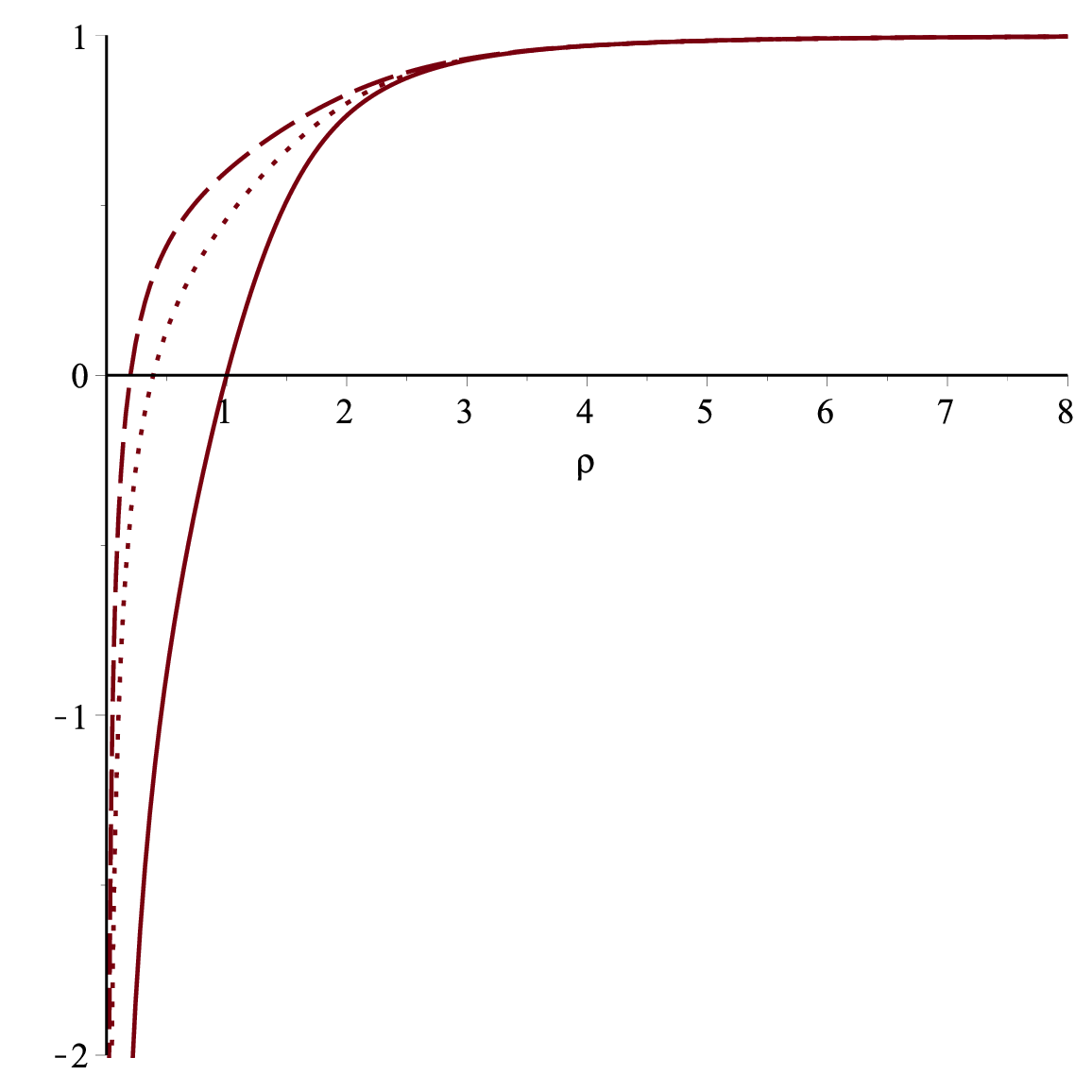}
    \includegraphics[width=0.3\textwidth]{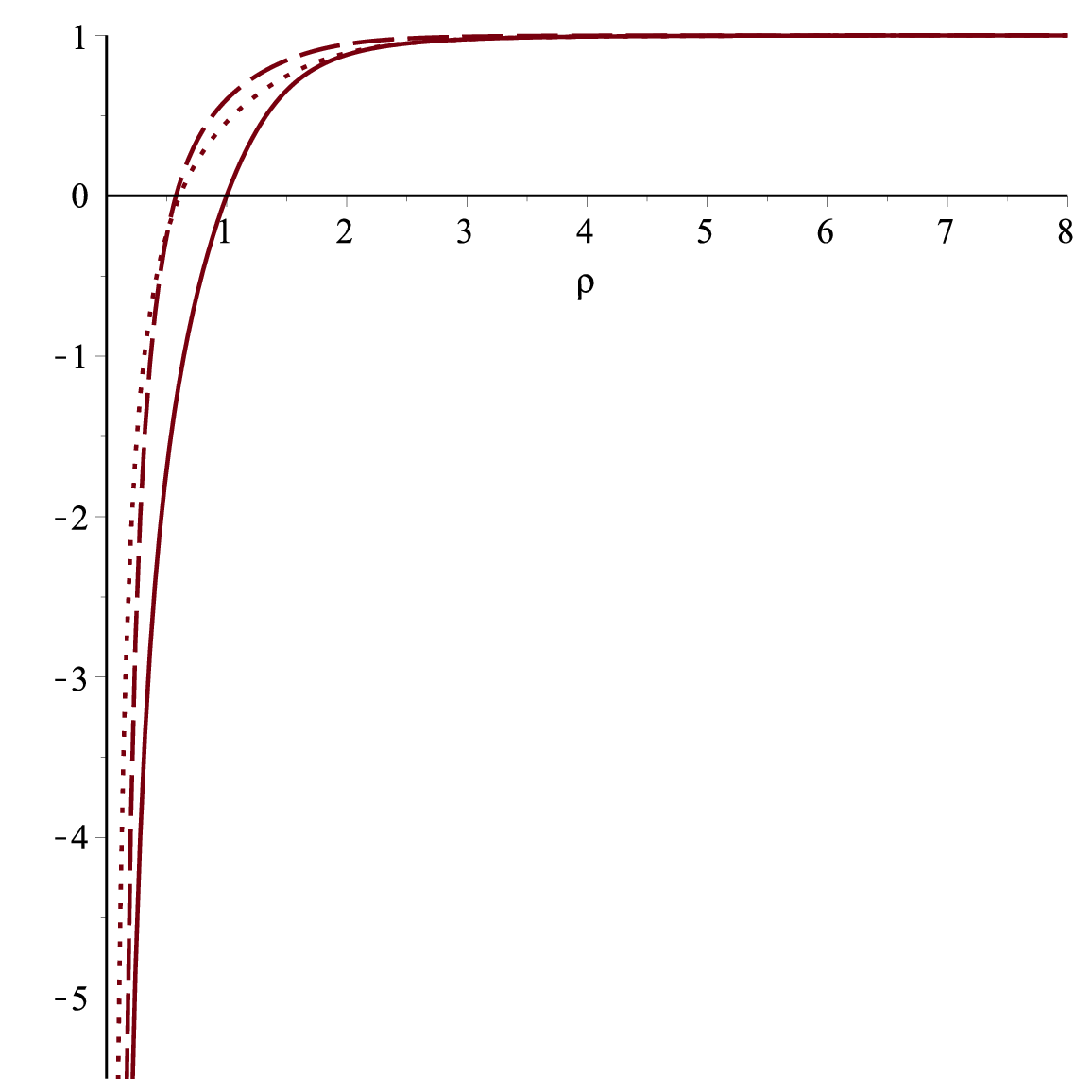}
    \caption{\label{figure01}
Plots of $f(\rho)$ as given by (\ref{frho}) for different values of $n$ and $\widehat{\alpha}$. Left panel: $n =3$, and $\widehat{\alpha} = 1$ (solid line), $1.9$ (dotted line), $2.1$ (dashed line). Central panel: $n =4$, and $\widehat{\alpha} = 1$ (solid line), $5$ (dotted line), $10$ (dashed line). Right panel: $n =5$, and $\widehat{\alpha} = 1$ (solid line), $5$ (dotted line), $10$ (dashed line).}
\end{figure}
From Figure~\ref{figure01}, we observe that the case $n = 3$ is unique in that it exhibits a nonvanishing event horizon for $0\leqslant\widehat{\alpha}<2$, an event horizon located at $\rho_h = 0$ when $\widehat{\alpha} = 0$ (see Table~\ref{table:event}), and no horizon for $\widehat{\alpha} > 2$. In contrast, for $n \geqslant 4$ and $\widehat{\alpha} > 0$, there is always a single event horizon, which approaches zero as $\widehat{\alpha}$ increases. In the following sections, we examine the QNMs associated with scalar, vector, and tensor perturbations. To this end, we reformulate the corresponding equations in the literature into a form suitable for the application of the SM, enabling the computation of the QNMs.

\begin{table}[htbp]
\centering
\caption{Typical numerical values of the rescaled event horizon $\rho_h$ for various values of $\widehat{\alpha}$ and $n$.}
\label{table:event}
\vspace*{1em}
\begin{tabular}{||c|c|c|c||}
\hline\hline
$\widehat{\alpha}$ & $n=3$        & $n=4$        & $n=5$       \\ [0.5ex]
\hline\hline
$0.0$              & $1.414213$   & $1.259921$   & $1.189207$  \\
$0.1$              & $1.378405$   & $1.233468$   & $1.168374$  \\
$0.2$              & $1.341641$   & $1.207040$   & $1.147931$  \\
$0.5$              & $1.224745$   & $1.128174$   & $1.089101$  \\
$1.0$              & $1$          & $1$          & $1$         \\
$1.5$              & $0.707107$   & $0.879615$   & $0.922378$  \\
$1.9$              & $0.316228$   & $0.791579$   & $0.868143$  \\
$2.0$              & $0$          & $0.770917$   & $0.855510$ \\ [1ex]
 \hline\hline 
 \end{tabular}
\end{table}

\begin{figure}[ht!] 
    \includegraphics[width=0.3\textwidth]{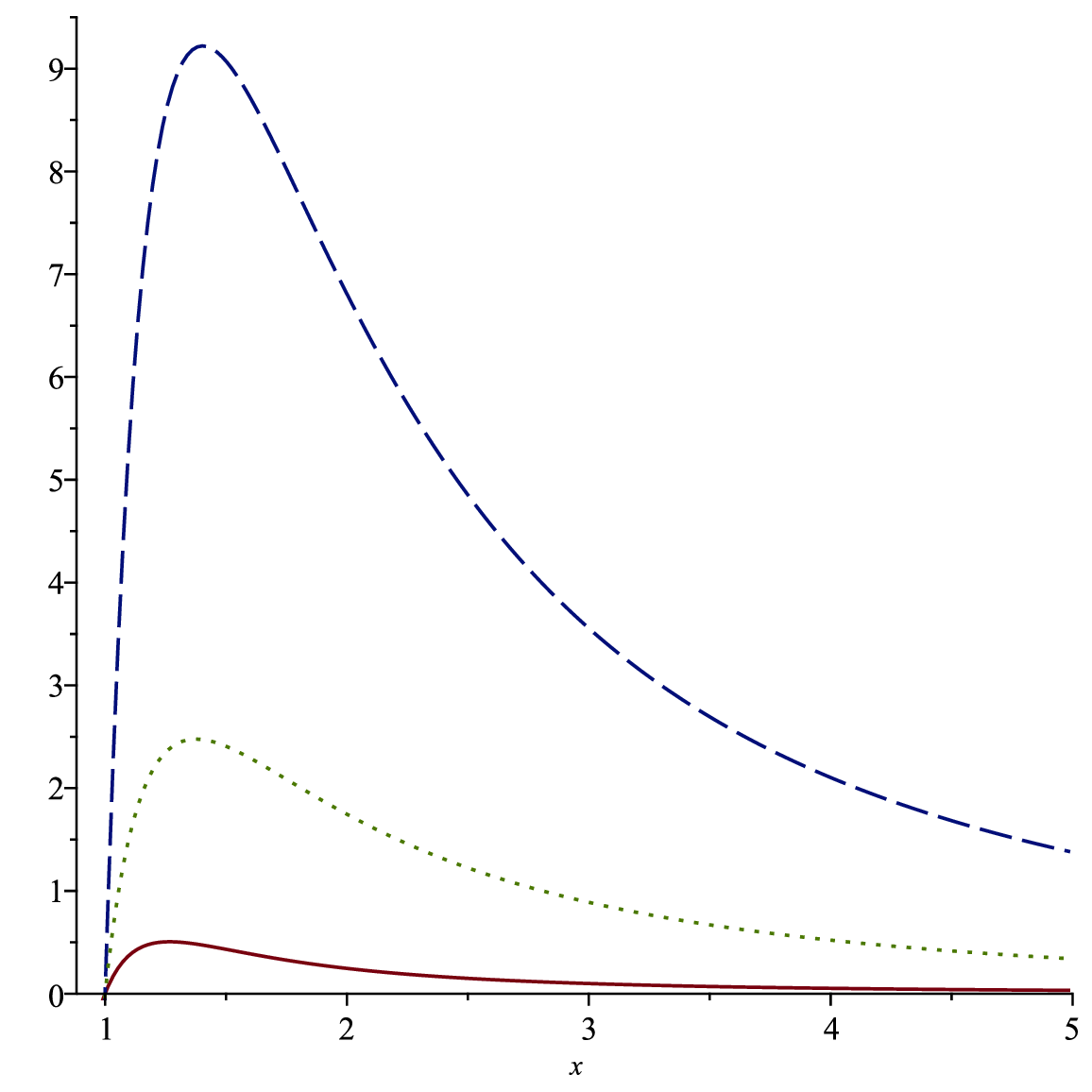}
    \includegraphics[width=0.3\textwidth]{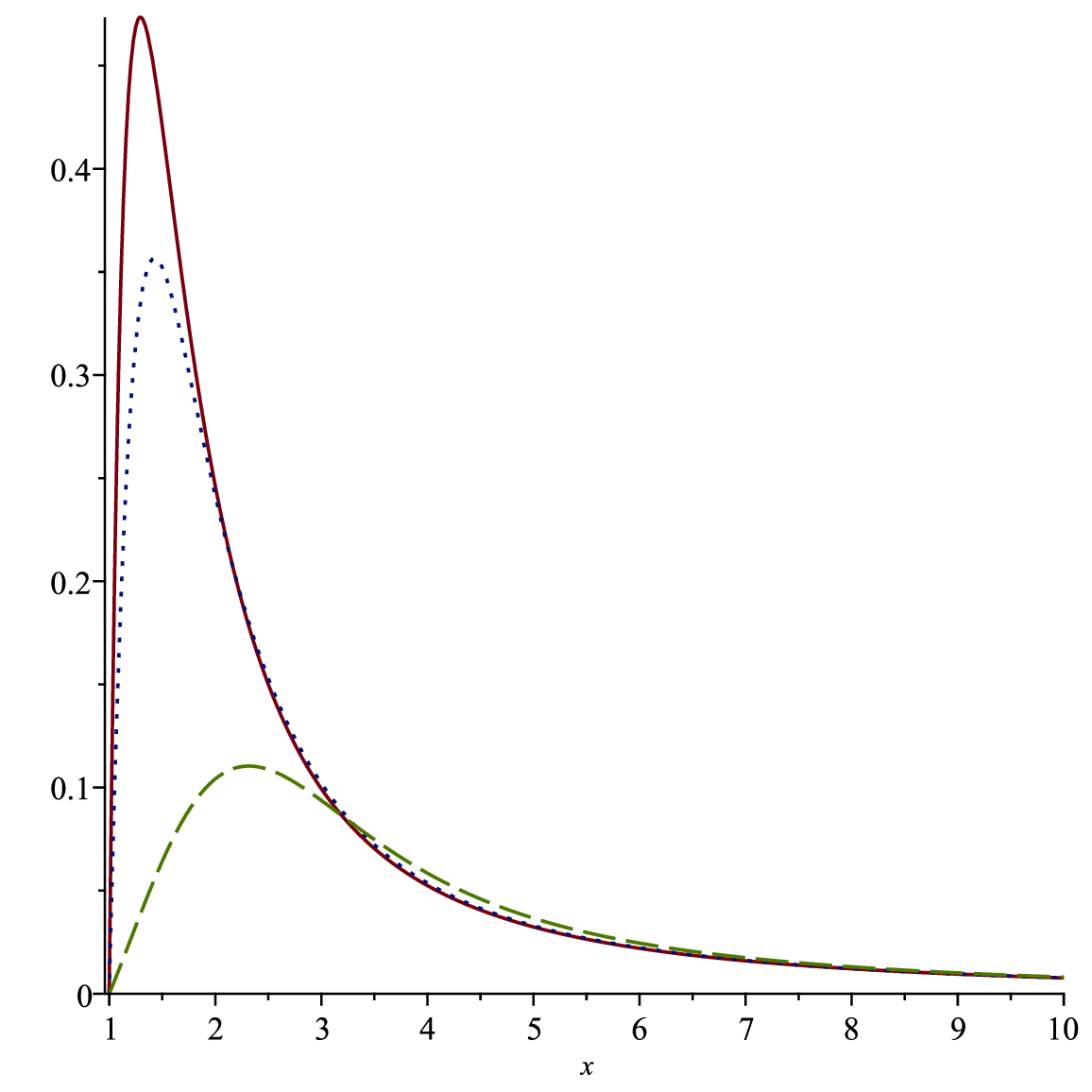}
    \includegraphics[width=0.3\textwidth]{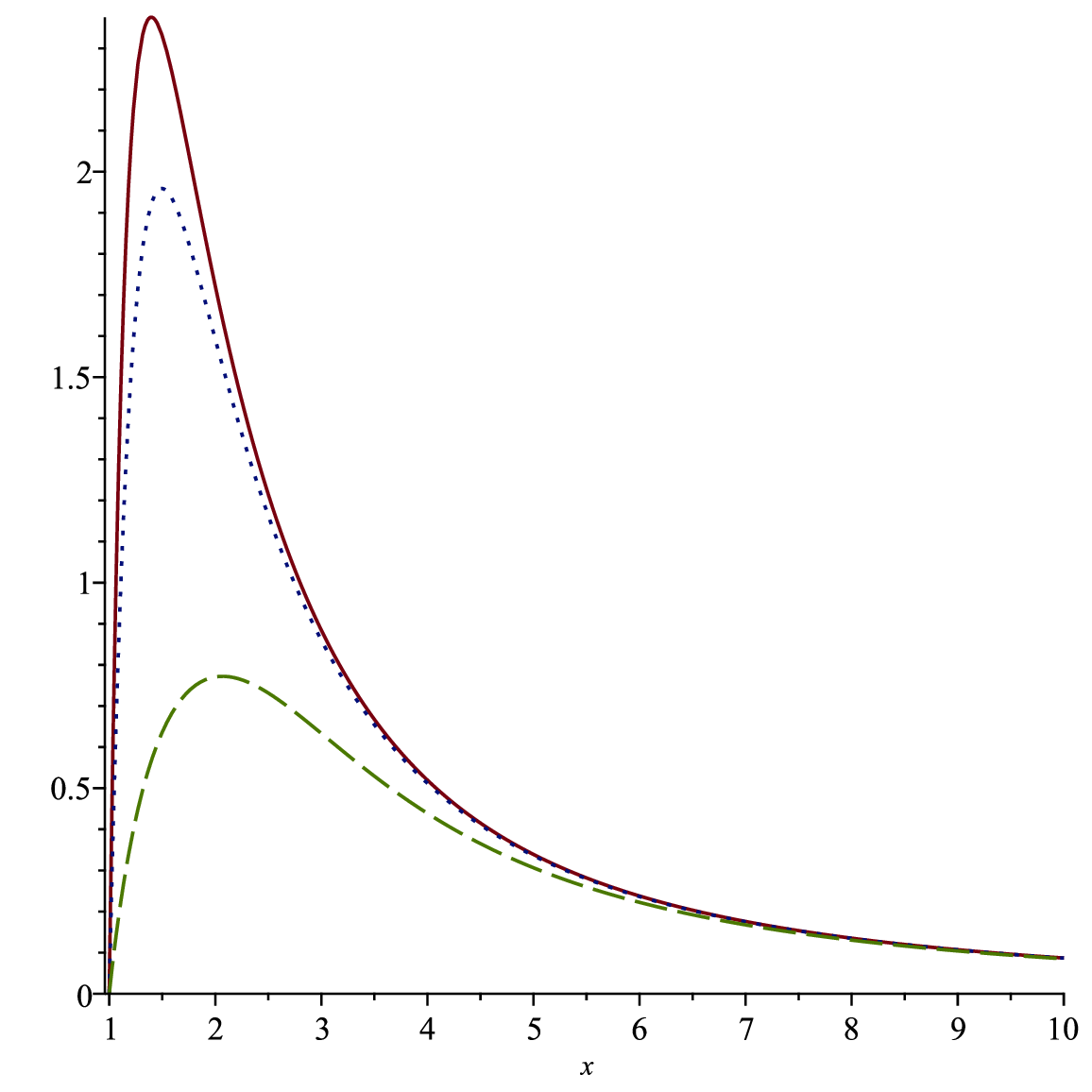}
    \caption{\label{figureUsn3}
Plots of the effective potential $\mathcal{U}_S(x)$ as a function of the rescaled variable $x$ (see equation \eqref{US}), for different values of $\widehat{\alpha}$ and $\ell_3$ in dimension $D=5$. {\bf{Left panel}}: $\widehat{\alpha} = 0$, with $\ell_3=0$ (solid line), $\ell_3=2$ (dotted line), and $\ell_3=5$ (dashed line). {\bf{Central panel}}: $\ell_3=0$, with $\widehat{\alpha}=0.1$ (solid line), $\widehat{\alpha}=0.5$ (dotted line), and $\widehat{\alpha}=1.5$ (dashed line). {\bf{Right panel}}: $\ell_3=2$, with $\widehat{\alpha}=0.1$ (solid line), $\widehat{\alpha}=0.5$ (dotted line), and $\widehat{\alpha}=1.5$ (dashed line).}
\end{figure}

\begin{figure}[ht!] 
    \includegraphics[width=0.3\textwidth]{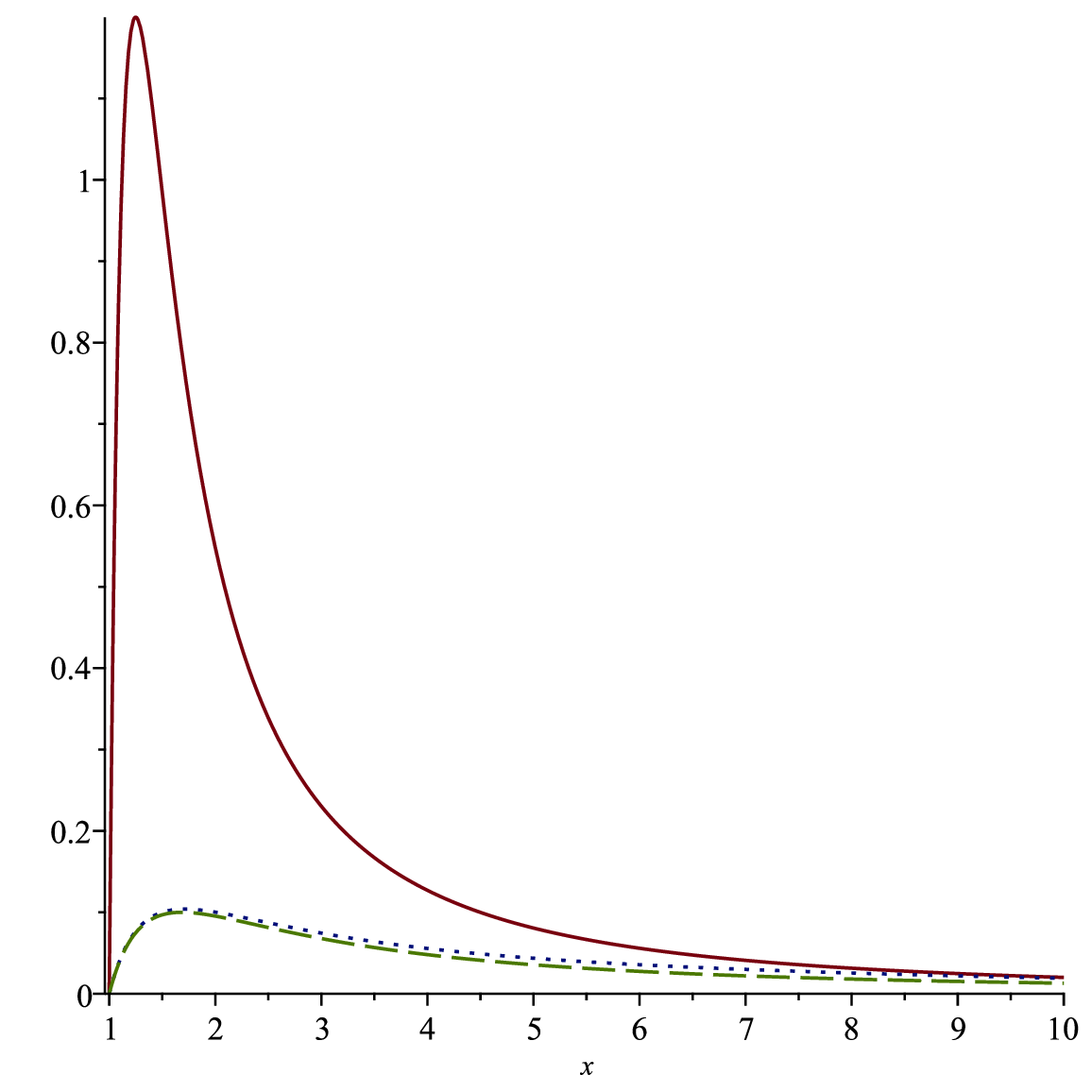}
    \includegraphics[width=0.3\textwidth]{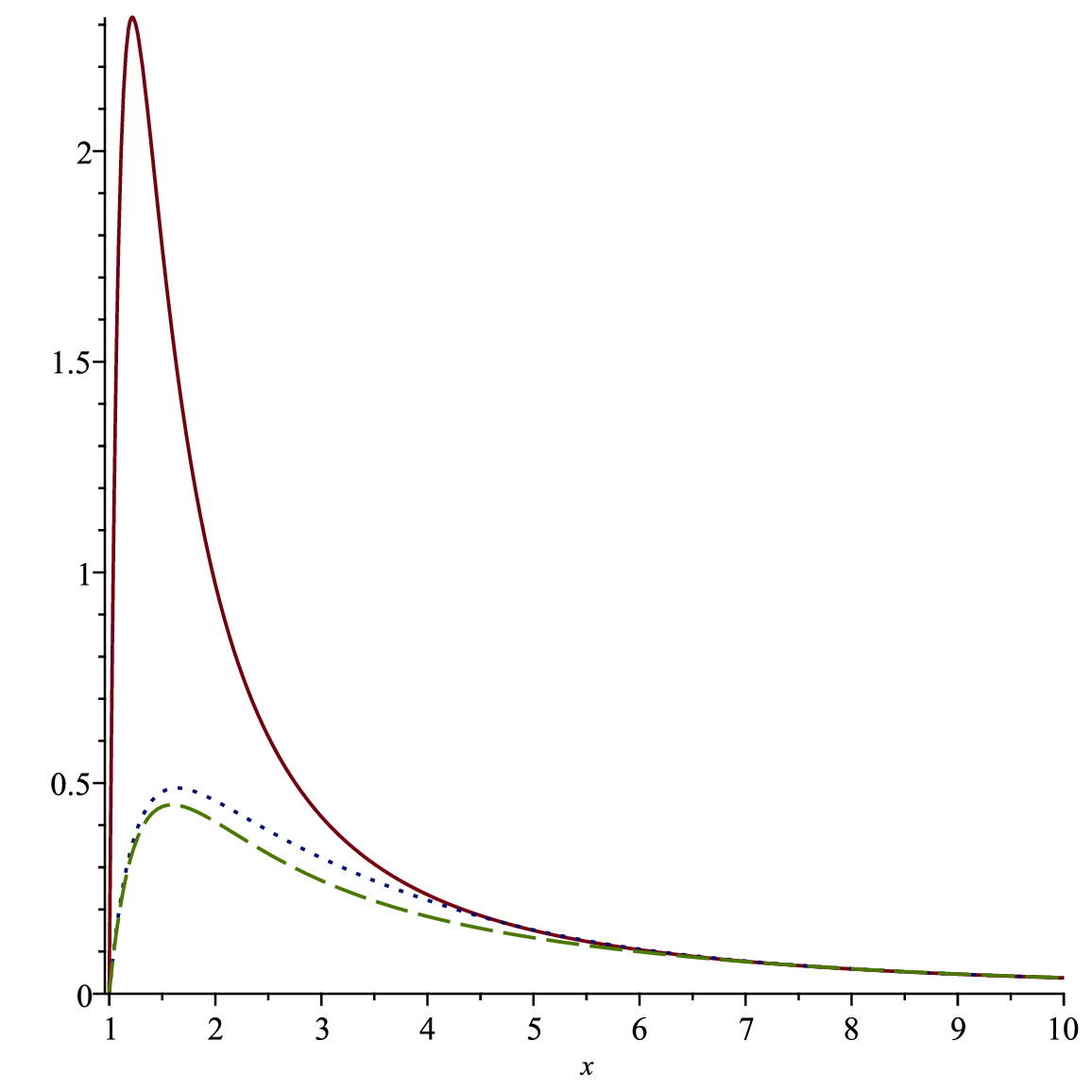}
    \includegraphics[width=0.3\textwidth]{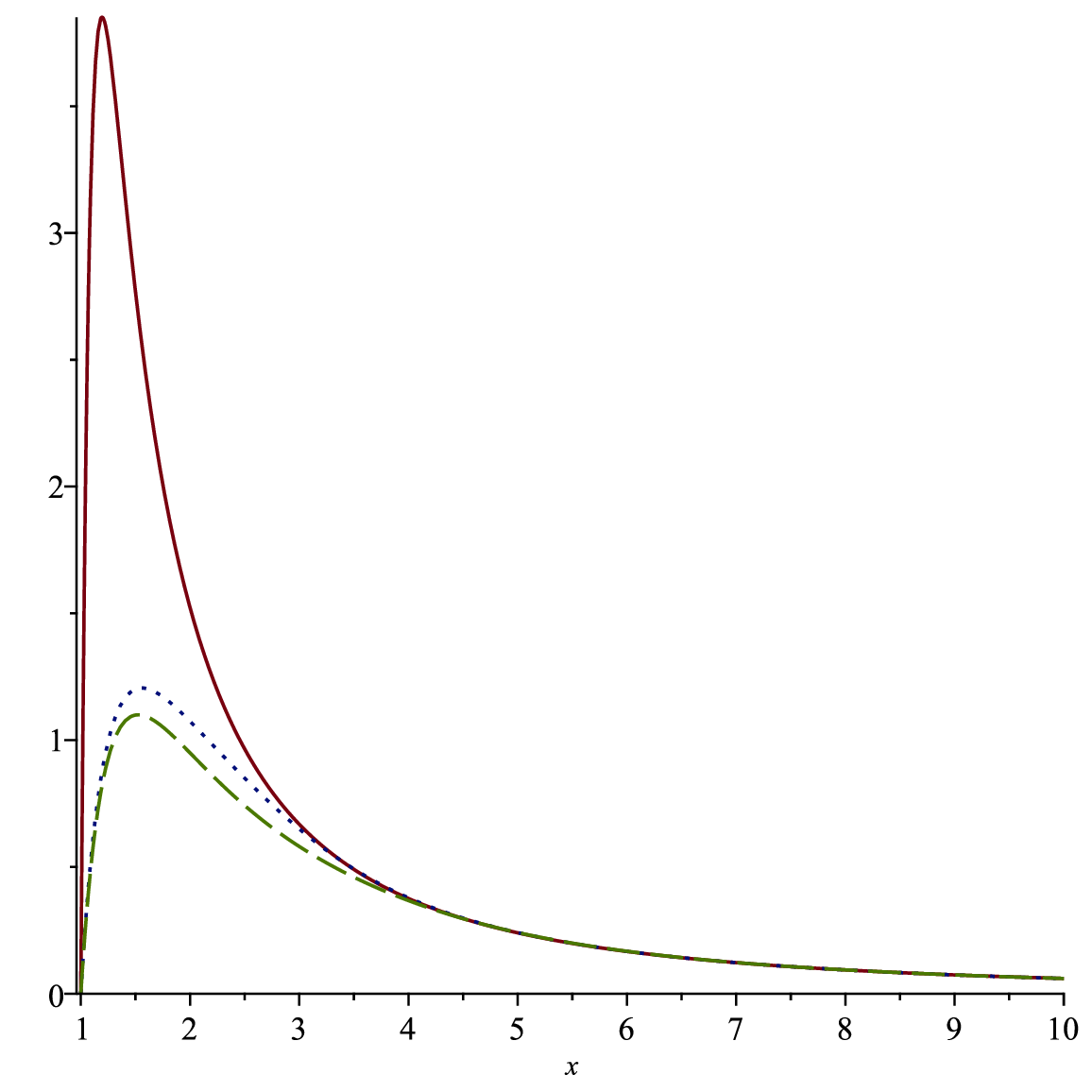}
    \caption{\label{figureUsn456}
Plots of the effective potential $\mathcal{U}_S(x)$ as a function of the rescaled variable $x$ (see equation \eqref{US}), for $\ell_n=0$ and $\widehat{\alpha} = 0.1$ (solid line), $\widehat{\alpha} = 10$ (dotted line), $\widehat{\alpha} = 20$ (dashed line). {\bf{Left panel}}: $D= 6$. {\bf{Central panel}}: $D=7$. {\bf{Right panel}}: $D=8$.}
\end{figure}

\begin{figure}[ht!] 
    \includegraphics[width=0.3\textwidth]{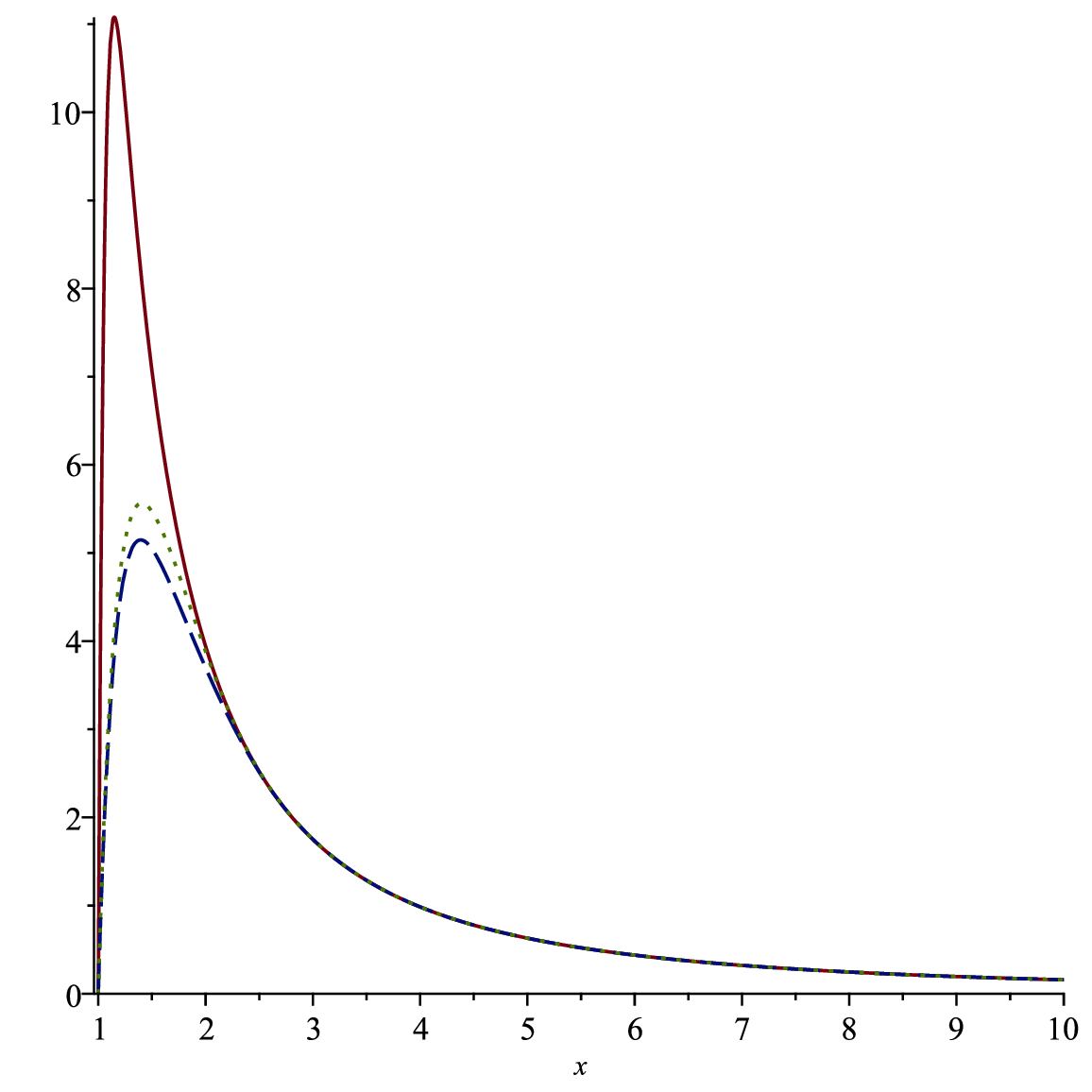}
    \includegraphics[width=0.3\textwidth]{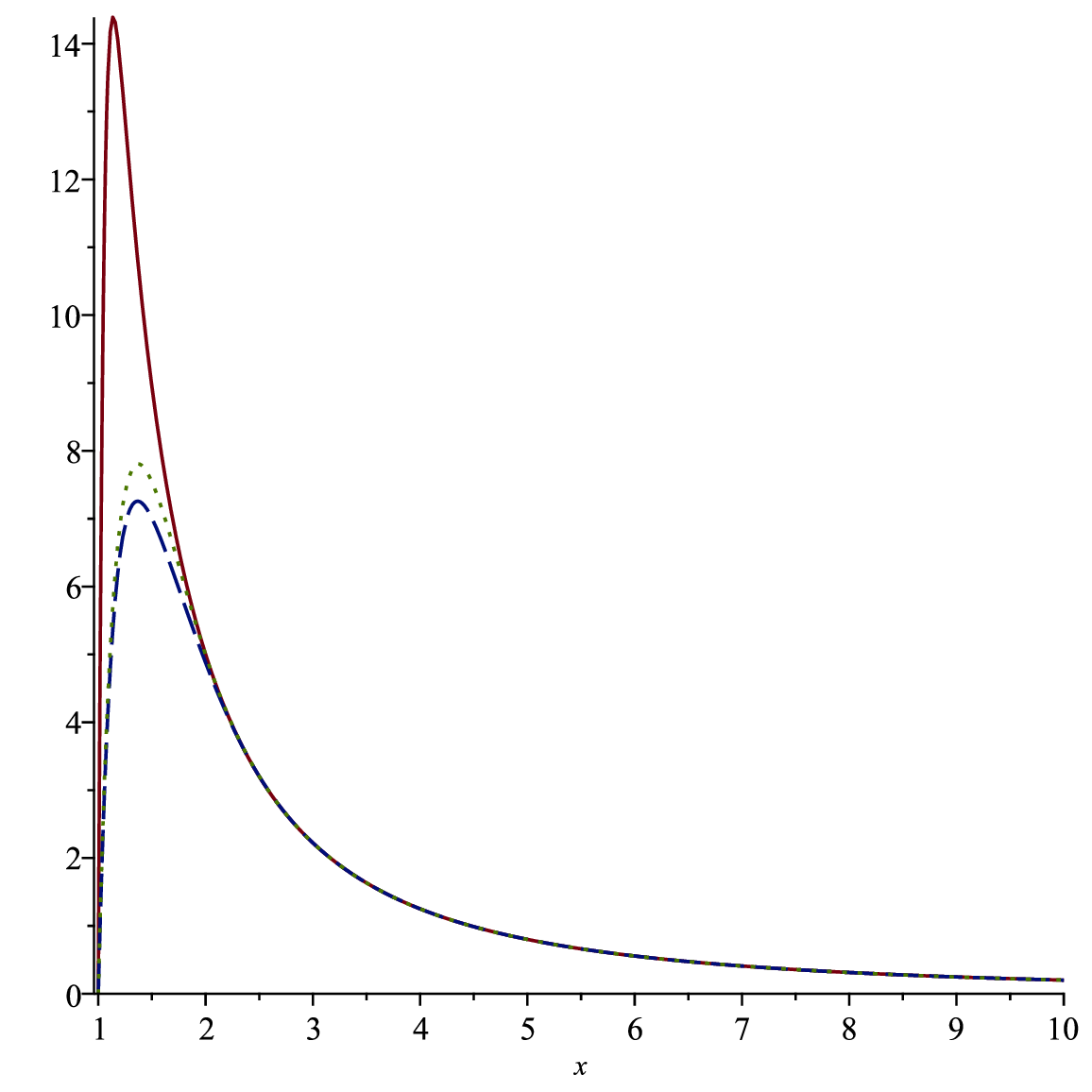}
    \includegraphics[width=0.3\textwidth]{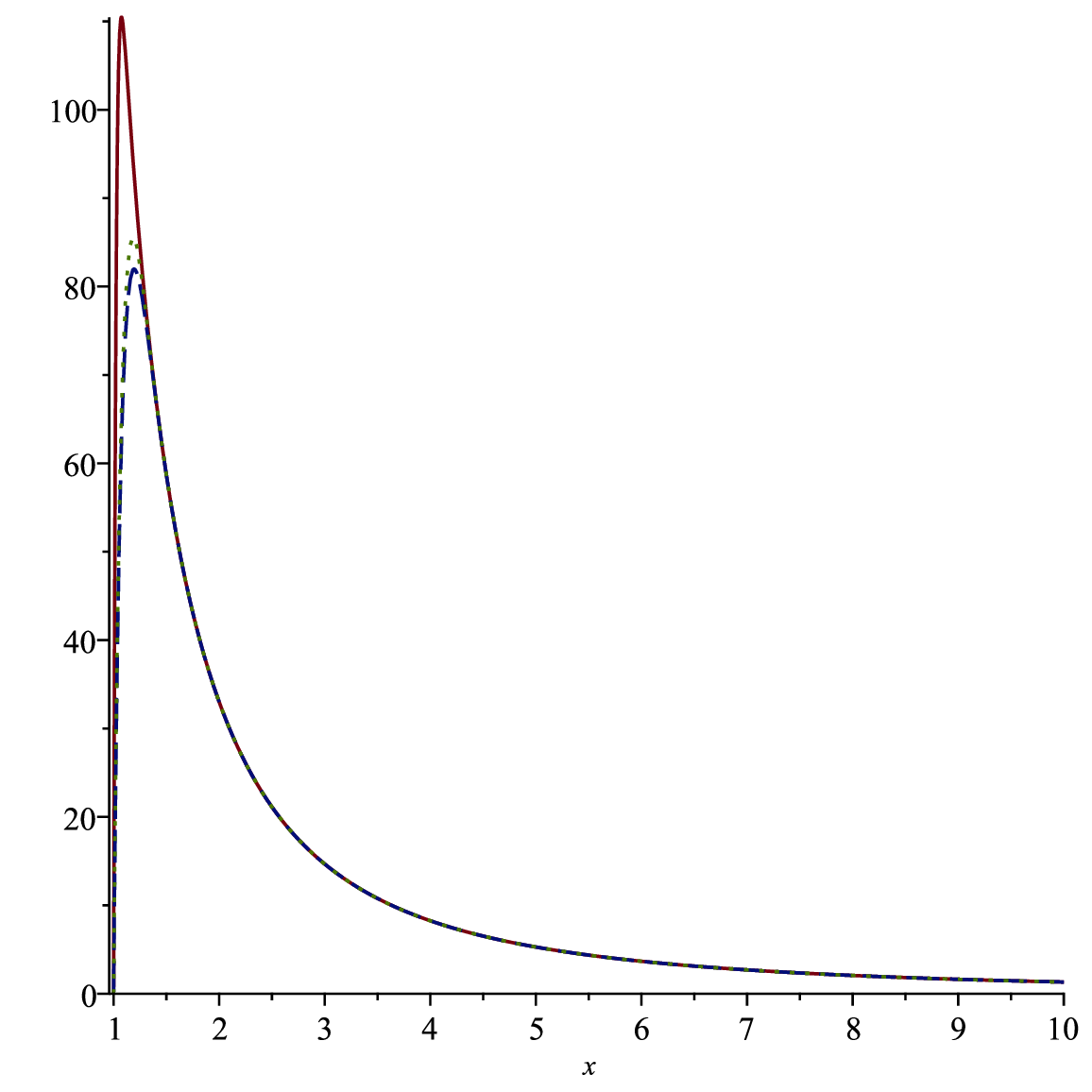}
    \caption{\label{figureUsn91024}
Plots of the effective potential $\mathcal{U}_S(x)$ as a function of the rescaled variable $x$ (see equation \eqref{US}), for $\ell_n=0$ and $\widehat{\alpha} = 0.1$ (solid line), $\widehat{\alpha} = 10$ (dotted line), $\widehat{\alpha} = 20$ (dashed line). {\bf{Left panel}}: $D= 11$. {\bf{Central panel}}: $D=12$. {\bf{Right panel}}: $D=26$.}
\end{figure}

In the case of scalar perturbations (spin $s = 0$), \cite{Iyer1989CQG} obtains the following equation governing the radial part of the perturbation
\begin{equation}\label{ODE0}
  f(r)\frac{d}{dr}\left(f(r)\frac{dR_S}{dr}\right)+\left[\omega^2-V_{S}(r)\right]R_S(r) = 0
\end{equation}
with $f(r)$ given by (\ref{f}) and potential 
\begin{equation}
  V_{S}(r) = f(r)\left[
  \frac{\ell_n(\ell_n+n-1)}{r^2}+\frac{n(n-2)}{4r^2}f(r)+\frac{n}{2r}\frac{df}{dr}
  \right].
\end{equation}
Here, $\ell_n$ denotes the total angular momentum of the perturbing field in dimension $D=n+2$ with $n \geqslant 3$. If we make use of \eqref{rescalings}, we can rewrite \eqref{ODE0} as 
\begin{equation}\label{ODE00}
    f(\rho)\frac{d}{d\rho}\left(f(\rho)\frac{dR_S}{d\rho}\right)+\left[\Omega^2-U_{S}(\rho)\right]R_S(\rho) = 0,\quad \Omega=\omega M^\frac{1}{n-1}
\end{equation}
with
\begin{equation}
  U_{S}(\rho) = f(\rho)\left[
  \frac{\ell_n(\ell_n+n-1)}{\rho^2}+\frac{n(n-2)}{4\rho^2}f(\rho)+\frac{n}{2\rho}\frac{df}{d\rho}.
  \right]
\end{equation}
and $f(\rho)$ given by \eqref{frho}. In the following analysis, we focus on computing the QNMs for the spectral problem stated in (\ref{ODE00}). For this purpose, we represent $\Omega$ as $\Omega = \Omega_R + i\Omega_I$, where $\Omega_I < 0$ ensures that the perturbation is dumped in time. The boundary conditions are set so that the radial field exhibits inward radiation at the event horizon and outward radiation at spatial infinity. This requires a detailed examination of the solution asymptotic behaviour in (\ref{ODE00}), both near the event horizon ($x \to 1^{+}$) and at large spatial distances ($x \to +\infty$). Moreover, to compute the QNMs using the SM, we must recast the differential equation in (\ref{ODE00}) and the appropriate boundary conditions over the compact interval $[-1,1]$. This adjustment is essential as the method expands the regular part of the eigenfunctions by means of Chebyshev polynomials. As a first step, we introduce the rescaling $x = \rho/\rho_h$, by means of which (\ref{ODE00}) becomes
\begin{equation}\label{ODE1}
  \frac{d^2 R_S}{dx^2}+p_S(x)\frac{dR_S}{dx}+q_S(x)R_S(x)=0   
\end{equation}
with 
\begin{equation}\label{pq}
  p_S(x)=\frac{1}{f(x)}\frac{df}{dx},\quad 
  q_S(x) = \frac{\rho_h^2 \Omega^2-\mathcal{U}_S(x)}{f^2(x)}
\end{equation}
and
\begin{equation}\label{US}
  \mathcal{U}_S(x)= f(x)\left[
  \frac{\ell_n(\ell_n+n-1)}{x^2}+\frac{n(n-2)}{4x^2}f(x)+\frac{n}{2x}\frac{df}{dx}
  \right],\quad
  f(x)=1-\frac{4\rho_h^{1-n}x^2}{x^{n+1}+\sqrt{x^{2n+2}+\kappa x^{n+1}}},\quad
  \kappa=\frac{8\widehat{\alpha}}{\rho_h^{n+1}}.
\end{equation}
From the central panel of Fig.~\ref{figureUsn3}, which depicts the effective potential $\mathcal{U}_S(x)$ in dimension $D=5$, we observe that for $\ell_3 = 0$, the potential decreases more gradually for $\widehat{\alpha} = 1.5$ compared to the cases $\widehat{\alpha} = 0.1$ and $\widehat{\alpha} = 0.5$ in the intermediate region $3.5 \lesssim x \lesssim 6$. This suggests that larger values of $\widehat{\alpha}$ result in a slower decay of the potential, which may have significant implications for the dynamics of test particles in this background. Moreover, from Figures~\ref{figureUsn456} and \ref{figureUsn91024}, we observe that for fixed $\widehat{\alpha}$ and $\ell_n$, the peak of the effective potential increases as the dimension of the underlying manifold increases.

In studying the asymptotic behaviour of the solutions to \eqref{ODE1} and applying the SM, it is useful to note that since $x=1$ is a simple root of the function $f(x)$, we have the following functional relation
\begin{equation}\label{frel}
4\rho_h^{1-n}=1+\sqrt{1+\kappa}.
\end{equation}
As a result, we can rewrite $f(x)$ in its final form as follows
\begin{equation}\label{fx}
f(x)=1-\frac{(1+\sqrt{1+\kappa})x^2}{x^{n+1}+\sqrt{x^{2n+2}+\kappa x^{n+1}}}.
\end{equation}
Moreover, with the help of \eqref{frel}, it is not difficult to verify that $f(x)$ admits the Taylor expansion around the point $x=1$
\begin{equation}\label{fpat1}
    f(x)=f^{'}(1)(x-1)+\mathcal{O}(x-1)^2,\quad
    f^{'}(1)=\frac{2(n-1)(1+\sqrt{1+\kappa})+\kappa(n-3)}{\sqrt{1+\kappa}(1+\sqrt{1+\kappa})}.
\end{equation}
where the prime denotes differentiation with respect to the variable $x$. Note that $f^{'}(1)>0$ for all $n\geqslant 3$. Finally, the asymptotic behaviour of the solutions to equation (\ref{ODE1}) can be deduced using the method outlined in \cite{Olver1994MAA}. For this purpose, we split our analysis by examining the behaviour of the radial field in two different regions.
\begin{enumerate}
\item 
{\underline{Asymptotic behaviour as $x\to 1^+$}}: Given that  
\begin{equation}  
    p_S(x) = \frac{1}{x-1} + \mathcal{O}(1), \quad  
    q_S(x) = \left(\frac{\rho_h \Omega}{f'(1)}\right)^2 \frac{1}{(x-1)^2} + \mathcal{O}((x-1)^{-1}),  
\end{equation}  
the point $x = 1$ is a simple pole for $p_S(x)$ and a second-order pole for $q_S(x)$. As a result, $x = 1$ qualifies as a regular singular point of \eqref{ODE1}, in accordance with Frobenius theory \cite{Ince1956}. Hence, we can construct solutions of the form
\begin{equation}
R_S(x) = (x-1)^\tau\sum_{m=0}^\infty a_m(x-1)^m,
\end{equation}
where the indicial equation determines $\tau$
\begin{equation}\label{indicial}
  \tau(\tau-1) + P_0\tau + Q_0 = 0
\end{equation}
with
\begin{equation}
  P_0 = \lim_{x\to 1}(x-1)p_S(x) =1 , \qquad
  Q_0=\lim_{x \to 1}(x-1)^2 q_S(x) = \left(\frac{\rho_h\Omega}{f^{'}(1)}\right)^2.
\end{equation}
The roots of (\ref{indicial}) are $\tau_\pm = \pm i\rho_h \Omega/f^{'}(1)$ and the correct QNM boundary condition at $x = 1$ reads
\begin{equation}\label{QNMBCz1}
  R_S\underset{{x\to 1^+}}{\longrightarrow} (x-1)^{-i \rho_h a\Omega}, \quad a=\frac{1}{f^{'}(1)}.
\end{equation}
Note that, as stated in the remark immediately following \eqref{fpat1}, the boundary condition above remains valid even in the case of a Schwarzschild--Tangherlini black hole.
\item 
{\underline{Asymptotic behaviour as $x\to +\infty$}}: We start by observing that in the limit of $x \to +\infty$
\begin{equation}\label{pqS}
    p_S(x) = \sum_{m=0}^\infty\frac{\mathfrak{f}_m}{x^m} = \mathcal{O}\left(\frac{1}{x^n}\right), \qquad
    q_S(x) = \sum_{m=0}^\infty\frac{\mathfrak{g}_m}{x^m}=\rho_h^2\Omega^2+\mathcal{O}\left(\frac{1}{x^2}\right).
\end{equation}
with $n\geqslant 3$. Given that at least one of the coefficients $\mathfrak{f}_0$, $\mathfrak{g}_0$, $\mathfrak{g}_1$ is nonzero, a formal asymptotic solution to (\ref{ODE1}) is represented by \cite{Olver1994MAA}
\begin{equation}\label{olvers}
    R_S^{(j)}(x) = x^{\mu_j}e^{\lambda_j x}\sum_{m=0}^\infty\frac{b_{m,j}}{x^m}, \qquad j \in \{1,2\},
\end{equation}
where $\lambda_1$, $\lambda_2$, $\mu_1$ and $\mu_2$ are the roots of the characteristic equations
\begin{equation}\label{chareqns}
   \lambda^2+\mathfrak{f}_0\lambda+\mathfrak{g}_0=0,\quad
   \mu_j=-\frac{\mathfrak{f}_1\lambda_j+\mathfrak{g}_1}{\mathfrak{f}_0+2\lambda_j}.
\end{equation}
A straightforward computation shows that $\lambda_\pm = \pm i \rho_h \Omega$. Since the scalar field must propagate as outward radiation in the limit $x \to +\infty$, we select the root $\lambda_+$. Hence, the QNM boundary condition at positive space-like infinity can be formulated as
\begin{equation}\label{QNMBCposinf}
    R_S\underset{{x\to +\infty}}{\longrightarrow} e^{i\rho_h\Omega x}.
\end{equation}
\end{enumerate}
At this point, we transform the radial function $R_S(x)$ into a new radial function $\Phi_S(x)$ such that the QNM boundary conditions are automatically implemented and  $\Phi_{S}(x)$ is regular at $x = 1$ and at space-like infinity. To this aim, we consider the transformation
\begin{equation}\label{Ansatz}
    R_S(x) = x^{i\rho_h a\Omega}(x-1)^{-i\rho_h a\Omega}e^{i\rho_h\Omega(x-1)} \Phi_S(x).
\end{equation}
If we multiply \eqref{ODE1} by $f^2(x)$ and substitute (\ref{Ansatz}) therein, we end up with the following ordinary differential equation for the radial eigenfunctions, namely
\begin{equation}\label{ODEznone}
    P_2(x)\Phi^{''}_S(x) + P_1(x)\Phi^{'}_S(x) + P_0(x)\Phi_S(x) = 0
\end{equation}
with
\begin{eqnarray}
    P_2(x)&=&x^2(x-1)^2 f^2(x),\label{P2}\\
    P_1(x)&=&x(x-1)f(x)\left[x(x-1)f^{'}(x)-2i\rho_h\Omega f(x)(x-x^2+a)\right],\label{P1}\\
    P_0(x)&=&-\Omega^2 Q_+(x)Q_{-}(x)+i\Omega f(x)L(x)-x^2(x-1)^2 \mathcal{U}_S(x),\label{P0}\\
    Q_\pm(x)&=&\rho_h\left[f(x)(a+x-x^2)\pm x(x-1)\right],\\
    L(x)&=&\rho_h\left[a(2x-1)f(x)-x(x-1)(a+x-x^2)f^{'}(x)\right].
\end{eqnarray}
Let us now introduce the transformation $x = 2/(1-y)$ mapping the point at infinity and the event horizon to $y = 1$ and $y = -1$, respectively. Furthermore, a dot denotes differentiation with respect to the new variable $y$. Then, equation \eqref{ODEznone} becomes
\begin{equation}\label{ODEynone}
    S_2(y)\ddot{\Phi}_{S}(y) + S_1(y)\dot{\Phi}_{S}(y) + S_0(y)\Phi_{S}(y) = 0,
\end{equation}
where
\begin{eqnarray}
  S_2(y) &=& (1+y)^2 f^2(y), \label{S2onone} \\
  S_1(y) &=& 2i\rho_h\Omega\frac{1+y}{(1-y)^2}f^2(y)\left[2(1+y)-a(1-y)^2\right]+(1+y)^2 f(y)\left[\dot{f}(y)-\frac{2f(y)}{1-y}\right], \label{S1onone}\\
  S_0(y) &=& \Omega^2\Sigma_2(y)+i\Omega\Sigma_1(y)+\Sigma_0(y) \label{S0onone}
\end{eqnarray}
with
\begin{eqnarray}
    \Sigma_2(y) &=&\frac{\rho_h^2}{(1-y)^4}\left\{4(1+y)^2-f^2(y)\left[a(1-y)^2-2(1+y)\right]^2\right\},\\
    \Sigma_1(y) &=&\frac{\rho_h f(y)}{1-y}\left\{af(y)(3+y)-\frac{1+y}{1-y}\left[a(1-y)^2-2(1+y)\right]\dot{f}(y)\right\},\\
    \Sigma_0(y) &=& -\frac{4(1+y)^2}{(1-y)^4}\mathcal{U}_S(y).\label{Sigma0}
\end{eqnarray}
Notice that we must also require that $\Phi_{S}(y)$ is regular at $y=\pm 1$. As a result of the transformation introduced above, we have 
\begin{eqnarray}
  f(y) &=& 1 - \frac{4(1+\sqrt{1+\kappa})(1-y)^{n-1}}{2^{n+1}\left(1+\sqrt{1+\kappa\left(\frac{1-y}{2}\right)^{n+1}}\right)},\label{fv1}\\
  \mathcal{U}_S(y) &=& \frac{(1-y)^2}{16}f(y)\left[n(n-2)f(y)+2n(1-y)\dot{f}(y)+4\ell_n(\ell_n+n-1)\right].\label{fv2}
\end{eqnarray}

\begin{table}%[ht]
\caption{Classification of the points $y=\pm 1$ for the relevant functions defined by \eqref{S2onone}-\eqref{S0onone}, \eqref{fv1}, and \eqref{fv2}. The abbreviations $z$ ord $n$ and $p$ ord $m$ stand for zero of order $n$ and pole of order $m$, respectively.}
\begin{center}
\begin{tabular}{ | c | c | c | c | c | c | c | c }
\hline
$y$  & $f(y)$  & $\mathcal{U}_S(y)$ & $S_2(y)$ & $S_1(y)$ & $S_0(y)$\\ \hline
$-1$ & z \mbox{ord} 1 & z \mbox{ord} 1 & z \mbox{ord} 4& z \mbox{ord} 3 & z \mbox{ord} 2 \\ \hline
$+1$ & $+1$           & z \mbox{ord} 2 & $+4$ & p \mbox{ord} 2 & p \mbox{ord} 2\\ \hline
\end{tabular}
\label{tableEinsnone}
\end{center}
\end{table}

Table~\ref{tableEinsnone} shows that the coefficients of the differential equation (\ref{ODEynone}) share a common zero of order $2$ at $y = -1$ while $y = 1$ is a pole of order $2$ for the coefficients $S_1$ and $S_0$. Hence, in order to apply the SM, we need to multiply (\ref{ODEynone}) by $(1-y)^2/(1+y)^2$. As a result, we end up with the following differential equation
\begin{equation}\label{ODEhynone}
    M_2(y)\ddot{\Phi}_{S}(y) + M_1(y)\dot{\Phi}_{S}(y) + M_0(y)\Phi_{S}(y) = 0,
\end{equation}
where
\begin{equation}\label{S210honone}
  M_2(y) = (1-y)^2 f^2(y), \qquad
  M_1(y) = i\Omega N_1(y)+N_0(y), \qquad
  M_0(y) = \Omega^2 C_2(y)+i\Omega C_1(y)+C_0(y)
\end{equation}
with
\begin{eqnarray}
    N_1(y) &=&\frac{2\rho_h f^2(y)^2}{1+y}\left[2(1 + y) - a(1 - y)^2\right], \quad
    N_0(y) = (1 - y)f(y)\left[(1 - y)\dot{f}(y)-2f(y)\right],\label{N0}\\
    C_2(y) &=&\frac{\rho_h^2}{(1-y^2)^2}\left\{4(1 + y)^2 - f^2(y)\left[a(1 - y)^2 - 2(1 + y)\right]^2\right\},\label{C2}\\
    C_1(y) &=&\frac{\rho_h f(y)(1 - y)}{(1 + y)^2}\left\{a f(y)(3 + y) - \frac{1 + y}{1-y}\left[a(1 - y)^2 - 2(1 + y)\dot{f}(y)\right]\right\},\label{C1}\\
    C_0(y) &=& -\frac{4\mathcal{U}_S(y)}{(1+y)(1-y)^2}.\label{C0}
\end{eqnarray}
It can be easily verified with Maple that
\begin{eqnarray}
    &&\lim_{y\to 1^{-}}M_2(y)=0=\lim_{y\to -1^{+}}M_2(y),\\
    &&\lim_{y\to 1^{-}}M_1(y)=4i\rho_h\Omega,\quad
    \lim_{y\to -1^{+}}M_1(y)=0,\\
    &&\lim_{y\to 1^{-}}M_0(y)=2a\rho_h\Omega^2-\frac{n}{4}(n-2)-\ell_n(\ell_n+n-1),\quad
     \lim_{y\to -1^{+}}M_0(y)=B_2\Omega^2,
\end{eqnarray}
where
\begin{equation}\label{B2}
B_2 =\rho_h^2\left\{1-\frac{a^2[\kappa(n-3)+2(n-1)(1+\sqrt{1+\kappa})]^2}{4(1+\kappa)\left(1+\sqrt{1+\kappa}\right)^2}\right\}.
\end{equation}
As a final step prior to implementing the SM, we rewrite the differential equation (\ref{ODEhynone}) in the following form
\begin{equation}\label{TSCH}
  L_0\left[\Phi_{S}, \dot{\Phi}_{S}, \ddot{\Phi}_{S}\right] +  i\Omega L_1\left[\Phi_{S}, \dot{\Phi}_{S}, \ddot{\Phi}_{S}\right] +  \Omega^2 L_2\left[\Phi_{S}, \dot{\Phi}_{S}, \ddot{\Phi}_{S}\right] = 0
\end{equation}
with
\begin{eqnarray}
L_0\left[\Phi_{S}, \dot{\Phi}_{S}, \ddot{\Phi}_{S}\right] &=& L_{00}(y)\Phi_{S} + L_{01}(y)\dot{\Phi}_{S} + L_{02}(y)\ddot{\Phi}_{S},\label{L0none}\\
L_1\left[\Phi_{S}, \dot{\Phi}_{S}, \ddot{\Phi}_{S}\right] &=&L_{10}(y)\Phi_{S} + L_{11}(y)\dot{\Phi}_{S}+L_{12}(y)\ddot{\Phi}_{S}, \label{L1none}\\
L_2\left[\Phi_{S}, \dot{\Phi}_{S}, \ddot{\Phi}_{S}\right]&=&L_{20}(y)\Phi_{S} +L_{21}(y)\dot{\Phi}_{S} + L_{22}(y)\ddot{\Phi}_{S}.\label{L2none}
\end{eqnarray}
Moreover, all the $L_{ij}$ terms appearing in (\ref{L0none})-(\ref{L2none}) exhibit a regular behavior at the endpoints $y = \pm 1$ with limits given as in Table~\ref{tableDreiScalar} .

\begin{table}%[ht]
\caption{Behaviour of the coefficients $L_{ij}$ at the endpoints of the interval $-1 \leqslant y \leqslant 1$ for the scalar perturbation of the Schwarzschild black hole with GB correction.}
\begin{center}
\begin{tabular}{ | c | c | c | c | c | c | c | c }
\hline
$(i,j)$ & $\displaystyle{\lim_{y\to -1^+}}L_{ij}$  & $L_{ij}$       & $\displaystyle{\lim_{y\to 1^-}}L_{ij}$  \\ \hline
$(0,0)$ &  $0$                                     & $C_0$          & $\frac{n}{2}-\frac{n^2}{4}-\ell_n(\ell_n+n-1)$\\ \hline
$(0,1)$ &  $0$                                     & $N_0$          & $0$\\ \hline
$(0,2)$ &  $0$                                     & $M_2$          & $0$\\ \hline 
$(1,0)$ &  $0$                                     & $C_1$          & $0$\\ \hline 
$(1,1)$ &  $0$                                     & $N_1$          & $4\rho_h$\\ \hline 
$(1,2)$ &  $0$                                     & $0$            & $0$\\ \hline 
$(2,0)$ &  $B_2$                                   & $C_2$          & $2a\rho_h^2$\\ \hline
$(2,1)$ &  $0$                                     & $0$            & $0$\\ \hline
$(2,2)$ &  $0$                                     & $0$            & $0$\\ \hline
\end{tabular}
\label{tableDreiScalar}
\end{center}
\end{table}

\section{Numerical method}

In order to solve the differential eigenvalue problem \eqref{TSCH} along with the corresponding frequencies $\Omega$, we have to discretize the differential operators $L_{j}[\cdot]$ with $j \in \{1,2,3\}$ defined in (\ref{L0none})-(\ref{L2none}) and in Table~\ref{tableDreiScalar}, respectively. Since our problem is posed on the finite interval $[-1, 1]$ without any boundary conditions, more precisely, we only require that the function $\Phi_S(y)$ be regular at $y = \pm 1$, then, it is natural to choose a Tchebyshev-type SM \cite{Trefethen2000, Boyd2000}. Namely, we are going to expand the function $y \mapsto \Phi_{S}(y)$ in the form of a truncated Tchebyshev series
\begin{equation}\label{eq:exp}
  \Phi_{S}(y)=\sum_{k=0}^{N} a_{k} T_k(y),
\end{equation}
where $N\ \in\ \mathbb{N}$ is kept as a numerical parameter, $\{a_{k}\}_{k=0}^{N}\ \subseteq\ \mathds{R}$, and $\{T_k(y)\}_{k=0}^{N}$ are the Tchebyshev polynomials of the first kind
\begin{equation}
    T_k: [-1, 1]\ \longrightarrow\ [-1, 1]\,, \qquad y\ \longmapsto\ \cos\,\bigl(k\arccos y\bigr)\,.
\end{equation}
After substituting expansion \eqref{eq:exp} into the differential equation \eqref{TSCH}, we obtain an eigenvalue problem with polynomial coefficients. In order to translate it into the realm of numerical linear algebra, we employ the collocation method \cite{Boyd2000}. Specifically, rather than insisting that the polynomial function in \( y \) is identically zero (a condition equivalent to having polynomial solutions for the differential problems as per equation \eqref{TSCH}), we impose a weaker requirement. This involves ensuring that the polynomial vanishes at \( N+1 \) strategically selected points. The number $N+1$ coincides exactly with the number of unknown coefficients $\{a_{k}\}_{k=0}^{N}$. For the collocation points, we implemented the Tchebyshev roots grid \cite{Fox1968}
\begin{equation}
  y_k= -\cos{\left(\frac{(2k+1)\pi}{2(n+1)}\right)},\quad k\in\{0, 1,\ldots,N\}.
\end{equation}
In our numerical codes, we also implemented the second option of the Tchebyshev extrema grid
\begin{equation*}
  y_k=-\cos{\left(\frac{k\pi}{n}\right)},\quad k\in\{0, 1,\ldots,N\}.
\end{equation*}
The users are free to choose their favourite collocation points. Notice that we used the roots grid in our computation, and in any case, the theoretical performance of the two available options is known to be absolutely comparable \cite{Fox1968, Boyd2000}.

Upon implementing the collocation method, we derive a classical matrix-based quadratic eigenvalue problem, as detailed in \cite{Tisseur2001}
\begin{equation}\label{eq:eig}
  (M_{0} + iM_{1}\Omega + M_{2}\Omega^2)\bf{a}_\kappa =\bf{0}.
\end{equation}
In this formulation, the square real matrices $M_{j}$, each of size $(N+1)\times(N+1)$ for $j=0,1,2$, represent the spectral discretizations of the operators $L_{j}[\cdot]$, respectively. The problem \eqref{eq:eig} is solved numerically with the \textsc{polyeig} function from \textsc{Matlab}. This polynomial eigenvalue problem yields \(2(N+1)\) potential values for the parameter \(\Omega\). To discern the physical values of \(\Omega\) that correspond to the black hole's QNM modes, we first overlap the root plots for various values of \(N\) in equation \eqref{eq:exp}, such as \(N \in \{250, 280, 300\}\). We then identify the consistent roots whose positions remain stable across these different \(N\) values. In order to reduce the rounding and other floating point errors, we performed all our computations with multiple precision arithmetic that is built in \textsc{Maple} and which is brought into \textsc{Matlab} by the \textsc{Advanpix} toolbox \cite{mct2015}. All numerical computations reported in this study have been performed with $300$ decimal digits of accuracy.

\section{Scalar quasinormal spectra}

In this section, we present a detailed numerical analysis of scalar QNMs in Schwarzschild--Gauss--Bonnet black holes across various spacetime dimensions. Several universal trends emerge from our study. First, we observe the systematic appearance of overdamped purely imaginary modes at moderate to strong GB couplings, forming towers of nearly equispaced frequencies along the imaginary axis in lower dimensions ($D\in\{5,6,7\}$), and isolated overdamped modes in higher dimensions ($D\in\{8,10,11,12,26\}$). Second, we detect non-monotonic behaviour in the real parts of higher overtone frequencies, which becomes increasingly prominent as the coupling parameter increases, pointing to a reorganisation of the QNM spectra. Third, the morphological structure of the QNM distributions in the complex plane evolves systematically with both dimension and coupling transitioning from Martini glass shapes at low coupling to 'coupe à glace', bottleneck, and eventually pyramidal or truncated 'coupe à glace' configurations at higher coupling strengths. Furthermore, the limitations of the traditional WKB and characteristic integration methods become evident, particularly in their failure to detect overdamped modes and in the frequent misidentification of higher overtones as fundamental modes at large coupling. In contrast, our SM consistently resolves these features with high accuracy across all examined dimensions and angular momenta. The results are organised dimension-by-dimension in the following subsections, where we highlight key spectral features, compare numerical methods, and provide a detailed mapping of the QNM landscape. In $D = 26$, for a black hole of mass $M$ and GB coupling $\widehat{\alpha} = 0.1$, the fundamental scalar QNM has a dimensionless frequency $\Omega = 9.4488 - 1.9916i$. Although this value is not directly comparable to the four-dimensional Schwarzschild case due to the different scaling of the mass parameter with spacetime dimensionality, the substantially larger dimensionless frequencies strongly suggest the presence of an amplification phenomenon in higher dimensions. For comparison, the corresponding dimensionless frequency in $D = 4$ is $\Omega_{\text{Sch}} = 0.1105 - 0.1049i$. The vector and tensor sectors analysed below display a similar, though somewhat less pronounced, amplification, with their fundamental modes in $D=26$ also greatly exceeding the corresponding four-dimensional Schwarzschild values in dimensionless units. To assess the potential detectability of these amplified modes, we convert the computed QNM frequencies into physical units (Hz) using the relation
\begin{equation}
  \nu = \frac{1}{2\pi} \left( \frac{c^3}{G M_\odot} \right)^p \frac{\Omega}{\eta^p}, \quad p = \frac{1}{n-1},
\end{equation}
where $\Omega$ is the dimensionless QNM frequency, $M$ is the black hole mass, $M_\odot$ is the solar mass, and $\eta=M/M_\odot$. For a stellar-mass black hole with $M = 10 M_\odot$, the mode $\Omega=9.4488-1.9916i$ corresponds to a frequency $\nu=2.315-0.488i$ Hz, while for a supermassive black hole with $M=10^9 M_\odot$, it reduces to $\nu=1.039-0.219i$ Hz. In the case of superstring theory ($D=10$) with the same GB coupling, the fundamental scalar mode is $\Omega=2.2676-0.8831i$, yielding physical frequencies of $\nu=1.488-0.580i$ Hz for a stellar black hole and $\nu=0.107-0.042i$ Hz for a supermassive black hole. Although these frequencies lie below the sensitivity threshold of current ground-based detectors such as LIGO and VIRGO (which have a lower cutoff near 20 Hz), and slightly exceed the upper limit of LISA's sensitivity band (which peaks near 1 Hz), they fall within the promising few-Hz to tens-of-Hz range targeted by proposed space-based interferometers like DECIGO  \cite{Kawamura2008JPCS, Kawamura2019IJMPD, Kawamura2021PTEP}. These results suggest, for the first time, that higher-dimensional corrections predicted by string theory could leave observable imprints on gravitational wave signals. Specifically, the amplification of QNM frequencies in string-theoretically motivated dimensions, such as $D\in\{10,26\}$, may bring the scalar ringdown phase of both stellar and supermassive black holes into the sensitivity window of next-generation detectors like DECIGO. This opens a potential empirical pathway to test string theory through gravitational wave spectroscopy.

\subsection{$D=5$}

\begin{figure}
    \centering
    \subfloat[$n=3$, $\ell_{3}=5$, $\widehat{\alpha}=0$ (Beer Glass)]{%
    \includegraphics[width=0.49\textwidth]{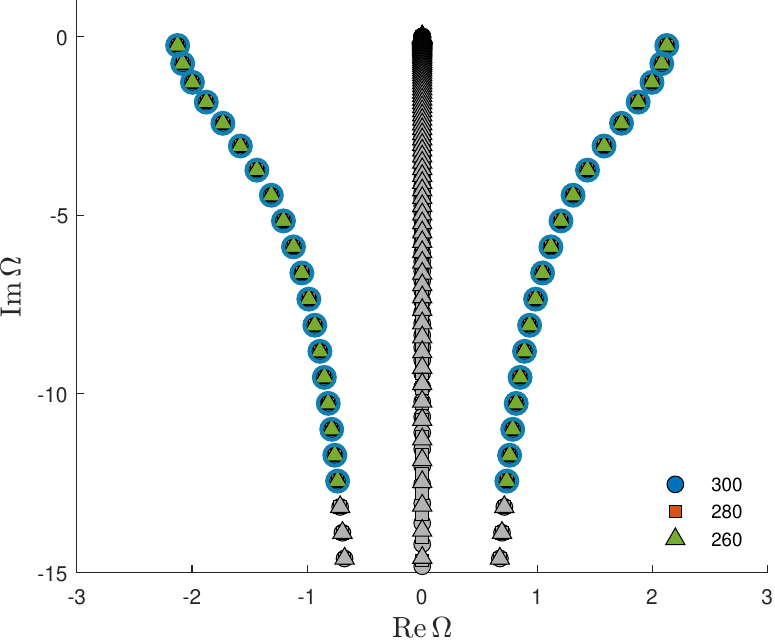}
    }
    \subfloat[$n=3$, $\ell_{3}=5$, $\widehat{\alpha}=1.9$ (Truncated Beer Glass)]{%
    \includegraphics[width=0.49\textwidth]{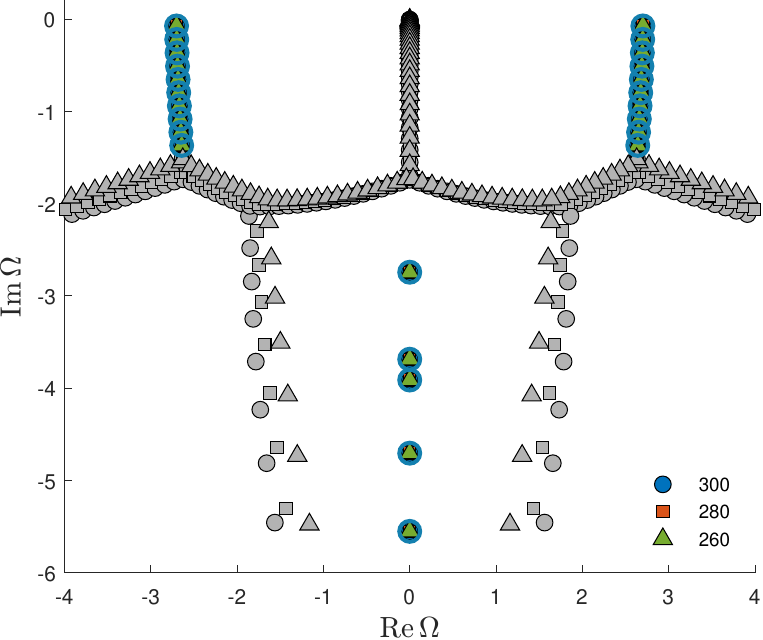}
    }
    \caption{Illustration of some distribution morphologies observed in the scalar perturbation spectrum. The left panel displays a typical beer glass configuration, while the right panel shows a truncated beer glass distribution. Here, $n$, $\ell_n$, and $\widehat{\alpha}$ denote the number of compactified dimensions, the angular momentum, and the coupling parameter, respectively.}
    \label{atlasQNMdistribscalirr}
\end{figure}

For scalar perturbations with $\ell_3 = 0$, overdamped QNMs first appear at moderate GB couplings $\widehat{\alpha} \geqslant 0.2$ and persist up to $\widehat{\alpha}=1.99$ (see Table~\ref{table:pi1}). These modes form a well-defined tower of nearly equispaced frequencies along the imaginary axis, with the spacing $\Delta\Omega$ exhibiting only weak dependence on $\widehat{\alpha}$ (see Tables~\ref{table:pi1} and \ref{table:pi2}). To quantitatively probe the dependence of $\Delta\Omega$ on angular momentum, we performed a linear regression analysis at small $\widehat{\alpha}$, finding strong linearity across $\ell_3$ with $R^2 > 0.97$ in all tested cases (see Figure~\ref{fig:scal-lin01}). These overdamped modes also emerge for higher angular momenta $\ell_3 \in {1, 2, 5}$ within the same coupling regime (see Table~\ref{table:pi2}). However, they are entirely absent in earlier WKB-based studies such as \cite{Iyer1989CQG, Konoplya2005PRDa}, highlighting the unique sensitivity of the SM in detecting such modes. At stronger coupling, specifically for $\widehat{\alpha} \geqslant 1.9$, we observe a restructuring of the spectrum. The overdamped modes progressively shift toward the origin of the complex frequency plane and begin to vanish. By $\widehat{\alpha} = 1.99$, only a single oscillatory mode remains, which stabilises to a limiting complex frequency as $\widehat{\alpha}$ increases further (see Table~\ref{table:1}). This transition is accompanied by a sharp reduction in the number of overdamped modes (Table~\ref{table:pi1}). Regarding numerical methods, Table~\ref{table:1} shows that both the SM and characteristic integration (CI) yield accurate results for $\ell_3 = 0$, whereas the sixth-order WKB approximation rapidly deteriorates for $\widehat{\alpha} \geqslant 0.5$. For $\ell_3 = 1$, the agreement between all methods improves, as expected in the $\ell > N$ regime. In particular, the WKB approximation becomes increasingly reliable with higher $\ell_3$, as seen in the example $\ell_3 = 5$, $\widehat{\alpha} = 1.9$ (Table~\ref{table:1a}). Nonetheless, neither WKB nor CI detects the overdamped tower observed in the range $0.2 \leqslant \widehat{\alpha} \leqslant 1.99$ for all tested angular momenta. Finally, for $0\leqslant \widehat{\alpha}\leqslant 0.5$ and $\ell_3\in\{0,1,2\}$, the QNM spectrum exhibits a characteristic Martini glass shape in the complex frequency plane (for a typical example see (a) in Fig.~\ref{atlasQNMdistribscal}). In contrast, for $\ell_3 = 5$, the distribution resembles a beer glass at $\widehat{\alpha}=0.1$ (for a typical example, see (a) in Fig.~\ref{atlasQNMdistribscalirr}), which gradually transitions into a truncated beer glass profile as $\widehat{\alpha}$ increases to $1.9$ (for a typical example, see (b) in Fig.~\ref{atlasQNMdistribscalirr}). Finally, we note that \cite{Konoplya2005PRDa} appears to overlook the earlier work of \cite{Iyer1989CQG}, as it contains no reference to it.

\subsection{$D=6$}

\begin{figure}
    \centering
    \subfloat[$n=10$, $\ell_{10}=0$, $\widehat{\alpha}=0.5$ (Martini)]{%
    \includegraphics[width=0.49\textwidth]{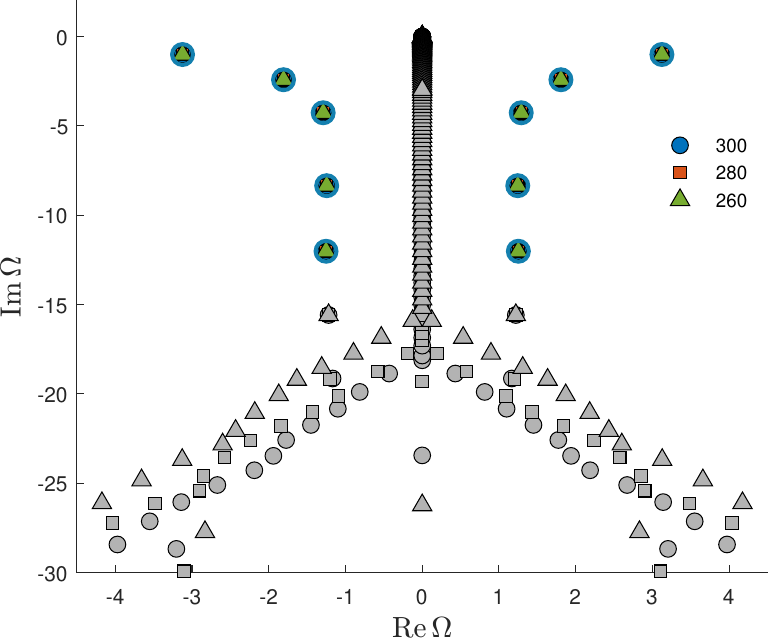}
    }
    \subfloat[$n=24$, $\ell_{24}=0$, $\widehat{\alpha}=0.2$ (Coupe)]{%
    \includegraphics[width=0.49\textwidth]{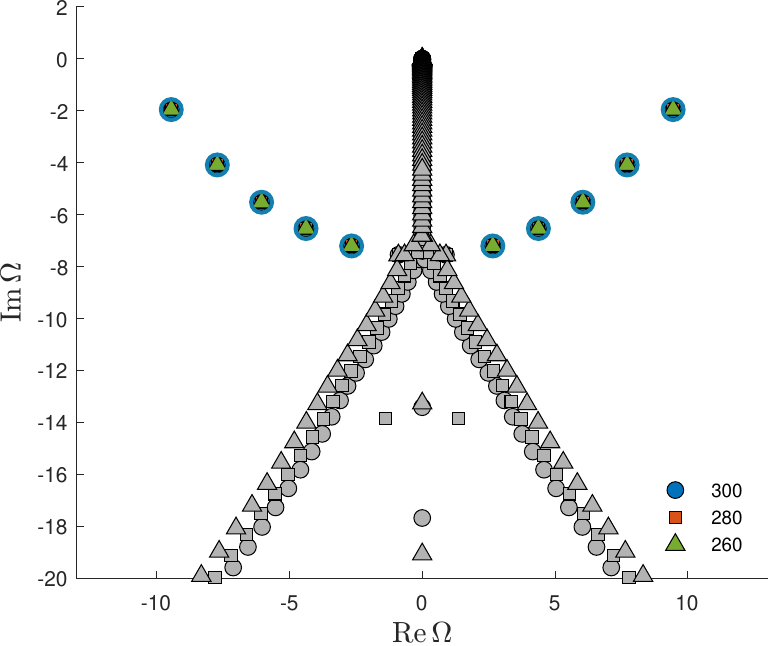}
    }\newline
    \subfloat[$n=9$, $\ell_{9}=1$, $\widehat{\alpha}=5.0$ (Bottleneck)]{%
    \includegraphics[width=0.49\textwidth]{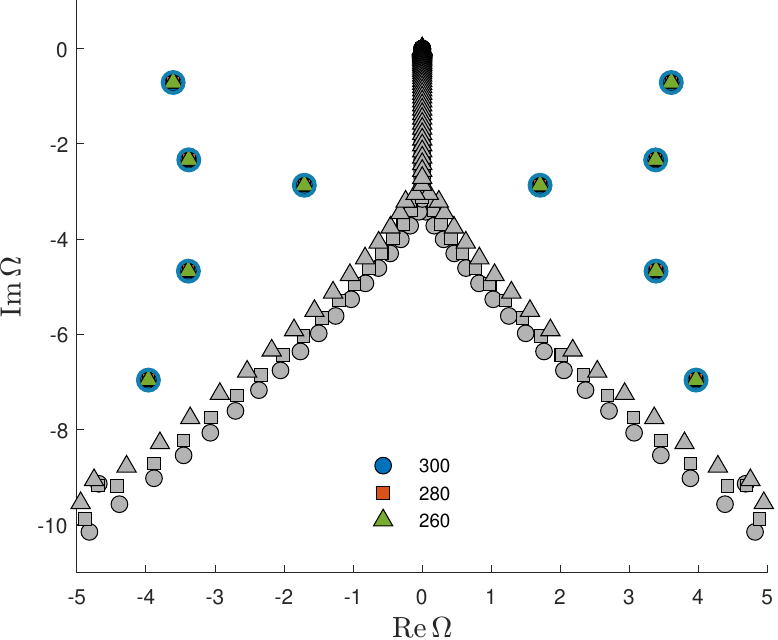}
    }
    \subfloat[$n=5$, $\ell_{5}=0$, $\widehat{\alpha}=5.0$ (Pyramid)]{%
    \includegraphics[width=0.49\textwidth]{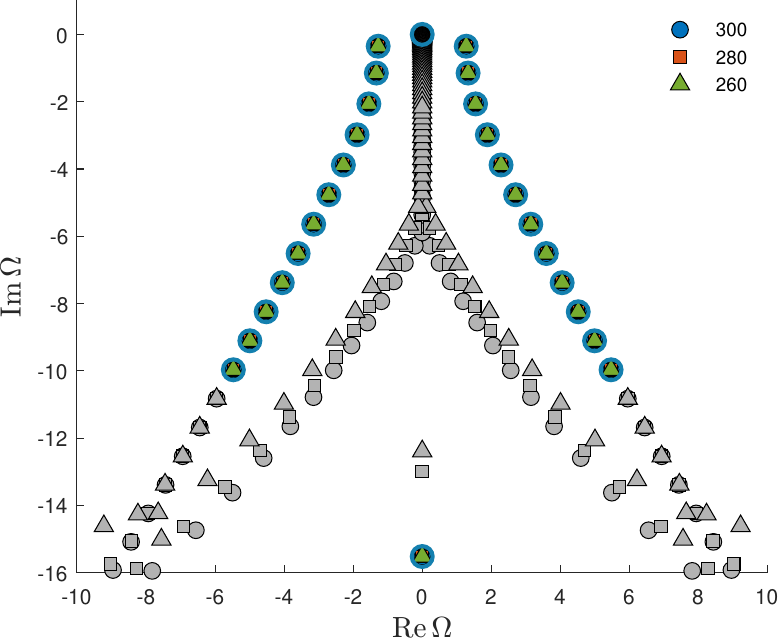}
    }
    \caption{The four main types of QNM distributions observed in the analysis of scalar perturbations. Here, $n$, $\ell_n$, and $\widehat{\alpha}$ denote the number of compactified dimensions, the angular momentum, and the coupling parameter, respectively.}
    \label{atlasQNMdistribscal}
\end{figure}

In the six-dimensional case, two central features emerge in the QNM spectrum. First, purely imaginary frequencies, are present for moderate couplings. For $\ell_4\in\{0,1\}$, they appear in the regime $0.2 \leqslant \widehat{\alpha} \leqslant 0.5$ (see Tables~\ref{table:pi3} and \ref{table:pi3}), while for $\ell_4=2$, they already emerge at $\widehat{\alpha}=0.1$ (see Table~\ref{table:pi5}). These modes form a nearly equispaced tower along the imaginary axis, similar to what we observed in the five-dimensional case. Our SM typically resolves up to fifteen such modes per configuration, but only a subset is reported here. Second, at larger couplings (e.g., $\widehat{\alpha}\geqslant 1.0$), we observe a non-monotonic behaviour in the real part of the QNMs. This pattern is clearly visible for $\ell_4\in\{0,5\}$ with $\widehat{\alpha}\geqslant 1.0$ (see Table~\ref{table:2} and \ref{table:2c})  and for $\ell_4=1$ with $\widehat{\alpha}\geqslant 5.0$ (see Table~\ref{table:2a}). These features indicate a qualitative reorganisation of the spectrum as the GB coupling increases. A further complication arises in identifying the fundamental mode when the WKB or CI technique is applied. For high $\widehat{\alpha}$ (e.g., $10$ and $20$), the sixth-order WKB approximation often misidentifies higher overtones as the fundamental. For example, we find that, at $\ell_4 = 0$, $\widehat{\alpha} = 10$, the WKB value $\Omega = 1.513624 - 0.45693i$ corresponds to our first overtone, not the fundamental (Table~\ref{table:2}, red vs. blue). At $\ell_4 = 0$, $\widehat{\alpha} = 20$, WKB predicts $\Omega = 2.961675 - 0.927128i$, which actually matches our fourth overtone. At $\ell_4 = 1$, the same misidentification occurs: $\Omega = 3.260415 - 0.498426i$ at $\widehat{\alpha} = 10$ is the third overtone, and $\Omega = 6.437139 - 0.991374i$ at $\widehat{\alpha} = 20$ is the fourth (Table~\ref{table:2a}). Similar discrepancies arise for $\ell_4 = 2$ where, at $\widehat{\alpha} = 10$ and $20$, WKB values coincide with the fifth overtone (Table~\ref{table:2b}). These consistent mismatches indicate that the sixth-order WKB method is not reliable for identifying the correct fundamental mode in this coupling regime. The CI method suffers from similar issues and, in some cases (e.g., $\ell_4 = 1$, $\widehat{\alpha} = 20$), fails to produce any result. Regarding the performance of approximation schemes we notice that, for $\ell_4 = 0$ and $\widehat{\alpha} \gtrsim 10$, both CI and WKB fail to yield trustworthy predictions. For $\ell_4 = 2$, sixth-order WKB gives inaccurate results already from $\widehat{\alpha} \gtrsim 5$, and finally, for $\ell_4 = 5$, the third-order WKB performs reasonably well in tracking the fundamental and first few overtones up to $\widehat{\alpha} \sim 5$, but it fails to detect the non-monotonic structure of higher overtones. Last but not least, for $\ell_4 \in \{0,1,2,5\}$, the QNM spectrum exhibits a 'coupe à glace' structure in the complex frequency plane for $0 \leqslant \widehat{\alpha} \leqslant 5$ (for a typical example see (b) in Fig.~\ref{atlasQNMdistribscal}), transitioning to a bottleneck-shaped distribution within the intermediate coupling range $1 \leqslant \widehat{\alpha} \leqslant 1.9$ (for a typical example, see (c) in Fig.~\ref{atlasQNMdistribscal}). For $\ell_4 = 0$, the distribution evolves into a pyramidal configuration when $5 \leqslant \widehat{\alpha} \leqslant 20$ (for a typical example, see (d) in Fig.~\ref{atlasQNMdistribscal}), a feature that persists at $ \widehat{\alpha} = 20$ also for $\ell_4 \in \{1,2\}$. In the coupling range $5 \leqslant \widehat{\alpha} \leqslant 10$, the spectrum for $\ell_4 = 1$ and $\ell_4 = 2$ retains a bottleneck-like shape.

\subsection{$D = 7$}

In this dimension, scalar perturbations exhibit two distinct phenomena. First, for low GB coupling ($0.1 \leqslant \widehat{\alpha} \leqslant 0.2$), a sequence of overdamped modes appears across all angular momenta $\ell_5\in\{0, 1, 2\}$ (see Tables~\ref{table:pi67} and \ref{table:pi78}). These modes are nearly equally spaced along the imaginary axis and disappear for larger $\widehat{\alpha}$. At stronger coupling, e.g., $\widehat{\alpha} = 10$ and $20$, only a single overdamped mode remains, at $\Omega = -21.7064i$ and $-30.6110i$, respectively. Second, non-monotonic behavior in the real part of the higher overtones becomes visible for $\widehat{\alpha} \geqslant 10$, again for all tested values of $\ell_5$ and it has been highlighted in bold in Tables~\ref{table:3a}, ~\ref{table:3b}, and \ref{table:3c}. This signals qualitative changes in the spectrum structure. As for the numerical methods, the sixth-order WKB approximation yields reliable estimates only for $\widehat{\alpha} \geqslant 10$, particularly for the fundamental QNM (see red-highlighted values). For smaller couplings, both WKB and CI become unreliable. In particular, CI fails to capture the variation in the real part of the QNM across $0.1 \leqslant \widehat{\alpha} \leqslant 0.5$, predicting instead a nearly constant value, contradicted by both WKB and SM. CI improves slightly at $\widehat{\alpha} \geqslant 10$, but neither WKB nor CI captures the overdamped towers or the non-monotonic overtones. Finally, for $\ell_5\in\{0,1,2\}$ and $0.1 \leqslant \widehat{\alpha} \leqslant 0.5$, the QNM spectrum displays a Martini glass configuration, which transitions into a 'coupe à glace' shape for $\ell_5 = 2$ at $\widehat{\alpha} = 0.5$. A pyramidal structure emerges for $\ell_5 = 0$ within the range $5 \leqslant \widehat{\alpha} \leqslant 20$, and similarly appears at $\widehat{\alpha} = 5$ for $\ell_5 \in \{1,2\}$. In the regime $10 \leqslant \widehat{\alpha} \leqslant 20$, the spectrum for $\ell_5 = 1$ and $\ell_5 = 2$ adopts a bottleneck-shaped distribution.

\subsection{$D = 8$}

In the scalar sector with $\ell_6 = 0$, we detect the presence of a single overdamped mode for increasing GB coupling, namely, $\Omega=-19.8069i$, $-24.5937i$, and $-30.8415i$ at $\widehat{\alpha}=5$, $10$, and $20.0$, respectively. These modes are not captured by WKB or CI approaches. Moreover, the sixth-order WKB predictions for the real part of the fundamental mode exhibit non-monotonic behaviour with increasing $\widehat{\alpha}$, but this appears to be a numerical artefact, as both CI and our SM provide a monotonic profile (contrast the red-highlighted values in Table~\ref{table:4a}). While the WKB method gains accuracy at higher $\widehat{\alpha}$, it still fails to resolve overtones. For $\ell_6 = 1$, the WKB method becomes progressively more accurate for the fundamental mode as $\widehat{\alpha}$ increases (see red-highlighted entries in Table~\ref{table:4b}). In contrast, the CI method is reasonably accurate both at small coupling ($\widehat{\alpha}=0.1$) and in the large-coupling regime ($\widehat{\alpha}\geqslant 5$). For $\ell_6=2$, the WKB approximation performs better for larger coupling ($\widehat{\alpha}\geqslant 10$), but fails to reproduce overtones (see Table~\ref{table:4c}). Overall, the WKB and CI methods are not always reliable. Moreover, non-monotonic behaviour in the real part of higher overtones emerges clearly for $\ell_6=0$ at $\widehat{\alpha}=5$ and $20$ (see bold entries in Table~\ref{table:4a}). This feature persists for $\ell_6=1$ across the range $0.5\leqslant\widehat{\alpha}\leqslant 20$ (Table~\ref{table:4b}), and becomes systematic for all tested values of $\widehat{\alpha}$ when $\ell_6=2$ (Table~\ref{table:4c}), indicating qualitative restructuring of the QNM spectrum. Finally, for $\ell_6\in\{0,1\}$ and $0.1\leqslant\widehat{\alpha}\leqslant 0.5$, the QNM spectrum exhibits a Martini glass distribution, while for $\ell_6=2$ in the same coupling range, the shape resembles a 'coupe à glace'. At $\widehat{\alpha}=5$, the spectrum transitions from a bottleneck profile for $\ell_6=0$ to a pyramidal structure for $\ell_6\in\{1,2\}$. At $\widehat{\alpha}=10$, the distribution becomes pyramidal for $\ell_6\in\{0,1\}$ and reverts to a bottleneck shape for $\ell_6=2$. For $\widehat{\alpha}=20$, a bottleneck pattern dominates across all tested values of $\ell_6$.

\subsection{$D=10$ (Superstring Theory)}

\begin{figure}
    \centering
    \subfloat[$n=8$, $\ell_{8}=0$, $\widehat{\alpha}=0.2$ (Deformed Coupe)]{%
    \includegraphics[width=0.49\textwidth]{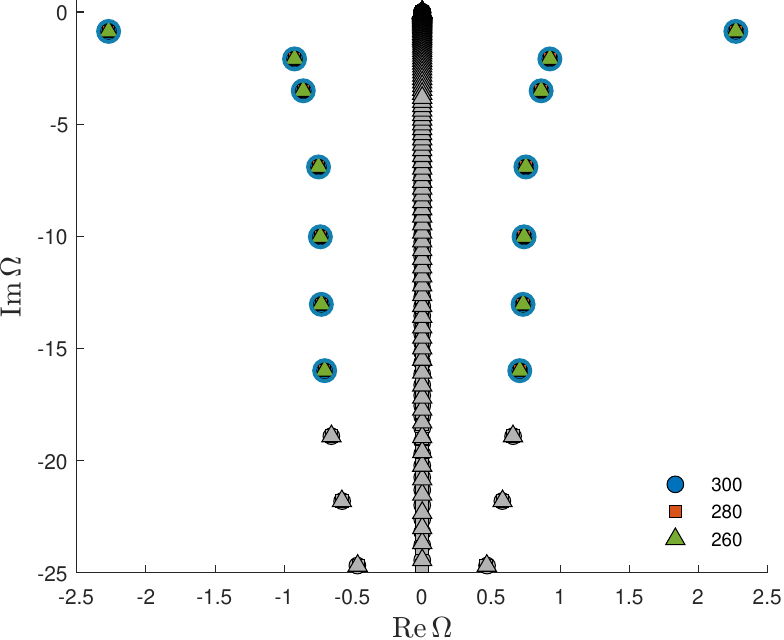}
    }
    \caption{Illustration of a typical deformed 'coupe à glace' distribution. Here, $n$, $\ell_n$, and $\widehat{\alpha}$ denote the number of compactified dimensions, the angular momentum, and the coupling parameter, respectively.}
    \label{defcoupel}
\end{figure}

In the ten-dimensional case, the scalar QNM spectrum exhibits several distinct features depending on the angular momentum index $\ell_8$ and the GB coupling $\widehat{\alpha}$. Purely imaginary QNMs are sparse but appear systematically for certain values of the coupling. For $\ell_8 = 0$, we detect a single overdamped mode at $\Omega=-46.0665i$ for $\widehat{\alpha} = 0.2$, at $\Omega=-30.7534i$ for $\widehat{\alpha}=5.0$, at $\Omega=-34.6687i$ for $\widehat{\alpha}=10$, and at $\Omega=-39.5350i$ for $\widehat{\alpha}=20$. For $\ell_8=1$, such modes appear only at $\Omega=-72.6316i$ for $\widehat{\alpha}=10$ and at $\Omega=-83.1977i$ for $\widehat{\alpha}=20$.  For $\ell_8=2$, no overdamped modes are found for $\widehat{\alpha}\leqslant 0.2$, but a single one emerges at $\Omega=-80.8597i$ for $\widehat{\alpha}=0.5$, at $\Omega=-113.4160i$ for $\widehat{\alpha}=5.0$, and at $\Omega=-129.8370i$ for $\widehat{\alpha}= 10$. Moreover, a non-monotonic trend in the real part of the higher overtones becomes apparent for $\ell_8=0$ within the range $0.5\leqslant\widehat{\alpha}\leqslant 20$, and extends more broadly for $\ell_8\in\{1, 2\}$, appearing already from $\widehat{\alpha}= 0.2$ onward (see bold entries in Table~\ref{table:5D10}). Regarding the morphology of the QNM spectrum, we identify a Martini glass distribution exclusively for $\ell_8 = 0$ in the narrow range $0 \leqslant \widehat{\alpha} \leqslant 0.1$, which transitions into a deformed 'coupe à glace' shape for $0.2\leqslant\widehat{\alpha}\leqslant 0.5$ (see Fig.~\ref{defcoupel} for a typical example). In contrast, for $\ell_8\in\{1, 2\}$, the QNMs exhibit a 'coupe à glace' distribution throughout $0.1\leqslant\widehat{\alpha}\leqslant 0.5$. At stronger coupling, $5 \leqslant \widehat{\alpha} \leqslant 20$, the spectrum adopts a bottleneck configuration across all tested angular momenta.

\subsection{$D=11$ (M-Theory)}

We observe the emergence of single overdamped modes across various configurations. Specifically, for $\ell_9 = 0$, overdamped QNMs appear at $\widehat{\alpha} \in \{ 5, 10, 20\}$ with frequencies $\Omega=-37.1819i$, $\Omega = -40.8974i$, and $\Omega = -45.5293i$, respectively. For $\ell_9 = 1$, overdamped modes emerge only at $\widehat{\alpha} = 10$ and $\widehat{\alpha}=20$, located at $\Omega=-75.8154i$ and $\Omega = -84.7772i$, respectively. For $\ell_9 = 2$, a single overdamped mode is detected already at $\widehat{\alpha} = 0.5$ with $\Omega = -88.7499i$, and persists at higher couplings. More precisely, we find $\Omega = -124.8810i$ for $\widehat{\alpha} = 10$, and $\Omega = -139.8890i$ for $\widehat{\alpha} = 20$. In addition to the appearance of overdamped modes, we observe non-monotonic behavior in the real part of the QNMs for $\ell_9\in\{0,1\}$ within the range $0.2\leqslant\widehat{\alpha}\leqslant 20$, and across all tested values of the coupling parameter for $\ell_9=2$ (see bold entries in Table~\ref{table:6D11}). The morphology of the QNM spectrum exhibits clear systematic trends. For $\ell_9 = 0$, a Martini glass distribution is observed in the regime $0.1 \leqslant \widehat{\alpha} \leqslant 0.5$, while in the same coupling range the spectrum transitions into a 'coupe à glace' distribution for $\ell_9 \in \{1,2\}$. At larger couplings ($5 \leqslant \widehat{\alpha} \leqslant 20$), a bottleneck distribution characterises the QNM spectrum uniformly across all tested values of $\ell_9$.

\subsection{$D=12$ (F-Theory)}

For scalar perturbations in twelve dimensions, overdamped modes begin to appear at moderate-to-strong GB coupling. Specifically, for $\ell_{10} = 0$, a single purely imaginary QNM is detected for $\widehat{\alpha} = 5$, $10$, and $20$, with frequencies $\Omega = -44.2085i$, $-47.7699i$, and $-52.2671i$, respectively. For $\ell_{10} = 1$, overdamped modes emerge only in the strong coupling regime. We find one such mode at $\Omega = -81.1465i$ for $\widehat{\alpha} = 10$ and another at $\Omega = -89.1708i$ for $\widehat{\alpha}=20$. For $\ell_{10} = 2$, the behavior is similar to the $\ell_{10} = 1$ case with overdamped modes identified at $\Omega = -125.7810i$ ($\widehat{\alpha} = 10$) and $\Omega = -138.4910i$ ($\widehat{\alpha} = 20$). We also observe non-monotonic behavior in the real part of the QNMs for $\ell_{10} = 0$ within the range $0.1\leqslant\widehat{\alpha}\leqslant 20$, and across all tested values of the coupling parameter for $\ell_{10} \in \{1,2\}$ (see bold entries in Table~\ref{table:8D12}). Regarding the morphology of the QNM spectrum, we observe that it mirrors the distribution patterns found in the $D = 11$ case, exhibiting the same transitions between Martini, coupe, and bottleneck structures across comparable ranges of the coupling parameter.

\subsection{$D=26$ (Bosonic String Theory)}

\begin{figure}
    \centering
    \subfloat[$n=24$, $\ell_{24}=0$, $\widehat{\alpha}=0.1$ (Coupe)]{%
    \includegraphics[width=0.49\textwidth]{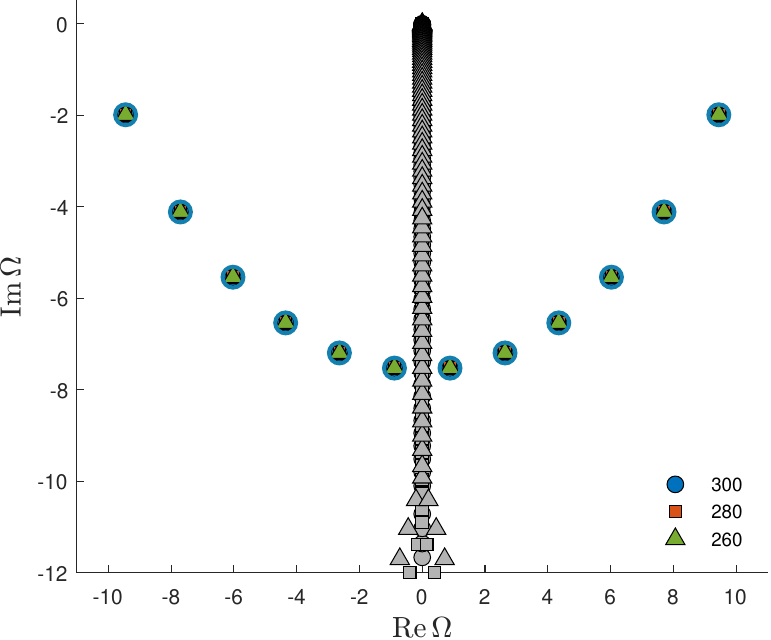}
    }
    \subfloat[$n=24$, $\ell_{24}=0$, $\widehat{\alpha}=0.5$ (Truncated Coupe)]{%
    \includegraphics[width=0.49\textwidth]{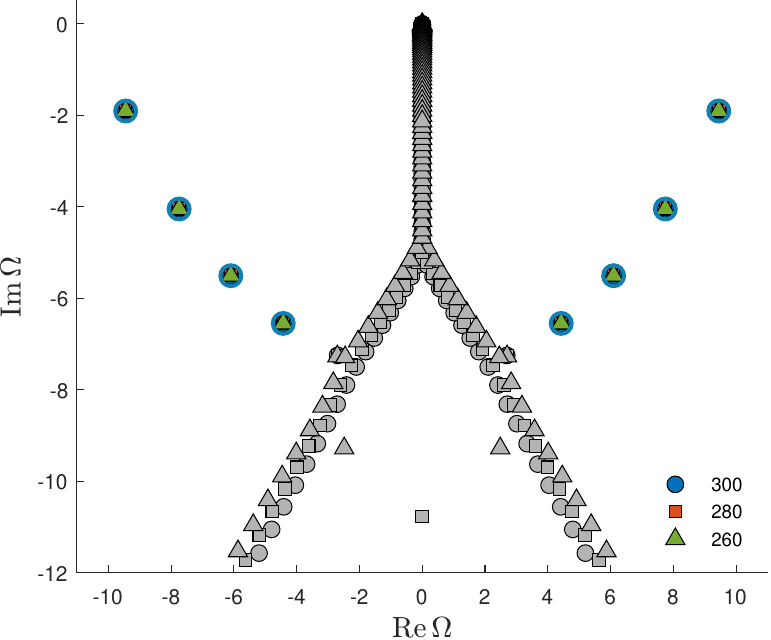}
    }
    \caption{Typical morphologies of QNM distribution observed in the scalar perturbation spectrum for $D=26$ (Bosonic String Theory). The left panel displays a typical coupe configuration, while the right panel shows a truncated coupe distribution. Here, $n$, $\ell_n$, and $\widehat{\alpha}$ denote the number of compactified dimensions, the angular momentum, and the coupling parameter, respectively.}
    \label{ctc26}
\end{figure}

\begin{figure}[H]
    \centering
    \includegraphics[width=0.8\linewidth]{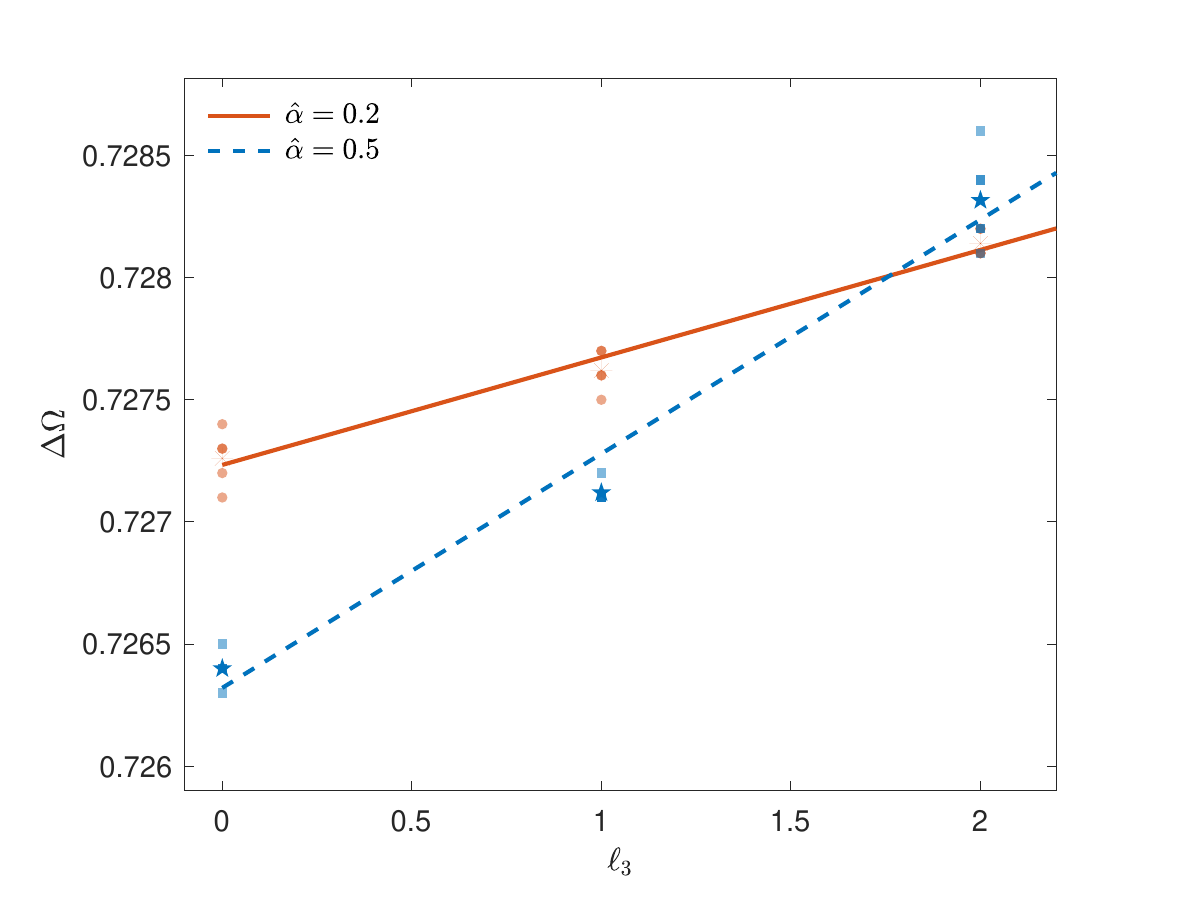}
    \caption{Linear regression analysis of the spectral gap values reported in Tables~\ref{table:pi1} and \ref{table:pi2}. The values of the linear regression determination coefficients for each considered value of the parameter $\widehat{\alpha}$ are as follows: $R^2 = 0.98910$ ($\widehat{\alpha}=0.2$), and $R^2 = 0.97980$ ($\widehat{\alpha}=0.5$).}
    \label{fig:scal-lin01}
\end{figure}

In this case, overdamped quasinormal modes appear across all tested angular momenta. For $\ell_{24}=0$, purely imaginary modes emerge at $\Omega=-202.9230i$, $-203.4280i$, and $-207.1870i$ for coupling parameters $\widehat{\alpha}\in\{5,10,20\}$, respectively. A similar pattern is observed for $\ell_{24} = 1$, with overdamped modes located at $\Omega=-241.4450i$, $-243.3330i$, and $-248.4790i$. For $\ell_{24}=2$, the purely imaginary QNMs shift further along the imaginary axis, taking the values $\Omega=-283.8540i$, $-287.2500i$, and $-293.9130i$ for the same sequence of couplings. At vanishing GB coupling, we observe that for each value of $\ell_{24}$, there are two distinguished QNMs emerging with significant separation from the main spectral branch (see Table~\ref{table:7D26}). These appear at $\Omega_1=0.8758-7.5000i$ and $\Omega_2=1.9261-13.8968i$ for $\ell_{24}= 0$; $\Omega_1=1.7509-8.1627i$ and $\Omega_2=1.8195-13.3975i$ for $\ell_{24}=1$; and $\Omega_1=0.9030-9.0166i$, $\Omega_2=1.6719-12.8802i$ for $\ell_{24}=2$. Regarding the morphology of the QNM distribution, we observe a coupe-shaped pattern in the range $0\leqslant\widehat{\alpha}\leqslant 0.2$ for all values of $\ell_{24}$ examined (for a typical examples, see (a) in Fig~\ref{ctc26}). This profile transitions into a truncated coupe configuration at $\widehat{\alpha}=0.5$, consistently across all tested angular momenta (for a typical example, see (b) in Fig~\ref{ctc26}). In the strong-coupling regime $5\leqslant\widehat{\alpha}\leqslant 20$, the spectrum simplifies significantly, exhibiting only two or three oscillatory modes accompanied by a single overdamped mode, as previously detailed at the beginning of this subsection. Finally, the QNM distribution morphology in $D = 26$ displays a markedly distinct structure compared to the cases $D\in\{10,11,12\}$. 

\begin{table}%[ht]
\centering
\caption{QNMs for scalar perturbations (spin $s = 0$) of a five-dimensional ($n=3$) Schwarzschild black hole with GB correction are presented in the table below for different values of $\widehat{\alpha}$ and $\ell_3=0$. The corresponding results are obtained through our SM, utilising $300$ polynomials with a precision of $300$ digits. In this context, $\Omega = \omega\sqrt{M}$ and $N$ represent the dimensionless frequency and the corresponding overtone, respectively. The notation 'N/A' indicates data not available, while 'SM' and 'CI' stand for Spectral Method and Characteristic Integration, respectively.}
\label{table:1}
\vspace*{1em}
\begin{tabular}{||c|c|c|c|c|c|c|c|c|c|c||}
\hline\hline
$\widehat{\alpha}$ $(D=5)$ & $\ell_3$ & $N$ & $\Omega$ \cite{Iyer1989CQG} (WKB 3rd order) & $\Omega$ \cite{Konoplya2005PRDa,Abdalla2005PRD} (WKB 6th order)& $\Omega$ \cite{Abdalla2005PRD} (CI) &$\Omega$ (SM) \\ [0.5ex]
\hline\hline
$0.0$   & $0$      & $0$ & \textcolor{red}{$0.3474-0.2906i$} & N/A          & N/A                            & \textcolor{red}{$0.3775-0.2711i$}\\
        &          & $1$ & N/A                               & N/A          & N/A                            & $0.2629-0.9342i$\\
        &          & $2$ & N/A                               & N/A          & N/A                            & $0.2124-1.6640i$\\
        &          & $3$ & N/A                               & N/A          & N/A                            & $0.1905-2.3881i$\\
        &          & $4$ & N/A                               & N/A          & N/A                            & $0.1783-3.1067i$\\
$0.1$   & $0$      & $0$ & N/A                               & \textcolor{red}{$0.389935-0.256159i$}&\textcolor{red}{$0.379-0.282i$} & \textcolor{red}{$0.3811-0.2643i$}\\
        &          & $1$ & N/A                               & N/A          & N/A                            & $0.2692-0.9042i$\\
        &          & $2$ & N/A                               & N/A          & N/A                            & $0.2106-1.6100i$\\
        &          & $3$ & N/A                               & N/A          & N/A                            & $0.1782-2.3155i$\\
        &          & $4$ & N/A                               & N/A          & N/A                            & $0.1551-3.0184i$\\
$0.2$   & $0$      & $0$ & N/A                               & \textcolor{red}{$0.396034-0.250548i$}&\textcolor{red}{$0.383-0.272i$}  & \textcolor{red}{$0.3842-0.2576i$}\\
        &          & $1$ & N/A                               & N/A          & N/A                            & $0.2746-0.8741i$\\
        &          & $2$ & N/A                               & N/A          & N/A                            & $0.2084-1.5569i$\\
        &          & $3$ & N/A                               & N/A          & N/A                            & $0.1666-2.2421i$\\
        &          & $4$ & N/A                               & N/A          & N/A                            & $0.1342-2.9224i$\\
$0.5$   & $0$      & $0$ & N/A                               & \textcolor{red}{$0.429741-0.208293i$}&\textcolor{red}{$0.391-0.246i$}  & \textcolor{red}{$0.3907-0.2381i$}\\
        &          & $1$ & N/A                               & N/A          & N/A                            & $0.2830-0.7882i$\\
        &          & $2$ & N/A                               & N/A          & N/A                            & $0.1796-1.3949i$\\
$1.0$   & $0$      & $0$ & \textcolor{red}{$0.3527-0.2118i$} & N/A          & N/A                            & \textcolor{red}{$0.3947-0.2098i$}\\
        &          & $1$ & N/A                               & N/A          & N/A                            & $0.2674-0.6707i$\\
        &          & $2$ & N/A                               & N/A          & N/A                            & $0.1372-1.2925i$\\
$1.9$   & $0$      & $0$ & \textcolor{red}{$0.3686-0.1696i$} & N/A          & N/A                            & \textcolor{red}{$0.4012-0.1666i$}\\
$1.99$  & $0$      & $0$ & N/A                             & N/A          & N/A                            & \textcolor{red}{$0.4013-0.1620i$}\\[1ex]
 \hline\hline 
 \end{tabular}
\end{table}

\begin{table}%[ht]
\centering
\caption{Scalar QNMs of a five-dimensional Schwarzschild black hole with GB correction are presented in the table below for different values of $\widehat{\alpha}$ and $\ell_3\in\{1,2,5\}$. The corresponding results are obtained through our SM, utilising $300$ polynomials with a precision of $300$ digits. In this context, $\Omega = \omega\sqrt{M}$ and $N$ represent the dimensionless frequency and the corresponding overtone, respectively. The notation 'N/A' indicates data not available, while 'SM' and 'CI' stand for Spectral Method and Characteristic Integration, respectively.}
\label{table:1a}
\vspace*{1em}
\begin{tabular}{||c|c|c|c|c|c|c|c|c|c|c||}
\hline\hline
$\widehat{\alpha}$ $(D=5)$ & $\ell_3$ & $N$ & $\Omega$ \cite{Iyer1989CQG} (WKB 3rd order) & $\Omega$ \cite{Konoplya2005PRDa,Abdalla2005PRD} (WKB 6th order)& $\Omega$ \cite{Abdalla2005PRD} (CI)& $\Omega$ (SM) \\ [0.5ex]
\hline\hline
$0.0$ & $1$  & $0$ & N/A                 & N/A         & N/A         & \textcolor{red}{$0.7184-0.2562i$}\\
      &      & $1$ & N/A                 & N/A         & N/A         & $0.6056-0.8186i$\\
      &      & $2$ & N/A                 & N/A         & N/A         & $0.4691-1.4914i$\\
      &      & $3$ & N/A                 & N/A         & N/A         & $0.3828-2.2180i$\\
      &      & $4$ & N/A                 & N/A         & N/A         & $0.3325-2.9502i$\\
$0.1$ & $1$  & $0$ & N/A                 & \textcolor{red}{$0.720423-0.255506i$} & \textcolor{red}{$0.7234-0.250i$} & \textcolor{red}{$0.7234-0.2506i$}\\
      &      & $1$ & N/A                 & N/A         & N/A         & $0.6174-0.7965i$\\
      &      & $2$ & N/A                 & N/A         & N/A         & $0.4836-1.4425i$\\
      &      & $3$ & N/A                 & N/A         & N/A         & $0.3896-2.1421i$\\
      &      & $4$ & N/A                 & N/A         & N/A         & $0.3265-2.8514i$\\
$0.2$ & $1$  & $0$ & N/A                 & \textcolor{red}{$0.723177-0.252684i$} & \textcolor{red}{$0.7284-0.245i$} &\textcolor{red}{$0.7285-0.2450i$}\\
      &      & $1$ & N/A                 & N/A         & N/A         & $0.6287-0.7751i$\\
      &      & $2$ & N/A                 & N/A         & N/A         & $0.4973-1.3954i$\\
      &      & $3$ & N/A                 & N/A         & N/A         & $0.3961-2.0687i$\\
      &      & $4$ & N/A                 & N/A         & N/A         & $0.3217-2.7559i$\\
$0.5$ & $1$  & $0$ & N/A                 & \textcolor{red}{$0.739205-0.236885i$} & \textcolor{red}{$0.7443-0.228i$} &\textcolor{red}{$0.7443-0.2284i$}\\
      &      & $1$ & N/A                 & N/A         & N/A         & $0.6604-0.7126i$\\
      &      & $2$ & N/A                 & N/A         & N/A         & $0.5369-1.2598i$\\
      &      & $3$ & N/A                 & N/A         & N/A         & $0.4209-1.8507i$\\
      &      & $4$ & N/A                 & N/A         & N/A         & $0.3210-2.4525i$\\
$0.0$ & $2$  & $0$ & N/A                 & N/A         & N/A         & \textcolor{red}{$1.0681-0.2528i$}\\
      &      & $1$ & N/A                 & N/A         & N/A         & $0.9848-0.7810i$\\
      &      & $2$ & N/A                 & N/A         & N/A         & $0.8441-1.3759i$\\
      &      & $3$ & N/A                 & N/A         & N/A         & $0.7032-2.0499i$\\
      &      & $4$ & N/A                 & N/A         & N/A         & $0.5982-2.7681i$\\
$0.1$ & $2$  & $0$ & N/A                 & \textcolor{red}{$1.07492-0.248021i$}  & N/A &\textcolor{red}{$1.0751-0.2476i$}\\
      &      & $1$ & N/A                 & N/A         & N/A         & $0.9969-0.7631i$\\
      &      & $2$ & N/A                 & N/A         & N/A         & $0.8638-1.3382i$\\
      &      & $3$ & N/A                 & N/A         & N/A         & $0.7250-1.9853i$\\
      &      & $4$ & N/A                 & N/A         & N/A         & $0.6144-2.6757i$\\
$0.2$ & $2$  & $0$ & N/A                 & \textcolor{red}{$1.08169-0.243352i$}  & N/A &\textcolor{red}{$1.0824-0.2423i$}\\
      &      & $1$ & N/A                 & N/A         & N/A        & $1.0091-0.7453i$\\
      &      & $2$ & N/A                 & N/A         & N/A         & $0.8829-1.3013i$\\
      &      & $3$ & N/A                 & N/A         & N/A         & $0.7467-1.9223i$\\
      &      & $4$ & N/A                 & N/A         & N/A         & $0.6314-2.5853i$\\
$0.5$ & $2$  & $0$ & N/A                 & \textcolor{red}{$1.10459-0.228154i$}  & N/A& \textcolor{red}{$1.1063-0.2261i$}\\
      &      & $1$ & N/A                 & N/A         & N/A         & $1.0460-0.6909i$\\
      &      & $2$ & N/A                 & N/A         & N/A         & $0.9396-1.1916i$\\
      &      & $3$ & N/A                 & N/A         & N/A         & $0.8148-1.7374i$\\
      &      & $4$ & N/A                 & N/A         & N/A         & $0.6955-2.3172i$\\
$0.0$ & $5$  & $0$ & \textcolor{red}{$2.125-0.2507i$}  & N/A    & N/A              & \textcolor{red}{$2.1250-0.2507i$}\\
      &      & $1$ & $2.081-0.7579i$     & N/A         & N/A         & $2.0814-0.7577i$\\
      &      & $2$ & $1.999-1.279i$      & N/A         & N/A         & $1.9964-1.2818i$\\
      &      & $3$ & $1.889-1.816i$      & N/A         & N/A         & $1.8760-1.8349i$\\
      &      & $4$ & $1.756-2.366i$      & N/A         & N/A         & $1.7316-2.4275i$\\
      &      & $5$ & $1.602-2.927i$      & N/A         & N/A         & $1.5796-3.0636i$\\
$1.0$ & $5$  & $0$ & \textcolor{red}{$2.300-0.1925i$}  & N/A  & N/A                & \textcolor{red}{$2.3005-0.1925i$}\\ 
      &      & $1$ & $2.279-0.5790i$     & N/A         & N/A         & $2.2799-0.5787i$\\ 
      &      & $2$ & $2.237-0.9694i$     & N/A         & N/A         & $2.2401-0.9686i$\\
      &      & $3$ & $2.177-1.365i$      & N/A         & N/A         & $2.1834-1.3644i$\\
      &      & $4$ & $2.100-1.767i$      & N/A         & N/A         & $2.1129-1.7686i$\\
      &      & $5$ & $2.008-2.174i$      & N/A         & N/A         & $2.0325-2.1828i$\\
$1.9$ & $5$  & $0$ & \textcolor{red}{$2.701-0.0730i$}  & N/A   & N/A               & \textcolor{red}{$2.7010-0.0730i$}\\
      &      & $1$ & $2.699-0.2188i$     & N/A         & N/A         & $2.6989-0.2190i$\\
      &      & $2$ & $2.696-0.3642i$     & N/A         & N/A         & $2.6951-0.3643i$\\
      &      & $3$ & $2.692-0.5091i$     & N/A         & N/A         & $2.6897-0.5090i$\\
      &      & $4$ & $2.685-0.6533i$     & N/A         & N/A         & $2.6833-0.6529i$\\
      &      & $5$ & $2.676-0.7965i$     & N/A         & N/A         & $2.6760-0.7939i$ \\ [1ex]
\hline\hline 
\end{tabular}
\end{table}

\begin{table}%[ht]
\centering
\caption{Purely imaginary QNMs for scalar perturbations of a five-dimensional ($n=3$) Schwarzschild black hole with GB correction are presented in the table below for different values of $\widehat{\alpha}$ and $\ell_3=0$. The corresponding results are obtained through our SM, utilising $300$ polynomials with a precision of $300$ digits. In this context, $\Omega = \omega\sqrt{M}$ and $N$ represent the dimensionless frequency and the corresponding overtone, respectively. The notation 'SM' stands for Spectral Method.}
\label{table:pi1}
\vspace*{1em}
\begin{tabular}{||c|c|c|c|c|c|c|c|c||}
\hline\hline
$\widehat{\alpha}$ $(D=5)$ & $\ell_3$ & $N$ & $\Omega$ (SM)    &$\Delta\Omega=\Omega_{N}-\Omega_{N+1}$\\ [0.5ex]
\hline\hline
$0.2$              & $0$      & $0$ & $0.0000-37.2370i$ & $0.7271i$\\
                   &          & $1$ & $0.0000-37.9641i$ & $0.7274i$\\
                   &          & $2$ & $0.0000-38.6915i$ & $0.7272i$\\
                   &          & $3$ & $0.0000-39.4187i$ & $0.7273i$\\
                   &          & $4$ & $0.0000-40.1460i$ & $0.7273i$\\
                   &          & $5$ & $0.0000-40.8733i$ & -        \\
$0.5$              & $0$      & $0$ & $0.0000-21.9462i$ & $0.7264i$\\
                   &          & $1$ & $0.0000-22.6762i$ & $0.7263i$\\
                   &          & $2$ & $0.0000-23.3989i$ & $0.7264i$\\
                   &          & $3$ & $0.0000-24.1253i$ & $0.7265i$\\
                   &          & $4$ & $0.0000-24.8518i$ & $0.7264i$\\
                   &          & $5$ & $0.0000-25.5782i$ & -        \\
$1.0$              & $0$      & $0$ & $0.0000-8.7628i$  & $0.7170i$\\
                   &          & $1$ & $0.0000-9.4798i$  & $0.7169i$\\
                   &          & $2$ & $0.0000-10.1967i$ & $0.7168i$\\
                   &          & $3$ & $0.0000-10.9135i$ & $0.7169i$\\
                   &          & $4$ & $0.0000-11.6304i$ & $0.7168i$\\
                   &          & $5$ & $0.0000-12.3472i$ & -        \\
$1.9$              & $0$      & $0$ & $0.0000-0.7947i$  & $0.7037i$\\
                   &          & $1$ & $0.0000-1.4984i$  & $0.7280i$\\
                   &          & $2$ & $0.0000-2.2264i$  & $0.7108i$\\
                   &          & $3$ & $0.0000-2.9372i$  & $0.7041i$\\
                   &          & $4$ & $0.0000-3.6413i$  & $0.7009i$\\
                   &          & $5$ & $0.0000-4.3422i$  & $0.6993i$\\
                   &          & $6$ & $0.0000-5.0415i$  & $0.6985i$\\
                   &          & $7$ & $0.0000-5.7400i$  & -        \\
$1.99$             & $0$      & $0$ & $0.0000-0.7034i$  & $0.7918i$\\
                   &          & $1$ & $0.0000-1.4952i$  & $0.7260i$\\
                   &          & $2$ & $0.0000-2.2212i$  & $0.7093i$\\
                   &          & $3$ & $0.0000-2.9305i$  & $0.6957i$\\
                   &          & $4$ & $0.0000-3.6262i$  & -        \\[1ex]
 \hline\hline 
 \end{tabular}
\end{table}

\begin{table}%[ht]
\centering
\caption{Purely imaginary QNMs for scalar perturbations of a five-dimensional ($n = 3$) Schwarzschild black hole with GB correction are presented in the table below for different values of $\widehat{\alpha}$ and $\ell_3\in\{1,2,5\}$. The corresponding results are obtained through our SM, utilising $300$ polynomials with a precision of $300$ digits. In this context, $\Omega = \omega\sqrt{M}$ and $N$ represent the dimensionless frequency and the corresponding overtone, respectively. The notation 'SM' stands for Spectral Method.}
\label{table:pi2}
\vspace*{1em}
\begin{tabular}{||c|c|c|c|c|c|c|c|c||}
\hline\hline
$\widehat{\alpha}$ $(D=5)$ & $\ell_3$ & $N$ & $\Omega$ (SM)    &$\Delta\Omega=\Omega_{N}-\Omega_{N+1}$\\ [0.5ex]
\hline\hline
$0.2$              & $1$      & $0$ & $0.0000-37.2203i$ & $0.7275i$\\
                   &          & $1$ & $0.0000-37.9478i$ & $0.7277i$\\
                   &          & $2$ & $0.0000-38.6755i$ & $0.7277i$\\
                   &          & $3$ & $0.0000-39.4031i$ & $0.7276i$\\
                   &          & $4$ & $0.0000-40.1307i$ & $0.7276i$\\
                   &          & $5$ & $0.0000-40.8583i$ & -        \\
$0.5$              & $1$      & $0$ & $0.0000-21.9234i$ & $0.7272i$\\
                   &          & $1$ & $0.0000-22.6506i$ & $0.7271i$\\
                   &          & $2$ & $0.0000-23.3777i$ & $0.7271i$\\
                   &          & $3$ & $0.0000-24.1048i$ & $0.7271i$\\
                   &          & $4$ & $0.0000-24.8319i$ & $0.7271i$\\
                   &          & $5$ & $0.0000-25.5589i$ & -        \\
$0.2$              & $2$      & $0$ & $0.0000-37.1922i$ & $0.7282i$\\
                   &          & $1$ & $0.0000-37.9204i$ & $0.7282i$\\
                   &          & $2$ & $0.0000-38.6486i$ & $0.7281i$\\
                   &          & $3$ & $0.0000-39.3767i$ & $0.7281i$\\
                   &          & $4$ & $0.0000-40.1048i$ & $0.7281i$\\
                   &          & $5$ & $0.0000-40.8329i$ & -        \\
$0.5$              & $2$      & $0$ & $0.0000-21.8848i$ & $0.7286i$\\
                   &          & $1$ & $0.0000-22.6134i$ & $0.7284i$\\
                   &          & $2$ & $0.0000-23.3418i$ & $0.7282i$\\
                   &          & $3$ & $0.0000-24.0700i$ & $0.7282i$\\
                   &          & $4$ & $0.0000-24.7982i$ & $0.7281i$\\
                   &          & $5$ & $0.0000-25.5263i$ & $0.7284i$\\
                   &          & $6$ & $0.0000-26.2542i$ & -        \\
$1.0$              & $5$      & $0$ & $0.0000-10.3638i$ & $0.7613i$\\
                   &          & $1$ & $0.0000-11.1251i$ & $0.7549i$\\
                   &          & $2$ & $0.0000-11.8800i$ & $0.7497i$\\
                   &          & $3$ & $0.0000-13.3751i$ & $0.7419i$\\
                   &          & $4$ & $0.0000-14.1170i$ & -        \\

$1.9$              & $5$      & $0$ & $0.0000-2.7429i$  & $0.9433i$\\
                   &          & $1$ & $0.0000-3.6862i$  & $0.2255i$\\
                   &          & $2$ & $0.0000-3.9117i$  & $0.7929i$\\
                   &          & $3$ & $0.0000-4.7046i$  & $0.8476i$\\
                   &          & $4$ & $0.0000-5.5522i$  & -        \\[1ex]
 \hline\hline 
 \end{tabular}
\end{table}

\begin{table}%[ht]
\centering
\caption{QNMs for scalar perturbations of a six-dimensional ($n=4$) Schwarzschild black hole with GB correction are presented in the table below for different values of $\widehat{\alpha}$ and $\ell_4=0$. The corresponding results are obtained through our SM, utilising $300$ polynomials with a precision of $300$ digits. In this context, $\Omega = \omega\sqrt[3]{M}$ and $N$ represent the dimensionless frequency and the corresponding overtone, respectively. The notation 'N/A' indicates data not available, while 'SM' and 'CI' stand for Spectral Method and Characteristic Integration, respectively.}
\label{table:2}
\vspace*{1em}
\begin{tabular}{||c|c|c|c|c|c|c|c|c|c|c||}
\hline\hline
$\widehat{\alpha}$ $(D=6)$ & $\ell_4$ & $N$ & $\Omega$ \cite{Iyer1989CQG} (WKB 3rd order)& $\Omega$ \cite{Konoplya2005PRDa,Abdalla2005PRD} (WKB 6th order) & $\Omega$ \cite{Abdalla2005PRD} (CI) & $\Omega$ (SM) \\ [0.5ex]
\hline\hline
$0.0$   & $0$ & $0$ & \textcolor{red}{$0.6410-0.4423i$} & N/A                                   & N/A                              & \textcolor{red}{$0.7059-0.4231i$} \\
        &     & $1$ & N/A                               & N/A                                   & N/A                              & $0.4368-1.5056i$ \\
        &     & $2$ & N/A                               & N/A                                   & N/A                              & $0.3291-2.7701i$ \\
        &     & $3$ & N/A                               & N/A                                   & N/A                              & $0.2931-4.0052i$ \\
        &     & $4$ & N/A                               & N/A                                   & N/A                              & $0.2751-5.2231i$ \\
$0.1$   & $0$ & $0$ & N/A                               & \textcolor{red}{$0.735854-0.402416i$} & \textcolor{red}{$0.7109-0.412i$} & \textcolor{red}{$0.7104-0.4124i$} \\
        &     & $1$ & N/A                               & N/A                                   & N/A                              & $0.4508-1.4503i$ \\
        &     & $2$ & N/A                               & N/A                                   & N/A                              & $0.3267-2.6743i$ \\
        &     & $3$ & N/A                               & N/A                                   & N/A                              & $0.2756-3.8796i$ \\
        &     & $4$ & N/A                               & N/A                                   & N/A                              & $0.2454-5.0710i$ \\
$0.2$   & $0$ & $0$ & N/A                               & \textcolor{red}{$0.748053-0.391049i$} & \textcolor{red}{$0.7144-0.402i$} & \textcolor{red}{$0.7141-0.4024i$} \\
        &     & $1$ & N/A                               & N/A                                   & N/A                              & $0.4643-1.4002i$ \\
        &     & $2$ & N/A                               & N/A                                   & N/A                              & $0.3301-2.5854i$ \\
        &     & $3$ & N/A                               & N/A                                   & N/A                              & $0.2734-3.7543i$ \\
        &     & $4$ & N/A                               & N/A                                   & N/A                              & $0.2368-4.8978i$ \\
$0.5$   & $0$ & $0$ & N/A                               & \textcolor{red}{$0.83530-0.304837i$}  & \textcolor{red}{$0.7225-0.375i$} & \textcolor{red}{$0.7223-0.3759i$} \\
        &     & $1$ & N/A                               & N/A                                   & N/A                              & $0.5001-1.2731i$ \\
        &     & $2$ & N/A                               & N/A                                   & N/A                              & $0.3438-2.3346i$ \\
        &     & $3$ & N/A                               & N/A                                   & N/A                              & $0.2362-3.3566i$ \\
        &     & $4$ & N/A                               & N/A                                   & N/A                              & N/A \\
$1.0$   & $0$ & $0$ & \textcolor{red}{$0.6794-0.3427i$} & N/A                                   & N/A                              & \textcolor{red}{$0.7335-0.3419i$} \\
        &     & $1$ & N/A                               & N/A                                   & N/A                              & $0.5508-1.1196i$ \\
        &     & $2$ & N/A                               & N/A                                   & N/A                              & $0.3786-2.0116i$ \\
        &     & $3$ & N/A                               & N/A                                   & N/A                              & ${\bf{0.2151-2.8724i}}$ \\
        &     & $4$ & N/A                               & N/A                                   & N/A                              & $0.3127-6.7200i$ \\
        &     & $5$ & N/A                               & N/A                                   & N/A                              & $0.3695-7.5772i$ \\
        &     & $6$ & N/A                               & N/A                                   & N/A                              & $0.4354-8.4371i$ \\
$1.9$   & $0$ & $0$ & \textcolor{red}{$0.7471-0.3124i$} & N/A                                   & N/A                              & \textcolor{red}{$0.7601-0.2972i$} \\
        &     & $1$ & N/A                               & N/A                                   & N/A                              & $0.6611-0.9427i$ \\
        &     & $2$ & N/A                               & N/A                                   & N/A                              & ${\bf{0.6111-1.6609i}}$ \\
        &     & $3$ & N/A                               & N/A                                   & N/A                              & $0.6455-2.3927i$ \\
        &     & $4$ & N/A                               & N/A                                   & N/A                              & $0.7168-3.1251i$ \\
        &     & $5$ & N/A                               & N/A                                   & N/A                              & $0.8030-3.8608i$ \\
        &     & $6$ & N/A                               & N/A                                   & N/A                              & $0.8985-4.6126i$ \\
$5.0$   & $0$ & $0$ & \textcolor{red}{$0.9067-0.1448i$} & \textcolor{red}{$0.906661-0.144820i$} & \textcolor{red}{$0.9526-0.226i$} & \textcolor{red}{$0.9530-0.2259i$} \\
        &     & $1$ & N/A                               & N/A                                   & N/A                              & $1.0685-0.7193i$ \\
        &     & $2$ & N/A                               & N/A                                   & N/A                              & $1.3048-1.2739i$ \\
        &     & $3$ & N/A                               & N/A                                   & N/A                              & $1.6103-1.8377i$ \\
        &     & $4$ & N/A                               & N/A                                   & N/A                              & $1.9499-2.3934i$ \\
        &     & $5$ & N/A                               & N/A                                   & N/A                              & $2.3078-2.9420i$ \\
$10.0$  & $0$ & $0$ & N/A                               & \textcolor{red}{$1.513624-0.456935i$} & \textcolor{red}{$1.5182-0.423i$} & \textcolor{red}{$1.1906-0.6477i$} \\
        &     & $1$ & N/A                               & N/A                                   & N/A                              & \textcolor{blue}{$1.5185-0.4740i$} \\
        &     & $2$ & N/A                               & N/A                                   & N/A                              & $1.7792-1.1836i$ \\
        &     & $3$ & N/A                               & N/A                                   & N/A                              & $2.2084-1.6844i$ \\
        &     & $4$ & N/A                               & N/A                                   & N/A                              & $2.6758-2.1894i$ \\
$20.0$  & $0$ & $0$ & N/A                               & \textcolor{red}{$2.961675-0.927128i$} & \textcolor{red}{$2.878-0.952i$}  & \textcolor{red}{$0.9504-0.8254i$} \\
        &     & $1$ & N/A                               & N/A                                   & N/A                              & $1.7268-1.2940i$ \\
        &     & $2$ & N/A                               & N/A                                   & N/A                              & $2.4919-1.6986i$ \\
        &     & $3$ & N/A                               & N/A                                   & N/A                              & \textcolor{blue}{$2.9611-0.9280i$} \\
        &     & $4$ & N/A                               & N/A                                   & N/A                              & $3.1454-2.1310i$ \\[1ex]
 \hline\hline 
 \end{tabular}
\end{table}

\begin{table}%[ht]
\centering
\caption{Purely imaginary QNMs for scalar perturbations of a six-dimensional ($n=4$) Schwarzschild black hole with GB correction are presented in the table below for different values of $\widehat{\alpha}$ and $\ell_4=0$. The corresponding results are obtained through our SM, utilising $300$ polynomia
ls with a precision of $300$ digits. In this context, $\Omega=\omega\sqrt[3]{M}$ and $N$ represent the dimensionless frequency and the corresponding overtone, respectively. The notation 'SM' stands for Spectral Method.}
\label{table:pi3}
\vspace*{1em}
\begin{tabular}{||c|c|c|c|c|c|c|c|c||}
\hline\hline
$\widehat{\alpha}$ $(D=6)$ & $\ell_4$ & $N$ & $\Omega$ (SM)    &$\Delta\Omega=\Omega_{N}-\Omega_{N+1}$\\ [0.5ex]
\hline\hline
$0.2$              & $0$      & $0$ & $0.0000-72.4541i$ & $1.2229i$\\
                   &          & $1$ & $0.0000-73.6670i$ & $1.2228i$\\
                   &          & $2$ & $0.0000-74.8998i$ & $1.2229i$\\
                   &          & $3$ & $0.0000-76.1227i$ & $1.2228i$\\
                   &          & $4$ & $0.0000-77.3455i$ & $1.2229i$\\
                   &          & $5$ & $0.0000-78.5684i$ & -        \\
$0.5$              & $0$      & $0$ & $0.0000-45.4215i$ & $1.2189i$\\
                   &          & $1$ & $0.0000-46.6404i$ & $1.2189i$\\
                   &          & $2$ & $0.0000-47.8593i$ & $1.2189i$\\
                   &          & $3$ & $0.0000-49.0782i$ & $1.2188i$\\
                   &          & $4$ & $0.0000-50.2970i$ & $1.2189i$\\
                   &          & $5$ & $0.0000-51.5159i$ & -        \\[1ex]
 \hline\hline 
 \end{tabular}
\end{table}

\begin{table}%[ht]
\centering
\caption{QNMs for scalar perturbations of a six-dimensional ($n=4$) Schwarzschild black hole with GB  correction are presented in the table below for different values of $\widehat{\alpha}$ and $\ell_4=1$. The corresponding results are obtained through our SM, utilising $300$ polynomials with a precision of $300$ digits. In this context, $\Omega = \omega\sqrt[3]{M}$ and $N$ represent the dimensionless frequency and the corresponding overtone, respectively. The notation 'N/A' indicates data not available, while 'SM' and 'CI' stand for Spectral Method and Characteristic Integration, respectively.}
\label{table:2a}
\vspace*{1em}
\begin{tabular}{||c|c|c|c|c|c|c|c|c|c|c||}
\hline\hline
$\widehat{\alpha}$ $(D=6)$& $\ell_4$ & $N$ & $\Omega$ \cite{Iyer1989CQG} (WKB 3rd order)& $\Omega$ \cite{Konoplya2005PRDa,Abdalla2005PRD} (WKB 6th order) & $\Omega$ \cite{Abdalla2005PRD} (CI) & $\Omega$ (SM) \\ [0.5ex]
\hline\hline
$0.0$   & $1$ & $0$ & N/A  & N/A                                   & N/A                            & \textcolor{red}{$1.1481-0.4042i$} \\
        &     & $1$ & N/A  & N/A                                   & N/A                            & $0.9076-1.3028i$ \\
        &     & $2$ & N/A  & N/A                                   & N/A                            & $0.6039-2.4756i$ \\
        &     & $3$ & N/A  & N/A                                   & N/A                            & $0.4625-3.7558i$ \\
        &     & $4$ & N/A  & N/A                                   & N/A                            & $0.3988-5.0139i$ \\
$0.1$   & $1$ & $0$ & N/A  & \textcolor{red}{$1.139007-0.415034i$} & \textcolor{red}{$1.153-0.395i$}& \textcolor{red}{$1.1536-0.3952i$} \\
        &     & $1$ & N/A  & N/A                                   & N/A                            & $0.9288-1.2667i$ \\
        &     & $2$ & N/A  & N/A                                   & N/A                            & $0.6293-2.3840i$ \\
        &     & $3$ & N/A  & N/A                                   & N/A                            & $0.4650-3.6204i$ \\
        &     & $4$ & N/A  & N/A                                   & N/A                            & $0.3779-4.8502i$ \\
$0.2$   & $1$ & $0$ & N/A  & \textcolor{red}{$1.136193-0.415706i$} & \textcolor{red}{$1.159-0.386i$}&\textcolor{red}{$1.1592-0.3867i$} \\
        &     & $1$ & N/A  & N/A                                   & N/A                            & $0.9484-1.2334i$ \\
        &     & $2$ & N/A  & N/A                                   & N/A                            & $0.6552-2.3022i$ \\
        &     & $3$ & N/A  & N/A                                   & N/A                            & $0.4766-3.4996i$ \\
        &     & $4$ & N/A  & N/A                                   & N/A                            & $0.3845-4.6995i$ \\
$0.5$   & $1$ & $0$ & N/A  & \textcolor{red}{$1.158635-0.391339i$} & \textcolor{red}{$1.176-0.363i$}& \textcolor{red}{$1.1762-0.3637i$} \\
        &     & $1$ & N/A  & N/A                                   & N/A                            & $1.0013-1.1462i$ \\
        &     & $2$ & N/A  & N/A                                   & N/A                            & $0.7402-2.0940i$ \\
        &     & $3$ & N/A  & N/A                                   & N/A                            & $0.5576-3.1639i$ \\
        &     & $4$ & N/A  & N/A                                   & N/A                            & $0.4582-4.2234i$ \\
$5.0$   & $1$ & $0$ & N/A  & \textcolor{red}{$1.790103-0.258382i$} & \textcolor{red}{$1.791-0.253i$}&\textcolor{red}{$1.7899-0.2521i$} \\
        &     & $1$ & N/A  & M/A                                   & N/A                            & ${\bf{1.7758-0.6602i}}$ \\
        &     & $2$ & N/A  & N/A                                   & N/A                            & $1.9729-1.0713i$ \\
        &     & $3$ & N/A  & N/A                                   & N/A                            & $2.2290-1.5812i$ \\
        &     & $4$ & N/A  & N/A                                   & N/A                            & $2.5218-2.1111i$ \\
$10.0$  & $1$ & $0$ & N/A  & \textcolor{red}{$3.260415-0.498426i$} & \textcolor{red}{$3.253-0.504i$}& \textcolor{red}{$1.5893-0.9926i$} \\
        &     & $1$ & N/A  & N/A                                   & N/A                            & $2.4471-1.3481i$ \\
        &     & $2$ & N/A  & N/A                                   & N/A                            & $3.1682-1.4396i$ \\
        &     & $3$ & N/A  & N/A                                   & N/A                            & \textcolor{blue}{$3.2604-0.4984i$} \\
        &     & $4$ & N/A  & N/A                                   & N/A                            & $3.3915-1.8209i$ \\
$20.0$  & $1$ & $0$ & N/A  & \textcolor{red}{$6.437139-0.991374i$} & N/A                            & \textcolor{red}{$1.2906-1.1358i$} \\
        &     & $1$ & N/A  & N/A                                   & N/A                            & $2.1185-1.6473i$ \\
        &     & $2$ & N/A  & N/A                                   & N/A                            & $2.9094-2.0685i$ \\
        &     & $3$ & N/A  & N/A                                   & N/A                            & $3.6843-2.4472i$ \\
        &     & $4$ & N/A  & N/A                                   & N/A                            & \textcolor{blue}{$6.4371-0.9914i$} \\[1ex]
 \hline\hline 
 \end{tabular}
\end{table}

\begin{table}%[ht]
\centering
\caption{Purely imaginary QNMs for scalar perturbations of a six-dimensional ($n=4$) Schwarzschild black hole with GB correction are presented in the table below for different values of $\widehat{\alpha}$ and $\ell_4=1$. The corresponding results are obtained through our SM, utilising $300$ polynomials with a precision of $300$ digits. In this context, $\Omega=\omega\sqrt[3]{M}$ and $N$ represent the dimensionless frequency and the corresponding overtone, respectively. The notation 'SM' stands for Spectral Method.}
\label{table:pi4}
\vspace*{1em}
\begin{tabular}{||c|c|c|c|c|c|c|c|c||}
\hline\hline
$\widehat{\alpha}$ $(D=6)$ & $\ell_4$ & $N$ & $\Omega$ (SM)    &$\Delta\Omega=\Omega_{N}-\Omega_{N+1}$\\ [0.5ex]
\hline\hline
$0.2$              & $1$      & $0$ & $0.0000-72.4393i$ & $1.2231i$\\
                   &          & $1$ & $0.0000-73.6624i$ & $1.2231i$\\
                   &          & $2$ & $0.0000-74.8855i$ & $1.2231i$\\
                   &          & $3$ & $0.0000-76.1086i$ & $1.2230i$\\
                   &          & $4$ & $0.0000-77.3316i$ & $1.2231i$\\
                   &          & $5$ & $0.0000-78.5547i$ & -        \\
$0.5$              & $1$      & $0$ & $0.0000-44.1822i$ & $1.2195i$\\
                   &          & $1$ & $0.0000-45.4077i$ & $1.2194i$\\
                   &          & $2$ & $0.0000-46.6211i$ & $1.2194i$\\
                   &          & $3$ & $0.0000-47.8405i$ & $1.2194i$\\
                   &          & $4$ & $0.0000-49.0599i$ & $1.2193i$\\
                   &          & $5$ & $0.0000-50.2792i$ & -        \\[1ex]
 \hline\hline 
 \end{tabular}
\end{table}

\begin{table}%[ht]
\centering
\caption{QNM modes for scalar perturbations of a six-dimensional ($n=4$) Schwarzschild black hole with GB  correction are presented in the table below for different values of $\widehat{\alpha}$ and $\ell_4=2$. The corresponding results are obtained through our SM, utilising $300$ polynomials with a precision of $300$ digits. In this context, $\Omega = \omega\sqrt[3]{M}$ and $N$ represent the dimensionless frequency and the corresponding overtone, respectively. The notation 'N/A' indicates data not available, while 'SM' stands for Spectral Method.}
\label{table:2b}
\vspace*{1em}
\begin{tabular}{||c|c|c|c|c|c|c|c|c||}
\hline\hline
$\widehat{\alpha}$ $(D=6)$ & $\ell_4$ & $N$ & $\Omega$ \cite{Iyer1989CQG} (WKB 3rd order)& $\Omega$ \cite{Konoplya2005PRDa} (WKB 6th order) & $\Omega$ (SM) \\ [0.5ex]
\hline\hline
$0.0$              & $2$      & $0$ & N/A                        & N/A                  & \textcolor{red}{$1.5966-0.3984i$} \\
                   &          & $1$ & N/A                        & N/A                  & $1.4176-1.2372i$ \\
                   &          & $2$ & N/A                        & N/A                  & $1.0940-2.2313i$ \\
                   &          & $3$ & N/A                        & N/A                  & $0.7986-3.4417i$ \\
                   &          & $4$ & N/A                        & N/A                  & $0.6339-4.7224i$ \\
$0.1$              & $2$      & $0$ & N/A                        & \textcolor{red}{$1.60284-0.39205i$}   & \textcolor{red}{$1.6039-0.3901i$} \\
                   &          & $1$ & N/A                        & N/A                  & $1.4360-1.2086i$ \\
                   &          & $2$ & N/A                        & N/A                  & $1.1309-2.1647i$ \\
                   &          & $3$ & N/A                        & N/A                  & $0.8336-3.3178i$ \\
                   &          & $4$ & N/A                        & N/A                  & $0.6455-4.5517i$ \\
$0.2$              & $2$      & $0$ & N/A                        & \textcolor{red}{$1.607508-0.387098i$} & \textcolor{red}{$1.6115-0.3822i$} \\
                   &          & $1$ & N/A                        & N/A                  & $1.4539-1.1815i$ \\
                   &          & $2$ & N/A                        & N/A                  & $1.1662-2.1036i$ \\
                   &          & $3$ & N/A                        & N/A                  & $0.8708-3.2055i$ \\
                   &          & $4$ & N/A                        & N/A                  & $0.6663-4.3984i$ \\
$0.5$              & $2$      & $0$ & N/A                        & \textcolor{red}{$1.62621-0.370114i$}  & \textcolor{red}{$1.6360-0.3599i$} \\
                   &          & $1$ & N/A                        & N/A                  & $1.5060-1.1066i$ \\
                   &          & $2$ & N/A                        & N/A                  & $1.2684-1.9432i$ \\
                   &          & $3$ & N/A                        & N/A                  & $1.0037-2.9167i$ \\
                   &          & $4$ & N/A                        & N/A                  & $0.8047-3.9876i$ \\
$5.0$              & $2$      & $0$ & N/A                        & \textcolor{red}{$2.619279-0.269015i$} & \textcolor{red}{$2.4506-0.8573i$} \\
                   &          & $1$ & N/A                        & N/A                  & $2.5815-0.9059i$ \\
                   &          & $2$ & N/A                        & N/A                  & \textcolor{blue}{$2.6188-0.2695i$} \\
                   &          & $3$ & N/A                        & N/A                  & $2.8415-1.4353i$ \\
                   &          & $4$ & N/A                        & N/A                  & $3.0977-1.9332i$ \\
$10.0$             & $2$      & $0$ & N/A                        & \textcolor{red}{$4.81752-0.505708i$}  & \textcolor{red}{$1.9944-1.2611i$} \\
                   &          & $1$ & N/A                        & N/A                  & $2.8943-1.6605i$ \\
                   &          & $2$ & N/A                        & N/A                  & $3.7412-1.9474i$ \\
                   &          & $3$ & N/A                        & N/A                  & $4.5300-2.1881i$ \\
                   &          & $4$ & N/A                        & N/A                  & $4.7656-1.5253i$ \\
                   &          & $5$ & N/A                        & N/A                  & \textcolor{blue}{$4.8175-0.5057i$} \\
                   &          & $6$ & N/A                        & N/A                  & $5.0049-2.5153i$ \\
                   &          & $7$ & N/A                        & N/A                  & $5.4233-2.9884i$ \\
$20.0$             & $2$      & $0$ & N/A                        & \textcolor{red}{$9.52384-1.00306i$}   & \textcolor{red}{$1.6229-1.3879i$} \\
                   &          & $1$ & N/A                        & N/A                  & $2.4919-1.9251i$ \\
                   &          & $2$ & N/A                        & N/A                  & $3.3004-2.3710i$ \\
                   &          & $3$ & N/A                        & N/A                  & $4.1033-2.7750i$ \\
                   &          & $4$ & N/A                        & N/A                  & $9.4270-3.0237i$ \\
                   &          & $5$ & N/A                        & N/A                  & \textcolor{blue}{$9.5238-1.0031i$} \\[1ex]
 \hline\hline 
 \end{tabular}
\end{table}

\begin{table}%[ht]
\centering
\caption{Purely imaginary QNMs for scalar perturbations of a six-dimensional ($n=4$) Schwarzschild black hole with GB correction are presented in the table below for different values of $\widehat{\alpha}$ and $\ell_4=2$. The corresponding results are obtained through our SM, utilising $300$ polynomials with a precision of $300$ digits. In this context, $\Omega=\omega\sqrt[3]{M}$ and $N$ represent the dimensionless frequency and the corresponding overtone, respectively. The notation 'SM' stands for Spectral Method.}
\label{table:pi5}
\vspace*{1em}
\begin{tabular}{||c|c|c|c|c|c|c|c|c||}
\hline\hline
$\widehat{\alpha}$ $(D=6)$ & $\ell_4$ & $N$ & $\Omega$ (SM)    &$\Delta\Omega=\Omega_{N}-\Omega_{N+1}$\\ [0.5ex]
\hline\hline
$0.1$              & $2$      & $0$ & $0.0000-61.5086i$ & $1.0983i$\\
                   &          & $1$ & $0.0000-62.6069i$ & $1.0984i$\\
                   &          & $2$ & $0.0000-63.7053i$ & $1.0980i$\\
                   &          & $3$ & $0.0000-64.8033i$ & $1.0979i$\\
                   &          & $4$ & $0.0000-65.9012i$ & $1.1018i$\\
                   &          & $5$ & $0.0000-67.0030i$ & -        \\
$0.2$              & $2$      & $0$ & $0.0000-72.4169i$ & $1.2235i$\\
                   &          & $1$ & $0.0000-73.6404i$ & $1.2234i$\\
                   &          & $2$ & $0.0000-74.8638i$ & $1.2235i$\\
                   &          & $3$ & $0.0000-76.0873i$ & $1.2234i$\\
                   &          & $4$ & $0.0000-77.3107i$ & $1.2234i$\\
                   &          & $5$ & $0.0000-78.5341i$ & -        \\
$0.5$              & $2$      & $0$ & $0.0000-45.3718i$ & $1.2202i$\\
                   &          & $1$ & $0.0000-46.5920i$ & $1.2202i$\\
                   &          & $2$ & $0.0000-47.8122i$ & $1.2201i$\\
                   &          & $3$ & $0.0000-49.0323i$ & $1.2200i$\\
                   &          & $4$ & $0.0000-50.2523i$ & $1.2200i$\\
                   &          & $5$ & $0.0000-51.4723i$ & -        \\[1ex]
 \hline\hline 
 \end{tabular}
\end{table}

\begin{table}%[ht]
\centering
\caption{QNMs for scalar perturbations of a six-dimensional ($n=4$) Schwarzschild black hole with GB correction are presented in the table below for different values of $\widehat{\alpha}$ and $\ell_4=5$. The corresponding results are obtained through our SM, utilising $300$ polynomials with a precision of $300$ digits. In this context, $\Omega = \omega\sqrt[3]{M}$ and $N$ represent the dimensionless frequency and the corresponding overtone, respectively. The notation 'N/A' indicates data not available, while 'SM' stands for Spectral Method.}
\label{table:2c}
\vspace*{1em}
\begin{tabular}{||c|c|c|c|c|c|c|c|c||}
\hline\hline
$\widehat{\alpha}$ $(D=6)$ & $\ell_4$ & $N$ & $\Omega$ \cite{Iyer1989CQG} (WKB 3rd order)& $\Omega$ \cite{Konoplya2005PRDa} (WKB 6th order) & $\Omega$ (SM) \\ [0.5ex]
\hline\hline
$0.0$              & $5$      & $0$ & \textcolor{red}{$2.950-0.3941i$}            & N/A                  & \textcolor{red}{$2.9503-0.3940i$} \\
                   &          & $1$ & $2.851-1.194i$             & N/A                  & $2.8529-1.1933i$ \\
                   &          & $2$ & $2.666-2.024i$             & N/A                  & $2.6583-2.0295i$ \\
                   &          & $3$ & $2.413-2.889i$             & N/A                  & $2.3735-2.9376i$ \\
                   &          & $4$ & $2.104-3.785i$             & N/A                  & $2.0294-3.9600i$ \\
                   &          & $5$ & $1.742-4.707i$             & N/A                  & $1.6942-5.1143i$ \\
$1.0$              & $5$      & $0$ & \textcolor{red}{$3.111-0.3208i$}            & N/A                  & \textcolor{red}{$3.1124-0.3209i$} \\
                   &          & $1$ & $3.062-0.9690i$            & N/A                  & $3.0655-0.9684i$ \\
                   &          & $2$ & $2.971-1.634i$             & N/A                  & $2.9742-1.6334i$ \\
                   &          & $3$ & $2.848-2.318i$             & N/A                  & $2.8441-2.3293i$ \\
                   &          & $4$ & $2.701-3.020i$             & N/A                  & $2.6854-3.0699i$ \\
                   &          & $5$ & $2.534-3.735i$             & N/A                  & $2.5139-3.8675i$ \\
                   &          & $6$ & N/A                        & N/A                  & $2.3530-4.7298i$ \\
                   &          & $7$ & N/A                        & N/A                  & $2.2377-5.6488i$ \\
                   &          & $8$ & N/A                        & N/A                  & $2.1860-6.5816i$ \\
                   &          & $9$ & N/A                        & N/A                  & ${\bf{2.1732-7.4994i}}$ \\
                   &          & $10$ & N/A                       & N/A                  & $2.1794-8.4028i$ \\
                   &          & $11$ & N/A                       & N/A                  & $2.1969-9.2963i$ \\
                   &          & $12$ & N/A                       & N/A                 & $2.2222-10.1824i$ \\
                   &          & $13$ & N/A                       & N/A                 & $2.2535-11.0629i$ \\
$1.9$              & $5$      & $0$ & \textcolor{red}{$3.328-0.2614i$}            & N/A                  & \textcolor{red}{$3.3268-0.2608i$} \\
                   &          & $1$ & $3.331-0.7920i$            & N/A                  & $3.3214-0.7888i$ \\
                   &          & $2$ & $3.346-1.342i$             & N/A                  & $3.3100-1.3350i$ \\
                   &          & $3$ & $3.384-1.917i$             & N/A                  & $3.2935-1.9084i$ \\
                   &          & $4$ & $3.455-2.512i$             & N/A                  & $3.2756-2.5152i$ \\
                   &          & $5$ & $3.562-3.123i$             & N/A                  & ${\bf{3.2638-3.1582i}}$ \\
                   &          & $6$ & N/A                        & N/A                  & $3.2678-3.8352i$ \\
                   &          & $7$ & N/A                        & N/A                  & ${\bf{0.5338-4.2539i}}$ \\
                   &          & $8$ & N/A                        & N/A                  & $3.2963-4.5382i$ \\
                   &          & $9$ & N/A                        & N/A                  & $3.3519-5.2551i$ \\
                   &          & $10$ & N/A                       & N/A                  & $3.4308-5.9762i$ \\
$5.0$              & $5$      & $0$ & \textcolor{red}{$5.014-0.2754i$}            & N/A                  & \textcolor{red}{$5.0145-0.2754i$} \\
                   &          & $1$ & $4.999-0.8270i$            & N/A                  & $4.9989-0.8271i$ \\
                   &          & $2$ & $4.969-1.381i$             & N/A                  & $4.9576-1.3895i$ \\
                   &          & $3$ & $4.926-1.939i$             & N/A                  & ${\bf{3.8876-1.5520i}}$ \\
                   &          & $4$ & $4.870-2.502i$             & N/A                  & $4.8324-1.7000i$ \\
                   &          & $5$ & $4.804-3.070i$             & N/A                  & $5.1533-2.0551i$ \\ 
                   &          & $6$ & N/A                        & N/A                  & $5.4040-2.5254i$ \\
                   &          & $7$ & N/A                        & N/A                  & $5.6707-3.0072i$ \\
                   &          & $8$ & N/A                        & N/A                  & ${\bf{0.5769-3.2918i}}$ \\
                   &          & $9$ & N/A                        & N/A                  & $5.9529-3.5019i$ \\
                   &          & $10$ & N/A                        & N/A                 & $6.2493-4.0060i$ \\[1ex]
 \hline\hline 
 \end{tabular}
\end{table}

\begin{table}%[ht]
\centering
\caption{QNMs for scalar perturbations with $\ell_5=0$ in a seven-dimensional ($n=5$) Schwarzschild black hole with GB correction are presented in the table below for different values of $\widehat{\alpha}$. The corresponding results are obtained through our SM, utilising $300$ polynomials with a precision of $300$ digits. In this context, $\Omega=\omega\sqrt[4]{M}$ while $N$ represents the corresponding overtone. The notation 'N/A' indicates data not available, while 'SM' and 'CI' stand for Spectral Method and Characteristic Integration, respectively.}
\label{table:3a}
\vspace*{1em}
\begin{tabular}{||c|c|c|c|c|c|c|c|c|c|c|c||}
\hline\hline
$\widehat{\alpha}$ $(D=7)$ & $\ell_5$ & $N$ & $\Omega$ \cite{Iyer1989CQG} (WKB 3rd order)& $\Omega$ \cite{Konoplya2005PRDa,Abdalla2005PRD} (WKB 6th order) & $\Omega$ \cite{Abdalla2005PRD} (CI) & $\Omega$ (SM) \\ [0.5ex]
\hline\hline
$0.0$   & $0$ & $0$ & N/A  & N/A                                 & N/A                            & \textcolor{red}{$1.0684-0.5599i$} \\
        &     & $1$ & N/A  & N/A                                 & N/A                            & $0.5747-2.0508i$ \\
        &     & $2$ & N/A  & N/A                                 & N/A                            & $0.4240-3.8992i$ \\
        &     & $3$ & N/A  & N/A                                 & N/A                            & $0.3836-5.6562i$ \\
        &     & $4$ & N/A  & N/A                                 & N/A                            & $0.3636-7.3809i$ \\
$0.1$   & $0$ & $0$ & N/A  & \textcolor{red}{$1.11738-0.546056i$}& \textcolor{red}{$1.092-0.523i$}& \textcolor{red}{$1.0730-0.5462i$} \\
        &     & $1$ & N/A  & N/A                                 & N/A                            & $0.5999-1.9685i$ \\
        &     & $2$ & N/A  & N/A                                 & N/A                            & $0.4224-3.7640i$ \\
        &     & $3$ & N/A  & N/A                                 & N/A                            & $0.3675-5.4807i$ \\
        &     & $4$ & N/A  & N/A                                 & N/A                            & $0.3394-7.1650i$ \\
        &     & $5$ & N/A  & N/A                                 & N/A                            & $0.3223-8.8302i$ \\
$0.2$   & $0$ & $0$ & N/A  & \textcolor{red}{$1.13699-0.529543i$}& \textcolor{red}{$1.092-0.520i$}& \textcolor{red}{$1.0768-0.5336i$} \\
        &     & $1$ & N/A  & N/A                                 & N/A                            & $0.6255-1.8980i$ \\
        &     & $2$ & N/A  & N/A                                 & N/A                            & $0.4355-3.6418i$ \\
        &     & $3$ & N/A  & N/A                                 & N/A                            & $0.3828-5.3032i$ \\
        &     & $4$ & N/A  & N/A                                 & N/A                            & $0.3520-6.9135i$ \\
        &     & $5$ & N/A  & N/A                                 & N/A                            & $0.3171-8.4892i$ \\
$0.5$   & $0$ & $0$ & N/A  & \textcolor{red}{$1.29469-0.395111i$}& \textcolor{red}{$1.092-0.493i$}& \textcolor{red}{$1.0856-0.5022i$} \\
        &     & $1$ & N/A  & N/A                                 & N/A                            & $0.6977-1.7330i$ \\
        &     & $2$ & N/A  & N/A                                 & N/A                            & $0.4938-3.3067i$ \\
        &     & $3$ & N/A  & N/A                                 & N/A                            & $0.4058-4.7757i$ \\
        &     & $4$ & N/A  & N/A                                 & N/A                            & $0.2849-6.1697i$ \\
        &     & $5$ & N/A  & N/A                                 & N/A                            & N/A \\
$5.0$   & $0$ & $0$ & N/A  & \textcolor{red}{$1.39823-0.574472i$}& \textcolor{red}{$1.275-0.351i$}& \textcolor{red}{$1.2749-0.3511i$} \\
        &     & $1$ & N/A  & N/A                                 & N/A                            & $1.3285-1.1481i$ \\
        &     & $2$ & N/A  & N/A                                 & N/A                            & $1.5450-2.0653i$ \\
        &     & $3$ & N/A  & N/A                                 & N/A                            & $1.8872-2.9830i$ \\
        &     & $4$ & N/A  & N/A                                 & N/A                            & $2.2828-3.8810i$ \\
        &     & $5$ & N/A  & N/A                                 & N/A                            & $2.7052-4.7657i$ \\
$10.0$  & $0$ & $0$ & N/A  & \textcolor{red}{$1.48053-0.385616i$}& \textcolor{red}{$1.515-0.358i$}& \textcolor{red}{$1.5166-0.3794i$} \\
        &     & $1$ & N/A  & N/A                                 & N/A                            & $1.5915-1.0871i$ \\
        &     & $2$ & N/A  & N/A                                 & N/A                            & $1.9504-1.8547i$ \\
        &     & $3$ & N/A  & N/A                                 & N/A                            & $2.4274-2.6536i$ \\
        &     & $4$ & N/A  & N/A                                 & N/A                            & $2.9544-3.4373i$ \\
        &     & $5$ & N/A  & N/A                                 & N/A                            & $3.5085-4.2078i$ \\
$20.0$  & $0$ & $0$ & N/A  & \textcolor{red}{$2.00368-0.56458i$} &\textcolor{red}{$2.001-0.567i$} & \textcolor{red}{$1.9997-0.5705i$} \\
        &     & $1$ & N/A  & N/A                                 & N/A                            & ${\bf{1.5834-1.1060i}}$ \\
        &     & $2$ & N/A  & N/A                                 & N/A                            & $2.2623-1.6901i$ \\
        &     & $3$ & N/A  & N/A                                 & N/A                            & $2.7932-2.3953i$ \\
        &     & $4$ & N/A  & N/A                                 & N/A                            & $3.3940-3.0918i$ \\
        &     & $5$ & N/A  & N/A                                 & N/A                            & $4.0217-3.7750i$ \\ [1ex]
 \hline\hline 
 \end{tabular}
\end{table}

\begin{table}%[ht]
\centering
\caption{QNMs for scalar perturbations with $\ell_5=1$ in a seven-dimensional ($n=5$) Schwarzschild black hole with GB correction are presented in the table below for different values of $\widehat{\alpha}$. The corresponding results are obtained through our SM, utilising $300$ polynomials with a precision of $300$ digits. In this context, $\Omega=\omega\sqrt[4]{M}$, while $N$ represents the corresponding overtone. The notation 'N/A' indicates data not available, while 'SM' and 'CI' stand for Spectral Method and Characteristic Integration, respectively.}
\label{table:3b}
\vspace*{1em}
\begin{tabular}{||c|c|c|c|c|c||c|c|c|c|c||}
\hline\hline
$\widehat{\alpha}$ $(D=7)$ & $\ell_5$ & $N$ & $\Omega$ \cite{Iyer1989CQG} (WKB 3rd order)& $\Omega$ \cite{Konoplya2005PRDa,Abdalla2005PRD} (WKB 6th order) & $\Omega$ \cite{Abdalla2005PRD} (CI) & $\Omega$ (SM) \\ [0.5ex]
\hline\hline
$0.0$   & $1$ & $0$ & N/A   & N/A                                    & N/A                            & \textcolor{red}{$1.5821-0.5391i$} \\
        &     & $1$ & N/A   & N/A                                    & N/A                            & $1.1711-1.7371i$ \\
        &     & $2$ & N/A   & N/A                                    & N/A                            & $0.6422-3.5121i$ \\
        &     & $3$ & N/A   & N/A                                    & N/A                            & $0.5040-5.3765i$ \\
        &     & $4$ & N/A   & N/A                                    & N/A                            & $0.4504-7.1591i$ \\
$0.1$   & $1$ & $0$ & N/A   & \textcolor{red}{$1.54573-0.577608i$}   & \textcolor{red}{$1.587-0.527i$} & \textcolor{red}{$1.5874-0.5274i$} \\
        &     & $1$ & N/A   & N/A                                    & N/A                            & $1.2049-1.6905i$ \\
        &     & $2$ & N/A   & N/A                                    & N/A                            & $0.6738-3.3715i$ \\
        &     & $3$ & N/A   & N/A                                    & N/A                            & $0.5004-5.1925i$ \\
        &     & $4$ & N/A   & N/A                                    & N/A                            & $0.4298-6.9446i$ \\
        &     & $5$ & N/A   & N/A                                    & N/A                            & $0.3953-8.6580i$ \\
$0.2$   & $1$ & $0$ & N/A   & \textcolor{red}{$1.53194-0.58701i$}    & \textcolor{red}{$1.530-0.517i$}&\textcolor{red}{$1.5927-0.5167i$} \\
        &     & $1$ & N/A   & N/A                                    & N/A                            & $1.2349-1.6493i$ \\
        &     & $2$ & N/A   & N/A                                    & N/A                            & $0.7105-3.2529i$ \\
        &     & $3$ & N/A   & N/A                                    & N/A                            & $0.5280-5.0340i$ \\
        &     & $4$ & N/A   & N/A                                    & N/A                            & $0.4716-6.7297i$ \\
        &     & $5$ & N/A   & N/A                                    & N/A                            & $0.4416-8.3606i$ \\
$0.5$   & $1$ & $0$ & N/A   & \textcolor{red}{$1.56277-0.554551i$}   & \textcolor{red}{$1.609-0.489i$}&\textcolor{red}{$1.6090-0.8491i$} \\
        &     & $1$ & N/A   & N/A                                    & N/A                            & $1.3123-1.5476i$ \\
        &     & $2$ & N/A   & N/A                                    & N/A                            & $0.8486-2.9695i$ \\
        &     & $3$ & N/A   & N/A                                    & N/A                            & $0.6786-4.5753i$ \\
        &     & $4$ & N/A   & N/A                                    & N/A                            & $0.6113-6.0721i$ \\
        &     & $5$ & N/A   & N/A                                    & N/A                            & $0.5453-7.5217i$ \\
$5.0$   & $1$ & $0$ & N/A   & \textcolor{red}{$2.01379-0.308316i$}   & \textcolor{red}{$1.982-0.337i$}&\textcolor{red}{$1.9807-0.3372i$} \\
        &     & $1$ & N/A   & N/A                                    & N/A                            & $2.0265-1.0471i$ \\
        &     & $2$ & N/A   & N/A                                    & N/A                            & $2.1739-1.8569i$ \\
        &     & $3$ & N/A   & N/A                                    & N/A                            & $2.4414-2.7364i$ \\
        &     & $4$ & N/A   & N/A                                    & N/A                            & $2.7908-3.6244i$ \\
        &     & $5$ & N/A   & N/A                                    & N/A                            & $3.1845-4.5035i$ \\
$10.0$  & $1$ & $0$ & N/A   & \textcolor{red}{$2.47824-0.423737i$}   & \textcolor{red}{$2.475-0.423i$}&\textcolor{red}{$2.4759-0.4217i$} \\
        &     & $1$ & N/A   & N/A                                    & N/A                            & ${\bf{2.3150-1.1177i}}$ \\
        &     & $2$ & N/A   & N/A                                    & N/A                            & $2.6105-1.6246i$ \\
        &     & $3$ & N/A   & N/A                                    & N/A                            & $3.0299-2.3867i$ \\
        &     & $4$ & N/A   & N/A                                    & N/A                            & $3.5005-3.1519i$ \\
        &     & $5$ & N/A   & N/A                                    & N/A                            & $4.0138-3.9138i$ \\
$20.0$  & $1$ & $0$ & N/A   & \textcolor{red}{$3.39094-0.59452i$}    & \textcolor{red}{$3.387-0.597i$}& \textcolor{red}{$3.3908-0.5944i$} \\
        &     & $1$ & N/A   & N/A                                    & N/A                            & ${\bf{2.0065-1.4164i}}$ \\
        &     & $2$ & N/A   & N/A                                    & N/A                            & $3.1392-1.6968i$ \\
        &     & $3$ & N/A   & N/A                                    & N/A                            & $3.4564-2.0625i$ \\
        &     & $4$ & N/A   & N/A                                    & N/A                            & $4.0062-2.7848i$ \\
        &     & $5$ & N/A   & N/A                                    & N/A                            & $4.5833-3.4564i$ \\[1ex]
 \hline\hline 
\end{tabular}
\end{table}

\begin{table}%[ht]
\centering
\caption{Purely imaginary QNMs for scalar perturbations of a seven-dimensional ($n=5$) Schwarzschild black hole with GB correction are presented in the table below for different values of $\widehat{\alpha}$ and $\ell_5\in\{0,1\}$. The corresponding results are obtained through our SM, utilising $300$ polynomials with a precision of $300$ digits. In this context, $\Omega=\omega\sqrt[4]{M}$ and $N$ represent the dimensionless frequency and the corresponding overtone, respectively. The notation 'SM' stands for Spectral Method.}
\label{table:pi67}
\vspace*{1em}
\begin{tabular}{||c|c|c|c|c|c|c|c|c||}
\hline\hline
$\widehat{\alpha}$ $(D=7)$ & $\ell_5$ & $N$ & $\Omega$ (SM)    &$\Delta\Omega=\Omega_{N}-\Omega_{N+1}$ & $\ell_5$ & $N$ & $\Omega$ (SM)    &$\Delta\Omega=\Omega_{N}-\Omega_{N+1}$\\ [0.5ex]
\hline\hline
$0.1$              & $0$      & $0$ & $0.0000-127.5750i$ & $1.7170i$ & $1$ & $0$ & $0.0000-129.2790i$ & $1.7170i$\\
                   &          & $1$ & $0.0000-129.2920i$ & $1.7180i$ &     & $1$ & $0.0000-130.9960i$ & $1.7180i$\\
                   &          & $2$ & $0.0000-131.0100i$ & $1.7170i$ &     & $2$ & $0.0000-132.7140i$ & $1.7170i$\\
                   &          & $3$ & $0.0000-132.7270i$ & $1.7170i$ &     & $3$ & $0.0000-134.4310i$ & $1.7140i$\\
                   &          & $4$ & $0.0000-134.4440i$ & $1.7180i$ &     & $4$ & $0.0000-136.1450i$ & $1.7210i$\\
                   &          & $5$ & $0.0000-136.1620i$ & -         &     & $5$ & $0.0000-137.8660i$ & -        \\
$0.2$              & $0$      & $0$ & $0.0000-107.3580i$ & $1.7240i$ & $1$ & $0$ & $0.0000-105.6200i$ & $1.7240i$\\
                   &          & $1$ & $0.0000-109.0820i$ & $1.7230i$ &     & $1$ & $0.0000-107.3440i$ & $1.7240i$\\
                   &          & $2$ & $0.0000-110.8050i$ & $1.7240i$ &     & $2$ & $0.0000-109.0680i$ & $1.7230i$\\
                   &          & $3$ & $0.0000-112.5290i$ & $1.7240i$ &     & $3$ & $0.0000-110.7910i$ & $1.7240i$\\
                   &          & $4$ & $0.0000-114.2530i$ & $1.7220i$ &     & $4$ & $0.0000-112.5150i$ & $1.7240i$\\
                   &          & $5$ & $0.0000-115.9760i$ & -         &     & $5$ & $0.0000-114.2390i$ & -        \\[1ex]
 \hline\hline 
 \end{tabular}
\end{table}

\begin{table}%[ht]
\centering
\caption{QNMs for scalar perturbations with $\ell_5=2$ in a seven-dimensional ($n=5$) Schwarzschild black hole with GB correction are presented in the table below for different values of $\widehat{\alpha}$. The corresponding results are obtained through our SM, utilising $300$ polynomials with a precision of $300$ digits. In this context, $\Omega=\omega\sqrt[4]{M}$ while $N$ represents the corresponding overtone. The notation 'N/A' indicates data not available, while 'SM' and 'CI' stand for Spectral Method and Characteristic Integration, respectively.}
\label{table:3c}
\vspace*{1em}
\begin{tabular}{||c|c|c|c|c|c|c|c|c||}
\hline\hline
$\widehat{\alpha}$ $(D=7)$ & $\ell_5$ & $N$ & $\Omega$ \cite{Iyer1989CQG} (WKB 3rd order)& $\Omega$ \cite{Konoplya2005PRDa} (WKB 6th order) & $\Omega$ (SM) \\ [0.5ex]
\hline\hline
$0.0$              & $2$      & $0$ & N/A                        & N/A                  & \textcolor{red}{$2.0995-0.5313i$} \\
                   &          & $1$ & N/A                        & N/A                  & $1.7972-1.6491i$ \\
                   &          & $2$ & N/A                        & N/A                  & $1.1782-3.0621i$ \\
                   &          & $3$ & N/A                        & N/A                  & $0.7479-4.9593i$ \\
                   &          & $4$ & N/A                        & N/A                  & $0.6020-6.8297i$ \\
$0.1$              & $2$      & $0$ & N/A                        & \textcolor{red}{$2.10314-0.525815i$}   & \textcolor{red}{$2.1063-0.5207i$} \\
                   &       & $1$ & N/A                        & N/A                   & $1.8226-1.6129i$ \\
                   &       & $2$ & N/A                        & N/A                   & $1.2430-2.9625i$ \\
                   &       & $3$ & N/A                        & N/A                   & $0.7830-4.7694i$ \\
                   &       & $4$ & N/A                        & N/A                   & $0.5958-6.6001i$ \\
                   &       & $5$ & N/A                        & N/A                   & $0.5039-8.3820i$ \\
$0.2$              & $2$      & $0$ & N/A                        & \textcolor{red}{$2.10106-0.524304i$}   & \textcolor{red}{$2.1134-0.5107i$} \\
                   &       & $1$ & N/A                        & N/A                   & $1.8464-1.5797i$ \\
                   &       & $2$ & N/A                        & N/A                   & $1.3029-2.8770i$ \\
                   &       & $3$ & N/A                        & N/A                   & $0.8267-4.6091i$ \\
                   &       & $4$ & N/A                        & N/A                   & $0.6339-6.4135i$ \\
                   &       & $5$ & N/A                        & N/A                   & $0.5740-8.1444i$ \\
$0.5$              & $2$      & $0$ & N/A                        & \textcolor{red}{$2.10473-0.51385i$}    & \textcolor{red}{$2.1362-0.4841i$} \\
                   &       & $1$ & N/A                        & N/A                   & $1.9134-1.4926i$ \\
                   &       & $2$ & N/A                        & N/A                   & $1.4683-2.6750i$ \\
                   &       & $3$ & N/A                        & N/A                   & $1.0309-4.2217i$ \\
                   &       & $4$ & N/A                        & N/A                   & $0.8885-5.8523i$ \\
                   &       & $5$ & N/A                        & N/A                   & $0.8446-7.3770i$ \\
$5.0$              & $2$      & $0$ & N/A                        & \textcolor{red}{$2.66996-0.330935i$}   & \textcolor{red}{$2.6715-0.3369i$} \\
                   &       & $1$ & N/A                        & N/A                   & $2.6955-1.0136i$ \\
                   &       & $2$ & N/A                        & N/A                   & $2.8053-1.7428i$ \\
                   &       & $3$ & N/A                        & N/A                   & $3.0143-2.5614i$ \\
                   &       & $4$ & N/A                        & N/A                   & $3.3070-3.4279i$ \\
                   &       & $5$ & N/A                        & N/A                   & $3.6587-4.3037i$ \\
$10.0$             & $2$      & $0$ & N/A                        & \textcolor{red}{$3.41504-0.433789i$}   & \textcolor{red}{$3.4149-0.4342i$} \\
                   &       & $1$ & N/A                        & N/A                   & $3.3681-1.3620i$ \\
                   &       & $2$ & N/A                        & N/A                   & ${\bf{2.9256-1.4641i}}$ \\
                   &       & $3$ & N/A                        & N/A                   & $3.6652-2.2046i$ \\
                   &       & $4$ & N/A                        & N/A                   & $4.0785-2.9396i$ \\
                   &       & $5$ & N/A                        & N/A                   & $4.5483-3.6886i$ \\
$20.0$             & $2$      & $0$ & N/A                        & \textcolor{red}{$4.69346-0.603341i$}   & \textcolor{red}{$4.6935-0.6033i$} \\
                   &       & $1$ & N/A                        & N/A                   & ${\bf{2.4192-1.6984i}}$ \\
                   &       & $2$ & N/A                        & N/A                   & $4.6048-1.8165i$ \\
                   &       & $3$ & N/A                        & N/A                   & ${\bf{3.6314-2.1340i}}$ \\
                   &       & $4$ & N/A                        & N/A                   & $4.6494-2.5982i$ \\
                   &       & $5$ & N/A                        & N/A                   & $5.2014-3.2038i$ \\ [1ex]
 \hline\hline 
\end{tabular}
\end{table}

\begin{table}%[ht]
\centering
\caption{Purely imaginary QNMs for scalar perturbations of a seven-dimensional ($n=5$) Schwarzschild black hole with GB correction are presented in the table below for different values of $\widehat{\alpha}$ and $\ell_5=2$. The corresponding results are obtained through our SM, utilising $300$ polynomials with a precision of $300$ digits. In this context, $\Omega=\omega\sqrt[4]{M}$ and $N$ represent the dimensionless frequency and the corresponding overtone, respectively. The notation 'SM' stands for Spectral Method.}
\label{table:pi78}
\vspace*{1em}
\begin{tabular}{||c|c|c|c|c|c|c|c|c||}
\hline\hline
$\widehat{\alpha}$ $(D=7)$ & $\ell_5$ & $N$ & $\Omega$ (SM)    &$\Delta\Omega=\Omega_{N}-\Omega_{N+1}$\\ [0.5ex]
\hline\hline
$0.1$              & $2$      & $0$ & $0.0000-130.9780i$ & $1.7170i$ \\
                   &          & $1$ & $0.0000-132.6950i$ & $1.7180i$ \\
                   &          & $2$ & $0.0000-134.4130i$ & $1.7180i$ \\
                   &          & $3$ & $0.0000-136.1310i$ & $1.7170i$ \\
                   &          & $4$ & $0.0000-137.8480i$ & $1.7180i$ \\
                   &          & $5$ & $0.0000-139.5660i$ & -         \\
$0.2$              & $2$      & $0$ & $0.0000-105.5990i$ & $1.8240i$ \\
                   &          & $1$ & $0.0000-107.3230i$ & $1.7250i$ \\
                   &          & $2$ & $0.0000-109.0480i$ & $1.7240i$ \\
                   &          & $3$ & $0.0000-110.7720i$ & $1.7240i$ \\
                   &          & $4$ & $0.0000-112.4960i$ & $1.7240i$ \\
                   &          & $5$ & $0.0000-114.2200i$ & -         \\[1ex]
 \hline\hline 
 \end{tabular}
\end{table}

\begin{table}%[ht]
\centering
\caption{QNMs for scalar perturbations in an eight-dimensional ($n = 6$) Schwarzschild black hole with GB correction are presented in the table below for different values of $\widehat{\alpha}$ and $\ell_6=0$. The corresponding results are obtained through our SM, utilising $300$ polynomials with a precision of $300$ digits. In this context, $\Omega$ is given by \eqref{ODE00} while $N$ represents the corresponding overtone. The notation 'N/A' indicates data not available, while 'SM' and 'CI' stand for Spectral Method and Characteristic Integration, respectively.}
\label{table:4a}
\vspace*{1em}
\begin{tabular}{||c|c|c|c|c|c|c|c|c|c|c||}
\hline\hline
$\widehat{\alpha}$ $(D=8)$ & $\ell_6$ & $N$ & $\Omega$ \cite{Iyer1989CQG} (WKB 3rd order)& $\Omega$ \cite{Konoplya2005PRDa,Abdalla2005PRD} (WKB 6th order) & $\Omega$ \cite{Abdalla2005PRD} (CI) & $\Omega$ (SM) \\ [0.5ex]
\hline\hline
$0.0$   & $0$ & $0$ & N/A  & N/A                                 & N/A                              & \textcolor{red}{$1.4528-0.6839i$} \\
        &     & $1$ & N/A  & N/A                                 & N/A                              & $0.6564-2.5984i$ \\
        &     & $2$ & N/A  & N/A                                 & N/A                              & $0.5099-5.0473i$ \\
        &     & $3$ & N/A  & N/A                                 & N/A                              & $0.4701-7.3260i$ \\
        &     & $4$ & N/A  & N/A                                 & N/A                              & $0.4503-9.5592i$ \\
$0.1$   & $0$ & $0$ & N/A  & \textcolor{red}{$1.51702-0.694245i$}&\textcolor{red}{$1.461-0.676i$}   & \textcolor{red}{$1.4572-0.6678i$} \\
        &     & $1$ & N/A  & N/A                                 & N/A                              & $0.6914-2.4814i$ \\
        &     & $2$ & N/A  & N/A                                 & N/A                              & $0.5112-4.8750i$ \\
        &     & $3$ & N/A  & N/A                                 & N/A                              & $0.4611-7.1012i$ \\
        &     & $4$ & N/A  & N/A                                 & N/A                              & $0.4390-9.2780i$ \\
        &     & $5$ & N/A  & N/A                                 & N/A                              & $0.4269-11.4270i$ \\
$0.2$   & $0$ & $0$ & N/A  & \textcolor{red}{$1.54463-0.673566i$}&\textcolor{red}{$1.463-0.658i$}   & \textcolor{red}{$1.4608-0.6534i$} \\
        &     & $1$ & N/A  & N/A                                 & N/A                              & $0.7302-2.3843i$ \\
        &     & $2$ & N/A  & N/A                                 & N/A                              & $0.5375-4.7204i$ \\
        &     & $3$ & N/A  & N/A                                 & N/A                              & $0.4972-6.8698i$ \\
        &     & $4$ & N/A  & N/A                                 & N/A                              & $0.4740-8.9472i$ \\
        &     & $5$ & N/A  & N/A                                 & N/A                              & $0.4417-10.9819i$ \\
$0.5$   & $0$ & $0$ & N/A  & \textcolor{red}{$1.7854-0.488086i$} &\textcolor{red}{$1.469-0.616$}    & \textcolor{red}{$1.4697-0.6182i$} \\
        &     & $1$ & N/A  & N/A                                 & N/A                              & $0.8500-2.1707i$ \\
        &     & $2$ & N/A  & N/A                                 & N/A                              & $0.6399-4.3000i$ \\
        &     & $3$ & N/A  & N/A                                 & N/A                              & $0.5731-6.2124i$ \\
        &     & $4$ & N/A  & N/A                                 & N/A                              & $0.4732-8.0467i$ \\
        &     & $5$ & N/A  & N/A                                 & N/A                              & $0.3062-9.8504i$ \\
$5.0$   & $0$ & $0$ & N/A  & \textcolor{red}{$1.47941-0.868859i$}&\textcolor{red}{$1.647-0.420i$}   & \textcolor{red}{$1.6481-0.4582i$} \\
        &     & $1$ & N/A  & N/A                                 & N/A                              & $1.6337-1.5232i$ \\
        &     & $2$ & N/A  & N/A                                 & N/A                              & ${\bf{1.8481-2.8025i}}$ \\
        &     & $3$ & N/A  & N/A                                 & N/A                              & $2.2552-4.0595i$ \\
        &     & $4$ & N/A  & N/A                                 & N/A                              & $2.7311-5.2803i$ \\
        &     & $5$ & N/A  & N/A                                 & N/A                              & $3.2409-6.4815i$ \\
$10.0$  & $0$ & $0$ & N/A  & \textcolor{red}{$1.800-0.410252i$}  & \textcolor{red}{$1.838-0.421i$}  & \textcolor{red}{$1.8420-0.4550i$} \\
        &     & $1$ & N/A  & N/A                                 & N/A                              & $1.9135-1.4231i$ \\
        &     & $2$ & N/A  & N/A                                 & N/A                              & $2.2759-2.5069i$ \\
        &     & $3$ & N/A  & N/A                                 & N/A                              & $2.8235-3.5943i$ \\
        &     & $4$ & N/A  & N/A                                 & N/A                              & $3.4415-4.6551i$ \\
        &     & $5$ & N/A  & N/A                                 & N/A                              & $4.0943-5.6966i$ \\
$20.0$  & $0$ & $0$ & N/A  & \textcolor{red}{$2.15426-0.556954i$}&\textcolor{red}{$2.154-0.544i$}   & \textcolor{red}{$2.1548-0.5466i$} \\
        &     & $1$ & N/A  & N/A                                 & N/A                              & ${\bf{2.0682-1.4079i}}$ \\
        &     & $2$ & N/A  & N/A                                 & N/A                              & $2.5713-2.2460i$ \\
        &     & $3$ & N/A  & N/A                                 & N/A                              & $3.2099-3.2004i$ \\
        &     & $4$ & N/A  & N/A                                 & N/A                              & $3.9110-4.1274i$ \\
        &     & $5$ & N/A  & N/A                                 & N/A                              & $4.6471-5.0360i$\\[1ex]
 \hline\hline 
 \end{tabular}
\end{table}

\begin{table}%[ht]
\centering
\caption{QNMs for scalar perturbations in an eight-dimensional ($n=6$) Schwarzschild black hole with GB correction are presented in the table below for different values of $\widehat{\alpha}$ and $\ell_6=1$. The corresponding results are obtained through our SM, utilising $300$ polynomials with a precision of $300$ digits. In this context, $\Omega$ is given by \eqref{ODE00} while $N$ represents the corresponding overtone. The notation 'N/A' indicates data not available, while 'SM' and 'CI' stand for Spectral Method and Characteristic Integration, respectively.}
\label{table:4b}
\vspace*{1em}
\begin{tabular}{||c|c|c|c|c|c|c|c|c|c|c||}
\hline\hline
$\widehat{\alpha}$ $(D=8)$ & $\ell_6$ & $N$ & $\Omega$ \cite{Iyer1989CQG} (WKB 3rd order)& $\Omega$ \cite{Konoplya2005PRDa,Abdalla2005PRD} (WKB 6th order) & $\Omega$ \cite{Abdalla2005PRD} (CI) & $\Omega$ (SM) \\ [0.5ex]
\hline\hline
$0.0$   & $1$  & $0$ & N/A  & N/A                                  & N/A                            & \textcolor{red}{$2.0203-0.6625i$} \\
        &      & $1$ & N/A  & N/A                                  & N/A                            & $1.3932-2.1077i$ \\
        &      & $2$ & N/A  & N/A                                  & N/A                            & $0.6514-4.6352i$ \\
        &      & $3$ & N/A  & N/A                                  & N/A                            & $0.5531-7.0457i$ \\
        &      & $4$ & N/A  & N/A                                  & N/A                            & $0.5119-9.3408i$ \\
$0.1$   & $1$  & $0$ & N/A  & \textcolor{red}{$1.93407-0.750046i$} &\textcolor{red}{$2.021-0.652i$} & \textcolor{red}{$2.0251-0.6487i$} \\
        &      & $1$ & N/A  & N/A                                  & N/A                            & $1.4447-2.0571i$ \\
        &      & $2$ & N/A  & N/A                                  & N/A                            & $0.6791-4.4516i$ \\
        &      & $3$ & N/A  & N/A                                  & N/A                            & $0.5537-6.8205i$ \\
        &      & $4$ & N/A  & N/A                                  & N/A                            & $0.5080-9.0718i$ \\
        &      & $5$ & N/A  & N/A                                  & N/A                            & $0.4895-11.2685i$ \\
$0.2$   & $1$  & $0$ & N/A  & \textcolor{red}{$1.90391-0.773947i$} & \textcolor{red}{$2.024-0.637i$}& \textcolor{red}{$2.0299-0.6362i$} \\
        &      & $1$ & N/A  & N/A                                  & N/A                            & $1.4885-2.0133i$ \\
        &      & $2$ & N/A  & N/A                                  & N/A                            & $0.7224-4.3027i$ \\
        &      & $3$ & N/A  & N/A                                  & N/A                            & $0.6108-6.6183i$ \\
        &      & $4$ & N/A  & N/A                                  & N/A                            & $0.5845-8.7782i$ \\
        &      & $5$ & N/A  & N/A                                  & N/A                            & $0.5636-10.8625i$ \\
$0.5$   & $1$  & $0$ & N/A  & \textcolor{red}{$1.94463-0.734591i$} &\textcolor{red}{$2.035-0.602i$} & \textcolor{red}{$2.0450-0.6049i$} \\
        &      & $1$ & N/A  & N/A                                  & N/A                            & $1.5954-1.9080i$ \\
        &      & $2$ & N/A  & N/A                                  & N/A                            & ${\bf{0.0496-2.1795i}}$ \\
        &      & $3$ & N/A  & N/A                                  & N/A                            & $0.9115-3.9392i$ \\
        &      & $4$ & N/A  & N/A                                  & N/A                            & $0.8235-6.0223i$ \\
        &      & $5$ & N/A  & N/A                                  & N/A                            & $0.7774-7.9419i$ \\
$5.0$   & $1$  & $0$ & N/A  & \textcolor{red}{$2.44511-0.455166i$} & \textcolor{red}{$2.348-0.434i$}& \textcolor{red}{$2.3503-0.4423i$} \\
        &      & $1$ & N/A  & N/A                                  & N/A                            & $2.3557-1.4053i$ \\
        &      & $2$ & N/A  & N/A                                  & N/A                            & ${\bf{0.2456-2.0932i}}$ \\
        &      & $3$ & N/A  & N/A                                  & N/A                            & $2.4697-2.5622i$ \\
        &      & $4$ & N/A  & N/A                                  & N/A                            & $2.7869-3.8057i$ \\
        &      & $5$ & N/A  & N/A                                  & N/A                            & $3.2197-5.0283i$ \\
$10.0$  & $1$  & $0$ & N/A  & \textcolor{red}{$2.6644-0.463966i$}  & \textcolor{red}{$2.670-0.458i$}& \textcolor{red}{$2.6748-0.4634i$} \\
        &      & $1$ & N/A  & N/A                                  & N/A                            & $2.6720-1.3521i$ \\
        &      & $2$ & N/A  & N/A                                  & N/A                            & ${\bf{0.2716-2.0029i}}$ \\
        &      & $3$ & N/A  & N/A                                  & N/A                            & $2.9233-2.2871i$ \\
        &      & $4$ & N/A  & N/A                                  & N/A                            & $3.3871-3.3304i$ \\
        &      & $5$ & N/A  & N/A                                  & N/A                            & $3.9495-4.3819i$ \\
$20.0$  & $1$  & $0$ & N/A  & \textcolor{red}{$3.2276-0.581024i$}  & \textcolor{red}{$3.209-0.590i$}& \textcolor{red}{$3.2264-0.5819i$} \\ 
        &      & $1$ & N/A  & N/A                                  & N/A                            & $2.7515-1.5914i$ \\
        &      & $2$ & N/A  & N/A                                  & N/A                            & ${\bf{0.2741-1.9110i}}$ \\
        &      & $3$ & N/A  & N/A                                  & N/A                            & $3.2695-1.9614i$ \\
        &      & $4$ & N/A  & N/A                                  & N/A                            & $3.8124-2.9294i$ \\
        &      & $5$ & N/A  & N/A                                  & N/A                            & $4.4499-3.8406i$ \\[1ex]
 \hline\hline 
 \end{tabular}
\end{table}

\begin{table}%[ht]
\centering
\caption{QNMs for scalar perturbations in an eight-dimensional ($n=6$) Schwarzschild black hole with GB correction are presented in the table below for different values of $\widehat{\alpha}$ and $\ell_6=2$. The corresponding results are obtained through our SM, utilising $300$ polynomials with a precision of $300$ digits. In this context, $\Omega$ is given by \eqref{ODE00} while $N$ represents the corresponding overtone. The notation 'N/A' indicates data not available. Moreover, 'SM' stands for Spectral Method.}
\label{table:4c}
\vspace*{1em}
\begin{tabular}{||c|c|c|c|c|c|c|c|c||}
\hline\hline
$\widehat{\alpha}$ $(D=8)$ & $\ell_6$ & $N$ & $\Omega$ \cite{Iyer1989CQG} (WKB 3rd order)& $\Omega$ \cite{Konoplya2005PRDa} (WKB 6th order) & $\Omega$ (SM) \\ [0.5ex]
\hline\hline
$0.0$              & $2$      & $0$ & N/A                        & N/A                  & \textcolor{red}{$2.5896-0.6534i$} \\
                   &          & $1$ & N/A                        & N/A                  & $2.1394-2.0137i$ \\
                   &          & $2$ & N/A                        & N/A                  & $\bf{0.2130-2.6968i}$ \\
                   &          & $3$ & N/A                        & N/A                  & $1.0364-3.9622i$ \\
                   &          & $4$ & N/A                        & N/A                  & $0.6912-6.6227i$ \\
$0.1$              & $2$      & $0$ & N/A                        & \textcolor{red}{$2.58782-0.651764i$}   & \textcolor{red}{$2.5956-0.6408i$} \\
                   &       & $1$ & N/A                        & N/A                   & $2.1726-1.9732i$ \\
                   &       & $2$ & N/A                        & N/A                   & ${\bf{0.2131-2.7109i}}$ \\
                   &       & $3$ & N/A                        & N/A                   & $1.1374-3.7986i$ \\
                   &       & $4$ & N/A                        & N/A                   & $0.7143-6.3853i$ \\
                   &       & $5$ & N/A                        & N/A                   & $0.6009-8.7593i$ \\
$0.2$              & $2$      & $0$ & N/A                        & \textcolor{red}{$2.57296-0.658318i$}   & \textcolor{red}{$2.6019-0.6292i$} \\
                   &       & $1$ & N/A                        & N/A                   & $2.2032-1.9368i$ \\
                   &       & $2$ & N/A                        & N/A                   & ${\bf{0.2090-2.7218i}}$ \\
                   &       & $3$ & N/A                        & N/A                   & $1.2327-3.6685i$ \\
                   &       & $4$ & N/A                        & N/A                   & $0.7707-6.2020i$ \\
                   &       & $5$ & N/A                        & N/A                   & $0.6994-8.5185i$ \\
$0.5$              & $2$      & $0$ & N/A                        & \textcolor{red}{$2.54771-0.665021i$}   & \textcolor{red}{$2.6225-0.5993i$} \\
                   &       & $1$ & N/A                        & N/A                   & $2.2859-1.8445i$ \\
                   &       & $2$ & N/A                        & N/A                   & ${\bf{0.1969-2.7472i}}$ \\
                   &       & $3$ & N/A                        & N/A                   & $1.4994-3.4052i$ \\
                   &       & $4$ & N/A                        & N/A                   & $1.0763-5.7052i$ \\
                   &       & $5$ & N/A                        & N/A                   & $1.0413-7.7553i$ \\
$5.0$              & $2$      & $0$ & N/A                        & \textcolor{red}{$3.06761-0.428616i$}   & \textcolor{red}{$3.0398-0.4355i$} \\
                   &       & $1$ & N/A                        & N/A                   & $3.0495-1.3498i$ \\
                   &       & $2$ & N/A                        & N/A                   & $3.1239-2.4011i$ \\
                   &       & $3$ & N/A                        & N/A                   & ${\bf{0.5656-2.6413i}}$ \\
                   &       & $4$ & N/A                        & N/A                   & $3.3466-3.5921i$ \\
                   &       & $5$ & N/A                        & N/A                   & $3.7131-4.8141i$ \\
$10.0$             & $2$      & $0$ & N/A                        & \textcolor{red}{$3.49737-0.47698i$}    & \textcolor{red}{$3.4987-0.4749i$} \\
                   &       & $1$ & N/A                        & N/A                   & $3.4277-1.3667i$ \\
                   &       & $2$ & N/A                        & N/A                   & $3.5774-2.1577i$ \\
                   &       & $3$ & N/A                        & N/A                   & ${\bf{0.6020-2.4805i}}$ \\
                   &       & $4$ & N/A                        & N/A                   & $3.9841-3.1265i$ \\
                   &       & $5$ & N/A                        & N/A                   & $4.4864-4.1593i$ \\
$20.0$             & $2$      & $0$ & N/A                        & \textcolor{red}{$4.26009-0.589931i$}   & \textcolor{red}{$4.2603-0.5901i$} \\ 
                   &       & $1$ & N/A                        & N/A                   & $4.1883-1.7815i$ \\
                   &       & $2$ & N/A                        & N/A                   & $3.2474-1.9277i$ \\
                   &       & $3$ & N/A                        & N/A                   & ${\bf{0.6022-2.3281i}}$ \\
                   &       & $4$ & N/A                        & N/A                   & $4.4564-2.7367i$ \\
                   &       & $5$ & N/A                        & N/A                   & $5.0328-3.6077i$ \\ [1ex]
 \hline\hline
 \end{tabular}
\end{table}

\begin{table}%[ht]
\centering
\caption{QNMs for scalar perturbations in a ten-dimensional ($n=8$) Schwarzschild black hole with GB correction are presented in the table below for different values of $\widehat{\alpha}$ and $\ell_8\in\{0,1,2\}$. The corresponding results are obtained through our SM, utilising $300$ polynomials with a precision of $300$ digits. In this context, $\Omega$ is given by \eqref{ODE00} while $N$ represents the corresponding overtone. The notation 'N/A' indicates data not available. Moreover, 'SM' stands for Spectral Method.}
\label{table:5D10}
\vspace*{1em}
\begin{tabular}{||c|c|c|c|c|c|c|c|c|c|c|c||}
\hline\hline
$\widehat{\alpha}$ $(D=10)$ & $\ell_8$ & $N$ & $\Omega$ (SM)& $\ell_8$ & $N$ & $\Omega$ (SM) & $\ell_8$ & $N$ & $\Omega$ (SM) \\ [0.5ex]
\hline\hline
$0.0$              & $0$      & $0$ & \textcolor{red}{$2.2639-0.9026i$} & $1$ & $0$ & \textcolor{red}{$2.9081-0.8820i$} & $2$ & $0$ & \textcolor{red}{$3.5513-0.8714i$} \\
                   &          & $1$ & $0.8783-2.0665i$  &     & $1$ & $1.7948-2.5200i$  &    & $1$ & $2.7543-2.5826i$ \\
                   &          & $2$ & $0.7843-3.8188i$  &     & $2$ & $0.9279-3.1682i$  &    & $2$ & $1.3872-3.2355i$ \\
                   &          & $3$ & $0.6787-7.3744i$  &     & $3$ & $0.7412-6.9869i$  &    & $3$ & $0.8077-6.3972i$ \\
                   &          & $4$ & $0.6418-10.6930i$ &     & $4$ & $0.6870-10.4287i$ &    & $4$ & $0.7353-10.0697i$ \\
$0.1$              & $0$      & $0$ & \textcolor{red}{$2.2676-0.8831i$} & $1$      & $0$ & \textcolor{red}{$2.9117-0.8648i$} & $2$      & $0$ & \textcolor{red}{$3.5555-0.8558i$}\\
                   &          & $1$ & $0.9024-2.0801i$ &       & $1$ & $1.8669-2.5125i$ &       & $1$ & $2.8031-2.5471i$\\
                   &          & $2$ & $0.8183-3.6508i$ &          & $2$ & $0.9659-3.0784i$ &          & $2$ & $1.4264-3.2201i$\\
                   &          & $3$ & $0.6928-7.1305i$ &          & $3$ & $0.7742-6.7391i$ &          & $3$ & $0.8780-6.1328i$\\
                   &          & $4$ & $0.6570-10.3673i$ &         & $4$ & $0.7171-10.1138i$ &         & $4$ & $0.7798-9.7621i$\\
                   &          & $5$ & $0.6476-13.5267i$ &         & $5$ & $0.7023-13.3422i$ &         & $5$ & $0.7572-13.0958i$\\
$0.2$              &  $0$        & $0$ & \textcolor{red}{$2.2706-0.8659i$} & $1$      & $0$ & \textcolor{red}{$2.9154-0.8497i$} & $2$      & $0$ & \textcolor{red}{$3.5603-0.8418i$}\\
                   &          & $1$ & $0.9242-2.0927i$ &         & $1$ & $1.9310-2.5000i$ &          & $1$ & $2.8467-2.5151i$\\
                   &          & $2$ & $0.8611-3.5130i$ &         & $2$ & $0.9949-3.0166i$ &          & $2$ & $1.4609-3.2099i$\\
                   &          & $3$ & $0.7506-6.9097i$ &         & $3$ & $0.8516-6.5374i$ &          & $3$ & $0.9686-5.9387i$\\
                   &          & $4$ & $0.7374-10.0251i$ &        & $4$ & ${\bf{0.8329-9.8026i}}$ &   & $4$ & ${\bf{0.9228-9.4938i}}$\\
                   &          & $5$ & $0.7302-13.0368i$ &        & $5$ & $0.8358-12.8842i$ &         & $5$ & $0.9415-12.6817i$\\
$0.5$              & $0$      & $0$ & \textcolor{red}{$2.2787-0.8250i$} & $1$      & $0$ & \textcolor{red}{$2.9277-0.8128i$} & $2$      & $0$ & \textcolor{red}{$3.5768-0.8067i$}\\
                   &          & $1$ & ${\bf{0.9741-2.1315i}}$ &       & $1$ & $2.0891-2.4547i$ &       & $1$ & $2.9595-2.4348i$\\
                   &          & $2$ & $1.0046-3.1910i$ &       & $2$ & ${\bf{1.0652-2.9042i}}$ &       & $2$ & $1.5479-3.1966i$\\
                   &          & $3$ & $0.9386-6.3160i$ &       & $3$ & $1.1457-6.0004i$ &       & $3$ & ${\bf{1.3537-5.4917i}}$\\
                   &          & $4$ & $0.9109-9.1081i$ &       & $4$ & $1.1456-8.9319i$ &       & $4$ & $1.3717-8.6887i$\\
                   &          & $5$ & $0.8456-11.8104i$ &     & $5$ & $1.1269-11.7017i$ &      & $5$ & $1.3886-11.5438i$\\
$5.0$              & $0$        & $0$ & \textcolor{red}{$2.4476-0.6445i$} & $1$      & $0$ & \textcolor{red}{$3.1764-0.6302i$} & $2$      & $0$ & \textcolor{red}{$3.8957-0.6207i$}\\
                   &          & $1$ & ${\bf{0.7811-2.1664}}i$ &   & $1$ & $3.0442-2.0470i$ &       & $1$ & $3.8070-1.9671i$\\
                   &          & $2$ & $2.2458-2.2003i$ &       & $2$ & ${\bf{1.2193-2.6400i}}$ &      & $2$ & $1.6770-3.1037i$\\
                   &          & $3$ & $2.4856-4.2517i$ &       & $3$ & $3.0780-3.9594i$ &       & $3$ & ${\bf{0.3383-3.4386i}}$\\
                   &          & $4$ & $3.0541-6.1700i$ &       & $4$ & $3.5604-5.9098i$ &       & $4$ & $3.7420-3.6950i$\\
                   &          & $5$ & $3.7158-8.0177i$ &       & $5$ & $4.1830-7.7726i$ &       & $5$ & $4.0864-5.6573i$\\
$10.0$             & $0$      & $0$ & \textcolor{red}{$2.6107-0.6174i$} & $1$      & $0$ & \textcolor{red}{$3.4063-0.6099i$} & $2$      & $0$ & \textcolor{red}{$4.1920-0.6078i$}\\
                   &       & $1$ & $2.5929-2.0453i$ &       & $1$ & $3.3913-1.9214i$ &       & $1$ & $4.1701-1.8599i$\\
                   &       & $2$ & ${\bf{0.7453-2.1266i}}$ &   & $2$ & ${\bf{1.1889-2.5787i}}$ &       & $2$ & ${\bf{1.6395-3.0061i}}$\\
                   &       & $3$ & $3.0036-3.7830i$ &       & $3$ & $3.6457-3.5169i$ &       & $3$ & $4.3358-3.3120i$\\
                   &       & $4$ & $3.7308-5.4533i$ &       & $4$ & $4.2675-5.1807i$ &       & $4$ & $4.8398-4.9366i$\\
                   &       & $5$ & $4.5609-7.0623i$ &      & $5$ & $5.0424-6.7954i$ &       & $5$ & $5.5465-6.5553i$\\
$20.0$             & $0$      & $0$ & \textcolor{red}{$2.8473-0.6445i$} & $1$      & $0$ & \textcolor{red}{$3.7470-0.6571i$} & $2$      & $0$ & \textcolor{red}{$4.6397-0.6679i$}\\
                   &       & $1$ & $2.8475-1.9445i$ &       & $1$ & $3.6576-1.8930i$ &       & $1$ & $4.4750-1.9186i$\\
                   &       & $2$ & ${\bf{0.7080-2.0926i}}$ &       & $2$ & ${\bf{1.1452-2.5168i}}$ &       & $2$ & ${\bf{1.5839-2.9108i}}$\\
                   &       & $3$ & $3.3656-3.3794i$ &       & $3$ & $4.0219-3.1379i$ &       & $3$ & $4.6955-2.9668i$\\
                   &       & $4$ & $4.1972-4.8220i$ &      & $4$ & $4.7588-4.5457i$ &       & $4$ & $5.3620-4.3080i$\\
                   &       & $5$ & $5.1297-6.2209i$ &       & $5$ & $5.6279-5.9423i$ &       & $5$ & $6.1625-5.6964i$\\[1ex]
 \hline\hline 
 \end{tabular}
\end{table}

\begin{table}%[ht]
\centering
\caption{QNMs for scalar perturbations in an eleven-dimensional ($n=9$) Schwarzschild black hole with GB correction are presented in the table below for different values of $\widehat{\alpha}$ and $\ell_9\in\{0,1,2\}$. The corresponding results are obtained through our SM, utilising $300$ polynomials with a precision of $300$ digits. In this context, $\Omega$ is given by \eqref{ODE00} while $N$ represents the corresponding overtone. The notation 'N/A' indicates data not available. Moreover, 'SM' stands for Spectral Method.}
\label{table:6D11}
\vspace*{1em}
\begin{tabular}{||c|c|c|c|c|c|c|c|c|c|c|c||}
\hline\hline
$\widehat{\alpha}$ $(D=11)$ & $\ell_9$ & $N$ & $\Omega$ (SM)& $\ell_9$ & $N$ & $\Omega$ (SM) & $\ell_9$ & $N$ & $\Omega$ (SM) \\ [0.5ex]
\hline\hline
$0.0$              & $0$      & $0$ & \textcolor{red}{$2.6840-1.0006i$} & $1$ & $0$ & \textcolor{red}{$3.3567-0.9807i$} & $2$ & $0$ & \textcolor{red}{$4.0275-0.9699i$} \\
                   &          & $1$ & $1.2951-2.2502i$  &     & $1$ & $2.1218-2.6135i$ &     & $1$ & $3.0588-2.7797i$ \\
                   &          & $2$ & $0.8695-4.4357i$  &     & $2$ & $1.1631-3.6541i$ &     & $2$ & $1.8259-3.5297i$ \\
                   &          & $3$ & $0.7638-8.5456i$  &     & $3$ & $0.8088-8.1719i$ &     & $3$ & $\bf{0.2592-4.1602i}$ \\
                   &          & $4$ & $0.7279-12.3835i$ &     & $4$ & $0.7635-12.1267i$ &    & $4$ & $0.8367-7.6450i$ \\
$0.1$              & $0$      & $0$ & \textcolor{red}{$2.6873-0.9797i$} & $1$      & $0$ & \textcolor{red}{$3.3597-0.9623i$} & $2$      & $0$ & \textcolor{red}{$4.0310-0.9531i$}\\
                   &       & $1$ & $1.3211-2.2548i$ &       & $1$ & $2.1728-2.6139i$ &       & $1$ & $3.1101-2.7528i$\\
                   &       & $2$ & $0.9126-4.2441i$ &       & $2$ & $1.2621-3.5479i$ &       & $2$ & $1.8736-3.4987i$\\
                   &       & $3$ & $0.7867-8.2660i$ &       & $3$ & $0.8510-7.8921i$ &       & $3$ & ${\bf{0.3071-4.1500i}}$\\
                   &       & $4$ & $0.7580-12.0060i$ &      & $4$ & $0.8118-11.7625i$ &      & $4$ & $0.9079-7.3555i$\\
                   &       & $5$ & $0.7551-15.6563i$ &      & $5$ & $0.8070-15.4786i$ &      & $5$ & ${\bf{0.8622-11.4407i}}$\\
$0.2$              & $0$   & $0$ & \textcolor{red}{$2.6901-0.9615i$} & $1$      & $0$ & \textcolor{red}{$3.3628-0.9461i$} & $2$      & $0$ & \textcolor{red}{$4.0350-0.9381i$}\\
                   &       & $1$ & $1.3444-2.2589i$ &       & $1$ & $2.2204-2.6128i$ &       & $1$ & $3.1562-2.7282i$\\
                   &       & $2$ & $0.9676-4.0899i$ &       & $2$ & $1.3334-3.4696i$ &       & $2$ & $1.9141-3.4742i$\\
                   &       & $3$ & $0.8615-8.0113i$ &       & $3$ & $0.9493-7.6590i$ &       & $3$ & ${\bf{0.3521-4.1566i}}$\\
                   &       & $4$ & ${\bf{0.8609-11.6076i}}$ &      & $4$ & ${\bf{0.9538-11.3938i}}$ &      & $4$ & $1.0219-7.1431i$\\
                   &       & $5$ & $0.8619-15.0871i$ &      & $5$ & $0.9665-14.9389i$ &      & $5$ & $1.0405-11.1125i$\\
$0.5$              & $0$   & $0$ & \textcolor{red}{$2.6976-0.9184i$} & $1$      & $0$ & \textcolor{red}{$3.3739-0.9070i$} & $2$      & $0$ & \textcolor{red}{$4.0497-0.9010i$}\\
                   &       & $1$ & $1.4004-2.2737i$ &       & $1$ & $2.3502-2.6072i$ &       & $1$ & $3.2764-2.6655i$\\
                   &       & $2$ & $1.1477-3.7419i$ &       & $2$ & $1.4785-3.3080i$ &       & $2$ & $2.0125-3.4235i$\\
                   &       & $3$ & $1.0911-7.3302i$ &       & $3$ & ${\bf{1.2901-7.0301i}}$ &       & $3$ & ${\bf{0.4990-4.1844i}}$\\
                   &       & $4$ & $1.0816-10.5611i$ &      & $4$ & $1.3126-10.3886i$ &      & $4$ & $1.4725-6.5960i$\\
                   &       & $5$ & $1.0324-13.6958i$ &      & $5$ & ${\bf{1.3066-13.5862i}}$ &     & $5$ & $1.5378-10.1617i$\\
$5.0$              & $0$   & $0$ & \textcolor{red}{$2.8631-0.7284i$} & $1$      & $0$ & \textcolor{red}{$3.6065-0.7152i$} & $2$      & $0$ & \textcolor{red}{$4.3414-0.7057i$}\\
                   &       & $1$ & ${\bf{1.2292-2.4393i}}$ &       & $1$ & $3.3801-2.3386i$ &       & $1$ & $4.1807-2.2501i$\\
                   &       & $2$ & $2.5333-2.5135i$ &       & $2$ & ${\bf{1.7068-2.8711i}}$ &       & $2$ & ${\bf{2.2240-3.2955i}}$\\
                   &       & $3$ & $2.8108-4.9765i$ &       & $3$ & $3.3858-4.6729i$ &       & $3$ & $4.0277-4.3699i$\\
                   &       & $4$ & $3.4685-7.2166i$ &       & $4$ & $3.9649-6.9593i$ &       & $4$ & $4.4786-6.7006i$\\
                   &       & $5$ & N/A              &       & $5$ & N/A              &       & $5$ & $5.1478-8.8984i$\\
$10.0$             &  $0$  & $0$ & \textcolor{red}{$3.0194-0.6939i$} & $1$      & $0$ & \textcolor{red}{$3.8165-0.6846i$} & $2$      & $0$ & \textcolor{red}{$4.6042-0.6797i$}\\
                   &       & $1$ & $2.9317-2.3384i$ &       & $1$ & $3.7567-2.1971i$ &       & $1$ & $4.5578-2.1199i$\\
                   &       & $2$ & ${\bf{1.1748-2.3985i}}$ &       & $2$ & ${\bf{1.6678-2.8200i}}$ &       & $2$ & ${\bf{2.1679-3.2240i}}$\\
                   &       & $3$ & $3.3779-4.4207i$ &       & $3$ & $4.0118-4.1352i$ &       & $3$ & $4.7025-3.8939i$\\
                   &       & $4$ & $4.2071-6.3808i$ &       & $4$ & $4.7341-6.1054i$ &       & $4$ & $5.2953-5.8484i$\\
                   &       & $5$ & N/A              &       & $5$ & N/A              &       & $5$ & $6.1185-7.7529i$\\
$20.0$             & $0$    & $0$ & \textcolor{red}{$3.2418-0.7062i$} & $1$      & $0$ & \textcolor{red}{$4.1187-0.7114i$} & $2$      & $0$ & \textcolor{red}{$4.9887-0.7182i$}\\
                   &       & $1$ & $3.2261-2.2106i$ &       & $1$ & $4.0544-2.1211i$ &       & $1$ & $4.8763-2.0967i$\\
                   &       & $2$ & ${\bf{1.1250-2.3707i}}$ &       & $2$ & ${\bf{1.6111-2.7724i}}$ &       & $2$ & ${\bf{2.0965-3.1507i}}$\\
                   &       & $3$ & $3.7824-3.9409i$ &       & $3$ & $4.4431-3.6834i$ &       & $3$ & $5.1398-3.4885i$\\
                   &       & $4$ & $4.7181-5.6367i$ &       & $4$ & $5.2718-5.3571i$ &       & $4$ & $5.8656-5.1072i$\\
                   &       & $5$ & N/A              &       & $5$ & $6.2651-6.9985i$ &       & $5$ & $6.7858-6.7439i$\\[1ex]
 \hline\hline 
 \end{tabular}
\end{table}

\begin{table}%[ht]
\centering
\caption{QNMs for scalar perturbations in a twelve-dimensional ($n=10$) Schwarzschild black hole with GB correction are presented in the table below for different values of $\widehat{\alpha}$ and $\ell_{10}\in\{0,1,2\}$. The corresponding results are obtained through our SM, utilising $300$ polynomials with a precision of $300$ digits. In this context, $\Omega$ is given by \eqref{ODE00} while $N$ represents the corresponding overtone. The notation 'N/A' indicates data not available. Moreover, 'SM' stands for Spectral Method.}
\label{table:8D12}
\vspace*{1em}
\begin{tabular}{||c|c|c|c|c|c|c|c|c|c|c|c||}
\hline\hline
$\widehat{\alpha}$ $(D=12)$ & $\ell_{10}$ & $N$ & $\Omega$ (SM)& $\ell_{10}$ & $N$ & $\Omega$ (SM) & $\ell_{10}$ & $N$ & $\Omega$ (SM) \\ [0.5ex]
\hline\hline
$0.0$        & $0$      & $0$ & \textcolor{red}{$3.1113-1.0924i$} & $1$ & $0$ & \textcolor{red}{$3.8080-1.0735i$} & $2$ & $0$ & \textcolor{red}{$4.5022-1.0627i$} \\
             &          & $1$ & $1.7045-2.4200i$  &     & $1$ & $2.4997-2.7239i$ &     & $1$ & $3.3882-2.9298i$ \\
             &          & $2$ & $0.9494-5.0500i$  &     & $2$ & $1.2535-4.0032i$ &     & $2$ & $2.2130-3.8302i$ \\
             &          & $3$ & $0.8496-9.7195i$  &     & $3$ & $\bf{0.4229-4.0965i}$ &     & $3$ & $\bf{0.6403-4.4577i}$ \\
             &          & $4$ & $0.8143-14.0766i$ &     & $4$ & $0.8833-9.3572i$ &     & $4$ & $0.8946-8.8735i$ \\
$0.1$        & $0$      & $0$ & \textcolor{red}{$3.1142-1.0704i$}  & $1$      & $0$ & \textcolor{red}{$3.8104-1.0540i$}  & $2$      & $0$ & \textcolor{red}{$4.5049-1.0448i$}\\
             &          & $1$ & $1.7302-2.4186i$  &       & $1$ & $2.5394-2.7211i$  &       & $1$ & $3.4361-2.9109i$\\
             &          & $2$ & $1.0000-4.8310i$  &       & $2$ & $1.4392-3.9249i$  &       & $2$ & $2.2736-3.7923i$\\
             &          & $3$ & $0.8822-9.4040i$  &       & $3$ & ${\bf{0.3309-4.0053i}}$  &       & $3$ & ${\bf{0.7029-4.4311i}}$\\
             &          & $4$ & ${\bf{0.8604-13.6467i}}$ &       & $4$ & $0.9364-9.0440i$  &       & $4$ & $0.9748-8.5562i$\\
             &          & $5$ & $0.8638-17.7876i$ &       & $5$ & ${\bf{0.9102-13.4108i}}$ &       & $5$ & $0.9550-13.1105i$\\
$0.2$        & $0$      & $0$ & \textcolor{red}{$3.1166-1.0514i$}  & $1$      & $0$ & \textcolor{red}{$3.8131-1.0369i$}  & $2$      & $0$ & \textcolor{red}{$4.5083-1.0290i$}\\
             &          & $1$ & $1.7528-2.4172i$  &       & $1$ & $2.5759-2.7187i$  &       & $1$ & $3.4799-2.8936i$\\
             &          & $2$ & $1.0670-4.6558i$  &       & $2$ & $1.5597-3.8592i$  &       & $2$ & $2.3240-3.7603i$\\
             &          & $3$ & $0.9744-9.1149i$  &       & $3$ & ${\bf{0.3051-3.9550i}}$  &       & $3$ & ${\bf{0.7562-4.4220i}}$\\
             &          & $4$ & ${\bf{0.9858-13.1919i}}$ &       & $4$ & $1.0555-8.7776i$  &       & $4$ & $1.1156-8.3157i$\\
             &          & $5$ & $0.9951-17.1391i$ &       & $5$ & $1.0773-12.9843i$ &       & $5$ & $1.1636-12.7214i$\\
$0.5$        & $0$      & $0$ & \textcolor{red}{$3.1235-1.0064i$}  & $1$      & $0$ & \textcolor{red}{$3.8230-0.9958i$}  & $2$      & $0$ & \textcolor{red}{$4.5213-0.9900i$}\\
             &          & $1$ & $1.8074-2.4178i$  &       & $1$ & $2.6739-2.7174i$  &       & $1$ & $3.5964-2.8504i$\\
             &          & $2$ & $1.2905-4.2657i$  &       & $2$ & $1.7958-3.7086i$  &       & $2$ & $2.4422-3.6871i$\\
             &          & $3$ & ${\bf{1.2451-8.3464i}}$  &       & $3$ & ${\bf{0.3455-3.8790i}}$  &       & $3$ & ${\bf{0.9034-4.4160i}}$\\
             &          & $4$ & $1.2533-12.0158i$ &       & $4$ & $1.4404-8.0573i$  &       & $4$ & $1.6194-7.6679i$\\
             &          & $5$ & ${\bf{1.2200-15.5829i}}$ &       & $5$ & $1.4816-11.8459i$ &       & $5$ & $1.7063-11.6299i$\\
$5.0$        & $0$      & $0$ & \textcolor{red}{$3.2861-0.8076i$}  & $1$      & $0$ & \textcolor{red}{$4.0430-0.7956i$}  & $2$      & $0$ & \textcolor{red}{$4.7923-0.7864i$}\\
             &          & $1$ & ${\bf{1.7066-2.6888i}}$  &       & $1$ & $3.7038-2.6116i$  &       & $1$ & $4.5454-2.5167i$\\
             &          & $2$ & $2.7957-2.8143i$  &       & $2$ & ${\bf{2.1930-3.0888i}}$  &       & $2$ & ${\bf{2.7533-3.4641i}}$\\
             &          & $3$ & $3.1391-5.7025i$  &       & $3$ & $3.7004-5.3940i$  &       & $3$ & $4.3153-5.0720i$\\
             &          & $4$ & N/A               &       & $4$ & N/A               &       & $4$ & $4.8805-7.7472i$\\
             &          & $5$ & N/A               &       & $5$ & N/A               &       & $5$ & N/A\\
$10.0$       & $0$      & $0$ & \textcolor{red}{$3.4377-0.7670i$}  & $1$      & $0$ & \textcolor{red}{$4.2394-0.7570i$}  & $2$      & $0$ & \textcolor{red}{$5.0326-0.7506i$}\\
             &          & $1$ & $3.2632-2.6217i$  &       & $1$ & $4.1185-2.4634i$  &       & $1$ & $4.9444-2.3742i$\\
             &          & $2$ & ${\bf{1.6231-2.6440i}}$  &       & $2$ & ${\bf{2.1521-3.0374i}}$  &       & $2$ & ${\bf{2.6954-3.4172i}}$\\
             &          & $3$ & $3.7552-5.0606i$  &       & $3$ & $4.3794-4.7610i$  &       & $3$ & $5.0622-4.4891i$\\
             &          & $4$ & N/A               &       & $4$ & $5.2095-7.0341i$  &       & $4$ & $5.7607-6.7664i$\\
             &          & $5$ & N/A               &       & $5$ & N/A               &       & $5$ & N/A\\
$20.0$       & $0$      & $0$ & \textcolor{red}{$3.6507-0.7683i$}  & $1$      & $0$ & \textcolor{red}{$4.5172-0.7690i$}  & $2$      & $0$ & \textcolor{red}{$5.3766-0.7721i$}\\
             &          & $1$ & $3.6008-2.4673i$  &       & $1$ & $4.4481-2.3551i$  &       & $1$ & $5.2822-2.3030i$\\
             &          & $2$ & ${\bf{1.5616-2.6213i}}$  &       & $2$ & ${\bf{2.0880-3.0019i}}$  &       & $2$ & ${\bf{2.6142-3.3643i}}$\\
             &          & $3$ & $4.2022-4.5032i$  &       & $3$ & $4.8629-4.2290i$  &       & $3$ & $5.5689-4.0068i$\\
             &          & $4$ & N/A               &       & $4$ & $5.7948-6.1707i$  &       & $4$ & $6.3811-5.9107i$\\
             &          & $5$ & N/A               &       & $5$ & N/A               &       & $5$ & N/A\\[1ex]
 \hline\hline 
 \end{tabular}
\end{table}

\begin{table}%[ht]
\centering
\caption{QNMs for scalar perturbations in a twenty-six-dimensional ($n=24$) Schwarzschild black hole with GB correction are presented in the table below for different values of $\widehat{\alpha}$ and $\ell_{24}\in\{0,1,2\}$. The corresponding results are obtained through our SM, utilising $300$ polynomials with a precision of $300$ digits. In this context, $\Omega$ is given by \eqref{ODE00} while $N$ represents the corresponding overtone. The notation 'N/A' indicates data not available. Moreover, 'SM' stands for Spectral Method.}
\label{table:7D26}
\vspace*{1em}
\begin{tabular}{||c|c|c|c|c|c|c|c|c|c|c|c||}
\hline\hline
$\widehat{\alpha}$ $(D=26)$ & $\ell_{24}$ & $N$ & $\Omega$ (SM)& $\ell_{24}$ & $N$ & $\Omega$ (SM) & $\ell_{24}$ & $N$ & $\Omega$ (SM) \\ [0.5ex]
\hline\hline
$0.0$        & $0$      & $0$ & \textcolor{red}{$9.4505-2.0207i$} & $1$ & $0$ & \textcolor{red}{$10.2970-2.0145i$} & $2$ & $0$ & \textcolor{red}{$11.1408-2.0097i$} \\
             &          & $1$ & $7.6887-4.1395i$  &     & $1$ & $8.5206-4.2539i$ &     & $1$ & $9.3566-4.3662i$ \\
             &          & $2$ & $6.0041-5.5551i$  &     & $2$ & $6.8362-5.7865i$ &     & $2$ & $7.6840-6.0071i$ \\
             &          & $3$ & $4.3221-6.5392i$  &     & $3$ & $5.1658-6.8965i$ &     & $3$ & $6.0435-7.2221i$ \\
             &          & $4$ & $2.6144-7.1800i$  &     & $4$ & $3.4765-7.6811i$ &     & $4$ & $4.3945-8.1116i$ \\
$0.1$        & $0$      & $0$ & \textcolor{red}{$9.4488-1.9916i$} & $1$      & $0$ & \textcolor{red}{$10.2942-1.9872i$}  & $2$      & $0$ & \textcolor{red}{$11.1375-1.9839i$}\\
             &          & $1$ & $7.7023-4.1158i$ &       & $1$ & $8.5345-4.2311i$   &       & $1$ & $9.3710-4.3439i$\\
             &          & $2$ & $6.0279-5.5419i$ &       & $2$ & $6.8625-5.7739i$   &       & $2$ & $7.7133-5.9942i$\\
             &          & $3$ & $4.3503-6.5407i$ &       & $3$ & $5.2004-6.8997i$   &       & $3$ & $6.0860-7.2241i$\\
             &          & $4$ & $2.6381-7.1978i$ &       & $4$ & $3.5121-7.7061i$   &       & $4$ & $4.4475-8.1350i$\\
             &          & $5$ & $0.8850-7.5290i$ &       & $5$ & $1.7737-8.2096i$   &       & $5$ & $2.7395-8.7668i$\\
$0.2$        & $0$      & $0$ & \textcolor{red}{$9.4475-1.9669i$} & $1$      & $0$ & \textcolor{red}{$10.2925-1.9638i$} & $2$       & $0$ & \textcolor{red}{$11.1355-1.9618i$}\\
             &          & $1$ & $7.7141-4.0958i$ &         & $1$ & $8.5469-4.2119i$ &         & $1$ & $9.3842-4.3252i$\\
             &          & $2$ & $6.0484-5.5310i$ &       & $2$ & $6.8854-5.7637i$ &         & $2$ & $7.7391-5.9840i$\\
             &          & $3$ & $4.3743-6.5424i$ &       & $3$ & $5.2303-6.9035i$ &         & $3$ & $6.1235-7.2271i$\\
             &          & $4$ & $2.6577-7.2132i$ &       & $4$ & $3.5421-7.7292i$ &         & $4$ & $4.4943-8.1575i$\\
             &          & $5$ & $0.8970-7.5723i$ &       & $5$ & $1.7907-8.2509i$ &         & $5$ & $2.7817-8.8179i$\\
$0.5$        & $0$      & $0$ & \textcolor{red}{$9.4471-1.9095i$} & $1$      & $0$ & \textcolor{red}{$10.2921-1.9091i$} & $2$       & $0$ & \textcolor{red}{$11.1354-1.9092i$}\\
             &          & $1$ & $7.7437-4.0512i$ &       & $1$ & $8.5794-4.1692i$ &         & $1$ & $9.4199-4.2838i$\\
             &          & $2$ & $6.0969-5.5095i$ &       & $2$ & $6.9417-5.7450i$ &         & $2$ & $7.8045-5.9662i$\\
             &          & $3$ & $4.4274-6.5519i$ &       & $3$ & $5.2999-6.9218i$ &         & $3$ & $6.2164-7.2464i$\\
             &          & $4$ & $2.7031-7.2712i$ &       & $4$ & $3.5990-7.7975i$ &         & $4$ & $4.6084-8.2365i$\\
             &          & $5$ & N/A              &       & $5$ & N/A              &         & $5$ & N/A\\
$5.0$        & $0$      & $0$ & \textcolor{red}{$9.5794-1.6404i$} & $1$      & $0$ & \textcolor{red}{$10.4403-1.6395i$} & $2$       & $0$ & \textcolor{red}{$11.2986-1.6381i$}\\
             &          & $1$ & $8.0098-3.9559i$ &        & $1$ & $8.8894-4.0894i$ &         & $1$ & $9.7786-4.2169i$\\
             &          & $2$ & $6.3535-5.6621i$ &       & $2$ & $7.2984-6.0078i$ &         & $2$ & $8.4011-6.3109i$\\
             &          & $3$ & N/A              &       & $3$ & N/A               &        & $3$ & N/A\\
             &          & $4$ & N/A              &       & $4$ & N/A                &       & $4$ & N/A\\
             &          & $5$ & N/A              &       & $5$ & N/A                &       & $5$ & N/A\\
$10.0$       & $0$      & $0$ & \textcolor{red}{$9.7109-1.5569i$} & $1$      & $0$ & \textcolor{red}{$10.5843-1.5537i$} & $2$       & $0$ & \textcolor{red}{$11.4545-1.5501i$}\\
             &          & $1$ & $8.1678-4.0167i$ &       & $1$ & $9.0807-4.1714i$ &         & $1$ & $10.0133-4.3198i$\\
             &          & $2$ & N/A              &       & $2$ & N/A               &        & $2$ & $9.3080-6.3600i$\\
             &          & $3$ & N/A              &       & $3$ & N/A               &        & $3$ & N/A\\
             &          & $4$ & N/A              &       & $4$ & N/A               &        & $4$ & N/A\\
             &          & $5$ & N/A              &       & $5$ & N/A               &        & $5$ & N/A\\
$20.0$       & $0$      & $0$ & \textcolor{red}{$9.8930-1.4970i$} & $1$      & $0$ & \textcolor{red}{$10.7832-1.4928i$} & $2$       & $0$ & \textcolor{red}{$11.6701-1.4884i$}\\
             &          & $1$ & $8.3463-4.1700i$ &       & $1$ & $9.3170-4.3799i$ &         & $1$ & $10.3601-4.5774i$\\
             &          & $2$ & N/A              &       & $2$ & N/A &                      & $2$ & $9.8708-5.8164i$\\
             &          & $3$ & N/A              &       & $3$ & N/A &                      & $3$ & N/A\\
             &          & $4$ & N/A              &       & $4$ & N/A &                      & $4$ & N/A\\
             &          & $5$ & N/A              &       & $5$ & N/A &                      & $5$ & N/A\\[1ex]
 \hline\hline 
 \end{tabular}
\end{table}

\section{Vector perturbations}

\begin{figure}[ht!] 
    \includegraphics[width=0.3\textwidth]{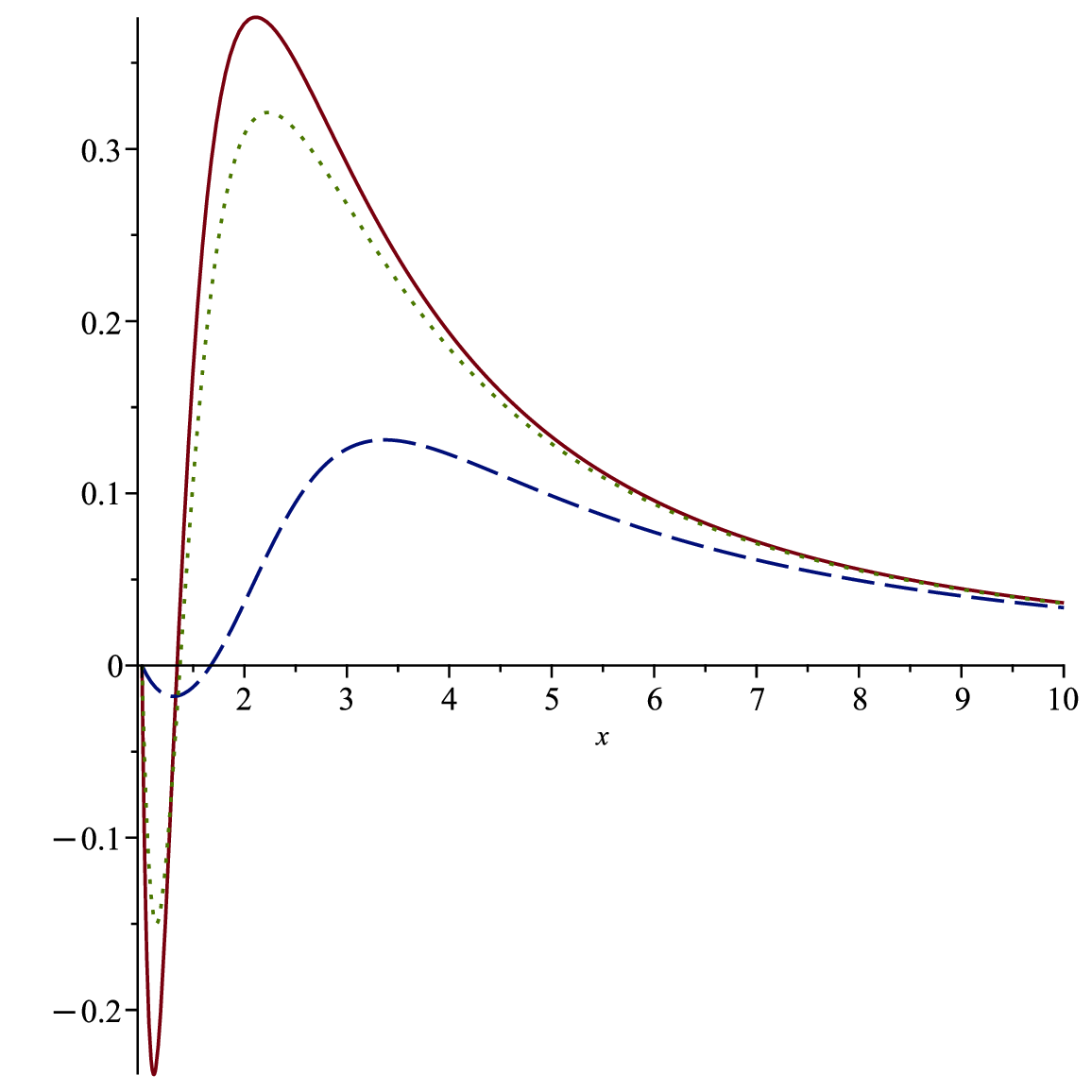}
    \includegraphics[width=0.3\textwidth]{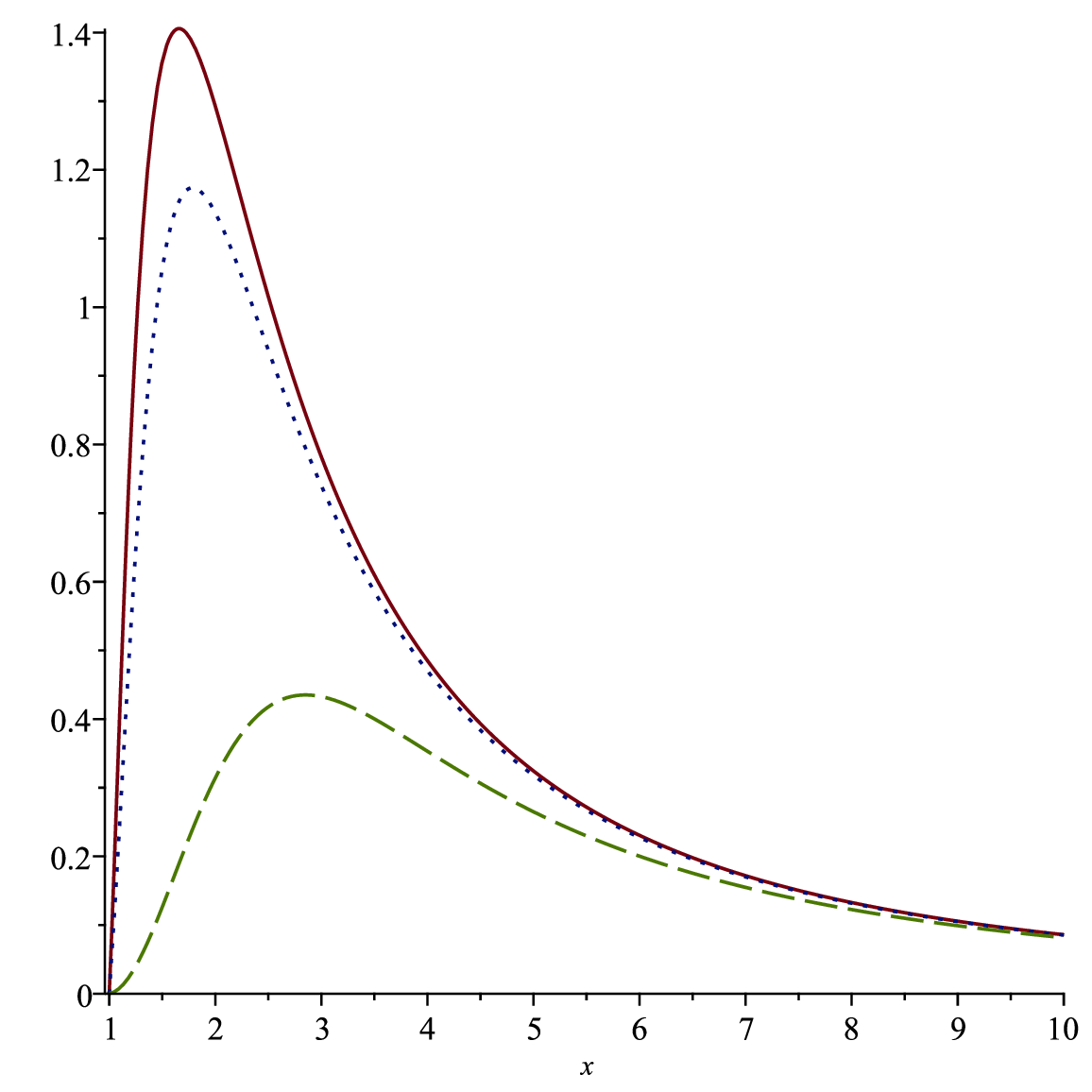}
    \caption{\label{figureUVD5}
Plots of the effective potential $\mathcal{U}_V(x)$ as a function of the rescaled variable $x$ (see equation \eqref{UV}), for different values of $\widehat{\alpha}$ and $\ell_3=1$. {\bf{Left panel}}: case $D=5$ and $\ell_3=1$, with $\widehat{\alpha} = 0.1$ (solid line), $\widehat{\alpha} = 0.5$  (dotted line), and $\widehat{\alpha} = 1.5$ (dashed line). {\bf{Right panel}}: case $D=5$ and $\ell_3=2$, with $\widehat{\alpha} = 0.1$ (solid line), $\widehat{\alpha} = 0.5$  (dotted line), and $\widehat{\alpha} = 1.5$ (dashed line)}
\end{figure}

\begin{figure}[ht!] 
    \includegraphics[width=0.3\textwidth]{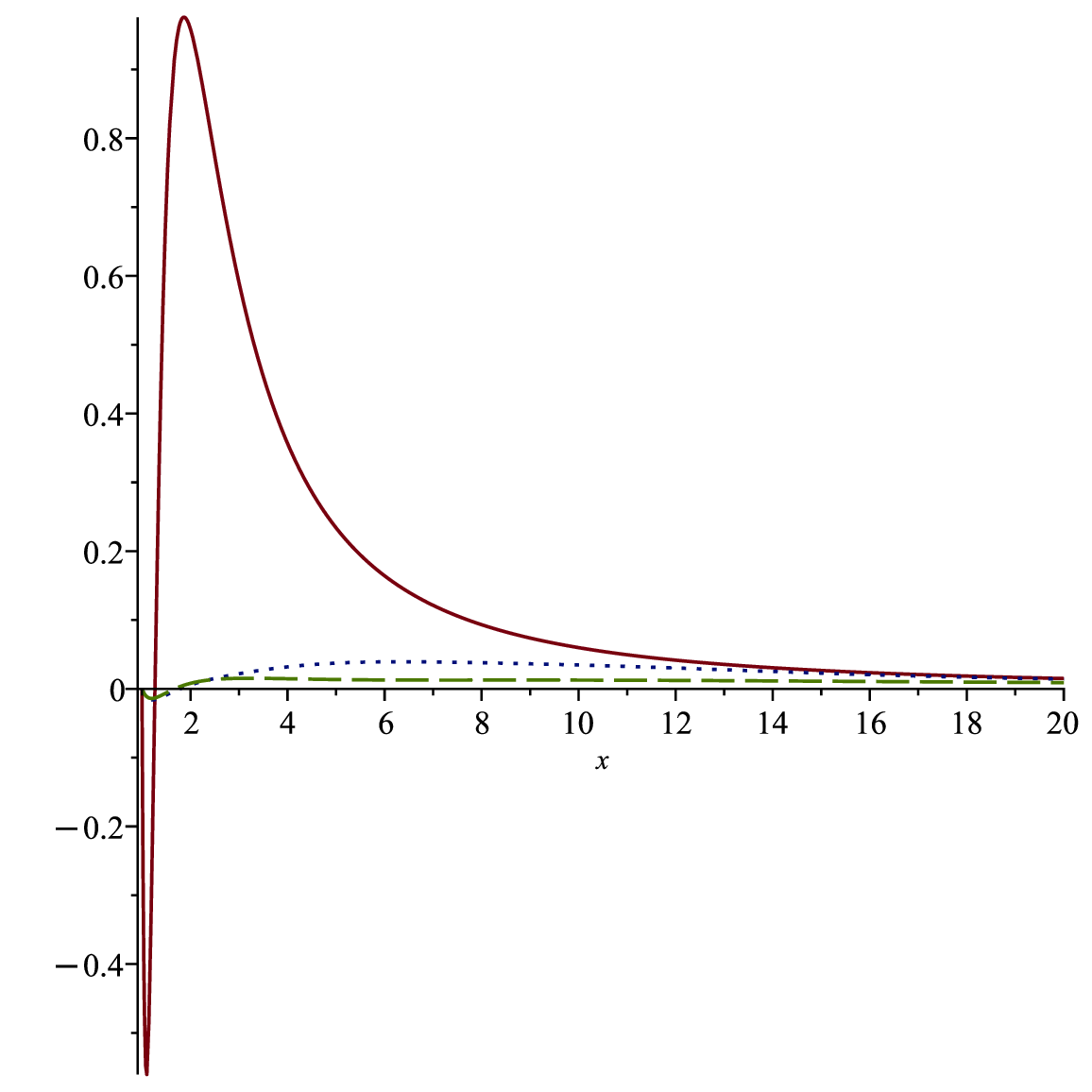}
    \includegraphics[width=0.3\textwidth]{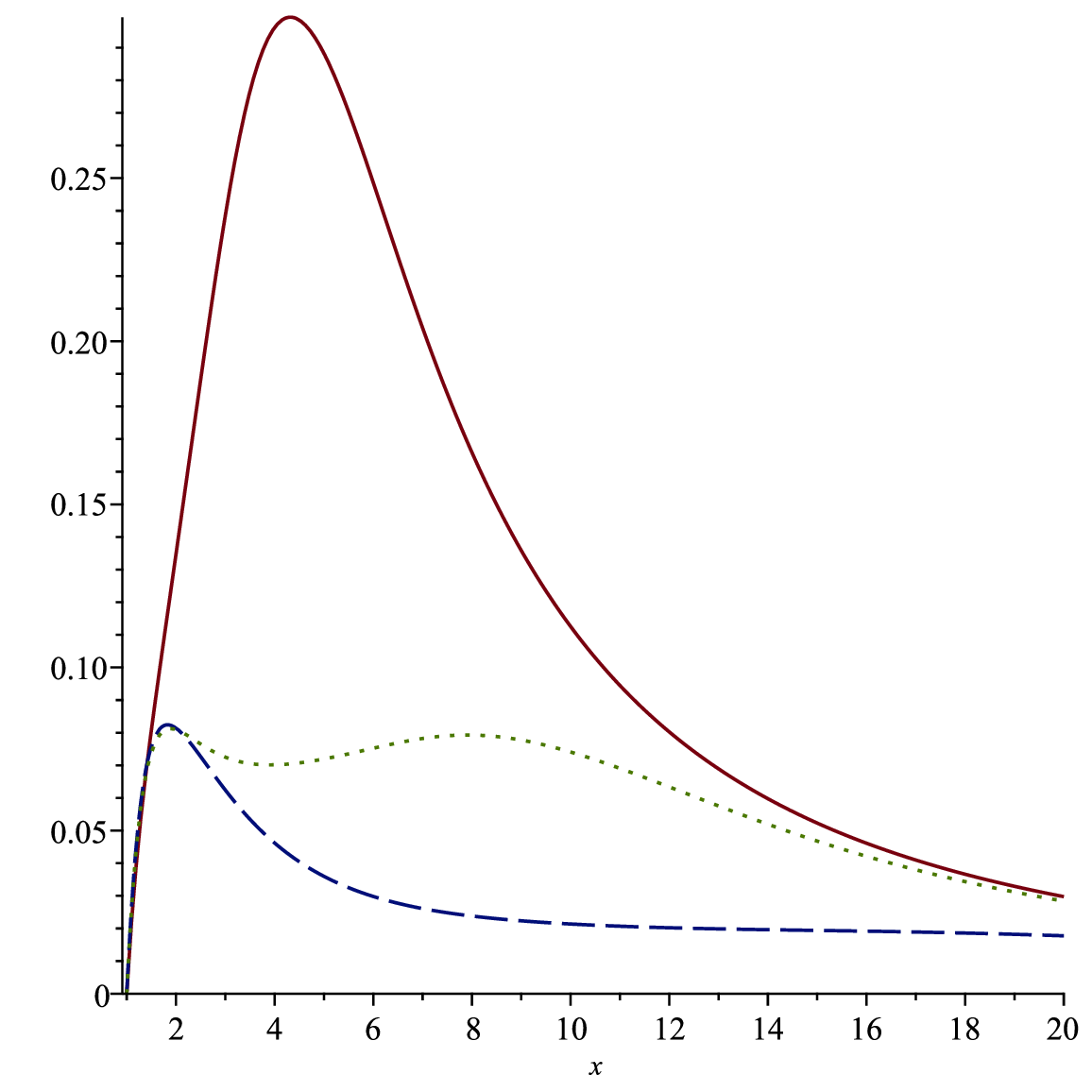}
    \caption{\label{figureUVD6}
Plots of the effective potential $\mathcal{U}_V(x)$ as a function of the rescaled variable $x$ (see equation \eqref{UV}), for different values of $\widehat{\alpha}$ and $\ell_n$. {\bf{Left panel}}: case $D=6$ and $\ell_4=1$, with $\widehat{\alpha} = 0.1$ (solid line), $\widehat{\alpha} = 10$  (dotted line), and $\widehat{\alpha} = 20$ (dashed line). {\bf{Right panel}}: $D=6$ and $\ell_4=2$, with $\widehat{\alpha} = 5$ (solid line), $\widehat{\alpha} = 10$  (dotted line), and $\widehat{\alpha} = 20$ (dashed line).}
\end{figure}

\begin{figure}[ht!] 
    \includegraphics[width=0.3\textwidth]{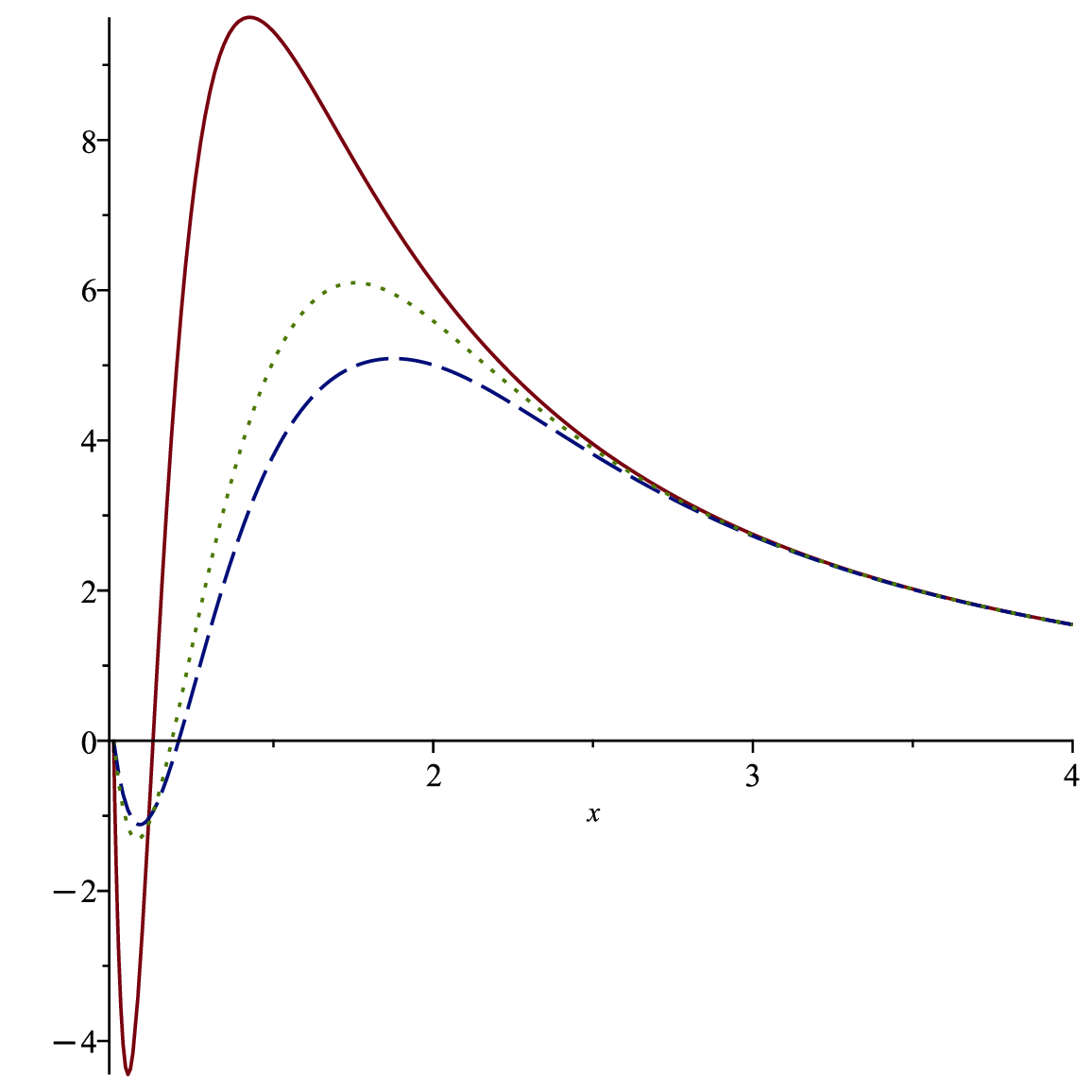}
    \includegraphics[width=0.3\textwidth]{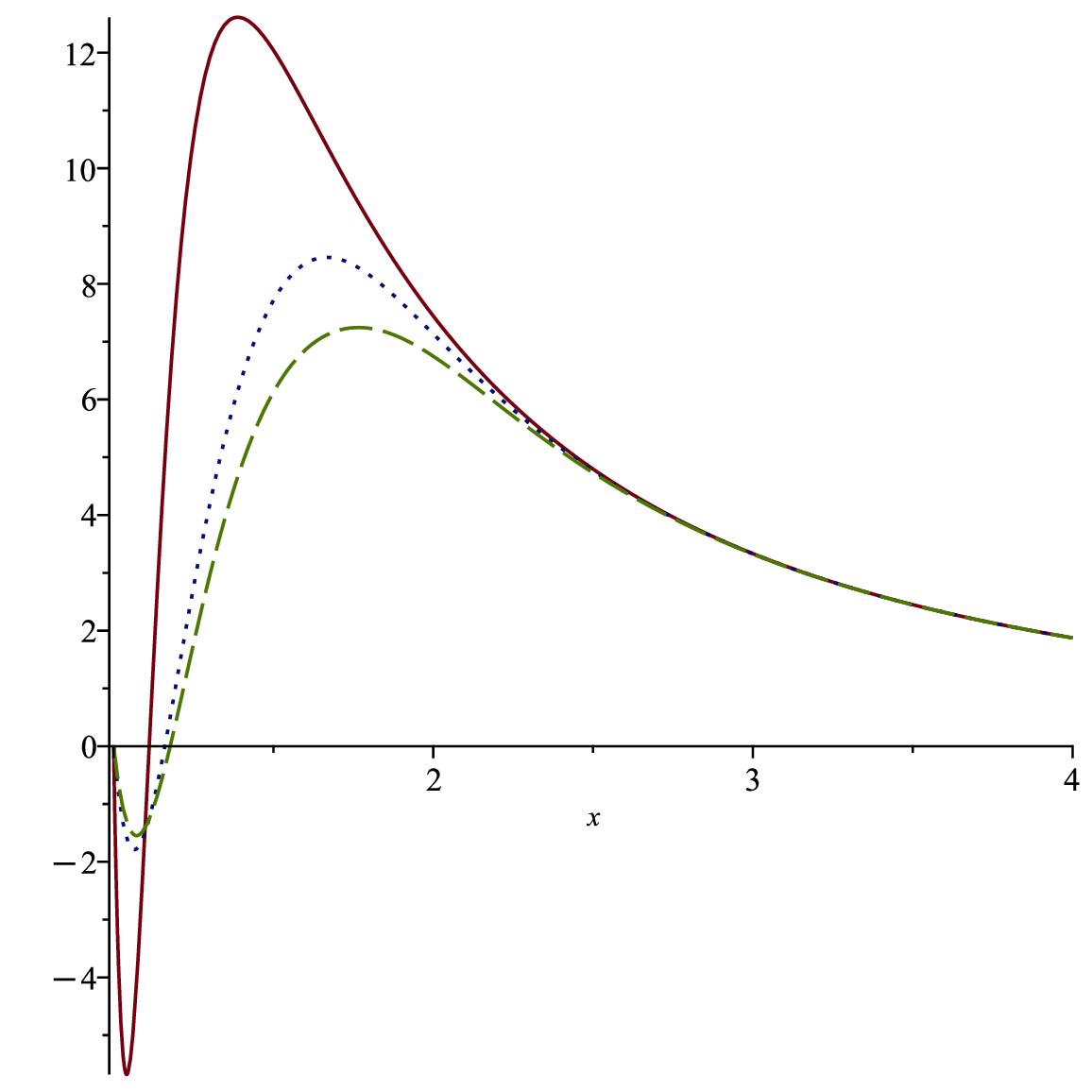}
    \includegraphics[width=0.3\textwidth]{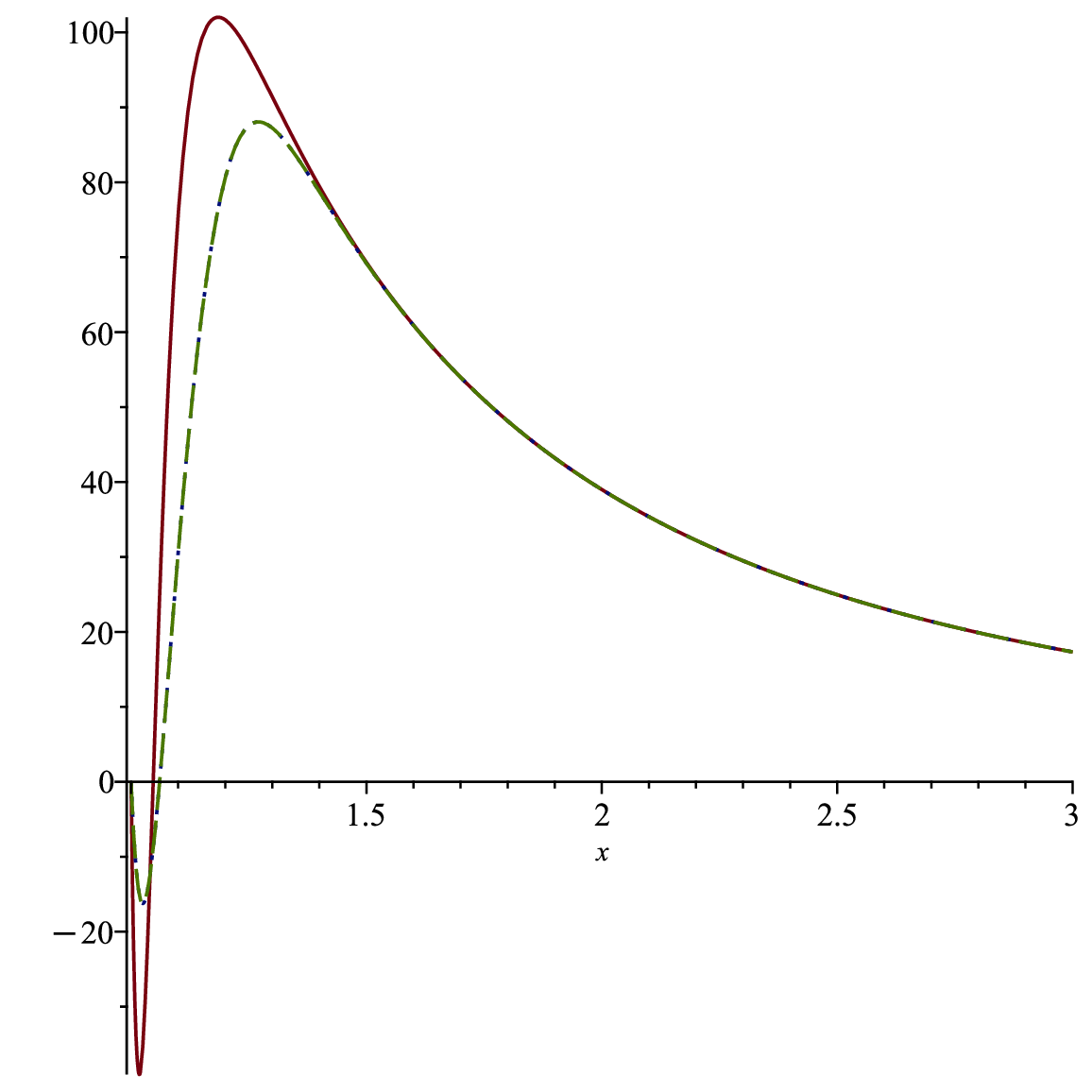}
    \caption{\label{figureUVD11_12_26}
Plots of the effective potential $\mathcal{U}_V(x)$ as a function of the rescaled variable $x$ (see equation \eqref{UV}), for different values of $\widehat{\alpha}$ and $\ell_n = 1$. {\bf{Left panel}}: case $D = 11$, with $\widehat{\alpha} = 0.1$ (solid line), $\widehat{\alpha} = 5$  (dotted line), and $\widehat{\alpha} = 10$ (dashed line). {\bf{Central panel}}: case $D = 12$, with $\widehat{\alpha} = 0.1$ (solid line), $\widehat{\alpha} = 5$  (dotted line), and $\widehat{\alpha} = 10$ (dashed line). {\bf{Right panel}}: $D = 26$, with $\widehat{\alpha} = 0.1$ (solid line), $\widehat{\alpha} = 5$  (dotted line), and $\widehat{\alpha} = 10$ (dashed line).}
\end{figure}

Vector perturbations of the Gauss--Bonnet-corrected Schwarzschild metric were analysed by \cite{Dotti2005CQG}. In geometrized units, and with $D=n+2$, the radial equation reads
\begin{equation}\label{ODEVP}
   f(r)\frac{d}{dr}\left(f(r)\frac{dR_V}{dr}\right)+\left[\omega^2-V_V(r)\right]R_V(r)=0
\end{equation}
with
\begin{eqnarray}
V_V(r)&=&q_V(r)+\frac{f(r)}{K_V(r)}\frac{d}{dr}\left(f(r)\frac{dK_V}{dr}\right),\quad
f(r)=1+\frac{r^2}{2\alpha}\left[1-\sqrt{1+\frac{8M\alpha}{r^{n+1}}}\right],\\
K_V(r)&=&\frac{1}{r^{\frac{n-2}{2}}\sqrt{r^2+\frac{\alpha}{(n-1)(n-2)}\left\{(n-3)[1-f(r)]-rf^{'}(r)\right\}}},\quad n\geqslant 3,\label{n3}\\
q_V(r)&=&\frac{(\ell_n-1)(\ell_n+n)f(r)}{2(n-1)r^2}\cdot\frac{(n-3)(\alpha^2\psi^2+2\alpha\psi)+2(n-1)}{(1+\alpha\psi)^2},
\end{eqnarray}
where the prime denotes differentiation with respect to $r$ and $\psi=\psi(r)$ is a solution of
\begin{equation}\label{psi}
n\psi^2+\frac{2n}{\alpha}\psi=\frac{4M}{\alpha r^{n+1}}.    
\end{equation}
Finally, by means of \eqref{psi}, we can bring $q_V(r)$ in its final form, namely
\begin{equation}
q_V(r)=\frac{(\ell_n-1)(\ell_n+n)f(r)}{(n-1)r^2}\cdot\frac{2\alpha M(n-3)+n(n-1)r^{n+1}}{nr^{n+1}+4\alpha M}.  
\end{equation}
In what follows, we will adopt the notation used in \cite{Konoplya2005PRDa, Abdalla2005PRD, Chakrabarti2007GRG} to facilitate the comparison of the numerical values of the QNMs with those reported in the cited literature. If we make use of the rescalings $\rho=r/M^\frac{1}{n-1}$ and $\widehat{\alpha}=\alpha/M^\frac{2}{n-1}$, we can rewrite \eqref{ODEVP} as
\begin{equation}\label{ODEV}
    f(\rho)\frac{d}{d\rho}\left(f(\rho)\frac{dR_V}{d\rho}\right)+\left[\Omega^2-U_{V}(\rho)\right]R_V(\rho) = 0,\quad \Omega=\omega M^\frac{1}{n-1}
\end{equation}
with
\begin{equation}
U_V(\rho)=q_V(\rho)+\frac{f(\rho)}{K_V(\rho)}\frac{d}{d\rho}\left(f(\rho)\frac{dK_V}{d\rho}\right),
\end{equation}
where
\begin{eqnarray}
K_V(\rho)&=&\frac{1}{\rho^{\frac{n-2}{2}}\sqrt{\rho^2+\frac{\widehat{\alpha}}{(n-1)(n-2)}\left\{(n-3)[1-f(\rho)]-\rho\frac{df}{d\rho}\right\}}},\\
q_V(\rho)&=&\frac{(\ell_n-1)(\ell_n+n)f(\rho)}{(n-1)\rho^2}\cdot
\frac{n(n-1)\rho^{n+1}+2\widehat{\alpha}(n-3)}{n\rho^{n+1}+4\widehat{\alpha}}.
\end{eqnarray}
Note that the explicit expression for $f(\rho)$ and the equation determining the event horizon are given by \eqref{frho} and \eqref{event}, respectively. Similarly to the scalar and tensorial cases, we introduce the rescaling $x = \rho/\rho_h$, by means of which our equation (\ref{ODEV}) becomes
\begin{equation}\label{ODEV1}
  \frac{d^2 R_V}{dx^2}+P_V(x)\frac{dR_V}{dx}+Q_V(x)R_V(x)=0   
\end{equation}
with 
\begin{equation}\label{pqV}
  P_V(x)=\frac{1}{f(x)}\frac{df}{dx},\quad 
  Q_V(x) = \frac{\rho_h^2 \Omega^2-\mathcal{U}_V(x)}{f^2(x)}
\end{equation}
and
\begin{equation}\label{UV}
  \mathcal{U}_V(x)= q_V(x)+\frac{f(x)}{K_V(x)}\frac{d}{dx}\left(f(x)\frac{dK_V}{dx}\right),
\end{equation}
\begin{eqnarray}
K_V(x)&=&\frac{1}{x^{\frac{n-2}{2}}\sqrt{\rho_h^2 x^2+\frac{\widehat{\alpha}}{(n-1)(n-2)}\left\{(n-3)[1-f(x)]-x\frac{df}{dx}\right\}}},\\
q_V(x)&=&\frac{(\ell_n-1)(\ell_n+n)f(x)}{(n-1)x^2}\cdot
\frac{n(n-1)\rho_h^{n+1}x^{n+1}+2\widehat{\alpha}(n-3)}{n\rho_h^{n+1}x^{n+1}+4\widehat{\alpha}}.
\end{eqnarray}
Here, the quantities $\kappa$ and $f(x)$ are defined as in \eqref{US}.  From Figure~\ref{figureUVD5}, we observe that in dimension $D=5$, the effective potential exhibits a markedly different behaviour compared to the corresponding scalar sector discussed above. Specifically, for $\ell_3 = 1$, a negative minimum emerges near the black hole event horizon for all tested values of the parameter $\widehat{\alpha}$. This scenario closely resembles the tensor sector in six dimensions discussed below, suggesting that similar instabilities, previously undetected for electromagnetic perturbations, might occur here as well. Nevertheless, we did not find any instability for the vector perturbations investigated. For $\ell_3 = 2$, the negative minimum disappears, and the effective potential becomes very similar to that of the Schwarzschild case. In $D = 6$, a negative minimum near the event horizon occurs exclusively for $\ell_4 = 1$ across the entire range of tested $\widehat{\alpha}$, again indicating potential instabilities. For $\ell_4 = 2$, no negative minimum is present. However, as \(\widehat{\alpha}\) increases, a distinct double-peak structure develops, rendering the WKB approximation unsuitable for the analysis of this particular configuration (see Figure~\ref{figureUVD6}). Additionally, in dimensions $D=7$ and $D=10$, we also detect a negative minimum near the event horizon for $\ell_n = 1$ and all values of $\widehat{\alpha}$ examined (ranging from $0.1$ to $20$). Finally, Figure~\ref{figureUVD11_12_26} illustrates the presence of a negative minimum near the horizon in higher dimensions $D\in\{11, 12, 26\}$ for all tested values of $\widehat{\alpha}$. Despite these minima, we have not observed any corresponding instabilities, in contrast to the tensor perturbation case in $D = 6$.

The asymptotic behaviour of the solutions to equation (\ref{ODEV1}) can be obtained using the method outlined in \cite{Olver1994MAA}. For this purpose, we split our analysis by examining the behaviour of the radial field in two different regions.
Concerning the asymptotic behaviour as $x\to 1^+$, we observe that  
\begin{equation}  
    P_V(x) = \frac{1}{x-1} + \mathcal{O}(1), \quad  
    Q_V(x) = \left(\frac{\rho_h \Omega}{f'(1)}\right)^2 \frac{1}{(x-1)^2} + \mathcal{O}((x-1)^{-1}).  
\end{equation}  
This signals that the point $x = 1$ is a simple pole for $P_V(x)$ and a second-order pole for $Q_V(x)$. Furthermore, the exponents associated with this singularity are identical to those found in the scalar case at the event horizon. Therefore, we can follow the same approach and conclude that the QNM boundary condition at $x = 1$ is given by
\begin{equation}\label{QNMBCz1V}
R_V\underset{{x\to 1^+}}{\longrightarrow} (x-1)^{-i \rho_h a\Omega},\quad
a=\frac{1}{f^{'}(1)}.
\end{equation}
Regarding the asymptotic behaviour as $x\to +\infty$, we start by observing that in the limit of $x \to +\infty$
\begin{equation}\label{pqVec}
    P_V(x) = \sum_{m=0}^\infty\frac{\mathfrak{f}_m}{x^m} = \mathcal{O}\left(\frac{1}{x^n}\right), \qquad
    Q_V(x) = \sum_{m=0}^\infty\frac{\mathfrak{g}_m}{x^m}=\rho_h^2\Omega^2+\mathcal{O}\left(\frac{1}{x^2}\right).
\end{equation}
with $n\geqslant 3$. Since this result matches the scalar case, we can apply the same approach and conclude that the QNM boundary condition at positive spatial infinity is given by
\begin{equation}\label{QNMBCposinfV}
    R_V\underset{{x\to +\infty}}{\longrightarrow} e^{i\rho_h\Omega x}.
\end{equation}
At this point, we transform the radial function $R_V(x)$ into a new radial function $\Phi_V(x)$ such that the QNM boundary conditions are automatically implemented and  $\Phi_{V}(x)$ is regular at $x = 1$ and at space-like infinity. To this aim, we consider the transformation
\begin{equation}\label{AnsatzV}
    R_V(x) = x^{i\rho_h a\Omega}(x-1)^{-i\rho_h a\Omega}e^{i\rho_h\Omega(x-1)} \Phi_V(x).
\end{equation}
Multiplying \eqref{ODEV1} by $f^2(x)$ and substituting \eqref{AnsatzV}, we obtain the following ordinary differential equation for the radial eigenfunctions
\begin{equation}\label{ODEznoneV}  
  P_{2V}(x) \Phi''_V(x) + P_{1V}(x) \Phi'_V(x) + P_{0V}(x) \Phi_V(x) = 0,  
\end{equation}  
where $P_{2V}(x) = P_2(x)$ and $P_{1V}(x) = P_1(x)$, with $P_1(x)$ and $P_2(x)$ given by \eqref{P1} and \eqref{P2}. The term $P_{0V}(x)$ is obtained from \eqref{P0} by replacing $\mathcal{U}_S(x)$ with $\mathcal{U}_V(x)$. Let us now introduce the transformation $x = 2/(1-y)$ mapping the point at infinity and the event horizon to $y = 1$ and $y = -1$, respectively. Furthermore, a dot denotes differentiation with respect to the new variable $y$. Then, equation \eqref{ODEznoneV} becomes
\begin{equation}\label{ODEynoneV}
    S_{2V}(y)\ddot{\Phi}_{V}(y) + S_{1V}(y)\dot{\Phi}_{V}(y) + S_{0V}(y)\Phi_{V}(y) = 0,
\end{equation}
where  $S_{2V}(y) = S_2(y)$ and $S_{1V}(y) = S_1(y)$, with $S_1(y)$ and $S_2(y)$ given by \eqref{S1onone} and \eqref{S2onone}. Also here, the term $S_{0V}(y)$ is obtained from \eqref{S0onone} by replacing $\mathcal{U}_S(y)$ with $\mathcal{U}_V(y)$ in \eqref{fv2}. Notice that we must also require that $\Phi_{V}(y)$ is regular at $y=\pm 1$. As a result of the transformation introduced above, $f(y)$ is still given by \eqref{fv1}, while the effective potential is
\begin{eqnarray}
\mathcal{U}_V(y)&=&q_V(y)+\frac{(1-y)^3 f(y)}{4K_V(y)}\left\{(1-y)\dot{f}(y)\dot{K}_V(y)+f(y)\left[(1-y)\ddot{K}_V(y)-2\dot{K}_V(y)\right]\right\},\label{f2vec}\\
K_V(y)&=&\frac{{(1-y)^\frac{n}{2}}}{2^{\frac{n}{2}-1}}\sqrt{4\rho_h^{2}+\frac{\widehat{\alpha}}{(n-1)(n-2)}(1-y)^2\left\{(n-3)[1-f(y)]-(1-y)\dot{f}(y)\right\}},\\
q_V(y)&=&\frac{(\ell_n-1)(\ell_n+n)(1-y)^2f(y)}{4(n-1)}\cdot\frac{n(n-1)(2\rho_h)^{n+1}+2\widehat{\alpha}(n-3)(1-y)^{n+1}}{n(2\rho_h)^{n+1}+4\widehat{\alpha}(1-y)^{n+1}}.
\end{eqnarray}

\begin{table}%[ht]
\caption{Classification of the points $y=\pm 1$ for the relevant functions entering in \eqref{fv1}, \eqref{ODEynoneV}, and \eqref{f2vec}. The abbreviations $z$ ord $n$ and $p$ ord $m$ stand for zero of order $n$ and pole of order $m$, respectively.}
\begin{center}
\begin{tabular}{ | c | c | c | c | c | c | c | c }
\hline
$y$  & $f(y)$  & $\mathcal{U}_V(y)$ & $S_{2V}(y)$ & $S_{1V}(y)$ & $S_{0V}(y)$\\ \hline
$-1$ & z \mbox{ord} 1 & z \mbox{ord} 1 & z \mbox{ord} 4& z \mbox{ord} 3 & z \mbox{ord} 2 \\ \hline
$+1$ & $+1$           & z \mbox{ord} 2 & $+4$ & p \mbox{ord} 2 & p \mbox{ord} 2\\ \hline
\end{tabular}
\label{tableEinsnoneV}
\end{center}
\end{table}

Table~\ref{tableEinsnoneV} shows that the coefficients of the differential equation (\ref{ODEynoneV}) share a common zero of order $2$ at $y = -1$ while $y = 1$ is a pole of order $2$ for the coefficients $S_{1V}(y)$ and $S_{0V}(y)$. Hence, in order to apply the spectral method, we need to multiply (\ref{ODEynoneV}) by $(1-y)^2/(1+y)^2$. As a result, we obtain the following differential equation 
\begin{equation}\label{MVec}  
  M_{2V}(y) \ddot{\Phi}_{V}(y) + M_{1V}(y) \dot{\Phi}_{V}(y) + M_{0V}(y) \Phi_{V}(y) = 0,  
\end{equation}  
where $M_{2V}(y) = M_2(y)$, $M_{1V}(y) = M_1(y)$, and $M_{0V}(y) = M_0(y)$, with $\mathcal{U}_S(y)$ replaced by $\mathcal{U}_V(y)$. We recall that $M_i(y)$, $N_0(y)$, $C_2(y)$, $C_1(y)$, and $C_0(y)$ were introduced in \eqref{S210honone} and \eqref{N0}--\eqref{C0}; in the vector sector one replaces $\mathcal{U}_S(y)$ by $\mathcal{U}_V(y)$ in \eqref{C0}. At this point, it can be easily verified with Maple that
\begin{eqnarray}
    &&\lim_{y\to 1^{-}}M_{2V}(y)=0=\lim_{y\to -1^{+}}M_{2V}(y),\\
    &&\lim_{y\to 1^{-}}M_{1V}(y)=4i\rho_h\Omega,\quad
    \lim_{y\to -1^{+}}M_{1V}(y)=0,\\
    &&\lim_{y\to 1^{-}}M_{0V}(y)=2a\rho_h^2\Omega^2-\frac{n}{4}(n+2)-(\ell_n-1)(\ell_n+n),\quad
    \lim_{y\to -1^{+}}M_{0V}(y)=\mathcal{B}\Omega^2,
\end{eqnarray}
where $\mathcal{B}$ is given by
\begin{equation}
\mathcal{B}=-\frac{\rho_h^2\left[\sqrt{1+\kappa}A_1+(1+\kappa)A_2\right]}{4(1+\kappa)\left[(\kappa^2+8\kappa+8)\sqrt{1+\kappa}+4\kappa^2+12\kappa+8\right]}
\end{equation}
with
\begin{eqnarray}
A_1&=&[\left(n-3\right)^{2} a^{2}-4]\kappa^{3}+[\left(18 n^{2}-68 n+58\right) a^{2}-36]\kappa^{2}+[\left(48 n^{2}-128 n+80\right) a^{2}-64]\kappa\nonumber\\
&&+32[\left(n-1\right)^{2} a^{2}-1],\\
A_2&=&2[\left(3n^{2}-14n+15\right) a^{2}-8]\kappa^{2}+16[2\left( n^{2}-3n+2\right) a^{2}-3] \kappa+32[\left(n-1\right)^{2} a^{2}-1].
\end{eqnarray}
As a final step prior to implementing the spectral method, we rewrite the differential equation \eqref{MVec} in the following form
\begin{equation}\label{TSCHV}
  L_{0V}\left[\Phi_{V}, \dot{\Phi}_{V}, \ddot{\Phi}_{V}\right] +  i\Omega L_{1V}\left[\Phi_{V}, \dot{\Phi}_{V}, \ddot{\Phi}_{V}\right] +  \Omega^2 L_{2V}\left[\Phi_{V}, \dot{\Phi}_{V}, \ddot{\Phi}_{V}\right] = 0
\end{equation}
with
\begin{eqnarray}
L_{0V}\left[\Phi_{V}, \dot{\Phi}_{V}, \ddot{\Phi}_{V}\right] &=& L_{00V}(y)\Phi_{V} + L_{01V}(y)\dot{\Phi}_{V} + L_{02V}(y)\ddot{\Phi}_{V},\label{L0noneV}\\
L_{1V}\left[\Phi_{V}, \dot{\Phi}_{V}, \ddot{\Phi}_{V}\right] &=&L_{10V}(y)\Phi_{V} + L_{11V}(y)\dot{\Phi}_{V}+L_{12V}(y)\ddot{\Phi}_{V}, \label{L1noneV}\\
L_{2V}\left[\Phi_{V}, \dot{\Phi}_{V}, \ddot{\Phi}_{V}\right]&=&L_{20V}(y)\Phi_{V} +L_{21V}(y)\dot{\Phi}_{V} + L_{22V}(y)\ddot{\Phi}_{V}.\label{L2noneV}
\end{eqnarray}
Moreover, all the $L_{ijV}$ terms appearing in (\ref{L0noneV})-(\ref{L2noneV}) exhibit a regular behavior at the endpoints $y = \pm 1$ with limits given as in Table~\ref{tableDreiVector}.

\begin{table}%[ht]
\caption{Behaviour of the coefficients $L_{ijV}$ at the endpoints of the interval $-1 \leqslant y \leqslant 1$ for the vector perturbation of the Schwarzschild black hole with GB correction.}
\begin{center}
\begin{tabular}{ | c | c | c | c | c | c | c | c }
\hline
$(i,j)$ & $\displaystyle{\lim_{y\to -1^+}}L_{ijV}$  & $L_{ijV}$       & $\displaystyle{\lim_{y\to 1^-}}L_{ijV}$  \\ \hline
$(0,0)$ &  $0$                                     & $C_{0V}$         & $-\frac{n}{4}(n+2))-(\ell_n-1)(\ell_n+n)$\\ \hline
$(0,1)$ &  $0$                                     & $N_{0V}$         & $0$\\ \hline
$(0,2)$ &  $0$                                     & $M_{2V}$         & $0$\\ \hline 
$(1,0)$ &  $0$                                     & $C_{1V}$         & $0$\\ \hline 
$(1,1)$ &  $0$                                     & $N_{1V}$         & $4\rho_h$\\ \hline 
$(1,2)$ &  $0$                                     & $0$              & $0$\\ \hline 
$(2,0)$ &  $\mathcal{B}$                           & $C_{2V}$         & $2a\rho_h^2$\\ \hline
$(2,1)$ &  $0$                                     & $0$              & $0$\\ \hline
$(2,2)$ &  $0$                                     & $0$              & $0$\\ \hline
\end{tabular}
\label{tableDreiVector}
\end{center}
\end{table} 

The numerical implementation of the resulting quadratic eigenvalue problem is the same as in the scalar sector and is described in Sec.~III.

\section{Vector quasinormal spectra}

In this section, we present a detailed analysis of the QNM spectra for vector-type perturbations of Schwarzschild--Gauss--Bonnet black holes across dimensions $D \in \{5, 6, 7, 8, 10, 11, 12, 26\}$. Several novel features emerge from our study. First, overdamped modes, either forming nearly equispaced towers or appearing as isolated modes, are observed systematically across all dimensions and angular momentum configurations, particularly at moderate to strong GB coupling. Second, a characteristic non-monotonic behaviour in the real parts of higher overtones is identified across multiple dimensions, signalling a complex reorganisation of the spectrum in the strong coupling regime. Third, the morphology of the QNM distributions exhibits a rich variety of structures, evolving from Martini and 'coupe à glace' shapes at small couplings to bottleneck and pyramidal profiles at stronger coupling. Notably, the Viking Snowman morphology is uncovered in $D = 8$ for specific configurations. Across all tested cases, the sixth-order WKB approximation systematically fails to capture the emergence of overdamped modes, overtone structures, and non-monotonic behaviours, particularly at strong coupling, thereby underscoring the superior resolving power of the SM.  In $D = 26$, for a black hole of mass $M$ and GB coupling $\widehat{\alpha} = 0.1$, the fundamental vector QNM exhibits a dimensionless frequency $\Omega = 9.4489 - 1.9918i$. Although this value is not directly comparable to its four-dimensional Schwarzschild counterpart due to the different scaling of the mass parameter with spacetime dimensionality, the substantially larger dimensionless frequency suggests a strong amplification phenomenon in higher dimensions. Specifically, the real part is approximately $35$ times larger than that of the corresponding vector mode in $D = 4$, where $\Omega_{\text{Sch}} = 0.2483-0.0925i$. The tensor sector discussed below exhibits a similar, though slightly less pronounced, amplification trend. Finally, to assess the observational prospects, we convert these QNM frequencies into physical units (Hz) using the relation
\begin{equation}
  \nu = \frac{1}{2\pi} \left( \frac{c^3}{G M_\odot} \right)^p \frac{\Omega}{\eta^p}, \quad p = \frac{1}{n-1},
\end{equation}
where $\Omega$ is the dimensionless QNM frequency, $M$ is the black hole mass, $M_\odot$ is the solar mass, and $\eta=M/M_\odot$. For a stellar-mass black hole with $M = 10 M_\odot$, the vector mode $\Omega = 9.4489 - 1.9918i$ corresponds to a frequency $\nu = 2.314-0.488i$ Hz, while for a supermassive black hole with $M = 10^9 M_\odot$, it becomes $\nu = 1.039-0.219i$ Hz.  For comparison, in the superstring theory case ($D = 10$) with the same GB coupling, the fundamental vector mode has $\Omega = 2.2682-0.8833i$, leading to physical frequencies of $\nu = 1.489-0.580i$ Hz for stellar black holes and $\nu = 0.107-0.042i$ Hz for supermassive black holes. Although these frequencies lie below the current sensitivity thresholds of ground-based detectors such as LIGO and VIRGO (with a lower limit near $20$ Hz) and exceed LISA’s upper frequency limit (peaking around 1 Hz), they fall within the few-Hz to tens-of-Hz band targeted by next-generation space-based detectors like DECIGO \cite{Kawamura2008JPCS, Kawamura2019IJMPD, Kawamura2021PTEP}. These results suggest, for the first time, that higher-dimensional corrections predicted by string theory could leave observable imprints on gravitational wave signals. Specifically, the amplification of QNM frequencies in string-theoretically motivated dimensions, such as $D\in\{10, 26\}$, could make the scalar and vector ringdown phases of stellar and supermassive black holes accessible to future gravitational wave observatories, thereby opening a promising pathway to empirically test string theory. Furthermore, an exact isospectrality between scalar monopole ($\ell = 0$) and vector dipole ($\ell = 1$) QNMs is observed at vanishing GB coupling. The scalar and vector frequencies coincide for the fundamental modes and the overtone towers, within numerical precision, for all $D \geq 5$. Its analytic origin lies in a Darboux-type factorisation of the corresponding Hamiltonians, which we establish rigorously in Appendix~\ref{AppendixA}. The isospectrality is lifted as soon as the GB coupling becomes nonzero, due to the emergence of spin-dependent corrections in the effective potentials. In contrast, no such degeneracy is observed in the tensor sector, indicating that this spectral pairing is specific to the scalar and vector sectors of Einstein gravity. The results are organised dimension by dimension in the following subsections, where we highlight the main spectral features, compare the performance of different numerical methods, and provide a detailed mapping of the QNM landscape.

\subsection{$D = 5$}

\begin{figure}
    \centering
    \subfloat[$n=10$, $\ell_{10}=1$, $\widehat{\alpha}=0.1$ (Martini)]{%
    \includegraphics[width=0.49\textwidth]{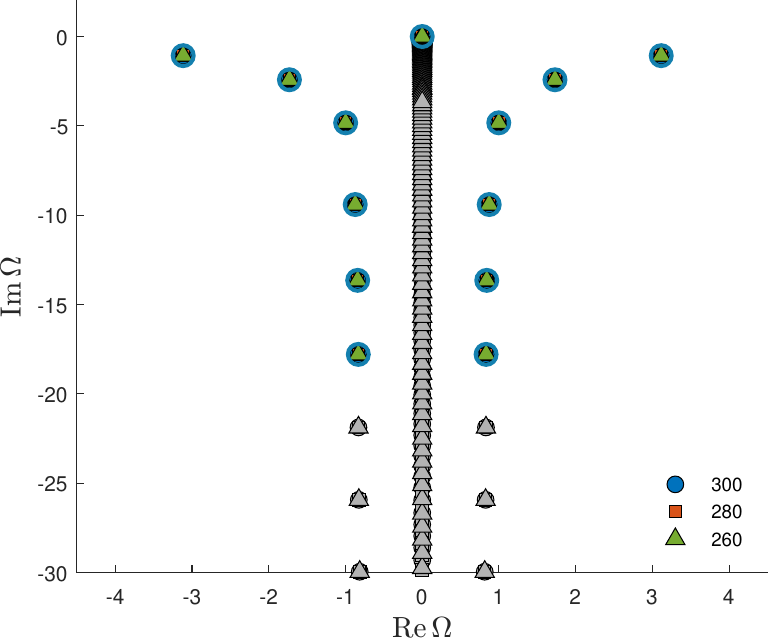}
    }
    \subfloat[$n=8$, $\ell_{8}=2$, $\widehat{\alpha}=0.5$ (Coupe)]{%
    \includegraphics[width=0.49\textwidth]{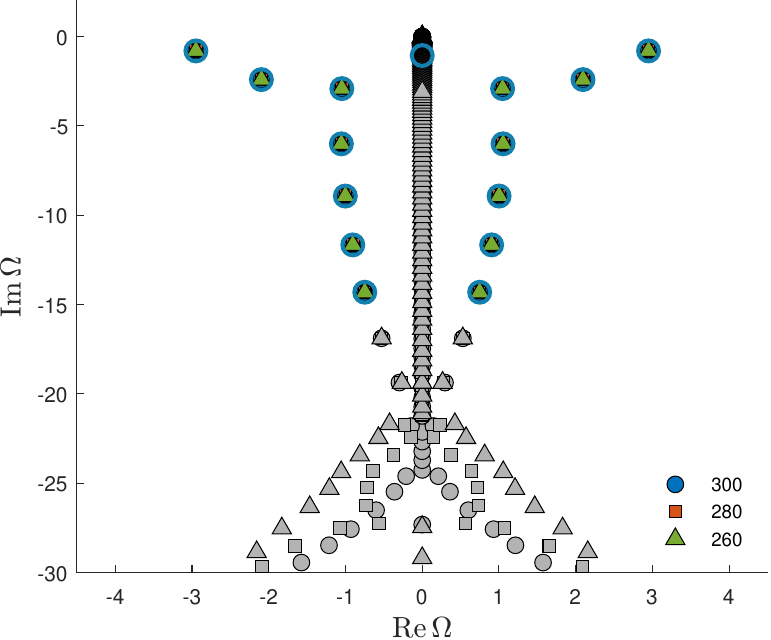}
    }\newline
    \subfloat[$n=9$, $\ell_{9}=1$, $\widehat{\alpha}=10$ (Bottleneck)]{%
    \includegraphics[width=0.49\textwidth]{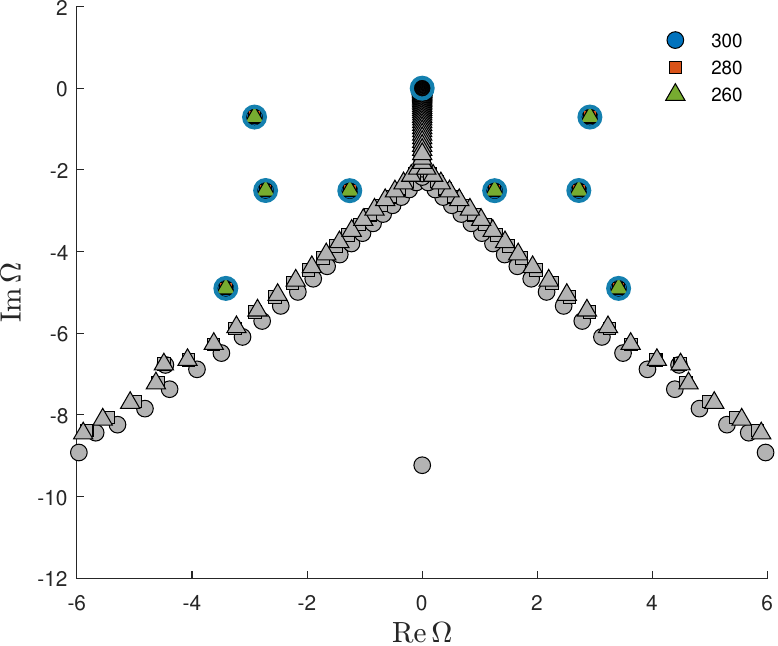}
    }
    \subfloat[$n=24$, $\ell_{24}=3$, $\widehat{\alpha}=0.5$ (Pyramid)]{%
    \includegraphics[width=0.49\textwidth]{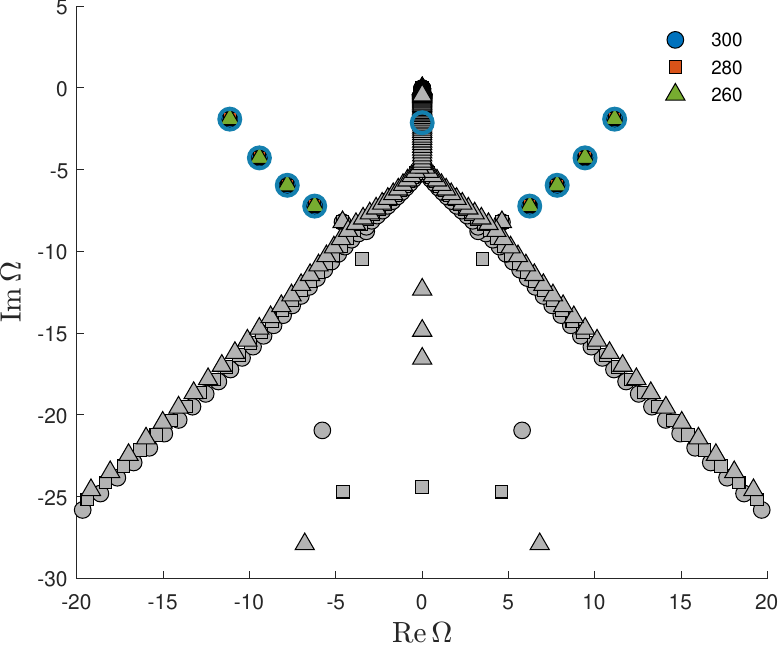}
    }
    \caption{The four main types of QNM distributions observed in the analysis of vector perturbations. Here, $n$, $\ell_n$, and $\widehat{\alpha}$ denote the number of compactified dimensions, the angular momentum, and the coupling parameter, respectively.}
    \label{atlasQNMdistribvec}
\end{figure}

In the five-dimensional case, the emergence of overdamped QNMs is observed across various values of the GB coupling $\widehat{\alpha}$ and angular momentum index $\ell_3$. For $\ell_3 = 1$, a single overdamped mode arises already at $\widehat{\alpha} = 0.1$, with frequency $\Omega=-16.4028i$. As the coupling increases to the range $0.6 \leqslant \widehat{\alpha} \leqslant 0.9$, a full tower of nearly equally spaced overdamped modes emerges (see Table~\ref{table:D5s1L1diffalpha}). This trend becomes more pronounced at stronger coupling. For instance, at $\widehat{\alpha} = 1.9$, we observe a distinguished overdamped mode at $\Omega = -0.1634i$, followed by several equally spaced modes (Table~\ref{table:piD5L123}). By $\widehat{\alpha} = 1.99$, the spectrum consists solely of overdamped modes, suggesting a highly stable configuration with purely exponential decay of perturbations. For $\ell_3 = 2$, overdamped modes are detected at all tested values of $\widehat{\alpha}$ except for $0.1$. At $\widehat{\alpha} = 1.0$, two distinguished overdamped modes appear at $\Omega_1 = -0.5379i$ and $\Omega_2 = -3.0056i$, followed by a sequence of equally spaced modes. At $\widehat{\alpha} = 1.9$, the spectrum features only two oscillatory QNMs, together with three distinguished overdamped modes at $\Omega_1 = -0.0370i$, $\Omega_2 = -0.1871i$, and $\Omega_3 = -0.3043i$, which are then followed by a tower of nearly equispaced purely imaginary modes. For $\ell_3 = 3$, the behaviour is analogous. At $\widehat{\alpha}=1.0$, we observe a distinguished overdamped mode at $\Omega = -0.9233i$, again followed by a sequence of overdamped frequencies (see Table~\ref{table:piD5L123}). At $\widehat{\alpha} = 1.9$, three distinguished modes occur at $\Omega_1=-0.0617i$, $\Omega_2=-0.2059i$, and $\Omega_3 = -0.3204i$, followed by several nearly equispaced overdamped modes. For $\ell_3 = 4$, overdamped modes are present for all tested couplings except $\widehat{\alpha}=0.1$. Notably, at $\widehat{\alpha}=0.5$, an isolated overdamped QNM is found at $\Omega=-8.8348i$, followed by a sequence of equally spaced imaginary frequencies. At $\widehat{\alpha}=1.9$, we detect three oscillatory QNMs together with two distinct groups of overdamped modes. By $\widehat{\alpha}=1.99$, only one oscillatory mode survives, accompanied by two well-separated groups of overdamped modes. In particular, we find two distinguished overdamped frequencies at $\Omega_1=-0.0049i$ and $\Omega_2=-0.0526i$, followed by three almost equally spaced overdamped modes (see Tables~\ref{table:piD5L4I} and \ref{table:piD5L4II}). We draw the reader's attention to the fact that the spectral gap parameter $\Delta\Omega$ for these overdamped modes has been measured and reported in Tables~\ref{table:D5s1L1diffalpha}, ~\ref{table:piD5L123}, \ref{table:piD5L4I}, and \ref{table:piD5L4II}. A linear regression analysis, displayed in Figure~\ref{fig:vec-lin01}, confirms a nearly linear dependence of $\Delta\Omega$ on the angular momentum $\ell_3$ for small values of $\widehat{\alpha}$, with determination coefficients $R^2$ exceeding $0.93$ across all considered cases. Furthermore, we observe a non-monotonic trend in the real part of the QNM frequencies for $\ell_3 = 1$ at $\widehat{\alpha} = 1.0$, and similarly for $\ell_3 = 4$ at $\widehat{\alpha} = 1.9$ (see bold entries in Tables~\ref{table:D5s1L1and2} and \ref{table:D5s1L3and4}). Moreover, \cite{Chakrabarti2007GRG} employed the sixth-order WKB approximation but did not report any values for the case $\widehat{\alpha} = 0$. In general, the WKB method underperforms compared to our SM for $\widehat{\alpha}\geqslant 0.2$ when $\ell_3 = 1$. It shows reasonable agreement for $\ell_3 = 2$ but it deteriorates for $\widehat{\alpha}\geqslant 0.5$ with $\ell_3 = 3$, and already for $\widehat{\alpha} \geqslant 0.1$ with $\ell_3 = 4$ (see Table~\ref{table:D5s1L1and2} and \ref{table:D5s1L3and4}). Furthermore, the WKB approximation fails to capture both overtones and the emergence of overdamped modes. Regarding the morphology of the QNM distribution, we observe a Martini glass structure (for a typical example, see (a) in Fig.~\ref{atlasQNMdistribvec}) across all tested values of $\ell_3$ within the range $0 \leqslant \widehat{\alpha} \leqslant 0.5$. At $\widehat{\alpha} = 1.0$, the distribution remains Martini-like for $\ell_3 = 1$, transitions to a bottleneck shape (for a typical examples see (c) in Fig.~\ref{atlasQNMdistribvec}) for $\ell_3 \in \{2, 3\}$, and assumes a 'coupe à glace' configuration (for a typical example, see (b) in Fig.~\ref{atlasQNMdistribvec}) for $\ell_3 = 4$. A qualitatively different regime sets in at stronger couplings $\widehat{\alpha}=1.9$ and $1.99$, where each value of $\ell_3$ yields only a few oscillatory QNMs, accompanied by a dominant sequence of overdamped modes.

\begin{figure}
    \centering
    \subfloat[$n=4$, $\ell_{4}=4$, $\widehat{\alpha}=0.5$ (Irregular Bottleneck)]{%
    \includegraphics[width=0.49\textwidth]{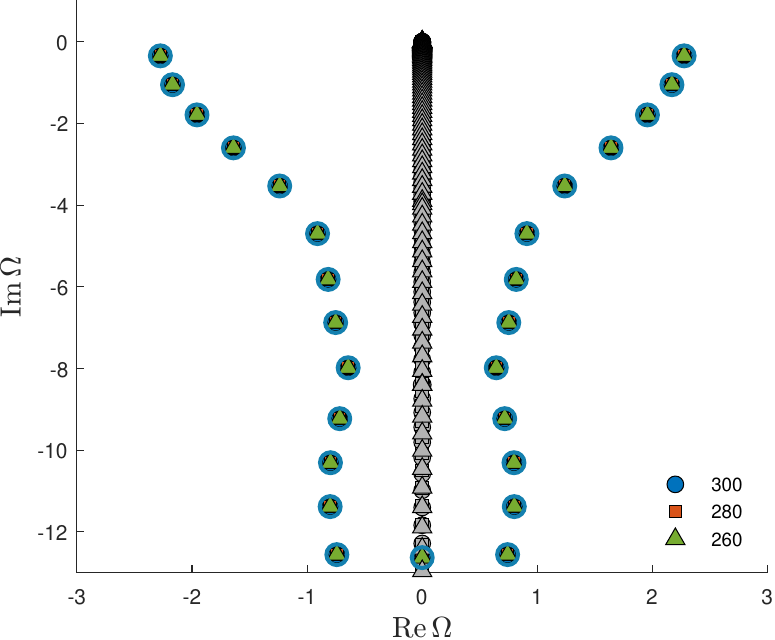}
    }
    \subfloat[$n=4$, $\ell_{4}=4$, $\widehat{\alpha}=10$ (Irregular Pyramid)]{%
    \includegraphics[width=0.49\textwidth]{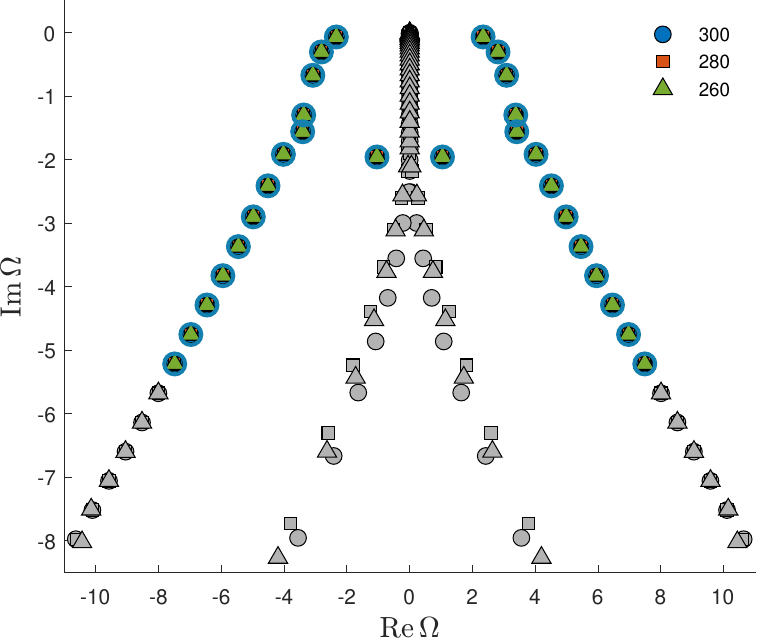}
    }
    \caption{Illustration of the irregular QNM distribution morphologies observed in the vectorial perturbation spectrum. The left panel displays a typical irregular bottleneck configuration, while the right panel shows an irregular pyramidal distribution. These deviations underline the richer spectral structure that emerges at specific values of the GB coupling and angular momentum. Here, $n$, $\ell_n$, and $\widehat{\alpha}$ denote the number of compactified dimensions, the angular momentum, and the coupling parameter, respectively.}
    \label{atlasQNMdistribvecirr}
\end{figure}

\subsection{$D=6$}

For $\ell_4 = 1$, a sequence of equally spaced overdamped modes appears in the regime $0.1 \leqslant \widehat{\alpha} \leqslant 0.5$ (see Table~\ref{table:piD6L12}). At $\widehat{\alpha} = 0.5$, this tower is accompanied by an isolated mode at $\Omega = -0.00013i$. When $\widehat{\alpha} = 1.0$, the spectrum contains only three oscillatory QNMs and a single overdamped mode at $\Omega = -29.3760i$ (see Table~\ref{table:D6s1L1and2}). For $\ell_4 = 2$, overdamped modes are again observed within $0.1 \leqslant \widehat{\alpha} \leqslant 0.5$. At $\widehat{\alpha} = 1.0$, an isolated overdamped mode appears at $\Omega = -0.8853i$, accompanied by two distinguished oscillatory modes at $\Omega_1 = 0.4724-12.8080i$ and $\Omega_2 = 0.6350-16.9672i$. At larger couplings, isolated overdamped modes persist such as $\Omega = -2.8811i$ for $\widehat{\alpha} = 5.0$, $\Omega = -6.2138i$ for $\widehat{\alpha} = 10$, $\Omega = -9.3803i$ for $\widehat{\alpha} = 15$, and $\Omega = -12.5263i$ for $\widehat{\alpha} = 20$. For $\ell_4 = 3$, overdamped modes emerge for $0.1 \leqslant \widehat{\alpha} \leqslant 0.2$ (see Table~\ref{table:D6s1L3and4}). At $\widehat{\alpha} = 1.0$, the spectrum contains four purely imaginary modes at $\Omega_1 = -12.2750i$, $\Omega_2 = -13.7852i$, $\Omega_3 = -16.4690i$, and $\Omega_4 = -17.7839i$. As $\widehat{\alpha}$ increases, isolated overdamped modes continue to appear. For instance, we find $\Omega = -36.3757i$ for $\widehat{\alpha} = 10$, $\Omega = -54.5868i$ for $\widehat{\alpha} = 15$, and $\Omega=-72.7900i$ for $\widehat{\alpha} = 20$. Finally, for $\ell_4=4$, overdamped modes are present within $0.1 \lesssim \widehat{\alpha} \lesssim 0.2$ (Table~\ref{table:piD6L4I}). At $\widehat{\alpha} = 0.5$, we detect two isolated purely imaginary modes at $\Omega_1 = -12.6318i$ and $\Omega_2 = -21.0513i$, in addition to the usual equispaced tower. Finally, in Table~\ref{table:piD6L12}, we report the measured spectral gap $\Delta\Omega$ for the overdamped QNMs. A linear regression analysis (see Figure~\ref{fig:vec-lin02}) confirms the hypothesis of a nearly linear dependence of $\Delta\Omega$ on $\ell_4$ for small $\widehat{\alpha}$. The correlation is remarkably strong, with determination coefficients $R^2 > 0.99$ across all tested values. We observe a clear non-monotonic behavior in the real part of the QNMs for $\ell_4 \in \{1,2,3\}$ across the full range $1 \leqslant  \widehat{\alpha} \leqslant 20$, while for $\ell_4 = 4$ this feature emerges in the narrower range $5 \leqslant  \widehat{\alpha} \leqslant 20$ (see bold entries in Tables~\ref{table:D6s1L1and2} and \ref{table:D6s1L3and4}). Turning to the sixth-order WKB method employed in \cite{Chakrabarti2007GRG}, we find that it performs poorly for $\ell_4 = 1$ in both the real and imaginary components of the QNMs. For $\ell_4 \in \{2,3\}$, WKB gives reasonable estimates only in the narrow range $0.1 \leqslant \widehat{\alpha} \leqslant 0.2$, but deteriorates significantly in accuracy for $\ell_4 = 4$ when $5 \leqslant \widehat{\alpha} \leqslant 20$. Moreover, the WKB approach fails to capture higher overtones and entirely misses the presence of overdamped modes, thus highlighting the superior resolving power of our SM. Regarding the morphology of the QNM spectrum, we find a 'coupe à glace' distribution at $\widehat{\alpha} = 0$ across all tested values of $\ell_4$, as well as in the range $0\leqslant \widehat{\alpha}\leqslant 0.2$ for $\ell_4 = 4$. A Martini distribution characterizes the range $0.1 \leqslant \widehat{\alpha} \leqslant 0.2$ for $\ell_4 \in \{1,2,3\}$. At $\widehat{\alpha} = 0.5$, this morphology persists for $\ell_4 \in \{1,2\}$, but transitions into a bottleneck structure for $\ell_4=3$, and an irregular bottleneck for $\ell_4 = 4$ (for a typical example, see panel (a) in Fig.~\ref{atlasQNMdistribvecirr}). The case $\widehat{\alpha} = 1.0$ is interesting because for $\ell_4\in\{1,2\}$, the spectrum features only a few oscillatory QNMs along with a single isolated overdamped mode. In contrast, for $\ell_4\in\{3,4\}$, the morphology evolves into a bottleneck distribution. Apart from the case $\ell_4 = 2$ at $\widehat{\alpha} = 10$, which retains a bottleneck shape, and $\ell_4 = 4$ at the same coupling, which exhibits an irregular pyramidal configuration (for a typical example, see panel (b) in Fig.~\ref{atlasQNMdistribvecirr}), the morphology becomes pyramidal (for a typical example, see (d) in Fig.~\ref{atlasQNMdistribvec}) in all remaining configurations with $\ell_4 \in \{1,2,3,4\}$ for $5 \leqslant \widehat{\alpha} \leqslant 20$.

\subsection{$D=7$}

For $\ell_5=1$ and within the range $0.1\leqslant\widehat{\alpha}\leqslant 0.2$, we observe the appearance of several equally spaced overdamped modes (see Table~\ref{table:piD7L123}). A similar pattern occurs for $\ell_5=2$ in the same coupling range. At $\widehat{\alpha}=0.5$, the spectrum features two isolated overdamped modes located at $\Omega_1=-49.0739i$ and $\Omega_2=-54.6814i$. As the coupling increases to $\widehat{\alpha}=1.0$, a single overdamped mode is observed at $\Omega =-0.9772i$. This trend continues at higher couplings, where isolated overdamped modes are found at $\Omega=-0.9130i$ for $\widehat{\alpha}=5$, $\Omega=-1.9924i$ for $\widehat{\alpha}=10$, $\Omega=-2.7739i$ for $\widehat{\alpha}=15$, and $\Omega=-3.3226i$ for $\widehat{\alpha}=20$. For $\ell_5=3$, a tower of nearly equispaced overdamped modes appears again in the low-coupling regime $0.1\leqslant \widehat{\alpha}\leqslant 0.2$ (see Table~\ref{table:piD7L123}). At $\widehat{\alpha}=0.5$, we find an isolated overdamped mode at $\Omega=-54.5279i$, and at $\widehat{\alpha}=5.0$ another one at $\Omega=-5.5968i$. Additional isolated overdamped modes are located at $\Omega=-8.7173i$ ($\widehat{\alpha}=10$), $\Omega=-10.8319i$ ($\widehat{\alpha}=15$), and $\Omega=-12.5697i$ ($\widehat{\alpha}=20$). For $\ell_5=4$, several equally spaced overdamped modes are observed in the regime $0.1\leqslant \widehat{\alpha}\leqslant 0.2$ (see Table~\ref{table:piD7L4}). At $\widehat{\alpha}=0.5$, the spectrum includes three isolated modes at $\Omega_1 = -19.5000i$, $\Omega_2 = -47.2340i$, and $\Omega_3 = -54.6650i$. As $\widehat{\alpha}$ increases, we identify isolated overdamped modes at $\Omega = -15.1297i$ for $\widehat{\alpha} = 5.0$, $\Omega = -21.8577i$ for $\widehat{\alpha} = 10$, $\Omega = -26.8747i$ for $\widehat{\alpha} = 15$, and $\Omega = -31.0744i$ for $\widehat{\alpha} = 20$. In Tables~\ref{table:piD7L123} and \ref{table:piD7L4}, we report the measured spectral gap $\Delta\Omega$ for the overdamped QNMs. To examine the dependence of $\Delta\Omega$ on angular momentum, we performed a linear regression analysis for small values of $\widehat{\alpha}$ (see Figure~\ref{fig:vec-lin03}). The resulting determination coefficients $R^2$ consistently exceed $0.89$, lending reasonable support to the hypothesis of a nearly linear dependence of the gap on $\ell_5$. Non-monotonic behaviour in the real part of the QNM spectrum is also evident in the seven-dimensional case. For $\ell_5\in\{1, 2, 3\}$, it appears consistently throughout the coupling range $1\leqslant\widehat{\alpha}\leqslant 20$, while for $\ell_5=4$, it becomes manifest only for stronger couplings, specifically $5\leqslant\widehat{\alpha}\leqslant 20$ (see bold entries in Tables~\ref{table:D7s1L1and2} and \ref{table:D7s1L3and4}). Although no bold entry is included in Table~\ref{table:D7s1L3and4} for $\ell_5=3$ at $\widehat{\alpha} = 1.0$, we note the presence of a non-monotonic mode at approximately $\Omega=0.4494 - 14.8999i$. This particular overtone lies beyond the fifth and was omitted from the table due to space constraints. In comparing the sixth-order WKB method employed by \cite{Chakrabarti2007GRG} with our spectral results, we find that WKB becomes entirely unreliable for $\ell_5=1$ in the range $0.5\leqslant\widehat{\alpha}\leqslant 20$. While it provides reasonable accuracy for the real part of the QNM at lower couplings ($0.1\leqslant\widehat{\alpha}\leqslant 0.2$), its predictions for the imaginary part remain unsatisfactory. For $\ell_5=2$, the WKB method shows modest agreement within $0.1\leqslant\widehat{\alpha}\leqslant 0.5$, but its accuracy deteriorates significantly as $\widehat{\alpha}$ increases. In the case of $\ell_5=3$, WKB yields reliable results only up to $\widehat{\alpha}=1.0$, beyond which its performance rapidly degrades, becoming completely unreliable for $5 \leqslant \widehat{\alpha} \leqslant 20$. For $\ell_5 = 4$, it performs adequately for $0.1 \leqslant \widehat{\alpha} \leqslant 0.5$, but fails entirely in the strong coupling regime $1\leqslant \widehat{\alpha} \leqslant 20$. Unsurprisingly, the WKB approximation does not capture higher overtones or the presence of overdamped modes, further underscoring the superiority of the SM in this context. Regarding the morphology of the QNM distributions, we observe a Martini glass shape for $\ell_5 \in \{1,2,3\}$ in the range $0 \leqslant \widehat{\alpha} \leqslant 0.5$, with two exceptions. More precisely, for $\ell_5 = 3$ at $\widehat{\alpha} = 0$, the distribution resembles a 'coupe à glace', while at $\widehat{\alpha} = 0.5$ it transitions into an irregular bottleneck. For $\ell_5 = 4$, the morphology is initially a 'coupe à glace' for $0 \leqslant \widehat{\alpha} \leqslant 0.2$, but becomes an irregular bottleneck at $\widehat{\alpha} = 0.5$. At $\widehat{\alpha} = 1.0$, the spectrum adopts a bottleneck structure for all values of $\ell_5$, except for $\ell_5 = 2$, where we observe an irregular bottleneck. When the coupling increases to $\widehat{\alpha} = 5$, a pyramidal distribution emerges for $\ell_5\in\{1,2\}$, while for $\ell_5 \in \{3,4\}$ the morphology becomes a bottleneck. In the strong coupling regime $10\leqslant\widehat{\alpha}\leqslant 20$, all tested values of $\ell_5$ exhibit a pyramidal configuration.

\begin{table}%[ht]
\centering
\caption{Isolated overdamped QNM modes for vector perturbations (spin $s = 1$) of an eight-dimensional ($n=6$) Schwarzschild black hole with GB correction are presented in the table below for different values of $\widehat{\alpha}$ and $\ell_6=2$. The corresponding results are obtained through our SM, utilising $300$ polynomials with a precision of $300$ digits. In this context, $\Omega$ represents the dimensionless frequency. The notation 'SM' stands for Spectral Method.}
\label{table:isolated}
\vspace*{1em}
\begin{tabular}{||c|c||}
\hline\hline
$\widehat{\alpha}$ & $\Omega$ (SM)   \\ [0.5ex]
\hline\hline
$0.1$              & $0.0000-1.0948i$\\
$0.2$              & $0.0000-1.0957i$ \\
$0.5$              & $0.0000-1.0839i$\\
$1.0$              & $0.0000-1.0041i$\\
$5.0$              & $0.0000-0.7798i$\\
$10$               & $0.0000-1.1518i$\\
$15$               & $0.0000-1.5504i$\\
$20$               & $0.0000-1.8755i$\\[1ex]
\hline\hline 
\end{tabular}
\end{table}
\subsection{$D=8$}
\begin{figure}
    \centering
    \subfloat[$n=6$, $\ell_{6}=4$, $\widehat{\alpha}=0.5$ (Viking Snowman)]{%
    \includegraphics[width=0.49\textwidth]{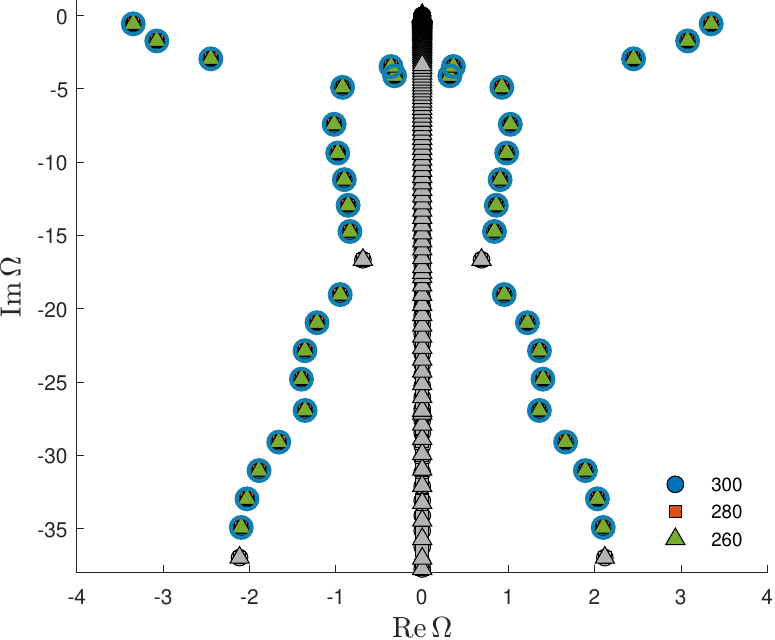}
    }
    \caption{Illustration of the Viking Snowman QNM distribution morphology observed in the vectorial perturbation spectrum. Here, $n$, $\ell_n$, and $\widehat{\alpha}$ denote the number of compactified dimensions, the angular momentum, and the coupling parameter, respectively.}
    \label{atlasQNMdistribvecVS}
\end{figure}

\begin{figure}[H]
    \centering
    \includegraphics[width=0.8\linewidth]{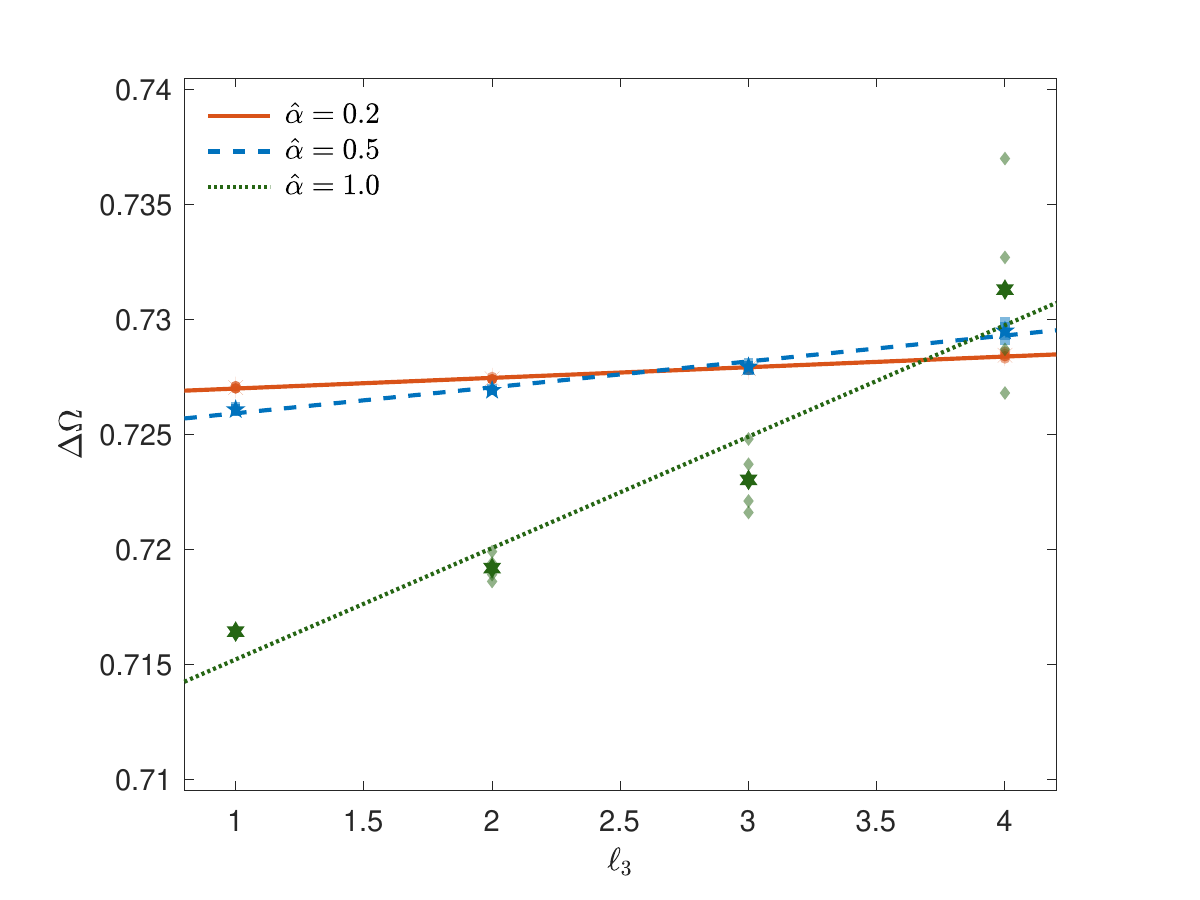}
    \caption{Linear regression analysis of the spectral gap values reported in Tables~\ref{table:piD5L123} and \ref{table:piD5L4I}. The values of the linear regression determinantion coefficients for each considered value of the parameter $\widehat{\alpha}$ are as follows: $R^2 = 0.990504$ ($\widehat{\alpha}=0.2$), $R^2 = 0.979058$ ($\widehat{\alpha}=0.5$), and $R^2 = 0.934907$ ($\widehat{\alpha}=1.0$).}
    \label{fig:vec-lin01}
\end{figure}

\begin{figure}[H]
    \centering
    \includegraphics[width=0.8\linewidth]{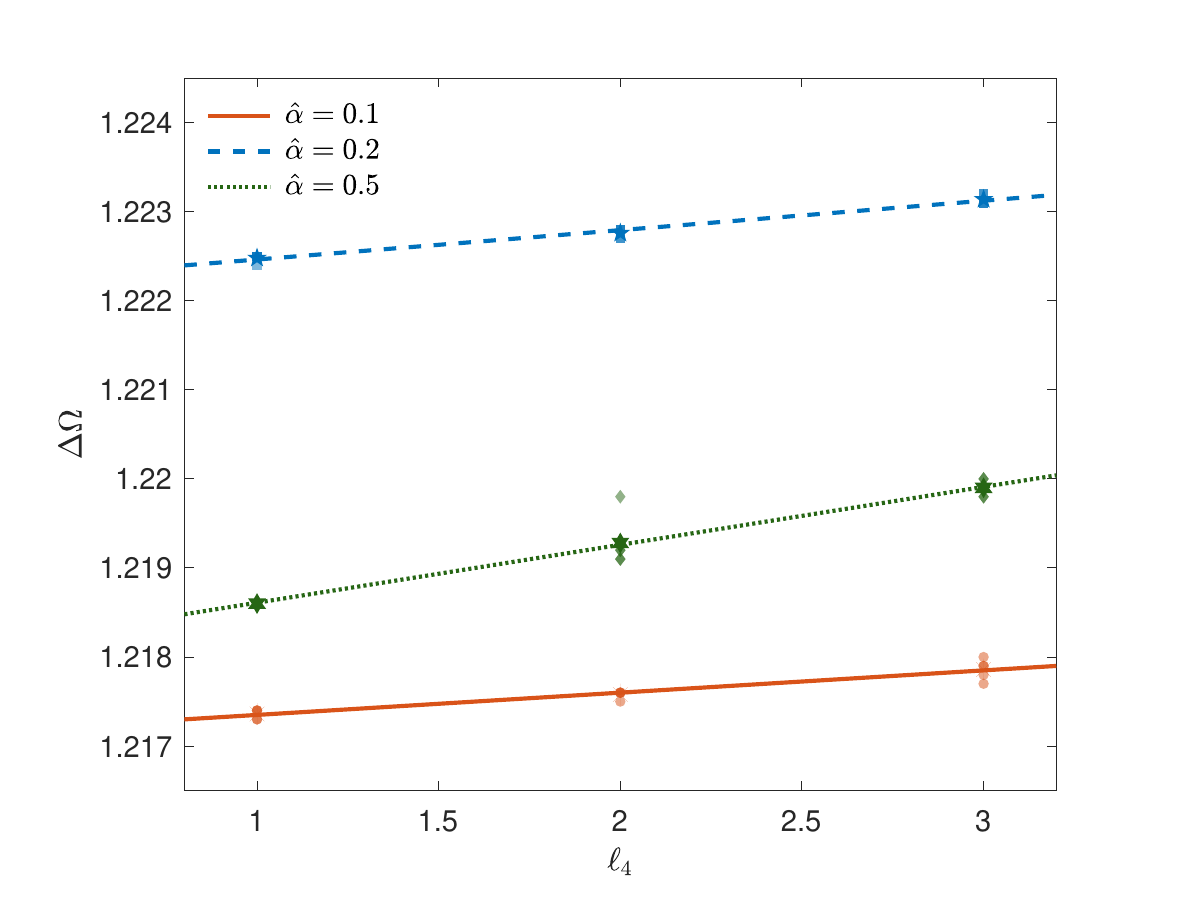}
    \caption{Linear regression analysis of the spectral gap values reported in Table~\ref{table:piD6L12}. The values of the linear regression determination coefficients for each considered value of the parameter $\widehat{\alpha}$ are as follows: $R^2 = 0.995223$ ($\widehat{\alpha}=0.1$), $R^2 = 0.992406$ ($\widehat{\alpha}=0.2$), and $R^2 = 0.999290$ ($\widehat{\alpha}=0.5$).}
    \label{fig:vec-lin02}
\end{figure}

\begin{figure}[H]
    \centering
    \includegraphics[width=0.8\linewidth]{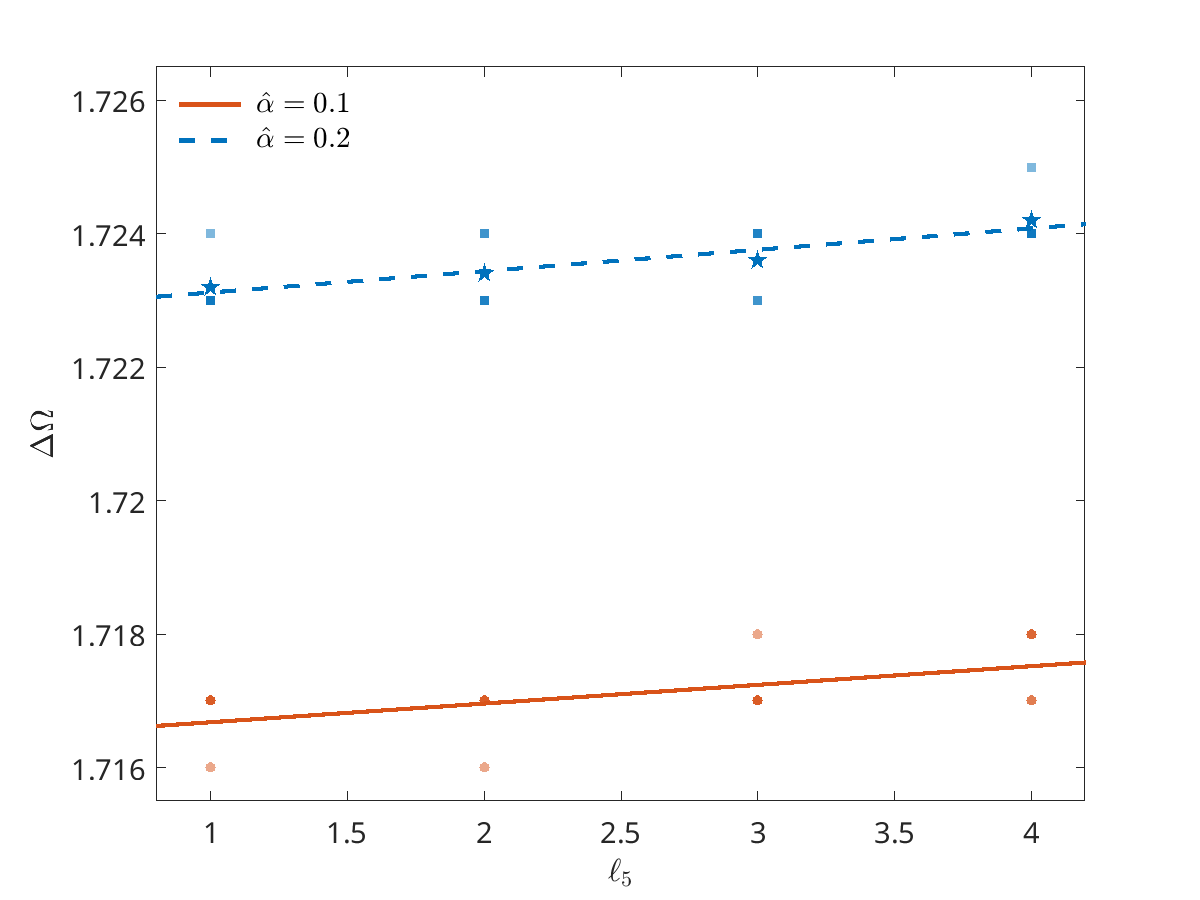}
    \caption{Linear regression analysis of the spectral gap values reported in Tables~\ref{table:piD7L123} and \ref{table:piD7L4}. The values of the linear regression determination coefficients for each considered value of the parameter $\widehat{\alpha}$ are as follows: $R^2 = 0.890909$ ($\widehat{\alpha}=0.1$), and $R^2 = 0.914286$ ($\widehat{\alpha}=0.2$).}
    \label{fig:vec-lin03}
\end{figure}

In the eight-dimensional case, we observe a single isolated overdamped mode at $\Omega=-1.0938i$ for $\ell_6=2$ when $\widehat{\alpha}=0$. For $\widehat{\alpha} = 0.1$, equally spaced overdamped modes emerge for $\ell_6\in\{1, 2\}$ (see Table~\ref{table:piD8L123}). The spacing of these modes appears largely independent of low values of $\ell_n$. For $\ell_6 = 2$, isolated overdamped modes are observed across various values of $\widehat{\alpha}$ (see Table~\ref{table:isolated}). We notice that the least damped mode occurs around $\widehat{\alpha} \approx 5.0$, with increased damping observed as $\widehat{\alpha}$ exceeds $10$. For $\ell_6 = 3$, isolated overdamped modes are detected within the range $0.5 \leqslant \widehat{\alpha} \leqslant 10$. Specifically, at $\widehat{\alpha} = 0.5$, modes occur at $\Omega = -59.5108i$ and $\Omega = -65.3882i$. At higher couplings, we find $\Omega = -4.5508i$ for $\widehat{\alpha} = 10$, $\Omega = -5.7320i$ for $\widehat{\alpha} = 15$, and $\Omega = -6.4925i$ for $\widehat{\alpha} = 20$. For $\ell_6 = 4$, two overdamped modes are observed at $\widehat{\alpha} = 0.5$, located at $\Omega = -58.9568i$ and $\Omega = -66.5611i$. As $\widehat{\alpha}$ increases, the overdamped modes shift to $\Omega = -7.3426i$ for $\widehat{\alpha} = 5.0$, $\Omega = -10.8719i$ for $\widehat{\alpha} = 10$, $\Omega = -12.7412i$ for $\widehat{\alpha} = 15$, and $\Omega = -14.1487i$ for $\widehat{\alpha} = 20$. Regarding the non-monotonic behavior of the real part of the QNMs, we observe this feature for $\ell_6\in\{1, 2, 4\}$ throughout the range $1 \leqslant \widehat{\alpha} \leqslant 20$, and for $\ell_6=3$, it persists across the entire interval $0 \leqslant \widehat{\alpha} \leqslant 20$ (see bold entries in Tables~\ref{table:D8s1L1and2} and \ref{table:D8s1L3and4}). A comparison with the sixth-order WKB results presented in \cite{Chakrabarti2007GRG} reveals several limitations. For $\ell_6=1$, the WKB method gives reasonably accurate results at small couplings ($0.1\leqslant \widehat{\alpha} \leqslant 0.2$), but its accuracy deteriorates further as $\widehat{\alpha}$ increases. For $\ell_6 = 2$, the approximation is acceptable within the range $0.1 \leqslant \widehat{\alpha} \leqslant 1$, but becomes unreliable for stronger couplings ($5 \leqslant \widehat{\alpha} \leqslant 20$). A similar trend is observed for $\ell_6 = 3$, where WKB shows acceptable agreement at small $\widehat{\alpha}$, yet fails completely in the higher coupling regime. For $\ell_6 = 4$, the method remains fairly reliable for $0.1 \leqslant \widehat{\alpha} \leqslant 1$, but again breaks down at larger values of $\widehat{\alpha}$. As in lower dimensions, the WKB method is unable to capture either the overdamped modes or the overtone structure of the spectrum. Regarding the morphology of the QNM spectrum, we observe a Martini distribution for $\ell_6\in\{1,2\}$ within the range $0\leqslant\widehat{\alpha}\leqslant 0.2$, which transitions into a 'coupe à glace' shape for higher angular momenta $\ell_6\in\{3,4\}$. A bottleneck configuration emerges uniformly across all tested values of $\ell_6$ in the intermediate coupling range $1\leqslant\widehat{\alpha}\leqslant 10$. The case $\widehat{\alpha}=0.5$ is particularly interesting: while a Martini morphology persists for $\ell_6=\{1,2\}$, it evolves into a bottleneck for $\ell_6=3$ and transforms into a distinctive Viking snowman distribution for $\ell_6 = 4$ (see Fig.~\ref{atlasQNMdistribvecVS}). At strong coupling ($15 \leqslant \widehat{\alpha} \leqslant 20$), the QNM spectrum retains a bottleneck structure for $\ell_6\in\{1,2\}$, but adopts a pyramidal configuration for $\ell_6\in\{3,4\}$.

\subsection{$D=10$ (Superstring Theory)}

For vanishing GB coupling, we observe isolated overdamped modes at $\Omega=-1.0611i$ for $\ell_8 = 2$ and $\Omega=-2.4419i$ for $\ell_8 = 3$. For $\ell_8=1$, a single overdamped mode appears at $\Omega=-67.5064i$ when $\widehat{\alpha}=0.5$. In the case of $\ell_8=2$, we detect one overdamped mode at $\Omega = -1.0640i$ for $\widehat{\alpha}=0.1$, another at $\Omega=-1.0662i$ for $\widehat{\alpha}=0.2$, and two modes at $\Omega=-1.0624i$ and $\Omega=-60.2992i$ for $\widehat{\alpha}=0.5$. Additional overdamped modes are found at $\Omega=-0.7370i$ for $\widehat{\alpha} = 5.0$, $\Omega=-0.8355i$ for $\widehat{\alpha}=10$, and $\Omega=-1.1105i$ for $\widehat{\alpha}=20$. Finally, for $\ell_8=3$, we identify isolated overdamped modes at $\Omega=-2.4483i$, $-2.4523i$, and $-2.4209i$ for $\widehat{\alpha} = 0.1$, $0.2$, and $0.5$, respectively. At higher couplings, we find $\Omega = -1.6473i$ for $\widehat{\alpha} = 5.0$, and $\Omega = -3.1400i$ for $\widehat{\alpha}=20$. Moreover, we observe non-monotonic behavior in the real part of the QNMs across different angular momenta: for $\ell_8=1$, this feature emerges in the range $5\leqslant\widehat{\alpha}\leqslant 20$; for $\ell_8 = 2$, it appears already at $\widehat{\alpha} \geqslant 0.5$; and for $\ell_8 = 3$, the non-monotonic trend is present throughout the entire interval $0.2\leqslant\widehat{\alpha}\leqslant 20$ (see bold entries in Table~\ref{table:D10s1L123}). Regarding the morphology of the QNM spectrum, we observe a 'coupe à glace' distribution for all tested values of $\ell_8$ within the range $0.1\leqslant\widehat{\alpha}\leqslant 0.5$, which transitions into a bottleneck configuration for $5\leqslant\widehat{\alpha}\leqslant 10$. The case $\widehat{\alpha}=0$ is somewhat exceptional: while a Martini-type distribution appears for $\ell_8 = 1$, it transforms into a coupe shape for $\ell_8\in\{2, 3\}$. At $\widehat{\alpha}=20$, the spectrum exhibits a pyramidal profile for $\ell_8\in\{1, 2\}$, which changes into a bottleneck distribution for $\ell_8 = 3$.

\subsection{$D=11$ (M-Theory)}

In the present case, for $\widehat{\alpha}=0$, we observe an isolated overdamped mode at $\Omega = -1.0516i$ for $\ell_9=2$. As the coupling increases, the overdamped sector becomes richer. Specifically, for $\ell_9=2$, we identify an isolated mode at $\Omega=-1.0546i$ for $\widehat{\alpha}=0.1$, another at $\Omega=-1.0547i$ for $\widehat{\alpha}=0.2$, and two overdamped modes at $\Omega=-1.0547i$ and $\Omega=-88.9332i$ when $\widehat{\alpha}=0.5$. At stronger coupling, we detect isolated overdamped modes at $\Omega=-0.7348i$ for $\widehat{\alpha}=5$, $\Omega=-0.7844i$ for $\widehat{\alpha}=10$, and $\Omega=-0.9829i$ for $\widehat{\alpha}=20$. For $\ell_9=3$, at $\widehat{\alpha}=0.5$, we detect a cluster of three overdamped modes located at $\Omega=-55.6172i$, $\Omega=-78.0753i$, and $\Omega=-91.8331i$. For higher coupling, we observe one isolated overdamped mode at $\Omega=-1.5967i$ for $\widehat{\alpha}=5$, another at $\Omega=-1.8467i$ for $\widehat{\alpha}=10$, and finally one at $\Omega=-2.5271i$ for $\widehat{\alpha}=20$. Concerning the non-monotonic behavior of the real part of the QNMs, we observe this feature for $\ell_9\in\{1,2\}$ within the range $5\leqslant\widehat{\alpha}\leqslant 20$, while for $\ell_9=3$, it is present throughout the entire coupling range $0\leqslant\widehat{\alpha}\leqslant 20$ (see bold entries in Table~\ref{table:D11s1L123}).  Finally, the morphology of the QNM spectrum in $D=11$ reveals distinct features compared to the $D=10$ case. For $\ell_9\in\{2, 3\}$, the distribution exhibits a 'coupe à glace' structure within the range $0\leqslant\widehat{\alpha}\leqslant 0.5$, while for all tested values of $\ell_9$, a bottleneck morphology emerges consistently for $10\leqslant\widehat{\alpha}\leqslant 20$. Interestingly, for $\ell_9=1$, the spectrum follows a coupe distribution at $\widehat{\alpha}= 0$, transitioning into a Martini shape for $0.1\leqslant\widehat{\alpha}\leqslant 0.5$. At intermediate coupling $\widehat{\alpha}=5$, the distribution becomes bottleneck-shaped for $\ell_9=1$, but  reverts to a Martini profile for $\ell_9\in\{2, 3\}$.

\subsection{$D=12$ (F-Theory)}

When $\widehat{\alpha} = 0$, we observe isolated overdamped modes at $\Omega=-1.0446i$ for $\ell_{10}=2$ and at $\Omega=-2.3328i$ for $\ell_{10}=3$. For $\ell_{10}=1$, a single overdamped mode appears at $\Omega=-101.4630i$ when $\widehat{\alpha}=0.5$. In the case $\ell_{10}=2$, we detect overdamped modes at $\Omega=-1.0476i$ for $\widehat{\alpha}=0.1$ and at $\Omega=-1.0499i$ for $\widehat{\alpha}=0.2$. At $\widehat{\alpha}=0.5$, two overdamped modes are observed at $\Omega=-105.5910i$ and $\Omega=-1.0485i$. As the coupling increases, further isolated overdamped modes appear at $\Omega=-0.7361i$, $\Omega=-0.7532i$, and $\Omega=-0.9035i$ for $\widehat{\alpha}=5$, $10$, and $20$, respectively. For $\ell_{10}=3$, an overdamped mode is found at $\Omega=-2.3392i$ when $\widehat{\alpha}=0.1$ and at $\Omega=-2.3435i$ for $\widehat{\alpha}=0.2$. At $\widehat{\alpha}=0.5$, we detect two modes at $\Omega=-108.7210i$ and $\Omega=-2.3287i$. Additional overdamped modes are present at $\Omega=-1.5727i$, $\Omega=-1.7075i$, and $\Omega=-2.1895i$ for $\widehat{\alpha}=5$, $10$, and $20$, respectively. Concerning the non-monotonic behavior in the real part of the QNMs, we observe that it emerges in the range \( 5 \leqslant \widehat{\alpha} \leqslant 20 \) for \( \ell_{10} = 1 \), while for \( \ell_{10} \in \{2, 3\} \), it persists throughout the entire coupling range \( 0 \leqslant \widehat{\alpha} \leqslant 20 \) (see bold entries in Table~\ref{table:D12s1L123}). Finally, the morphology of the QNM spectrum exhibits a Martini distribution for $\ell_{10}\in\{1, 2\}$ in the range $0.1\leqslant\widehat{\alpha}\leqslant 0.2$, which transitions into a bottleneck structure for all tested values of $\ell_{10}$ when $10\leqslant\widehat{\alpha}\leqslant 20$. At $\widehat{\alpha}= 0$, the distribution assumes a coupe shape across all angular momenta, and for $\ell_{10}=3$, this coupe profile persists up to $\widehat{\alpha}=0.5$. Interestingly, at $\widehat{\alpha}=5$, the morphology is Martini-like for $\ell_{10}=1$, transitions to a coupe distribution for $\ell_{10}=2$, and then shifts to a pyramidal structure for $\ell_{10}=3$. Thus, the spectrum reveals a subtle dependence on both the angular momentum and the GB coupling.

\subsection{$D=26$ (Bosonic String Theory)}

For $\widehat{\alpha}= 0$, isolated overdamped modes appear at $\Omega=-1.0149i$ for $\ell_{24}=2$ and $\Omega=-2.1196i$ for $\ell_{24}=3$. When $\ell_{24}=2$, additional isolated modes are observed at $\Omega=-1.0166i$ and $\Omega=-1.0179i$ for $\widehat{\alpha}=0.1$ and $0.2$, respectively. This trend continues with $\Omega=-1.0184i$ at $\widehat{\alpha}=0.5$, followed by a gradual shift toward weaker damping with $\Omega=-0.7934i$, $-0.6877i$, and $-0.6661i$ for $\widehat{\alpha}=5$, $10$, and $20$, respectively. For $\ell_{24}=3$, isolated overdamped modes are found at $\Omega=-2.1232i$ and $\Omega=-2.1259i$ for $\widehat{\alpha}=0.1$ and $0.2$, respectively. At $\widehat{\alpha}=0.5$, an overdamped mode occurs at $\Omega=-2.1256i$, while at $\widehat{\alpha}=5$ its value reads $\Omega=-1.6193i$. Finally, the spectrum continues to evolve with purely imaginary modes at $\Omega=-1.4111i$ for $\widehat{\alpha}=10$ and $\Omega=-1.3850i$ for $\widehat{\alpha}=20$. Concerning the non-monotonic behavior in the real part of the QNMs, we find a striking contrast with the patterns observed in all other dimensions: it manifests exclusively at $\widehat{\alpha}=0$, consistently across all tested values of $\ell_{24}$ (see bold entries in Table~\ref{table:D26s1L123}). Regarding the morphology of the QNM spectrum, we observe a truncated coupe shape for all values of $\ell_{24}$ at $\widehat{\alpha}=0$, which transitions into a full coupe distribution in the range $0.1\leqslant\widehat{\alpha}\leqslant 0.2$. A pyramidal structure emerges only at $\widehat{\alpha}=0.5$ for $\ell_{24}\in\{2,3\}$, and at $\widehat{\alpha}=5$ for $\ell_{24}=5$. Furthermore, for $5\leqslant\widehat{\alpha}\leqslant 20$ and $\ell_{24}\in\{1,3\}$, the spectrum contains only a small number of oscillatory modes, accompanied by isolated overdamped modes. A similar pattern is also observed for $\ell_{24}=2$ in the range $10\leqslant\widehat{\alpha}\leqslant 20 $.

\begin{table}%[ht]
\centering
\caption{QNM modes for vector perturbations (spin $s = 1$) of a five-dimensional ($n=3$) Schwarzschild black hole with GB correction are presented in the table below for different values of $\widehat{\alpha}$ and $\ell_3\in\{1,2\}$. The corresponding results are obtained through our SM, utilising $300$ polynomials with a precision of $300$ digits. In this context, $\Omega$ and $N$ represent the dimensionless frequency and the corresponding overtone, respectively. The notation 'N/A' indicates data not available, while 'SM' stands for Spectral Method.}
\label{table:D5s1L1and2}
\vspace*{1em}
\begin{tabular}{||c|c|c|c|c|c|c|c|c|c|c|c|c|c||}
\hline\hline
$\widehat{\alpha}$ $(D=5)$ & $\ell_3$ & $N$ & $\Omega$ (WKB) \cite{Chakrabarti2007GRG} & $\Omega$ (SM) & $\ell_3$ & $N$ & $\Omega$ (WKB) \cite{Chakrabarti2007GRG} & $\Omega$ (SM) \\ [0.5ex]
\hline\hline
$0.0$   & $1$  & $0$ & N/A                                 & \textcolor{red}{$0.3775-0.2711i$}  & $2$  & $0$ & N/A                                  &\textcolor{red}{$0.8019-0.2316i$}\\
        &      & $1$ & N/A                                 & $0.2629-0.9352i$                   &      & $1$ & N/A                                  & $0.6699-0.7227i$\\
        &      & $2$ & N/A                                 & $0.2124-1.6640i$                   &      & $2$ & N/A                                  & $0.3839-1.3609i$\\
        &      & $3$ & N/A                                 & $0.1905-2.3881i$                   &      & $3$ & N/A                                  & $0.3081-2.2068i$\\
        &      & $4$ & N/A                                 & $0.1784-3.1069i$                   &      & $4$ & N/A                                  & $0.2825-2.9556i$\\
$0.1$   & $1$  & $0$ & \textcolor{red}{$0.38852-0.2349i$}  & \textcolor{red}{$0.3851-0.2644i$}  & $2$  & $0$ & \textcolor{red}{$0.80908-0.22980i$}  & \textcolor{red}{$0.8096-0.2293i$} \\
        &      & $1$ & N/A                                 & $0.2763-0.9016i$                   &      & $1$ & N/A                                  & $0.6838-0.7158i$ \\            
        &      & $2$ & N/A                                 & $0.2199-1.6010i$                   &      & $2$ & N/A                                  & $0.4359-1.3533i$ \\ 
        &      & $3$ & N/A                                 & $0.1880-2.2973i$                   &      & $3$ & N/A                                  & $0.3527-2.1278i$ \\          
        &      & $4$ & N/A                                 & $0.1626-2.9889i$                   &      & $4$ & N/A                                  & $0.3102-2.8455i$ \\           
$0.2$   & $1$  & $0$ & \textcolor{red}{$0.38845-0.223310i$}& \textcolor{red}{$0.3919-0.2580i$}  & $2$  & $0$ & \textcolor{red}{$0.81282-0.22383i$}  & \textcolor{red}{$0.8170-0.2269i$} \\
        &      & $1$ & N/A                                 & $0.2857-0.8699i$                   &      & $1$ & N/A                                  & $0.6966-0.7077i$ \\           
        &      & $2$ & N/A                                 & $0.2186-1.5403i$                   &      & $2$ & N/A                                  & $0.4735-1.3261i$ \\ 
        &      & $3$ & N/A                                 & $0.1689-2.2079i$                   &      & $3$ & N/A                                  & $0.3768-2.0517i$ \\           
        &      & $4$ & N/A                                 & $0.1197-2.8725i$                   &      & $4$ & N/A                                  & $0.3143-2.7369i$ \\  
$0.5$   & $1$  & $0$ & \textcolor{red}{$0.38801-0.22691i$} & \textcolor{red}{$0.4078-0.2414i$}  & $2$  & $0$ & \textcolor{red}{$0.82668-0.20401i$}  & \textcolor{red}{$0.8371-0.2201i$} \\
        &      & $1$ & N/A                                 & $0.2962-0.7868i$                   &      & $1$ & N/A                                  & $0.7294-0.6819i$ \\            
        &      & $2$ & N/A                                 & $0.1695-1.3746i$                   &      & $2$ & N/A                                  & $0.5460-1.2322i$ \\ 
        &      & $3$ & N/A                                 & N/A                                &      & $3$ & N/A                                  & $0.4080-1.8362i$ \\            
        &      & $4$ & N/A                                 & N/A                                &      & $4$ & N/A                                  & $0.2922-2.4163i$ \\ 
        &      & $5$ & N/A                                 & N/A                                &      & $5$ & N/A                                  & $0.1446-2.9901i$ \\
$1.0$   & $1$  & $0$ & \textcolor{red}{$0.38612-0.21348i$} & \textcolor{red}{$0.4262-0.2207i$}  & $2$  & $0$ & \textcolor{red}{$0.86280-0.16695i$}  & \textcolor{red}{$0.8669-0.2099i$} \\        
        &      & $1$ & N/A                                 & $0.2765-0.6929i$                   &      & $1$ & N/A                                  & $0.7766-0.6427i$ \\            
        &      & $2$ & N/A                                 & $0.1432-1.2565i$                   &      & $2$ & N/A                                  & $0.6156-1.1101i$ \\ 
        &      & $3$ & N/A                                 & N/A                                &      & $3$ & N/A                                  & $0.4418-1.6132i$ \\            
        &      & $4$ & N/A                                 & N/A                                &      & $4$ & N/A                                  & ${\bf{0.3594-2.1302i}}$ \\ 
        &      & $5$ & N/A                                 & N/A                                &      & $5$ & N/A                                  & $0.3674-2.7250i$ \\
$1.9$   & $1$  & $0$ & N/A                                 & \textcolor{red}{$0.4540-0.1914i$}  & $2$  & $0$ & N/A                                  & \textcolor{red}{$0.9101-0.1970i$} \\
        &      & $1$ & N/A                                 & N/A                                &      & $1$ & N/A                                  & $0.8374-0.6065i$ \\           
$1.99$  & $1$  & $0$ & N/A                                 & N/A                                & $2$  & $0$ & N/A                                  & \textcolor{red}{$0.9135-0.1962i$} \\[1ex]
 \hline\hline 
 \end{tabular}
\end{table}

\begin{table}%[ht]
\centering
\caption{QNM modes for vector perturbations (spin $s = 1$) of a five-dimensional ($n = 3$) Schwarzschild black hole with GB correction are presented in the table below for different values of $\widehat{\alpha}$ and $\ell_3\in\{3,4\}$. The corresponding results are obtained through our SM, utilising $300$ polynomials with a precision of $300$ digits. In this context, $\Omega$ and $N$ represent the dimensionless frequency and the corresponding overtone, respectively. The notation 'N/A' indicates data not available, while 'SM' stands for Spectral Method.}
\label{table:D5s1L3and4}
\vspace*{1em}
\begin{tabular}{||c|c|c|c|c|c|c|c|c|c|c|c|c|c||}
\hline\hline
$\widehat{\alpha}$ $(D=5)$ & $\ell_3$ & $N$ & $\Omega$ (WKB) \cite{Chakrabarti2007GRG} & $\Omega$ (SM) & $\ell_3$ & $N$ & $\Omega$ (WKB)\cite{Chakrabarti2007GRG} & $\Omega$ (SM) \\ [0.5ex]
\hline\hline
$0.0$   & $3$  & $0$ & N/A                                 &\textcolor{red}{$1.2200-0.2361i$}                & $4$  & $0$ & N/A                                  &\textcolor{red}{$1.6126-0.2404i$}\\
        &      & $1$ & N/A                                 & $1.1442-0.7212i$                                &      & $1$ & N/A                                  & $1.5559-0.7295i$\\
        &      & $2$ & N/A                                 & $0.9996-1.2484i$                                &      & $2$ & N/A                                  & $1.4462-1.2444i$\\
        &      & $3$ & N/A                                 & $0.8121-1.8454i$                                &      & $3$ & N/A                                  & $1.2950-1.8037i$\\
        &      & $4$ & N/A                                 & $0.6221-2.5133i$                                &      & $4$ & N/A                                  & $1.1253-2.4197i$\\
$0.1$   & $3$  & $0$ & \textcolor{red}{$1.22732-0.23143i$} & \textcolor{red}{$1.2271-0.2330i$}  & $4$  & $0$ & \textcolor{red}{$1.62271-0.23563i$}  & \textcolor{red}{$1.6196-0.2369i$} \\
        &      & $1$ & N/A                                 & $1.1535-0.7110i$                   &      & $1$ & N/A                                  & $1.5647-0.7181i$ \\                
        &      & $2$ & N/A                                 & $1.0121-1.2271i$                   &      & $2$ & N/A                                  & $1.4581-1.2225i$ \\ 
        &      & $3$ & N/A                                 & $0.8235-1.8066i$                   &      & $3$ & N/A                                  & $1.3098-1.7664i$ \\              
        &      & $4$ & N/A                                 & $0.6159-2.4558i$                   &      & $4$ & N/A                                  & $1.1393-2.3607i$ \\                
$0.2$   & $3$  & $0$ & \textcolor{red}{$1.23556-0.22612i$} & \textcolor{red}{$1.2339-0.2301i$}  & $4$  & $0$ & \textcolor{red}{$1.63381-0.23061i$}  & \textcolor{red}{$1.6264-0.2334i$} \\
        &      & $1$ & N/A                                 & $1.1621-0.7009i$                   &      & $1$ & N/A                                  & $1.5729-0.7070i$ \\                
        &      & $2$ & N/A                                 & $1.0235-1.2059i$                   &      & $2$ & N/A                                  & $1.4688-1.2011i$ \\ 
        &      & $3$ & N/A                                 & $0.8346-1.7683i$                   &      & $3$ & N/A                                  & $1.3229-1.7302i$ \\              
        &      & $4$ & N/A                                 & $0.6147-2.4025i$                   &      & $4$ & N/A                                  & $1.1514-2.3040i$ \\  
$0.5$   & $3$  & $0$ & \textcolor{red}{$1.26382-0.20983i$} & \textcolor{red}{$1.2526-0.2217i$}  & $4$  & $0$ & \textcolor{red}{$1.67107-0.21492i$}  & \textcolor{red}{$1.6449-0.2238i$} \\
        &      & $1$ & N/A                                 & $1.1850-0.6729i$                   &      & $1$ & N/A                                  & $1.5947-0.6763i$ \\                
        &      & $2$ & N/A                                 & $1.0530-1.1497i$                   &      & $2$ & N/A                                  & $1.4962-1.1438i$ \\ 
        &      & $3$ & N/A                                 & $0.8682-1.6751i$                   &      & $3$ & N/A                                  & $1.3543-1.6370i$ \\              
        &      & $4$ & N/A                                 & $0.6811-2.2775i$                   &      & $4$ & N/A                                  & $1.1788-2.1671i$ \\ 
        &      & $5$ & N/A                                 & $0.5619-2.8956i$                   &      & $5$ & N/A                                  & $0.9897-2.7479i$ \\
$1.0$   & $3$  & $0$ & \textcolor{red}{$1.32628-0.18150i$} & \textcolor{red}{$1.2793-0.2103i$}  & $4$  & $0$ & \textcolor{red}{$1.75070-0.18540i$}  & \textcolor{red}{$1.6702-0.2112i$} \\             
        &      & $1$ & N/A                                 & $1.2193-0.6376i$                   &      & $1$ & N/A                                  & $1.6253-0.6380i$ \\                
        &      & $2$ & N/A                                 & $1.1023-1.0862i$                   &      & $2$ & N/A                                  & $1.5362-1.0783i$ \\ 
        &      & $3$ & N/A                                 & $0.9389-1.5687i$                   &      & $3$ & N/A                                  & $1.4051-1.5422i$ \\              
        &      & $4$ & N/A                                 & $0.7658-2.0782i$                   &      & $4$ & N/A                                  & $1.2407-2.0365i$ \\ 
        &      & $5$ & N/A                                 & $0.6660-2.6134i$                   &      & $5$ & N/A                                  & $1.0711-2.5566i$ \\
$1.9$   & $3$  & $0$ & N/A                                 & \textcolor{red}{$1.3133-0.2005i$}  & $4$  & $0$ & N/A                                  & \textcolor{red}{$1.6978-0.2021i$} \\
        &      & $1$ & N/A                                 & $1.2677-0.6120i$                   &      & $1$ & N/A                                  & $1.6641-0.6135i$ \\                
        &      & $2$ & N/A                                 & $1.1719-1.0561i$                   &      & $2$ & N/A                                  & $1.5943-1.0464i$ \\ 
        &      & $3$ & N/A                                 & $0.9514-1.5150i$                   &      & $3$ & N/A                                  & ${\bf{0.7433-2.1182i}}$ \\        
$1.99$  & $3$  & $0$ & N/A                                 & \textcolor{red}{$1.3156-0.2001i$}  & $4$  & $0$ & N/A                                   & \textcolor{red}{$1.6993-0.2020i$} \\[1ex]
 \hline\hline 
 \end{tabular}
\end{table}

\begin{table}%[ht]
\centering
\caption{QNM modes for vector perturbations (spin $s = 1$) of a five-dimensional ($n = 3$) Schwarzschild black hole with GB correction are presented in the table below for different values of $\widehat{\alpha}$ and $\ell_3=1$. The corresponding results are obtained through our SM, utilising $300$ polynomials with a precision of $300$ digits. In this context, $\Omega$ and $N$ represent the dimensionless frequency and the corresponding overtone, respectively. The notation 'N/A' indicates data not available, while 'SM' stands for Spectral Method. Moreover, $\Omega_o$ indicates the overdamped modes and $\Delta\Omega=\Omega_N-\Omega_{N+1}$.}
\label{table:D5s1L1diffalpha}
\vspace*{1em}
\begin{tabular}{||c|c|c|c|c|c|c|c|c|c|c|c|c|c||}
\hline\hline
$\widehat{\alpha}$ $(D=5)$ & $\ell_3$ & $N$ & $\Omega$ (SM) & $\Omega_o$ & $\Delta\Omega$\\ [0.5ex]
\hline\hline
$0.6$                      & $1$      & $0$ & \textcolor{red}{$0.4120-0.2366i$} & $0.0000-17.7791i$ & $0.7245i$\\
                           &          & $1$ & $0.2952-0.7630i$                  & $0.0000-18.5036i$ & $0.7245i$\\                
                           &          & $2$ & $0.1313-1.3284i$                  & $0.0000-19.2281i$ & $0.7246i$\\ 
                           &          & $3$ & $0.0946-2.0467i$                  & $0.0000-19.9527i$ & $0.7245i$\\
                           &          & $4$ & $0.0834-2.6222i$                  & $0.0000-20.6772i$ & $0.7246i$\\ 
                           &          & $5$ & N/A                               & $0.0000-21.4018i$ & -\\
$0.7$                      & $1$      & $0$ & \textcolor{red}{$0.4159-0.2321i$} & $0.0000-14.8405i$ & $0.7226i$\\
                           &          & $1$ & $0.2923-0.7414i$                  & $0.0000-15.5631i$ & $0.7227i$\\                
                           &          & $2$ & N/A                               & $0.0000-16.2858i$ & $0.7227i$\\ 
                           &          & $3$ & N/A                               & $0.0000-17.0085i$ & $0.7227i$\\  
                           &          & $4$ & N/A                               & $0.0000-17.7312i$ & $0.7228i$\\ 
                           &          & $5$ & N/A                               & $0.0000-18.4540i$ & -\\
$0.8$                      & $1$      & $0$ & \textcolor{red}{$0.4195-0.2280i$} & $0.0000-11.9143i$ & $0.7205i$\\
                           &          & $1$ & $0.2878-0.7221i$                  & $0.0000-12.6348i$ & $0.7206i$\\                
                           &          & $2$ & $0.1192-1.3590i$                  & $0.0000-13.3554i$ & $0.7206i$\\ 
                           &          & $3$ & $0.0941-1.8864i$                  & $0.0000-14.0760i$ & $0.7207i$\\ 
                           &          & $4$ & N/A                               & $0.0000-14.7967i$ & $0.7207i$\\ 
                           &          & $5$ & N/A                               & $0.0000-15.5174i$ & -\\
$0.9$                      & $1$      & $0$ & \textcolor{red}{$0.4229-0.2242i$} & $0.0000-10.4396i$ & $0.7185i$\\
                           &          & $1$ & $0.2822-0.7058i$                  & $0.0000-11.1581i$ & $0.7185i$\\                
                           &          & $2$ & $0.1405-1.3099i$                  & $0.0000-11.8766i$ & $0.7186i$\\ 
                           &          & $3$ & N/A                               & $0.0000-12.5952i$ & $0.7186i$\\
                           &          & $4$ & N/A                               & $0.0000-13.3138i$ & $0.7187i$\\ 
                           &          & $5$ & N/A                               & $0.0000-14.0325i$ & -\\[1ex]
 \hline\hline 
 \end{tabular}
\end{table}

\begin{table}%[ht]
\centering
\caption{Overdamped QNM modes for vector perturbations (spin $s = 1$) of a five-dimensional ($n=3$) Schwarzschild black hole with GB correction are presented in the table below for different values of $\widehat{\alpha}$ and $\ell_3\in\{1,2,3\}$. The corresponding results are obtained through our SM, utilising $300$ polynomials with a precision of $300$ digits. In this context, $\Omega$ and $N$ represent the dimensionless frequency and the corresponding overtone, respectively. The notation 'SM' stands for Spectral Method. Moreover, $\Delta\Omega=\Omega_{N}-\Omega_{N+1}$.}
\label{table:piD5L123}
\vspace*{1em}
\begin{tabular}{||c|c|c|c|c|c|c|c|c|c|c|c|c||}
\hline\hline
$\widehat{\alpha}$ $(D=5)$ & $\ell_3$ & $N$ & $\Omega$ (SM)    &$\Delta\Omega$ & $\ell_3$ & $N$ & $\Omega$ (SM)    &$\Delta\Omega$ & $\ell_3$ & $N$ & $\Omega$ (SM)    &$\Delta\Omega$\\ [0.5ex]
\hline\hline
$0.2$              & $1$      & $0$ & $0.0000-37.4920i$ & $0.7271i$ & $2$ & $0$ & $0.0000-37.4719i$ & $0.7275i$ & $3$ & $0$ & $0.0000-38.1719i$ & $0.7279i$\\
                   &          & $1$ & $0.0000-38.2191i$ & $0.7270i$ &     & $1$ & $0.0000-38.1994i$ & $0.7274i$ &     & $1$ & $0.0000-38.8998i$ & $0.7279i$\\
                   &          & $2$ & $0.0000-38.9461i$ & $0.7270i$ &     & $2$ & $0.0000-38.9268i$ & $0.7274i$ &     & $2$ & $0.0000-39.6277i$ & $0.7278i$\\
                   &          & $3$ & $0.0000-39.6731i$ & $0.7271i$ &     & $3$ & $0.0000-39.6542i$ & $0.7274i$ &     & $3$ & $0.0000-40.3555i$ & $0.7279i$\\
                   &          & $4$ & $0.0000-40.4002i$ & $0.7270i$ &     & $4$ & $0.0000-40.3816i$ & $0.7274i$ &     & $4$ & $0.0000-41.0834i$ & $0.7278i$\\
                   &          & $5$ & $0.0000-41.1272i$ & -         &     & $5$ & $0.0000-41.1090i$ & -         &     & $5$ & $0.0000-41.8112i$ & -        \\
$0.5$              & $1$      & $0$ & $0.0000-19.9995i$ & $0.7260i$ & $2$ & $0$ & $0.0000-22.1507i$ & $0.7270i$ & $3$ & $0$ & $0.0000-22.1131i$ & $0.7281i$\\
                   &          & $1$ & $0.0000-20.7255i$ & $0.7261i$ &     & $1$ & $0.0000-22.8777i$ & $0.7269i$ &     & $1$ & $0.0000-22.8412i$ & $0.7280i$\\
                   &          & $2$ & $0.0000-21.4516i$ & $0.7260i$ &     & $2$ & $0.0000-23.6046i$ & $0.7269i$ &     & $2$ & $0.0000-23.5692i$ & $0.7280i$\\
                   &          & $3$ & $0.0000-22.1776i$ & $0.7261i$ &     & $3$ & $0.0000-24.3315i$ & $0.7269i$ &     & $3$ & $0.0000-24.2972i$ & $0.7278i$\\
                   &          & $4$ & $0.0000-22.9037i$ & $0.7262i$ &     & $4$ & $0.0000-25.0584i$ & $0.7269i$ &     & $4$ & $0.0000-25.0250i$ & $0.7278i$\\
                   &          & $5$ & $0.0000-23,6299i$ & -         &     & $5$ & $0.0000-25.7853i$ & -         &     & $5$ & $0.0000-25.7528i$ & -        \\
$1.0$              & $1$      & $0$ & $0.0000-8.9728i$  & $0.7163i$ & $2$ & $0$ & $0.0000-9.6382i$  & $0.7199i$ & $3$ & $0$ & $0.0000-9.5666i$  & $0.7248i$\\
                   &          & $1$ & $0.0000-9.6891i$  & $0.7164i$ &     & $1$ & $0.0000-10.3581i$ & $0.7194i$ &     & $1$ & $0.0000-10.2914i$ & $0.7237i$\\
                   &          & $2$ & $0.0000-10.4055i$ & $0.7164i$ &     & $2$ & $0.0000-11.0775i$ & $0.7191i$ &     & $2$ & $0.0000-11.0151i$ & $0.7229i$\\
                   &          & $3$ & $0.0000-11.1219i$ & $0.7165i$ &     & $3$ & $0.0000-11.7966i$ & $0.7189i$ &     & $3$ & $0.0000-11.7380i$ & $0.7221i$\\
                   &          & $4$ & $0.0000-11.8384i$ & $0.7165i$ &     & $4$ & $0.0000-12.5155i$ & $0.7186i$ &     & $4$ & $0.0000-12.4601i$ & $0.7216i$\\
                   &          & $5$ & $0.0000-12.5549i$ & -         &     & $5$ & $0.0000-13.2341i$ & -         &     & $5$ & $0.0000-13.1817i$ & -       \\
$1.9$              & $1$      & $0$ & $0.0000-0.6330i$  & $0.5351i$ & $2$ & $0$ & $0.0000-0.9004i$  & $0.7145i$ & $3$ & $0$ & $0.0000-1.1205i$  & $0.9450i$\\
                   &          & $1$ & $0.0000-1.1681i$  & $0.5834i$ &     & $1$ & $0.0000-1.6149i$  & $0.6704i$ &     & $1$ & $0.0000-2.0655i$  & $0.7586i$\\
                   &          & $2$ & $0.0000-1.7515i$  & $0.6797i$ &     & $2$ & $0.0000-2.2853i$  & $0.7169i$ &     & $2$ & $0.0000-2.8241i$  & $0.7466i$\\
                   &          & $3$ & $0.0000-2.4312i$  & $0.6961i$ &     & $3$ & $0.0000-3.0022i$  & $0.7185i$ &     & $3$ & $0.0000-3.5707i$  & $0.7359i$\\
                   &          & $4$ & $0.0000-3.1273i$  & $0.6984i$ &     & $4$ & $0.0000-3.7207i$  & $0.7137i$ &     & $4$ & $0.0000-4.3066i$  & $0.7267i$\\
                   &          & $5$ & $0.0000-3.8257i$  & -         &     & $5$ & $0.0000-4.4344i$  & -         &     & $5$ & $0.0000-5.0333i$  & -       \\
$1.99$             & $1$      & $0$ & $0.0000-0.6294i$  & $0.5260i$ & $2$ & $0$ & $0.0000-0.8978i$  & $0.7062i$ & $4$ & $0$ & $0.0000-1.1201i$  & $0.9384i$\\
                   &          & $1$ & $0.0000-1.1554i$  & $0.5857i$ &     & $1$ & $0.0000-1.6040i$  & $0.6715i$ &     & $1$ & $0.0000-2.0585i$  & $0.7576i$\\
                   &          & $2$ & $0.0000-1.7411i$  & $0.6797i$ &     & $2$ & $0.0000-2.2755i$  & $0.7185i$ &     & $2$ & $0.0000-2.8161i$  & $0.7447i$\\
                   &          & $3$ & $0.0000-2.4208i$  & $0.6962i$ &     & $3$ & $0.0000-2.9940i$  & -         &     & $3$ & $0.0000-3.56801i$ & -       \\
                   &          & $4$ & $0.0000-3.1170i$  & -         &     & $4$ & -                 & -         &     & $4$ & -                 & -\\
[1ex]
 \hline\hline 
 \end{tabular}
\end{table}

\begin{table}%[ht]
\centering
\caption{Overdamped QNM modes for vector perturbations (spin $s = 1$) of a five-dimensional ($n = 3$) Schwarzschild black hole with GB correction are presented in the table below for different values of $\widehat{\alpha}$ and $\ell_3=4$. The corresponding results are obtained through our SM, utilising $300$ polynomials with a precision of $300$ digits. In this context, $\Omega$ and $N$ represent the dimensionless frequency and the corresponding overtone, respectively. The notation 'SM' stands for Spectral Method. Moreover, $\Delta\Omega=\Omega_{N}-\Omega_{N+1}$.}
\label{table:piD5L4I}
\vspace*{1em}
\begin{tabular}{||c|c|c|c|c|c|c|c|c|c|c|c|c||}
\hline\hline
$\widehat{\alpha}$ $(D=5)$ & $\ell_3$ & $N$ & $\Omega$ (SM)    &$\Delta\Omega$\\ [0.5ex]
\hline\hline
$0.2$              & $4$      & $0$ & $0.0000-38.8650i$ & $0.7286i$ \\
                   &          & $1$ & $0.0000-39.5936i$ & $0.7284i$ \\
                   &          & $2$ & $0.0000-40.3220i$ & $0.7285i$ \\
                   &          & $3$ & $0.0000-41.0505i$ & $0.7284i$ \\
                   &          & $4$ & $0.0000-41.7789i$ & $0.7283i$ \\
                   &          & $5$ & $0.0000-42.5072i$ & -         \\
$0.5$              & $4$      & $0$ & $0.0000-21.3346i$ & $0.7299i$ \\
                   &          & $1$ & $0.0000-22.0645i$ & $0.7297i$ \\
                   &          & $2$ & $0.0000-22.7492i$ & $0.7295i$ \\
                   &          & $3$ & $0.0000-23.5237i$ & $0.7293i$ \\
                   &          & $4$ & $0.0000-24.2530i$ & $0.7291i$ \\
                   &          & $5$ & $0.0000-24.9821i$ & -         \\
$1.0$              & $4$      & $0$ & $0.0000-1.2432i$  & $0.7593i$ \\
                   &          & $1$ & $0.0000-2.0025i$  & $0.7370i$ \\
                   &          & $2$ & $0.0000-2.7395i$  & $0.7268i$ \\
                   &          & $3$ & $0.0000-3.4663i$  & $0.7287i$ \\
                   &          & $4$ & $0.0000-4.1950i$  & $0.7327i$ \\
                   &          & $5$ & $0.0000-4.9277i$  & $0.7370i$ \\
                   &          & $6$ & $0.0000-5.6647i$  & -         \\
                   &          & $7$ & $0.0000-9.4740i$  & $0.7311i$ \\
                   &          & $8$ & $0.0000-10.2051i$ & $0.7293i$ \\
                   &          & $9$ & $0.0000-10.9344i$ & $0.7278i$ \\
                   &          & $10$& $0.0000-11.6622i$ & $0.7265i$ \\
                   &          & $11$& $0.0000-12.3887i$ & $0.7255i$ \\
                   &          & $12$& $0.0000-13.1142i$ & -         \\[1ex]
 \hline\hline 
 \end{tabular}
\end{table}

\begin{table}%[ht]
\centering
\caption{Overdamped QNM modes for vector perturbations (spin $s = 1$) of a five-dimensional ($n = 3$) Schwarzschild black hole with GB correction are presented in the table below for different values of $\widehat{\alpha}$ and $\ell_3=4$. The corresponding results are obtained through our SM, utilising $300$ polynomials with a precision of $300$ digits. In this context, $\Omega$ and $N$ represent the dimensionless frequency and the corresponding overtone, respectively. The notation 'SM' stands for Spectral Method. Moreover, $\Delta\Omega=\Omega_{N}-\Omega_{N+1}$.}
\label{table:piD5L4II}
\vspace*{1em}
\begin{tabular}{||c|c|c|c|c|c|c|c|c|c|c|c|c||}
\hline\hline
$\widehat{\alpha}$ $(D=5)$ & $\ell_3$ & $N$ & $\Omega$ (SM)    &$\Delta\Omega$\\ [0.5ex]
\hline\hline
$1.9$              & $4$      & $0$ & $0.0000-0.0834i$  & $0.1403i$ \\
                   &          & $1$ & $0.0000-0.2237i$  & $0.1138i$ \\
                   &          & $2$ & $0.0000-0.3375i$  & $0.0991i$ \\
                   &          & $3$ & $0.0000-0.4366i$  & -         \\
                   &          & $4$ & $0.0000-1.2850i$  & $1.2225i$ \\
                   &          & $5$ & $0.0000-2.5075i$  & $0.8555i$ \\
                   &          & $6$ & $0.0000-3.3630i$  & $0.7733i$ \\
                   &          & $7$ & $0.0000-4.1363i$  & $0.7508i$ \\
                   &          & $8$ & $0.0000-4.8871i$  & $0.7378i$ \\
                   &          & $9$& $0.0000- 5.6249i$  & $0.7289i$ \\
                   &          & $10$& $0.0000-6.3538i$  & -         \\
$1.99$             & 4        & $0$ & $0.0000-0.0070i$  & $0.0469i$ \\
                   &          & $1$ & $0.0000-0.0539i$  & $0.03262i$ \\
                   &          & $2$ & $0.0000-0.0901i$  & -         \\
                   &          & $3$ & $0.0000-1.2855i$  & $1.2217$ \\
                   &          & $4$ & $0.0000-2.5072i$  & -        \\[1ex]
 \hline\hline 
 \end{tabular}
\end{table}

\begin{table}%[ht]
\centering
\caption{QNM modes for vector perturbations (spin $s = 1$) of a six-dimensional ($n = 4$) Schwarzschild black hole with GB correction are presented in the table below for different values of $\widehat{\alpha}$ and $\ell_4\in\{1,2\}$. The corresponding results are obtained through our SM, utilising $300$ polynomials with a precision of $300$ digits. In this context, $\Omega$ and $N$ represent the dimensionless frequency and the corresponding overtone, respectively. The notation 'N/A' indicates data not available, while 'SM' stands for Spectral Method.}
\label{table:D6s1L1and2}
\vspace*{1em}
\begin{tabular}{||c|c|c|c|c|c|c|c|c|c|c|c|c|c||}
\hline\hline
$\widehat{\alpha}$ $(D=6)$ & $\ell_4$ & $N$ & $\Omega$ (WKB) \cite{Chakrabarti2007GRG} & $\Omega$ (SM) & $\ell_4$ & $N$ & $\Omega$ (WKB) \cite{Chakrabarti2007GRG} & $\Omega$ (SM) \\ [0.5ex]
\hline\hline
$0.0$   & $1$  & $0$ & N/A                                 & \textcolor{red}{$0.7059-0.4231i$}  & $2$  & $0$ & N/A                                  &\textcolor{red}{$1.2101-0.3763i$}\\
        &      & $1$ & N/A                                 & $0.4368-1.5056i$                   &      & $1$ & N/A                                  & $0.9172-1.2027i$\\
        &      & $2$ & N/A                                 & $0.3291-2.7701i$                   &      & $2$ & N/A                                  & $0.5328-2.4564i$\\
        &      & $3$ & N/A                                 & $0.2931-4.0052i$                   &      & $3$ & N/A                                  & $0.4220-3.7658i$\\
        &      & $4$ & N/A                                 & $0.2751-5.2230i$                   &      & $4$ & N/A                                  & $0.3725-5.0272i$\\
$0.1$   & $1$  & $0$ & \textcolor{red}{$0.71811-0.37339i$}  & \textcolor{red}{$0.7127-0.4126i$} & $2$  & $0$ & \textcolor{red}{$1.22692-0.37114i$} & \textcolor{red}{$1.2178-0.3697i$} \\
        &      & $1$ & N/A                                 & $0.4547-1.4484i$                   &      & $1$ & N/A                                  & $0.9470-1.1777i$ \\                
        &      & $2$ & N/A                                 & $0.3288-2.6669i$                   &      & $2$ & N/A                                  & $0.5710-2.3581i$ \\ 
        &      & $3$ & N/A                                 & $0.2719-3.8657i$                   &      & $3$ & N/A                                  & $0.4315-3.6209i$ \\              
        &      & $4$ & N/A                                 & $0.2316-5.0507i$                   &      & $4$ & N/A                                  & $0.3545-4.8467i$ \\                
$0.2$   & $1$  & $0$ & \textcolor{red}{$0.71772-0.37044i$} & \textcolor{red}{$0.7183-0.4031i$}  & $2$  & $0$ & \textcolor{red}{$1.22918-0.36377i$}  & \textcolor{red}{$1.2248-0.3638i$} \\
        &      & $1$ & N/A                                 & $0.4687-1.3976i$                   &      & $1$ & N/A                                  & $0.9712-1.1540i$ \\                
        &      & $2$ & N/A                                 & $0.3238-2.5724i$                   &      & $2$ & N/A                                  & $0.6011-2.2717i$ \\ 
        &      & $3$ & N/A                                 & $0.2446-3.7330i$                   &      & $3$ & N/A                                  & $0.4359-3.4897i$ \\              
        &      & $4$ & N/A                                 & $0.1805-4.8736i$                   &      & $4$ & N/A                                  & $0.3367-4.6759i$ \\  
$0.5$   & $1$  & $0$ & \textcolor{red}{$0.71640-0.36036i$} & \textcolor{red}{$0.7303-0.3797i$}  & $2$  & $0$ & \textcolor{red}{$1.23772-0.33942i$}  & \textcolor{red}{$1.2433-0.3487i$} \\
        &      & $1$ & N/A                                 & $0.4958-1.2744i$                   &      & $1$ & N/A                                  & $1.0275-1.0942i$ \\                
        &      & $2$ & N/A                                 & $0.2939-2.3209i$                   &      & $2$ & N/A                                  & $0.6867-2.0621i$ \\ 
        &      & $3$ & N/A                                 & $0.1262-3.3274i$                   &      & $3$ & N/A                                  & $0.4787-3.1312i$ \\              
        &      & $4$ & N/A                                 & N/A                                &      & $4$ & N/A                                  & $0.3433-4.1439i$ \\ 
        &      & $5$ & N/A                                 & N/A                                &      & $5$ & N/A                                  & $0.2340-5.0954i$ \\
$1.0$   & $1$  & $0$ & \textcolor{red}{$0.71389-0.33903i$} & \textcolor{red}{$0.7445-0.3523i$}  & $2$  & $0$ & \textcolor{red}{$1.26100-0.29391i$}  & \textcolor{red}{$1.2703-0.3305i$}\\ 
        &      & $1$ & N/A                                 & $0.5254-1.1355i$                   &      & $1$ & N/A                                  & $1.1008-1.0264i$ \\                
        &      & $2$ & N/A                                 & ${\bf{0.2613-1.9939i}}$                   &      & $2$ & N/A                                  & $0.8252-1.8429i$ \\ 
        &      & $3$ & N/A                                 & $0.5573-8.4672i$                   &      & $3$ & N/A                                  & $0.5746-2.7225i$ \\              
        &      & $4$ & N/A                                 & $0.6983-9.3233i$                   &      & $4$ & N/A                                  & ${\bf{0.3523-3.6880i}}$ \\ 
        &      & $5$ & N/A                                 & $0.8129-10.1803i$                  &      & $5$ & N/A                                  & $0.5076-4.6522i$ \\
        &      & $6$ & N/A                                 & $0.9138-11.0344i$                  &      & $5$ & N/A                                  & $0.5252-5.6823i$ \\
        &      & $7$ & N/A                                 & $1.0081-11.8876i$                  &      & $5$ & N/A                                  & $0.6789-6.7619i$ \\
$5.0$   & $1$  & $0$ & \textcolor{red}{$0.95379-0.07595i$} & \textcolor{red}{$0.8840-0.2552i$}  & $2$  & $0$ & \textcolor{red}{$1.79853-0.29350i$}  & \textcolor{red}{$1.3459-0.2818i$} \\
        &      & $1$ & N/A                                 & $0.9675-0.8138i$                   &      & $1$ & N/A                                  & ${\bf{1.1691-0.6457i}}$ \\                
        &      & $2$ & N/A                                 & $1.2548-1.3939i$                   &      & $2$ & N/A                                  & $1.4288-1.0692i$ \\ 
        &      & $3$ & N/A                                 & ${\bf{0.3085-1.4238i}}$            &      & $3$ & N/A                                  & $1.7257-1.6945i$ \\              
        &      & $4$ & N/A                                 & $1.6622-1.9179i$                   &      & $4$ & N/A                                  & ${\bf{0.5939-1.7131i}}$ \\ 
        &      & $5$ & N/A                                 & $2.0833-2.4328i$                   &      & $5$ & N/A                                  & $2.1129-2.2815i$ \\
$10.0$  & $1$  & $0$ & N/A                                 & \textcolor{red}{$1.0295-0.2396i$}  & $2$  & $0$ & N/A                                  & \textcolor{red}{$1.4621-0.1130i$} \\
        &      & $1$ & N/A                                 & $1.2742-0.7356i$                   &      & $1$ & N/A                                  & $1.7172-0.4718i$ \\                
        &      & $2$ & N/A                                 & ${\bf{0.2085-1.0907i}}$            &      & $2$ & N/A                                  & $2.0116-0.9705i$ \\ 
        &      & $3$ & N/A                                 & $1.7397-1.2110i$                   &      & $3$ & N/A                                  & ${\bf{0.4581-1.3637i}}$ \\         
        &      & $4$ & N/A                                 & $2.2626-1.6758i$                   &      & $4$ & N/A                                  & $2.3873-1.4450i$ \\ 
        &      & $5$ & N/A                                 & $2.7966-2.1334i$                   &      & $5$ & N/A                                  & $2.8667-1.9076i$ \\
$15.0$  & $1$  & $0$ & N/A                                 & \textcolor{red}{$1.1255-0.2644i$}  & $2$  & $0$ & N/A                                  & \textcolor{red}{$1.5878-0.1867i$} \\
        &      & $1$ & N/A                                 & $1.3915-0.7140i$                   &      & $1$ & N/A                                  & $2.0267-0.5161i$ \\                
        &      & $2$ & N/A                                 & ${\bf{0.1714-0.9519i}}$            &      & $2$ & N/A                                  & $2.2398-0.9495i$ \\ 
        &      & $3$ & N/A                                 & $1.9220-1.1528i$                   &      & $3$ & N/A                                  & ${\bf{0.4008-1.2117i}}$ \\     
        &      & $4$ & N/A                                 & $2.4833-1.6034i$                   &      & $4$ & N/A                                  & $2.6117-1.3176i$ \\ 
        &      & $5$ & N/A                                 & $3.0619-2.0430i$                   &      & $5$ & N/A                                  & $3.1903-1.7511i$ \\      
$20.0$  & $1$  & $0$ & N/A                                 & \textcolor{red}{$1.1808-0.3114i$}  & $2$  & $0$ & N/A                                  & \textcolor{red}{$1.5911-0.2597i$} \\
        &      & $1$ & N/A                                 & $1.4454-0.6997i$                   &      & $1$ & N/A                                  & $2.1209-0.7932i$ \\                
        &      & $2$ & N/A                                 & ${\bf{0.1513-0.8731i}}$            &      & $2$ & N/A                                  & $2.5111-1.0279i$ \\ 
        &      & $3$ & N/A                                 & $2.0187-1.1258i$                   &      & $3$ & N/A                                  & $2.7643-1.0398i$ \\              
        &      & $4$ & N/A                                 & $2.5870-1.5659i$                   &      & $4$ & N/A                                  & ${\bf{0.3657-1.1207i}}$ \\ 
        &      & $5$ & N/A                                 & $3.1857-1.9997i$                   &      & $5$ & N/A                                  & $3.3826-1.6629i$ \\[1ex]
 \hline\hline 
 \end{tabular}
\end{table}

\begin{table}%[ht]
\centering
\caption{Overdamped QNM modes for vector perturbations (spin $s = 1$) of a six-dimensional ($n = 4$) Schwarzschild black hole with GB correction are presented in the table below for different values of $\widehat{\alpha}$ and $\ell_4\in\{1,2,3\}$. The corresponding results are obtained through our SM, utilising $300$ polynomials with a precision of $300$ digits. In this context, $\Omega$ and $N$ represent the dimensionless frequency and the corresponding overtone, respectively. The notation 'SM' stands for Spectral Method. Moreover, $\Delta\Omega=\Omega_{N}-\Omega_{N+1}$.}
\label{table:piD6L12}
\vspace*{1em}
\begin{tabular}{||c|c|c|c|c|c|c|c|c|c|c|c|c||}
\hline\hline
$\widehat{\alpha}$ $(D=6)$ & $\ell_4$ & $N$ & $\Omega$ (SM)    &$\Delta\Omega$ & $\ell_4$ & $N$ & $\Omega$ (SM)    &$\Delta\Omega$ & $\ell_4$ & $N$ & $\Omega$ (SM)    &$\Delta\Omega$\\ [0.5ex]
\hline\hline
$0.1$              & $1$      & $0$ & $0.0000-88.3814i$ & $1.2173i$ & $2$ & $0$ & $0.0000-88.3650i$ & $1.2176i$ & $3$ & $0$ & $0.0000-88.3432i$ & $1.2180i$\\
                   &          & $1$ & $0.0000-89.5987i$ & $1.2174i$ &     & $1$ & $0.0000-89.5826i$ & $1.2175i$ &     & $1$ & $0.0000-89.5612i$ & $1.2177i$\\
                   &          & $2$ & $0.0000-90.8161i$ & $1.2173i$ &     & $2$ & $0.0000-90.8001i$ & $1.2176i$ &     & $2$ & $0.0000-90.7789i$ & $1.2179i$\\
                   &          & $3$ & $0.0000-92.0334i$ & $1.2174i$ &     & $3$ & $0.0000-92.0177i$ & $1.2176i$ &     & $3$ & $0.0000-91.9968i$ & $1.2178i$\\
                   &          & $4$ & $0.0000-93.2508i$ & $1.2174i$ &     & $4$ & $0.0000-93.2353i$ & $1.2176i$ &     & $4$ & $0.0000-93.2146i$ & $1.2179i$\\
                   &          & $5$ & $0.0000-94.4682i$ & -         &     & $5$ & $0.0000-94.4529i$ & -         &     & $5$ & $0.0000-94.4325i$ & -        \\
$0.2$              & $1$      & $0$ & $0.0000-70.3680i$ & $1.2225i$ & $2$ & $0$ & $0.0000-71.5728i$ & $1.2227i$ & $3$ & $0$ & $0.0000-71.5492i$ & $1.2231i$\\
                   &          & $1$ & $0.0000-71.5905i$ & $1.2224i$ &     & $1$ & $0.0000-72.7955i$ & $1.2228i$ &     & $1$ & $0.0000-72.7723i$ & $1.2232i$\\
                   &          & $2$ & $0.0000-72.8129i$ & $1.2225i$ &     & $2$ & $0.0000-74.0183i$ & $1.2228i$ &     & $2$ & $0.0000-73.9955i$ & $1.2231i$\\
                   &          & $3$ & $0.0000-74.0354i$ & $1.2225i$ &     & $3$ & $0.0000-75.2411i$ & $1.2227i$ &     & $3$ & $0.0000-75.2186i$ & $1.2231i$\\
                   &          & $4$ & $0.0000-75.2579i$ & $1.2225i$ &     & $4$ & $0.0000-76.4638i$ & $1.2228i$ &     & $4$ & $0.0000-76.4417i$ & $1.2232i$\\
                   &          & $5$ & $0.0000-76.4804i$ & -         &     & $5$ & $0.0000-77.6866i$ & -         &     & $5$ & $0.0000-77.6649i$ & -        \\
$0.5$              & $1$      & $0$ & $0.0000-44.5101i$ & $1.2186i$ & $2$ & $0$ & $0.0000-45.7056i$ & $1.2198i$ & $3$ & $0$ & $0.0000-45.6749i$ & $1.2200i$\\
                   &          & $1$ & $0.0000-45.7287i$ & $1.2186i$ &     & $1$ & $0.0000-46.9248i$ & $1.2192i$ &     & $1$ & $0.0000-46.8949i$ & $1.2200i$\\
                   &          & $2$ & $0.0000-46.9473i$ & $1.2186i$ &     & $2$ & $0.0000-48.1440i$ & $1.2192i$ &     & $2$ & $0.0000-48.1149i$ & $1.2199i$\\
                   &          & $3$ & $0.0000-48.1659i$ & $1.2186i$ &     & $3$ & $0.0000-49.3632i$ & $1.2191i$ &     & $3$ & $0.0000-49.3348i$ & $1.2198i$\\
                   &          & $4$ & $0.0000-49.3845i$ & $1.2186i$ &     & $4$ & $0.0000-50.5823i$ & $1.2191i$ &     & $4$ & $0.0000-50.5546i$ & $1.2198i$\\
                   &          & $5$ & $0.0000-50.6031i$ & -         &     & $5$ & $0.0000-51.8014i$ & -         &     & $5$ & $0.0000-51.7744i$ & -       \\[1ex]
 \hline\hline 
 \end{tabular}
\end{table}

\begin{table}%[ht]
\centering
\caption{QNM modes for vector perturbations (spin $s = 1$) of a six-dimensional ($n=4$) Schwarzschild black hole with GB correction are presented in the table below for different values of $\widehat{\alpha}$ and $\ell_4 \in \{3,4\}$. The corresponding results are obtained through our SM, utilising $300$ polynomials with a precision of $300$ digits. In this context, $\Omega$ and $N$ represent the dimensionless frequency and the corresponding overtone, respectively. The notation 'N/A' indicates data not available, while 'SM' stands for Spectral Method.}
\label{table:D6s1L3and4}
\vspace*{1em}
\begin{tabular}{||c|c|c|c|c|c|c|c|c|c|c|c|c|c||}
\hline\hline
$\widehat{\alpha}$ $(D=6)$ & $\ell_4$ & $N$ & $\Omega$ (WKB) \cite{Chakrabarti2007GRG} & $\Omega$ (SM) & $\ell_4$ & $N$ & $\Omega$ (WKB) \cite{Chakrabarti2007GRG} & $\Omega$ (SM) \\ [0.5ex]
\hline\hline
$0.0$   & $3$  & $0$ & N/A                                 & \textcolor{red}{$1.7367-0.3714i$}  & $4$  & $0$ & N/A                                  &\textcolor{red}{$2.2412-0.3750i$}\\
        &      & $1$ & N/A                                 & $1.5633-1.1361i$                   &      & $1$ & N/A                                  & $2.1136-1.1397i$\\
        &      & $2$ & N/A                                 & $1.1950-1.9878i$                   &      & $2$ & N/A                                  & $1.8548-1.9558i$\\
        &      & $3$ & N/A                                 & $0.5567-3.0840i$                   &      & $3$ & N/A                                  & $1.4740-2.8821i$\\
        &      & $4$ & N/A                                 & $0.4345-4.7265i$                   &      & $4$ & N/A                                  & $1.0375-3.9787i$\\
$0.1$   & $3$  & $0$ & \textcolor{red}{$1.74514-0.36568i$} & \textcolor{red}{$1.7440-0.3653i$}  & $4$  & $0$ & \textcolor{red}{$2.25087-0.36801i$} & \textcolor{red}{$2.2487-0.3689i$} \\
        &      & $1$ & N/A                                 & $1.5790-1.1167i$                   &      & $1$ & N/A                                  & $2.1265-1.1202i$ \\                
        &      & $2$ & N/A                                 & $1.2308-1.9486i$                   &      & $2$ & N/A                                  & $1.8793-1.9179i$ \\ 
        &      & $3$ & N/A                                 & $0.6461-3.0228i$                   &      & $3$ & N/A                                  & $1.5142-2.8131i$ \\              
        &      & $4$ & N/A                                 & $0.4717-4.5354i$                   &      & $4$ & N/A                                  & $1.0809-3.8604i$ \\                
$0.2$   & $3$  & $0$ & \textcolor{red}{$1.75199-0.35737i$} & \textcolor{red}{$1.7509-0.3598i$}  & $4$  & $0$ & \textcolor{red}{$2.26119-0.36020i$}  & \textcolor{red}{$2.2557-0.3632i$} \\
        &      & $1$ & N/A                                 & $1.5930-1.0984i$                   &      & $1$ & N/A                                  & $2.1383-1.1018i$ \\                
        &      & $2$ & N/A                                 & $1.2621-1.9105i$                   &      & $2$ & N/A                                  & $1.9012-1.8823i$ \\ 
        &      & $3$ & N/A                                 & $0.7234-2.9400i$                   &      & $3$ & N/A                                  & $1.5506-2.7492i$ \\              
        &      & $4$ & N/A                                 & $0.4942-4.3652i$                   &      & $4$ & N/A                                  & $1.1249-3.7523i$ \\  
$0.5$   & $3$  & $0$ & \textcolor{red}{$1.77616-0.33261i$} & \textcolor{red}{$1.7696-0.3454i$}  & $4$  & $0$ & \textcolor{red}{$2.29598-0.33755i$}  & \textcolor{red}{$2.2751-0.3482i$} \\
        &      & $1$ & N/A                                 & $1.6288-1.0518i$                   &      & $1$ & N/A                                  & $2.1694-1.0545i$ \\                
        &      & $2$ & N/A                                 & $1.3392-1.8180i$                   &      & $2$ & N/A                                  & $1.9568-1.7937i$ \\ 
        &      & $3$ & N/A                                 & $0.9135-2.7664i$                   &      & $3$ & N/A                                  & $1.6400-2.6006i$ \\              
        &      & $4$ & N/A                                 & $0.6790-3.9234i$                   &      & $4$ & N/A                                  & $1.2381-3.5343i$ \\ 
        &      & $5$ & N/A                                 & $0.5762-4.9755i$                   &      & $5$ & N/A                                  & $0.9094-4.7001i$ \\
$1.0$   & $3$  & $0$ & \textcolor{red}{$1.83089-0.29695i$} & \textcolor{red}{$1.7965-0.3282i$}  & $4$  & $0$ & \textcolor{red}{$2.36771-0.30442i$}  & \textcolor{red}{$2.3018-0.3303i$} \\     
        &      & $1$ & N/A                                 & $1.6789-1.0002i$                   &      & $1$ & N/A                                  & $2.2122-1.0009i$ \\                
        &      & $2$ & N/A                                 & $1.4430-1.7300i$                   &      & $2$ & N/A                                  & $2.0296-1.7045i$ \\ 
        &      & $3$ & N/A                                 & $1.1319-2.5760i$                   &      & $3$ & N/A                                  & $1.7531-2.4723i$ \\              
        &      & $4$ & N/A                                 & $0.8715-3.4901i$                   &      & $4$ & N/A                                  & $1.4244-3.3402i$ \\ 
        &      & $5$ & N/A                                 & ${\bf{0.7921-4.4895i}}$                   &      & $5$ & N/A                                  & $1.1909-4.2998i$ \\
        &      & $6$ & N/A                                 & $0.8520-5.5120i$                   &      & $6$ & N/A                                  & $1.1593-5.3347i$ \\
$5.0$   & $3$  & $0$ & \textcolor{red}{$2.60496-0.31368i$} & \textcolor{red}{$1.8343-0.3219i$}  & $4$  & $0$ & \textcolor{red}{$3.35548-0.32099i$}  & \textcolor{red}{$2.3208-0.3176i$} \\
        &      & $1$ & N/A                                 & ${\bf{1.5730-0.4114i}}$            &      & $1$ & N/A                                  & ${\bf{1.9341-0.3505i}}$ \\              
        &      & $2$ & N/A                                 & $1.9523-0.9139i$                   &      & $2$ & N/A                                  & $2.3927-0.8403i$ \\ 
        &      & $3$ & N/A                                 & $2.1504-1.4401i$                   &      & $3$ & N/A                                  & $2.5712-1.3151i$ \\              
        &      & $4$ & N/A                                 & $2.4298-1.9953i$                   &      & $4$ & N/A                                  & $2.8011-1.8444i$ \\ 
        &      & $5$ & N/A                                 & ${\bf{0.9260-2.0451i}}$            &      & $5$ & N/A                                  & $3.0854-2.3813i$ \\
        &      & $6$ & N/A                                 & $2.7626-2.5421i$                   &      & $6$ & N/A                                  & ${\bf{1.2811-2.3873i}}$ \\
$10.0$  & $3$  & $0$ & N/A                                 & \textcolor{red}{$1.9220-0.0803i$}  & $4$  & $0$ & N/A                                  & \textcolor{red}{$2.3378-0.0695i$} \\
        &      & $1$ & N/A                                 & $2.2869-0.3474i$                   &      & $1$ & N/A                                  & $2.8130-0.3052i$ \\                
        &      & $2$ & N/A                                 & $2.5580-0.8189i$                   &      & $2$ & N/A                                  & $3.0863-0.6727i$ \\ 
        &      & $3$ & N/A                                 & $2.7920-1.3177i$                   &      & $3$ & N/A                                  & $3.3780-1.2977i$ \\              
        &      & $4$ & N/A                                 & ${\bf{0.7425-1.6589i}}$            &      & $4$ & N/A                                  & $3.4105-1.5609i$ \\ 
        &      & $5$ & N/A                                 & $3.1791-1.6605i$                   &      & $5$ & N/A                                  & $4.0243-1.9165i$ \\
        &      & $6$ & N/A                                 & $3.6861-2.1146i$                   &      & $6$ & N/A                                  & ${\bf{1.0430-1.9583i}}$ \\
$15.0$  & $3$  & $0$ & N/A                                 & \textcolor{red}{$1.9738-0.1707i$}  & $4$  & $0$ & N/A                                  & \textcolor{red}{$2.3533-0.1465i$} \\
        &      & $1$ & N/A                                 & $2.6043-0.6596i$                   &      & $1$ & N/A                                  & $2.9690-0.6529i$ \\                
        &      & $2$ & N/A                                 & $3.2023-0.7208i$            &      & $2$ & N/A                                  & ${\bf{4.1912-0.7410i}}$ \\ 
        &      & $3$ & N/A                                 & ${\bf{2.7537-1.3483i}}$            &      & $3$ & N/A                                  & $3.9207-1.2184i$ \\              
        &      & $4$ & N/A                                 & $3.5906-1.4480i$                   &      & $4$ & N/A                                  & $3.0791-1.5910i$ \\ 
        &      & $5$ & N/A                                 & ${\bf{0.6602-1.4854i}}$            &      & $5$ & N/A                                  & ${\bf{0.9336-1.7614i}}$ \\   
        &      & $6$ & N/A                                 & $4.1459-1.9605i$                   &      & $6$ & N/A                                  & $4.6047-1.7638i$ \\
        &      & $7$ & N/A                                 & $4.6435-2.4611i$                   &      & $7$ & N/A                                  & $5.0681-2.3036i$ \\      
$20.0$  & $3$  & $0$ & N/A                                 & \textcolor{red}{$1.9516-0.2209i$}  & $4$  & $0$ & N/A                                  & \textcolor{red}{$2.3153-0.1887i$} \\
        &      & $1$ & N/A                                 & $2.5706-0.8367i$                   &      & $1$ & N/A                                  & $2.9372-0.7909i$ \\                
        &      & $2$ & N/A                                 & $4.2139-0.9554i$            &      & $2$ & N/A                                  & ${\bf{5.5937-0.9751i}}$ \\ 
        &      & $3$ & N/A                                 & ${\bf{2.6195-1.3705i}}$            &      & $3$ & N/A                                  & $3.8784-1.4090i$ \\              
        &      & $4$ & N/A                                 & $3.6359-1.3748i$                   &      & $4$ & N/A                                  & ${\bf{2.9018-1.6083i}}$ \\ 
        &      & $5$ & N/A                                 & $4.4095-1.9394i$                   &      & $5$ & N/A                                  & $4.7724-1.8968i$ \\[1ex]
 \hline\hline 
 \end{tabular}
\end{table}

\begin{table}%[ht]
\centering
\caption{Overdamped QNM modes for vector perturbations (spin $s = 1$) of a six-dimensional ($n = 4$) Schwarzschild black hole with GB correction are presented in the table below for different values of $\widehat{\alpha}$ and $\ell_4 = 4$. The corresponding results are obtained through our SM, utilising $300$ polynomials with a precision of $300$ digits. In this context, $\Omega$ and $N$ represent the dimensionless frequency and the corresponding overtone, respectively. The notation 'SM' stands for Spectral Method. Moreover, $\Delta\Omega=\Omega_{N}-\Omega_{N+1}$.}
\label{table:piD6L4I}
\vspace*{1em}
\begin{tabular}{||c|c|c|c|c|c|c|c|c|c|c|c|c||}
\hline\hline
$\widehat{\alpha}$ $(D=6)$ & $\ell_4$ & $N$ & $\Omega$ (SM)    &$\Delta\Omega$\\ [0.5ex]
\hline\hline
$0.1$              & $4$      & $0$ & $0.0000-89.5344i$ & $1.2179i$ \\
                   &          & $1$ & $0.0000-90.7532i$ & $1.2183i$ \\
                   &          & $2$ & $0.0000-91.9706i$ & $1.2182i$ \\
                   &          & $3$ & $0.0000-93.1888i$ & $1.2182i$ \\
                   &          & $4$ & $0.0000-94.4070i$ & $1.2182i$ \\
                   &          & $5$ & $0.0000-95.6252i$ & -         \\
$0.2$              & $4$      & $0$ & $0.0000-71.5197i$ & $1.2236i$ \\
                   &          & $1$ & $0.0000-72.7433i$ & $1.2237i$ \\
                   &          & $2$ & $0.0000-73.9670i$ & $1.2236i$ \\
                   &          & $3$ & $0.0000-75.1906i$ & $1.2235i$ \\
                   &          & $4$ & $0.0000-76.4141i$ & $1.2236i$ \\
                   &          & $5$ & $0.0000-77.6377i$ & -         \\
$0.5$              & $4$      & $0$ & $0.0000-46.8576i$ & $1.2209i$ \\
                   &          & $1$ & $0.0000-48.0785i$ & $1.2207i$ \\
                   &          & $2$ & $0.0000-49.2992i$ & $1.2207i$ \\
                   &          & $3$ & $0.0000-50.5199i$ & $1.2206i$ \\
                   &          & $4$ & $0.0000-51.7405i$ & $1.2205i$ \\
                   &          & $5$ & $0.0000-52.9610i$ & -        \\
$1.0$              &          & $0$ & $0.0000-2.5884i$  & $9.6916$ \\
                   &          & $1$ & $0.0000-12.2800i$ & $4.0952i$ \\
                   &          & $2$ & $0.0000-16.3752i$ & $1.8850i$ \\
                   &          & $3$ & $0.0000-18.2602i$ & $2.2079i$ \\
                   &          & $4$ & $0.0000-20.4681i$ & - \\[1ex]
 \hline\hline 
 \end{tabular}
\end{table}

\begin{table}%[ht]
\centering
\caption{QNM modes for vector perturbations (spin $s = 1$) of a seven-dimensional ($n = 5$) Schwarzschild black hole with GB correction are presented in the table below for different values of $\widehat{\alpha}$ and $\ell_5\in\{1,2\}$. The corresponding results are obtained through our SM, utilising $300$ polynomials with a precision of $300$ digits. In this context, $\Omega$ and $N$ represent the dimensionless frequency and the corresponding overtone, respectively. The notation 'N/A' indicates data not available, while 'SM' stands for Spectral Method.}
\label{table:D7s1L1and2}
\vspace*{1em}
\begin{tabular}{||c|c|c|c|c|c|c|c|c|c|c|c|c|c||}
\hline\hline
$\widehat{\alpha}$ $(D=7)$ & $\ell_5$ & $N$ & $\Omega$ (WKB) \cite{Chakrabarti2007GRG} & $\Omega$ (SM) & $\ell_5$ & $N$ & $\Omega$ (WKB) \cite{Chakrabarti2007GRG} & $\Omega$ (SM) \\ [0.5ex]
\hline\hline
$0.0$   & $1$  & $0$ & N/A                                 & \textcolor{red}{$1.0684-0.5599i$}  & $2$  & $0$ & N/A                                  &\textcolor{red}{$1.6267-0.5149i$}\\
        &      & $1$ & N/A                                 & $0.5747-2.0508i$                   &      & $1$ & N/A                                  & $1.1627-1.6568i$\\
        &      & $2$ & N/A                                 & $0.4240-3.8992i$                   &      & $2$ & N/A                                  & $0.5888-3.5185i$\\
        &      & $3$ & N/A                                 & $0.3836-5.6562i$                   &      & $3$ & N/A                                  & $0.4803-5.3913i$\\
        &      & $4$ & N/A                                 & $0.3638-7.3810i$                   &      & $4$ & N/A                                  & $0.4362-7.1721i$\\
$0.1$   & $1$  & $0$ & \textcolor{red}{$1.08031-0.50041i$} & \textcolor{red}{$1.0746-0.5464i$}  & $2$  & $0$ &\textcolor{red}{$1.64991-0.50074i$}   & \textcolor{red}{$1.6337-0.5049i$} \\
        &      & $1$ & N/A                                 & $0.6025-1.9665i$                   &      & $1$ & N/A                                  & $1.2043-1.6168i$ \\                
        &      & $2$ & N/A                                 & $0.4202-3.7573i$                   &      & $2$ & N/A                                  & $0.6225-3.3690i$ \\ 
        &      & $3$ & N/A                                 & $0.3568-5.4698i$                   &      & $3$ & N/A                                  & $0.4739-5.1969i$ \\              
        &      & $4$ & N/A                                 & $0.3167-7.1516i$                   &      & $4$ & N/A                                  & $0.4034-6.9403i$ \\                
$0.2$   & $1$  & $0$ & \textcolor{red}{$1.07951-0.49660i$} & \textcolor{red}{$1.0795-0.5344i$}  & $2$  & $0$ & \textcolor{red}{$1.65098-0.49298i$}  & \textcolor{red}{$1.6401-0.4960i$} \\
        &      & $1$ & N/A                                 & $0.6269-1.8950i$                   &      & $1$ & N/A                                  & $1.2381-1.5815i$ \\                
        &      & $2$ & N/A                                 & $0.4201-3.6307i$                   &      & $2$ & N/A                                  & $0.6546-3.2424i$ \\ 
        &      & $3$ & N/A                                 & $0.3420-5.2898i$                   &      & $3$ & N/A                                  & $0.4812-5.0259i$ \\              
        &      & $4$ & N/A                                 & $0.2849-6.9000i$                   &      & $4$ & N/A                                  & $0.4043-6.7117i$ \\  
$0.5$   & $1$  & $0$ & \textcolor{red}{$1.07711-0.48362i$} & \textcolor{red}{$0.6870-1.7322i$}  & $2$  & $0$ & \textcolor{red}{$1.65571-0.46751i$}  & \textcolor{red}{$1.6573-0.4745i$} \\
        &      & $1$ & N/A                                 & $0.4291-3.2966i$                   &      & $1$ & N/A                                  & $1.3167-1.4979i$ \\                
        &      & $2$ & N/A                                 & $0.2856-4.7464i$                   &      & $2$ & N/A                                  & $0.7723-2.9439i$ \\ 
        &      & $3$ & N/A                                 & N/A                                &      & $3$ & N/A                                  & $0.5646-4.5416i$ \\              
        &      & $4$ & N/A                                 & N/A                                &      & $4$ & N/A                                  & $0.4425-5.9893i$ \\ 
        &      & $5$ & N/A                                 & N/A                                &      & $5$ & N/A                                  & $0.3234-7.3377i$ \\
$1.0$   & $1$  & $0$ & \textcolor{red}{$1.07361-0.45689i$} & \textcolor{red}{$1.1041-0.4732i$}  & $2$  & $0$ & \textcolor{red}{$1.67071-0.42051i$}  & \textcolor{red}{$1.6835-0.4493i$} \\     
        &      & $1$ & N/A                                 & $0.7655-1.5632i$                   &      & $1$ & N/A                                  & $1.4162-1.4074i$ \\                
        &      & $2$ & N/A                                 & $0.4788-2.8991i$                   &      & $2$ & N/A                                  & $0.9785-2.6333i$ \\ 
        &      & $3$ & N/A                                 & ${\bf{0.2872-4.1043i}}$                   &      & $3$ & N/A                                  & $0.7149-3.9897i$ \\              
        &      & $4$ & N/A                                 & $0.5421-9.3577i$                   &      & $4$ & N/A                                  & ${\bf{0.5289-5.1870i}}$ \\ 
        &      & $5$ & N/A                                 & $0.6405-10.6761i$                  &      & $5$ & N/A                                  & $0.6556-8.0575i$ \\
$5.0$   & $1$  & $0$ & \textcolor{red}{$1.33952-0.45084i$} & \textcolor{red}{$1.2252-0.3775i$}  & $2$  & $0$ & \textcolor{red}{$2.03929-0.32006i$}  & \textcolor{red}{$1.7865-0.3865i$} \\
        &      & $1$ & N/A                                 & $1.2179-1.2569i$                   &      & $1$ & N/A                                  & $1.6867-1.1980i$ \\                
        &      & $2$ & N/A                                 & ${\bf{0.4277-2.1909i}}$            &      & $2$ & N/A                                  & $1.6766-2.1553i$ \\ 
        &      & $3$ & N/A                                 & $1.4696-2.3232i$                   &      & $3$ & N/A                                  & ${\bf{0.8388-2.3986i}}$ \\              
        &      & $4$ & N/A                                 & $1.9499-3.2000i$                   &      & $4$ & N/A                                  & $2.0779-3.1035i$ \\ 
        &      & $5$ & N/A                                 & $2.4673-4.0131i$                   &      & $5$ & N/A                                  & $2.5709-3.9365i$ \\
$10.0$  & $1$  & $0$ & \textcolor{red}{$1.46092-0.36981i$} & \textcolor{red}{$1.3269-0.3588i$}  & $2$  & $0$ & \textcolor{red}{$2.37045-0.45111i$}  & \textcolor{red}{$1.8105-0.3268i$} \\
        &      & $1$ & N/A                                 & $1.4818-1.1688i$                   &      & $1$ & N/A                                  & $1.8338-0.9780i$ \\                
        &      & $2$ & N/A                                 & ${\bf{0.3654-1.6483i}}$            &      & $2$ & N/A                                  & $2.1235-1.7707i$ \\ 
        &      & $3$ & N/A                                 & $1.9553-1.9755i$                   &      & $3$ & N/A                                  & ${\bf{0.7297-1.8889i}}$ \\              
        &      & $4$ & N/A                                 & $2.5569-2.6984i$                   &      & $4$ & N/A                                  & $2.6645-2.5583i$ \\ 
        &      & $5$ & N/A                                 & $3.1854-3.4103i$                   &      & $5$ & N/A                                  & $3.2779-3.2978i$ \\
$15.0$  & $1$  & $0$ & \textcolor{red}{$1.68586-0.47334i$} & \textcolor{red}{$1.3835-0.3538i$}  & $2$  & $0$ & \textcolor{red}{$2.53501-0.45766i$}  & \textcolor{red}{$1.8430-0.2609i$} \\
        &      & $1$ & N/A                                 & $1.6017-1.1012i$                   &      & $1$ & N/A                                  & $2.0060-0.8550i$ \\                
        &      & $2$ & N/A                                 & ${\bf{0.3421-1.4359i}}$            &      & $2$ & N/A                                  & $2.3559-1.6027i$ \\ 
        &      & $3$ & N/A                                 & $2.1441-1.8080i$                   &      & $3$ & N/A                                  & ${\bf{0.6778-1.6825i}}$ \\              
        &      & $4$ & N/A                                 & $2.7920-2.4756i$                   &      & $4$ & N/A                                  & $2.9013-2.3145i$ \\ 
        &      & $5$ & N/A                                 & $3.4603-3.1355i$                   &      & $5$ & N/A                                  & $3.5526-2.9983i$ \\      
$20.0$  & $1$  & $0$ & \textcolor{red}{$1.87237-0.54573i$} & \textcolor{red}{$1.4215-0.3513i$}  & $2$  & $0$ & \textcolor{red}{$2.62139-0.34334i$}  & \textcolor{red}{$1.8933-0.2243i$} \\
        &      & $1$ & N/A                                 & $1.6639-1.0558i$                   &      & $1$ & N/A                                  & $2.1208-0.8061i$ \\                
        &      & $2$ & N/A                                 & ${\bf{0.3274-1.3164i}}$            &      & $2$ & N/A                                  & $2.4734-1.5161i$ \\ 
        &      & $3$ & N/A                                 & $2.2403-1.7034i$                   &      & $3$ & N/A                                  & ${\bf{0.6434-1.5617i}}$ \\              
        &      & $4$ & N/A                                 & $2.9103-2.3379i$                   &      & $4$ & N/A                                  & $3.0181-2.1580i$ \\ 
        &      & $5$ & N/A                                 & $3.5980-2.9638i$                   &      & $5$ & N/A                                  & $3.6953-2.8039i$ \\[1ex]
 \hline\hline 
 \end{tabular}
\end{table}

\begin{table}%[ht]
\centering
\caption{QNM modes for vector perturbations (spin $s = 1$) of a seven-dimensional ($n=5$) Schwarzschild black hole with GB correction are presented in the table below for different values of $\widehat{\alpha}$ and $\ell_5\in\{3,4\}$. The corresponding results are obtained through our SM, utilising $300$ polynomials with a precision of $300$ digits. In this context, $\Omega$ and $N$ represent the dimensionless frequency and the corresponding overtone, respectively. The notation 'N/A' indicates data not available, while 'SM' stands for Spectral Method.}
\label{table:D7s1L3and4}
\vspace*{1em}
\begin{tabular}{||c|c|c|c|c|c|c|c|c|c|c|c|c|c||}
\hline\hline
$\widehat{\alpha}$ $(D=7)$ & $\ell_5$ & $N$ & $\Omega$ (WKB) \cite{Chakrabarti2007GRG} & $\Omega$ (SM) & $\ell_5$ & $N$ & $\Omega$ (WKB)  \cite{Chakrabarti2007GRG} & $\Omega$ (SM) \\ [0.5ex]
\hline\hline
$0.0$   & $3$  & $0$ & N/A                                 & \textcolor{red}{$2.2178-0.5000i$}  & $4$  & $0$ & N/A                                  &\textcolor{red}{$2.7959-0.4999i$}\\
        &      & $1$ & N/A                                 & $1.9145-1.5270i$                   &      & $1$ & N/A                                  & $2.5728-1.5166i$\\
        &      & $2$ & N/A                                 & $1.1297-2.7074i$                   &      & $2$ & N/A                                  & $2.0858-2.6004i$\\
        &      & $3$ & N/A                                 & $0.5581-4.9840i$                   &      & $3$ & N/A                                  & $1.2324-3.9048i$\\
        &      & $4$ & N/A                                 & $0.5062-6.8780i$                   &      & $4$ & N/A                                  & $0.3962-6.4431i$\\
$0.1$   & $3$  & $0$ & \textcolor{red}{$2.22978-0.43349i$} & \textcolor{red}{$2.2248-0.4914i$}  & $4$  & $0$ & \textcolor{red}{$2.80549-0.49205i$}  & \textcolor{red}{$2.8031-0.4916i$} \\
        &      & $1$ & N/A                                 & $1.9381-1.5002i$                   &      & $1$ & N/A                                  & $2.5905-1.4907i$ \\                
        &      & $2$ & N/A                                 & $1.2221-2.6548i$                   &      & $2$ & N/A                                  & $2.1299-2.5520i$ \\ 
        &      & $3$ & N/A                                 & $0.5922-4.7687i$                   &      & $3$ & N/A                                  & $1.3303-3.8064i$ \\              
        &      & $4$ & N/A                                 & $0.4942-6.6324i$                   &      & $4$ & N/A                                  & $0.4390-6.1368i$ \\                
$0.2$   & $3$  & $0$ & \textcolor{red}{$2.23449-0.48374i$} & \textcolor{red}{$2.2314-0.4836i$}  & $4$  & $0$ & \textcolor{red}{$2.81365-0.48223i$}  & \textcolor{red}{$2.8099-0.4841i$} \\
        &      & $1$ & N/A                                 & $1.9588-1.4757i$                   &      & $1$ & N/A                                  & $2.6065-1.4671i$ \\                
        &      & $2$ & N/A                                 & $1.2969-2.6029i$                   &      & $2$ & N/A                                  & $2.1688-2.5077i$ \\ 
        &      & $3$ & N/A                                 & $0.6253-4.5863i$                   &      & $3$ & N/A                                  & $1.4176-3.7209i$ \\              
        &      & $4$ & N/A                                 & $0.5125-6.4226i$                   &      & $4$ & N/A                                  & $0.4706-5.8693i$ \\  
$0.5$   & $3$  & $0$ & \textcolor{red}{$2.25186-0.45407i$} & \textcolor{red}{$2.2498-0.4644i$}  & $4$  & $0$ & \textcolor{red}{$2.84194-0.45386i$}  & \textcolor{red}{$2.8293-0.4649i$} \\
        &      & $1$ & N/A                                 & $2.0113-1.4158i$                   &      & $1$ & N/A                                  & $2.6488-1.4083i$ \\                
        &      & $2$ & N/A                                 & $1.4674-2.8466i$                   &      & $2$ & N/A                                  & $2.2662-2.4033i$ \\ 
        &      & $3$ & N/A                                 & $0.8415-4.1524i$                   &      & $3$ & N/A                                  & $1.6231-3.5520i$ \\              
        &      & $4$ & N/A                                 & $0.7120-5.7812i$                   &      & $4$ & N/A                                  & $0.9345-5.3495i$ \\ 
        &      & $5$ & N/A                                 & $0.6239-7.2289i$                   &      & $5$ & N/A                                  & $0.8849-7.0128i$ \\
$1.0$   & $3$  & $0$ & \textcolor{red}{$2.29356-0.40943i$} & \textcolor{red}{$2.2777-0.4419i$}  & $4$  & $0$ & \textcolor{red}{$2.90198-0.41485i$}  & \textcolor{red}{$2.8577-0.4425i$} \\               
        &      & $1$ & N/A                                 & $2.0830-1.3515i$                   &      & $1$ & N/A                                  & $2.7074-1.3439i$ \\                
        &      & $2$ & N/A                                 & $1.6693-2.3799i$                   &      & $2$ & N/A                                  & $2.3890-2.3045i$ \\ 
        &      & $3$ & N/A                                 & $1.1909-3.7247i$                   &      & $3$ & N/A                                  & $1.8786-3.4222i$ \\              
        &      & $4$ & N/A                                 & $0.9470-5.0624i$                   &      & $4$ & N/A                                  & $1.4136-4.8269i$ \\ 
        &      & $5$ & N/A                                 & $0.8438-6.3180i$                   &      & $5$ & N/A                                  & $1.2577-6.1677i$ \\
$5.0$   & $3$  & $0$ & \textcolor{red}{$2.80530-0.34825i$} & \textcolor{red}{$2.3408-0.3963i$}  & $4$  & $0$ & \textcolor{red}{$3.54531-0.35999i$}  & \textcolor{red}{$2.8896-0.4060i$} \\
        &      & $1$ & N/A                                 & $2.1394-1.1910i$                   &      & $1$ & N/A                                  & ${\bf{2.5074-1.2385i}}$ \\                
        &      & $2$ & N/A                                 & ${\bf{1.9407-1.6642i}}$                   &      & $2$ & N/A                                  & $2.6669-1.3776i$ \\ 
        &      & $3$ & N/A                                 & $2.2005-2.5990i$                   &      & $3$ & N/A                                  & $2.8123-2.3472i$ \\              
        &      & $4$ & N/A                                 & ${\bf{1.2783-2.6890i}}$            &      & $4$ & N/A                                  & ${\bf{1.7272-3.0145i}}$ \\ 
        &      & $5$ & N/A                                 & $2.6253-3.5940i$                   &      & $5$ & N/A                                  & $3.0666-3.2932i$ \\
$10.0$  & $3$  & $0$ & \textcolor{red}{$3.23719-0.47697i$} & \textcolor{red}{$2.2812-0.2838i$}  & $4$  & $0$ & \textcolor{red}{$4.06418-0.48967i$}  & \textcolor{red}{$2.7528-0.2457i$} \\
        &      & $1$ & N/A                                 & $2.3902-0.7363i$                   &      & $1$ & N/A                                  & $2.9524-0.6282i$ \\                
        &      & $2$ & N/A                                 & $2.6471-1.4474i$                   &      & $2$ & N/A                                  & $3.2018-1.3187i$ \\ 
        &      & $3$ & N/A                                 & ${\bf{1.1132-2.1779i}}$            &      & $3$ & N/A                                  & $3.4689-2.0525i$ \\              
        &      & $4$ & N/A                                 & $2.9902-2.2208i$                   &      & $4$ & N/A                                  & ${\bf{1.5026-2.4789i}}$ \\ 
        &      & $5$ & N/A                                 & $3.4772-2.9801i$                   &      & $5$ & N/A                                  & $3.8647-2.7682i$ \\
$15.0$  & $3$  & $0$ & \textcolor{red}{$3.40873-0.51161i$} & \textcolor{red}{$2.3476-0.1726i$}  & $4$  & $0$ & \textcolor{red}{$4.24763-0.53548i$}  & \textcolor{red}{$2.8519-0.1253i$} \\
        &      & $1$ & N/A                                 & $2.6119-0.6653i$                   &      & $1$ & N/A                                  & $3.1757-0.5608i$ \\                
        &      & $2$ & N/A                                 & $2.8987-1.3742i$                   &      & $2$ & N/A                                  & $3.4774-1.2500i$ \\ 
        &      & $3$ & N/A                                 & ${\bf{1.0302-1.9594i}}$            &      & $3$ & N/A                                  & $3.7009-1.9550i$ \\              
        &      & $4$ & N/A                                 & $3.2395-2.0449i$                   &      & $4$ & N/A                                  & ${\bf{1.3889-2.2435i}}$ \\ 
        &      & $5$ & N/A                                 & $3.7857-2.6820i$                   &      & $5$ & N/A                                  & $4.1417-2.4733i$ \\      
$20.0$  & $3$  & $0$ & \textcolor{red}{$3.46675-0.48649i$} & \textcolor{red}{$2.4240-0.1580i$}  & $4$  & $0$ & \textcolor{red}{$4.30050-0.53964i$}  & \textcolor{red}{$2.9234-0.1439i$} \\
        &      & $1$ & N/A                                 & $2.7528-0.6243i$                   &      & $1$ & N/A                                  & $3.3518-0.5121i$ \\                
        &      & $2$ & N/A                                 & $3.0501-1.3527i$                   &      & $2$ & N/A                                  & $3.6830-1.2188i$ \\ 
        &      & $3$ & N/A                                 & ${\bf{0.9758-1.8286i}}$            &      & $3$ & N/A                                  & $3.7182-1.9529i$ \\              
        &      & $4$ & N/A                                 & $3.3280-1.9246i$                   &      & $4$ & N/A                                  & ${\bf{1.3149-2.1004i}}$ \\ 
        &      & $5$ & N/A                                 & $3.9567-2.4744i$                   &      & $5$ & N/A                                  & $4.3192-2.2177i$ \\[1ex]
 \hline\hline 
 \end{tabular}
\end{table}

\begin{table}%[ht]
\centering
\caption{Overdamped QNM modes for vector perturbations (spin $s = 1$) of a seven-dimensional ($n = 5$) Schwarzschild black hole with GB correction are presented in the table below for different values of $\widehat{\alpha}$ and $\ell_5\in\{1,2,3\}$. The corresponding results are obtained through our SM, utilising $300$ polynomials with a precision of $300$ digits. In this context, $\Omega$ and $N$ represent the dimensionless frequency and the corresponding overtone, respectively. The notation 'SM' stands for Spectral Method. Moreover, $\Delta\Omega=\Omega_{N}-\Omega_{N+1}$.}
\label{table:piD7L123}
\vspace*{1em}
\begin{tabular}{||c|c|c|c|c|c|c|c|c|c|c|c|c||}
\hline\hline
$\widehat{\alpha}$ $(D=7)$ & $\ell_5$ & $N$ & $\Omega$ (SM)    &$\Delta\Omega$ & $\ell_5$ & $N$ & $\Omega$ (SM)    &$\Delta\Omega$ & $\ell_5$ & $N$ & $\Omega$ (SM)    &$\Delta\Omega$\\ [0.5ex]
\hline\hline
$0.1$    & $1$      & $0$ & $0.0000-131.489i$ & $1.716i$ & $2$ & $0$ & $0.0000-133.191i$ & $1.716i$ & $3$ & $0$ & $0.0000-134.888i$ & $1.718i$\\
         &          & $1$ & $0.0000-133.205i$ & $1.717i$ &     & $1$ & $0.0000-134.907i$ & $1.717i$ &     & $1$ & $0.0000-136.606i$ & $1.717i$\\
         &          & $2$ & $0.0000-134.922i$ & $1.717i$ &     & $2$ & $0.0000-136.624i$ & $1.717i$ &     & $2$ & $0.0000-138.323i$ & $1.717i$\\
         &          & $3$ & $0.0000-136.639i$ & $1.717i$ &     & $3$ & $0.0000-138.341i$ & $1.717i$ &     & $3$ & $0.0000-140.040i$ & $1.717i$\\
         &          & $4$ & $0.0000-138.356i$ & $1.717i$ &     & $4$ & $0.0000-140.058i$ & $1.717i$ &     & $4$ & $0.0000-141.757i$ & $1.717i$\\
         &          & $5$ & $0.0000-140.073i$ & -        &     & $5$ & $0.0000-141.775i$ & -        &     & $5$ & $0.0000-143.474i$ & -\\
$0.2$    & $1$      & $0$ & $0.0000-107.786i$ & $1.723i$ & $2$ & $0$ & $0.0000-111.216i$ & $1.724i$ & $3$ & $0$ & $0.0000-111.196i$ & $1.723i$\\
         &          & $1$ & $0.0000-109.509i$ & $1.723i$ &     & $1$ & $0.0000-112.940i$ & $1.723i$ &     & $1$ & $0.0000-112.919i$ & $1.724i$\\
         &          & $2$ & $0.0000-111.232i$ & $1.724i$ &     & $2$ & $0.0000-114.663i$ & $1.724i$ &     & $2$ & $0.0000-114.643i$ & $1.724i$\\
         &          & $3$ & $0.0000-112.956i$ & $1.723i$ &     & $3$ & $0.0000-116.387i$ & $1.723i$ &     & $3$ & $0.0000-116.367i$ & $1.723i$\\
         &          & $4$ & $0.0000-114.679i$ & $1.723i$ &     & $4$ & $0.0000-118.110i$ & $1.723i$ &     & $4$ & $0.0000-118.090i$ & $1.724i$\\
         &          & $5$ & $0.0000-116.402i$ & -        &     & $5$ & $0.0000-119.833i$ & -        &     & $5$ & $0.0000-119.814i$ & -\\[1ex]
 \hline\hline 
 \end{tabular}
\end{table}

\begin{table}%[ht]
\centering
\caption{Overdamped QNM modes for vector perturbations (spin $s = 1$) of a seven-dimensional ($n = 5$) Schwarzschild black hole with GB correction are presented in the table below for different values of $\widehat{\alpha}$ and $\ell_5=4$. The corresponding results are obtained through our SM, utilising $300$ polynomials with a precision of $300$ digits. In this context, $\Omega$ and $N$ represent the dimensionless frequency and the corresponding overtone, respectively. The notation 'SM' stands for Spectral Method. Moreover, $\Delta\Omega=\Omega_{N}-\Omega_{N+1}$.}
\label{table:piD7L4}
\vspace*{1em}
\begin{tabular}{||c|c|c|c|c|c|c|c|c|c|c|c|c||}
\hline\hline
$\widehat{\alpha}$ $(D=7)$ & $\ell_5$ & $N$ & $\Omega$ (SM)    &$\Delta\Omega$ \\ [0.5ex]
\hline\hline
$0.1$              & $4$      & $0$ & $0.0000-134.865i$ & $1.718i$ \\
                   &          & $1$ & $0.0000-136.583i$ & $1.717i$ \\
                   &          & $2$ & $0.0000-138.300i$ & $1.718i$ \\
                   &          & $3$ & $0.0000-140.018i$ & $1.717i$ \\
                   &          & $4$ & $0.0000-141.735i$ & $1.718i$ \\
                   &          & $5$ & $0.0000-143.453i$ & -        \\
$0.2$              & $4$      & $0$ & $0.0000-111.170i$ & $1.724i$ \\
                   &          & $1$ & $0.0000-112.894i$ & $1.725i$ \\
                   &          & $2$ & $0.0000-114.619i$ & $1.724i$ \\
                   &          & $3$ & $0.0000-116.343i$ & $1.724i$ \\
                   &          & $4$ & $0.0000-118.067i$ & $1.724i$ \\
                   &          & $5$ & $0.0000-119.791i$ & -        \\[1ex]
 \hline\hline 
 \end{tabular}
\end{table}

\begin{table}%[ht]
\centering
\caption{QNM modes for vector perturbations (spin $s = 1$) of an eight-dimensional ($n = 6$) Schwarzschild black hole with GB correction are presented in the table below for different values of $\widehat{\alpha}$ and $\ell_6\in\{1,2\}$. The corresponding results are obtained through our SM, utilising $300$ polynomials with a precision of $300$ digits. In this context, $\Omega$ and $N$ represent the dimensionless frequency and the corresponding overtone, respectively. The notation 'N/A' indicates data not available, while 'SM' stands for Spectral Method.}
\label{table:D8s1L1and2}
\vspace*{1em}
\begin{tabular}{||c|c|c|c|c|c|c|c|c|c|c|c|c|c||}
\hline\hline
$\widehat{\alpha}$ $(D=8)$ & $\ell_6$ & $N$ & $\Omega$ (WKB)\cite{Chakrabarti2007GRG} & $\Omega$ (SM) & $\ell_6$ & $N$ & $\Omega$ (WKB) \cite{Chakrabarti2007GRG} & $\Omega$ (SM) \\ [0.5ex]
\hline\hline
$0.0$   & $1$  & $0$ & N/A                                 & \textcolor{red}{$1.4528-0.6839i$}  & $2$  & $0$ & N/A                                  &\textcolor{red}{$2.0535-0.6431i$}\\
        &      & $1$ & N/A                                 & $0.6564-2.5984i$                   &      & $1$ & N/A                                  & $1.3863-2.0360i$\\
        &      & $2$ & N/A                                 & $0.5099-5.0473i$                   &      & $2$ & N/A                                  & $0.6175-4.6529i$\\
        &      & $3$ & N/A                                 & $0.4701-7.3260i$                   &      & $3$ & N/A                                  & $0.5397-7.0592i$\\
        &      & $4$ & N/A                                 & $0.4501-9.5593i$                   &      & $4$ & N/A                                  & $0.5040-9.3512i$\\
$0.1$   & $1$  & $0$ & \textcolor{red}{$1.46373-0.61715i$} & \textcolor{red}{$1.4583-0.6680i$}  & $2$  & $0$ &\textcolor{red}{$2.07904-0.61909i$}   & \textcolor{red}{$2.0596-0.6304i$} \\
        &      & $1$ & N/A                                 & $0.6930-2.4787i$                   &      & $1$ & N/A                                  & $1.4431-1.9918i$ \\                
        &      & $2$ & N/A                                 & $0.5062-4.8692i$                   &      & $2$ & N/A                                  & $0.6418-4.4624i$ \\ 
        &      & $3$ & N/A                                 & $0.4466-7.0930i$                   &      & $3$ & N/A                                  & $0.5313-6.8239i$ \\              
        &      & $4$ & N/A                                 & $0.4119-9.2688i$                   &      & $4$ & N/A                                  & $0.4790-9.0679i$ \\                
$0.2$   & $1$  & $0$ & \textcolor{red}{$1.46250-0.61271i$} & \textcolor{red}{$1.4627-0.6542i$}  & $2$  & $0$ & \textcolor{red}{$2.07920-0.61118i$}  & \textcolor{red}{$2.0652-0.6194i$} \\
        &      & $1$ & N/A                                 & $0.7296-2.3799i$                   &      & $1$ & N/A                                  & $1.4887-1.9530i$ \\                
        &      & $2$ & N/A                                 & $0.5163-4.7118i$                   &      & $2$ & N/A                                  & $0.6741-4.3056i$ \\ 
        &      & $3$ & N/A                                 & $0.4511-6.8621i$                   &      & $3$ & N/A                                  & $0.5598-6.6128i$ \\              
        &      & $4$ & N/A                                 & $0.4029-8.9397i$                   &      & $4$ & N/A                                  & $0.5100-8.7675i$ \\  
$0.5$   & $1$  & $0$ & \textcolor{red}{$1.45897-0.59754i$} & \textcolor{red}{$1.4725-0.6216i$}  & $2$  & $0$ & \textcolor{red}{$2.08109-0.58516i$}  & \textcolor{red}{$2.0807-0.5929i$} \\
        &      & $1$ & N/A                                 & $0.8343-2.1659i$                   &      & $1$ & N/A                                  & $1.5926-1.8616i$ \\                
        &      & $2$ & N/A                                 & $0.5668-4.2957i$                   &      & $2$ & N/A                                  & ${\bf{0.0603-2.1879i}}$ \\ 
        &      & $3$ & N/A                                 & $0.4444-6.1910i$                   &      & $3$ & N/A                                  & $0.8215-3.9264i$ \\              
        &      & $4$ & N/A                                 & $0.2902-7.9540i$                   &      & $4$ & N/A                                  & $0.6910-5.9996i$ \\ 
        &      & $5$ & N/A                                 & N/A                                &      & $5$ & N/A                                  & $0.5788-7.8732i$ \\
$1.0$   & $1$  & $0$ & \textcolor{red}{$1.45410-0.56680i$} & \textcolor{red}{$1.4863-0.5847i$}  & $2$  & $0$ & \textcolor{red}{$2.09037-0.53716i$}  & \textcolor{red}{$2.1056-0.5621i$} \\
        &      & $1$ & N/A                                 & $0.9740-1.9636i$                   &      & $1$ & N/A                                  & $1.7186-1.7619i$ \\                
        &      & $2$ & N/A                                 & $0.6698-3.8238i$                   &      & $2$ & N/A                                  & ${\bf{0.1072-2.2098i}}$ \\ 
        &      & $3$ & N/A                                 & ${\bf{0.5045-5.4760i}}$                   &      & $3$ & N/A                                  & $1.0786-3.5055i$ \\              
        &      & $4$ & N/A                                 & $0.7617-13.9224i$                  &      & $4$ & N/A                                  & $0.8816-5.3383i$ \\ 
        &      & $5$ & N/A                                 & $0.9749-15.7398i$                  &      & $5$ & N/A                                  & $0.6808-6.9272i$ \\
$5.0$   & $1$  & $0$ & \textcolor{red}{$1.66619-0.49783i$} & \textcolor{red}{$1.6021-0.4810i$}  & $2$  & $0$ & \textcolor{red}{$2.41661-0.43291i$}  & \textcolor{red}{$2.2195-0.4874i$} \\
        &      & $1$ & N/A                                 & $1.5070-1.6282i$                   &      & $1$ & N/A                                  & $2.0768-1.5441i$ \\                
        &      & $2$ & N/A                                 & ${\bf{0.4638-3.0049i}}$            &      & $2$ & N/A                                  & $2.0035-2.9483i$ \\ 
        &      & $3$ & N/A                                 & $1.7435-3.1604i$                   &      & $3$ & N/A                                  & ${\bf{0.9707-3.0617i}}$ \\     
        &      & $4$ & N/A                                 & $2.3332-4.4126i$                   &      & $4$ & N/A                                  & $2.5253-4.2975i$ \\ 
        &      & $5$ & N/A                                 & $2.9438-5.5254i$                   &      & $5$ & N/A                                  & $3.1056-5.4491i$ \\
$10.0$  & $1$  & $0$ & \textcolor{red}{$1.79779-0.39821i$} & \textcolor{red}{$1.6955-0.4576i$}  & $2$  & $0$ & \textcolor{red}{$2.66779-0.49225i$}  & \textcolor{red}{$2.2574-0.4454i$} \\
        &      & $1$ & N/A                                 & $1.7777-1.5459i$                   &      & $1$ & N/A                                  & $2.2337-1.4102i$ \\                
        &      & $2$ & N/A                                 & ${\bf{0.4919-2.1762i}}$            &      & $2$ & N/A                                  & ${\bf{0.9865-2.3598i}}$ \\ 
        &      & $3$ & N/A                                 & $2.2879-2.7255i$                   &      & $3$ & N/A                                  & $2.5128-2.5670i$ \\              
        &      & $4$ & N/A                                 & $2.9864-3.7148i$                   &      & $4$ & N/A                                  & $3.1486-3.6018i$ \\ 
        &      & $5$ & N/A                                 & $3.7371-4.6674i$                   &      & $5$ & N/A                                  & $3.8734-4.5722i$ \\
$15.0$  & $1$  & $0$ & \textcolor{red}{$1.94080-0.49012i$} & \textcolor{red}{$1.7456-0.4510i$}  & $2$  & $0$ & \textcolor{red}{$2.83381-0.55573i$}  & \textcolor{red}{$2.2710-0.4088i$} \\
        &      & $1$ & N/A                                 & $1.9054-1.4695i$                   &      & $1$ & N/A                                  & $2.3264-1.2950i$ \\                
        &      & $2$ & N/A                                 & ${\bf{0.5251-1.8796i}}$            &      & $2$ & N/A                                  & ${\bf{0.9732-2.1052i}}$ \\ 
        &      & $3$ & N/A                                 & $2.4960-2.4813i$                   &      & $3$ & N/A                                  & $2.7174-2.3194i$ \\              
        &      & $4$ & N/A                                 & $3.2518-3.3774i$                   &      & $4$ & N/A                                  & $3.4070-3.2526i$ \\ 
        &      & $5$ & N/A                                 & $4.0482-4.2578i$                   &      & $5$ & N/A                                  & $4.1835-4.1493i$ \\      
$20.0$  & $1$  & $0$ & \textcolor{red}{$2.07360-0.54533i$} & \textcolor{red}{$1.7756-0.4458i$}  & $2$  & $0$ & \textcolor{red}{$2.92724-0.55988i$}  & \textcolor{red}{$2.2804-0.3752i$} \\
        &      & $1$ & N/A                                 & $1.9751-1.4062i$                   &      & $1$ & N/A                                  & $2.4016-1.2047i$ \\                
        &      & $2$ & N/A                                 & ${\bf{0.5360-1.7253i}}$            &      & $2$ & N/A                                  & ${\bf{0.9526-1.9610i}}$ \\ 
        &      & $3$ & N/A                                 & $2.6059-2.3225i$                   &      & $3$ & N/A                                  & $2.8358-2.1618i$ \\              
        &      & $4$ & N/A                                 & $3.3893-3.1649i$                   &      & $4$ & N/A                                  & $3.5409-3.0318i$ \\ 
        &      & $5$ & N/A                                 & $4.2059-3.9965i$                   &      & $5$ & N/A                                  & $4.3426-3.8779i$ \\[1ex]
 \hline\hline 
 \end{tabular}
\end{table}

\begin{table}%[ht]
\centering
\caption{QNM modes for vector perturbations (spin $s = 1$) of an eight-dimensional ($n=6$) Schwarzschild black hole with GB correction are presented in the table below for different values of $\widehat{\alpha}$ and $\ell_6\in\{3,4\}$. The corresponding results are obtained through our SM, utilising $300$ polynomials with a precision of $300$ digits. In this context, $\Omega$ and $N$ represent the dimensionless frequency and the corresponding overtone, respectively. The notation 'N/A' indicates data not available, while 'SM' stands for Spectral Method.}
\label{table:D8s1L3and4}
\vspace*{1em}
\begin{tabular}{||c|c|c|c|c|c|c|c|c|c|c|c|c|c||}
\hline\hline
$\widehat{\alpha}$ $(D=8)$ & $\ell_6$ & $N$ & $\Omega$ (WKB) \cite{Chakrabarti2007GRG} & $\Omega$ (SM) & $\ell_6$ & $N$ & $\Omega$ (WKB) \cite{Chakrabarti2007GRG} & $\Omega$ (SM) \\ [0.5ex]
\hline\hline
$0.0$   & $3$  & $0$ & N/A                                 & \textcolor{red}{$2.6870-0.6222i$}  & $4$  & $0$ & N/A                                  &\textcolor{red}{$3.3160-0.6175i$}\\
        &      & $1$ & N/A                                 & $2.2323-1.8888i$                   &      & $1$ & N/A                                  & $2.9767-1.8642i$\\
        &      & $2$ & N/A                                 & ${\bf{0.5712-2.8883i}}$            &      & $2$ & N/A                                  & $2.1592-3.1525i$\\
        &      & $3$ & N/A                                 & $0.5900-3.8767i$                   &      & $3$ & N/A                                  & $0.4557-3.2968i$\\
        &      & $4$ & N/A                                 & $0.5919-6.6837i$                   &      & $4$ & N/A                                  & $0.4094-6.1338i$\\
$0.1$   & $3$  & $0$ & \textcolor{red}{$2.70311-0.61275i$} & \textcolor{red}{$2.6933-0.6113i$}  & $4$  & $0$ & \textcolor{red}{$3.32693-0.60901i$}  & \textcolor{red}{$3.3225-0.6073i$} \\
        &      & $1$ & N/A                                 & $2.2642-1.8570i$                   &      & $1$ & N/A                                  & $2.9995-1.8338i$ \\                
        &      & $2$ & N/A                                 & ${\bf{0.5730-3.0115i}}$            &      & $2$ & N/A                                  & $2.2320-3.1042i$ \\ 
        &      & $3$ & N/A                                 & $0.7034-3.5547i$                   &      & $3$ & N/A                                  & $0.4494-3.3298i$ \\              
        &      & $4$ & N/A                                 & $0.6044-6.4335i$                   &      & $4$ & N/A                                  & ${\bf{0.2657-4.3958i}}$ \\    
        &      & $5$ & N/A                                 & $0.5389-8.7904i$                   &      & $5$ & N/A                                  & $0.4553-5.8138i$ \\ 
$0.2$   & $3$  & $0$ & \textcolor{red}{$2.70603-0.60251i$} & \textcolor{red}{$2.6993-0.6016i$}  & $4$  & $0$ & \textcolor{red}{$3.33289-0.59806i$}  & \textcolor{red}{$3.3289-0.5982i$} \\
        &      & $1$ & N/A                                 & $2.2918-1.8283i$                   &      & $1$ & N/A                                  & $3.0201-1.8064i$ \\                
        &      & $2$ & N/A                                 & ${\bf{0.4117-3.0990i}}$            &      & $2$ & N/A                                  & $2.2952-3.0602i$ \\ 
        &      & $3$ & N/A                                 & $0.9568-3.3061i$                   &      & $3$ & N/A                                  & $0.4358-3.3670i$ \\              
        &      & $4$ & N/A                                 & $0.6425-6.2348i$                   &      & $4$ & N/A                                  & ${\bf{0.3451-4.3048i}}$ \\  
        &      & $5$ & N/A                                 & $0.5995-8.5269i$                   &      & $5$ & N/A                                  & $0.4760-5.5293i$ \\  
$0.5$   & $3$  & $0$ & \textcolor{red}{$2.71766-0.57078i$} & \textcolor{red}{$2.7168-0.5781i$}  & $4$  & $0$ & \textcolor{red}{$3.35437-0.56581i$}  & \textcolor{red}{$3.3476-0.5754i$} \\
        &      & $1$ & N/A                                 & $2.3615-1.7596i$                   &      & $1$ & N/A                                  & $3.0746-1.7397i$ \\                
        &      & $2$ & N/A                                 & ${\bf{0.2375-2.9489i}}$            &      & $2$ & N/A                                  & $2.4485-2.9604i$ \\ 
        &      & $3$ & N/A                                 & $1.3914-3.1482i$                   &      & $3$ & N/A                                  & $0.3614-3.4821i$ \\              
        &      & $4$ & N/A                                 & $0.8961-5.6881i$                   &      & $4$ & N/A                                  & ${\bf{0.3201-4.1150i}}$ \\ 
        &      & $5$ & N/A                                 & $0.8284-7.6993i$                   &      & $5$ & N/A                                  & $0.9225-4.9021i$ \\
$1.0$   & $3$  & $0$ & \textcolor{red}{$2.74804-0.52034i$} & \textcolor{red}{$2.7447-0.5507i$}  & $4$  & $0$ & \textcolor{red}{$3.40253-0.51999i$}  & \textcolor{red}{$3.3765-0.5489i$} \\  
        &      & $1$ & N/A                                 & $2.4549-1.6857i$                   &      & $1$ & N/A                                  & $3.1497-1.6673i$ \\                
        &      & $2$ & N/A                                 & ${\bf{0.2884-2.9111i}}$            &      & $2$ & N/A                                  & $2.6364-2.8714i$ \\ 
        &      & $3$ & N/A                                 & $1.7631-3.0326i$                   &      & $3$ & N/A                                  & ${\bf{0.5394-3.6586i}}$ \\              
        &      & $4$ & N/A                                 & $1.2517-5.0710i$                   &      & $4$ & N/A                                  & $1.7114-4.5587i$ \\ 
        &      & $5$ & N/A                                 & $1.0459-6.7974i$                   &      & $5$ & N/A                                  & $1.4235-6.5709i$ \\
$5.0$   & $3$  & $0$ & \textcolor{red}{$3.15756-0.41819i$} & \textcolor{red}{$2.8322-0.4939i$}  & $4$  & $0$ & \textcolor{red}{$3.90098-0.42763i$}  & \textcolor{red}{$3.4380-0.5008i$} \\
        &      & $1$ & N/A                                 & $2.6560-1.5120i$                   &      & $1$ & N/A                                  & $3.2433-1.5136i$ \\                
        &      & $2$ & N/A                                 & ${\bf{0.3255-2.5965i}}$            &      & $2$ & N/A                                  & $2.8116-2.3929i$ \\ 
        &      & $3$ & N/A                                 & $2.3483-2.6654i$                   &      & $3$ & N/A                                  & ${\bf{0.6598-3.0381i}}$ \\              
        &      & $4$ & N/A                                 & ${\bf{1.5167-3.2992i}}$            &      & $4$ & N/A                                  & $2.7526-3.4619i$ \\ 
        &      & $5$ & N/A                                 & $2.6457-4.1497i$                   &      & $5$ & N/A                                  & ${\bf{2.1196-3.5654i}}$ \\
        &      & $6$ & N/A                                 & $3.2187-5.3450i$                   &      & $6$ & N/A                                  & $3.2765-4.9050i$ \\ 
$10.0$  & $3$  & $0$ & \textcolor{red}{$3.51637-0.51655i$} & \textcolor{red}{$2.8093-0.4375i$}  & $4$  & $0$ & \textcolor{red}{$4.34217-0.52963i$}  & \textcolor{red}{$3.3557-0.4398i$} \\
        &      & $1$ & N/A                                 & ${\bf{2.7068-1.2351i}}$                   &      & $1$ & N/A                                  & ${\bf{3.2491-1.0759i}}$ \\                
        &      & $2$ & N/A                                 & $2.8247-2.1568i$                   &      & $2$ & N/A                                  & $3.4715-1.8993i$ \\ 
        &      & $3$ & N/A                                 & ${\bf{1.4745-2.6340i}}$            &      & $3$ & N/A                                  & ${\bf{0.6191-2.7234i}}$ \\              
        &      & $4$ & N/A                                 & $3.2943-3.2940i$                   &      & $4$ & N/A                                  & $1.9480-2.9274i$ \\ 
        &      & $5$ & N/A                                 & $3.9772-4.3542i$                   &      & $5$ & N/A                                  & $3.7780-2.9507i$ \\
$15.0$  & $3$  & $0$ & \textcolor{red}{$3.70743-0.58013i$} & \textcolor{red}{$2.7878-0.3603i$}  & $4$  & $0$ & \textcolor{red}{$4.55712-0.59556i$}  & \textcolor{red}{$3.3049-0.3157i$} \\
        &      & $1$ & N/A                                 & $2.8687-1.0449i$                   &      & $1$ & N/A                                  & $3.4779-0.8968i$ \\                
        &      & $2$ & N/A                                 & $3.1381-1.9635i$                   &      & $2$ & N/A                                  & $3.7470-1.7779i$ \\ 
        &      & $3$ & N/A                                 & ${\bf{1.4154-2.3770i}}$            &      & $3$ & N/A                                  & ${\bf{1.8517-2.6608i}}$ \\              
        &      & $4$ & N/A                                 & $3.6283-2.9493i$                   &      & $4$ & N/A                                  & $4.0936-2.7056i$ \\ 
        &      & $5$ & N/A                                 & $4.3228-3.9020i$                   &      & $5$ & N/A                                  & $4.6677-3.6087i$ \\      
$20.0$  & $3$  & $0$ & \textcolor{red}{$3.80401-0.59800i$} & \textcolor{red}{$2.7976-0.2925i$}  & $4$  & $0$ & \textcolor{red}{$4.65663-0.62039i$}  & \textcolor{red}{$3.3278-0.2285i$} \\
        &      & $1$ & N/A                                 & $2.9970-0.9603i$                   &      & $1$ & N/A                                  & $3.6126-0.8275i$ \\                
        &      & $2$ & N/A                                 & $3.3015-1.8650i$                   &      & $2$ & N/A                                  & $3.9117-1.7092i$ \\ 
        &      & $3$ & N/A                                 & ${\bf{1.3681-2.2272i}}$            &      & $3$ & N/A                                  & ${\bf{1.7818-2.5015i}}$ \\              
        &      & $4$ & N/A                                 & $3.7859-2.7463i$                   &      & $4$ & N/A                                  & $4.2295-2.5565i$ \\ 
        &      & $5$ & N/A                                 & $4.5033-3.6146i$                   &      & $5$ & N/A                                  & $4.8488-3.3354i$ \\[1ex]
 \hline\hline 
 \end{tabular}
\end{table}

\begin{table}%[ht]
\centering
\caption{Overdamped QNM modes for vector perturbations (spin $s = 1$) of an eight-dimensional ($n=6$) Schwarzschild black hole with GB correction are presented in the table below for different values of $\widehat{\alpha}$ and $\ell_6\in\{1,2,3\}$. The corresponding results are obtained through our SM, utilising $300$ polynomials with a precision of $300$ digits. In this context, $\Omega$ and $N$ represent the dimensionless frequency and the corresponding overtone, respectively. The notation 'SM' stands for Spectral Method. Moreover, $\Delta\Omega=\Omega_{N}-\Omega_{N+1}$.}
\label{table:piD8L123}
\vspace*{1em}
\begin{tabular}{||c|c|c|c|c|c|c|c|c|c|c|c|c||}
\hline\hline
$\widehat{\alpha}$ $(D=8)$ & $\ell_6$ & $N$ & $\Omega$ (SM)    &$\Delta\Omega$ & $\ell_6$ & $N$ & $\Omega$ (SM)    &$\Delta\Omega$ \\ [0.5ex]
\hline\hline
$0.1$    & $1$      & $0$ & $0.0000-169.844i$ & $2.217i$ & $2$ & $0$ & $0.0000-169.829i$ & $2.218i$ \\
         &          & $1$ & $0.0000-172.061i$ & $2.218i$ &     & $1$ & $0.0000-172.047i$ & $2.218i$ \\
         &          & $2$ & $0.0000-174.279i$ & $2.218i$ &     & $2$ & $0.0000-174.265i$ & $2.217i$ \\
         &          & $3$ & $0.0000-176.497i$ & $2.217i$ &     & $3$ & $0.0000-176.482i$ & $2.218i$ \\
         &          & $4$ & $0.0000-178.714i$ & $2.217i$ &     & $4$ & $0.0000-178.700i$ & $2.217i$ \\
         &          & $5$ & $0.0000-180.931i$ & -        &     & $5$ & $0.0000-180.917i$ & -        \\[1ex]
 \hline\hline 
 \end{tabular}
\end{table}

\begin{table}%[ht]
\centering
\caption{QNM modes for vector perturbations (spin $s = 1$) of a ten-dimensional ($n=8$) Schwarzschild black hole with GB correction are presented in the table below for different values of $\widehat{\alpha}$ and $\ell_{8}\in\{1,2,3\}$. The corresponding results are obtained through our SM, utilising $300$ polynomials with a precision of $300$ digits. In this context, $\Omega$ is given by \eqref{ODEV} while $N$ represents the corresponding overtone. The notation 'N/A' indicates data not available. Moreover, 'SM' stands for Spectral Method.}
\label{table:D10s1L123}
\vspace*{1em}
\begin{tabular}{||c|c|c|c|c|c|c|c|c|c|c|c||}
\hline\hline
$\widehat{\alpha}$ $(D=10)$ & $\ell_{8}$ & $N$ & $\Omega$ (SM)& $\ell_{8}$ & $N$ & $\Omega$ (SM) & $\ell_{8}$ & $N$ & $\Omega$ (SM) \\ [0.5ex]
\hline\hline
$0.0$   & $1$  & $0$ & \textcolor{red}{$2.2639-0.9026i$}   & $2$   & $0$ & \textcolor{red}{$2.9286-0.8698i$}  & $3$  & $0$ & \textcolor{red}{$3.6180-0.8457i$}\\
        &      & $1$ & $0.8783-2.0665i$                    &       & $1$ & $1.8223-2.4704i$                   &      & $1$ & $2.8289-2.4809i$\\
        &      & $2$ & $0.7843-3.8188i$                    &       & $2$ & $0.8862-3.1823i$                   &      & $2$ & $1.4201-3.1725i$\\
        &      & $3$ & $0.6787-7.3744i$                    &       & $3$ & $0.7291-7.0017i$                   &      & $3$ & $0.7108-6.4907i$\\
        &      & $4$ & $0.6418-10.6930i$                   &       & $4$ & $0.6821-10.4373i$                  &      & $4$ & $0.7065-10.1126i$\\
$0.1$   & $1$  & $0$ & \textcolor{red}{$2.2682-0.8833i$}   & $2$   & $0$ & \textcolor{red}{$2.9330-0.8533i$}  & $3$  & $0$ & \textcolor{red}{$3.6226-0.8312i$}\\
        &      & $1$ & $0.9023-2.0810i$                    &       & $1$ & $1.8909-2.4639i$                   &      & $1$ & $2.8745-2.4501i$\\
        &      & $2$ & $0.8170-3.6483i$                    &       & $2$ & $0.9346-3.0895i$                   &      & $2$ & $1.4702-3.1633i$\\
        &      & $3$ & $0.6848-7.1265i$                    &       & $3$ & $0.7555-6.7491i$                   &      & $3$ & $0.7627-6.2172i$\\
        &      & $4$ & $0.6391-10.3628i$                   &       & $4$ & $0.6965-10.1151i$                  &      & $4$ & $0.7373-9.7950i$\\
        &      & $5$ & $0.6175-13.5229i$                   &       & $5$ & $0.6708-13.3405i$                  &      & $5$ & $0.7162-13.1110i$\\
$0.2$   & $1$  & $0$ & \textcolor{red}{$2.2717-0.8667i$}   & $2$   & $0$ & \textcolor{red}{$2.9371-0.8391i$}  & $3$  & $0$ & \textcolor{red}{$3.6272-0.8186i$}\\
        &      & $1$ & $0.9240-2.0944i$                    &       & $1$ & $1.9510-2.4522i$                   &      & $1$ & $2.9141-2.4224i$\\
        &      & $2$ & $0.8544-3.5087i$                    &       & $2$ & $0.9695-3.0260i$                   &      & $2$ & $1.5125-3.1575i$\\
        &      & $3$ & $0.7240-6.9057i$                    &       & $3$ & $0.8156-6.5424i$                   &      & $3$ & ${\bf{0.8333-6.0127i}}$\\
        &      & $4$ & $0.6873-10.0242i$                   &       & $4$ & $0.7790-9.8025i$                   &      & $4$ & $0.8484-9.5153i$\\
        &      & $5$ & $0.6556-13.0360i$                   &       & $5$ & $0.7576-12.8822i$                  &      & $5$ & $0.8515-12.6888i$\\
$0.5$   & $1$  & $0$ & \textcolor{red}{$2.2798-0.8280i$}   & $2$   & $0$ & \textcolor{red}{$2.9493-0.8055i$}  & $3$  & $0$ & \textcolor{red}{$3.6419-0.7882i$}\\
        &      & $1$ & $0.9754-2.1341i$                    &       & $1$ & $2.0955-2.4105i$                   &      & $1$ & $3.0128-2.3553i$\\
        &      & $2$ & $0.9732-3.1891i$                    &       & $2$ & ${\bf{1.0476-2.9138i}}$                   &      & $2$ & $1.6126-3.1564i$\\
        &      & $3$ & $0.8586-6.3212i$                    &       & $3$ & $1.0536-6.0049i$                   &      & $3$ & ${\bf{0.1596-3.8346i}}$\\
        &      & $4$ & $0.7738-9.1007i$                    &       & $4$ & $1.0019-8.9271i$                   &      & $4$ & $1.1658-5.5268i$\\
        &      & $5$ & $0.6336-11.7495i$                   &       & $5$ & $0.9047-11.6607i$                  &      & $5$ & $1.2062-8.6937i$\\
$5.0$   & $1$  & $0$ & \textcolor{red}{$2.4054-0.6633i$}   & $2$   & $0$ & \textcolor{red}{$3.0941-0.6680i$}  & $3$  & $0$ & \textcolor{red}{$3.7822-0.6727i$}\\
        &      & $1$ & ${\bf{0.7879-2.2455i}}$             &       & $1$ & $2.7965-2.1561i$                   &      & $1$ & $3.5058-2.1021i$\\
        &      & $2$ & $2.0818-2.2908i$                    &       & $2$ & ${\bf{1.2359-2.7535i}}$            &      & $2$ & ${\bf{1.7808-3.3063i}}$\\
        &      & $3$ & $2.3140-4.7262i$                    &       & $3$ & $2.6132-4.4944i$                   &      & $3$ & $2.9512-4.1743i$\\
        &      & $4$ & $3.1603-6.7531i$                    &       & $4$ & $3.3937-6.6016i$                   &      & $4$ & $3.6309-6.4381i$\\
        &      & $5$ & N/A                                 &       & $5$ & N/A                                &      & $5$ & $4.4318-8.2851i$\\
$10.0$  & $1$  & $0$ & \textcolor{red}{$2.4960-0.6294i$}   & $2$   & $0$ & \textcolor{red}{$3.1476-0.6296i$}  & $3$  & $0$ & \textcolor{red}{$3.7953-0.6317i$}\\
        &      & $1$ & $2.4095-2.2040i$                    &       & $1$ & $3.0069-2.0494i$                   &      & $1$ & $3.6152-1.9497i$\\
        &      & $2$ & ${\bf{0.7815-2.2883i}}$             &       & $2$ & ${\bf{1.3776-2.8658i}}$            &      & $2$ & ${\bf{2.1103-3.2474i}}$\\
        &      & $3$ & $3.0321-4.1857i$                    &       & $3$ & $3.3176-3.9998i$                   &      & $3$ & $3.6110-3.7501i$\\
        &      & $4$ & $3.9529-5.7559i$                    &       & $4$ & $4.1809-5.6323i$                   &      & $4$ & $4.4011-5.4778i$\\
        &      & $5$ & $4.9463-7.1954i$                    &       & $5$ & $5.1313-7.1036i$                   &      & $5$ & $5.3158-6.9864i$\\
$20.0$  & $1$  & $0$ & \textcolor{red}{$2.5763-0.6120i$}   & $2$   & $0$ & \textcolor{red}{$3.1772-0.5813i$}  & $3$  & $0$ & \textcolor{red}{$3.7691-0.5544i$}\\
        &      & $1$ & $2.6657-2.0604i$                    &       & $1$ & $3.1884-1.8907i$                   &      & $1$ & $3.7369-1.7094i$\\
        &      & $2$ & ${\bf{1.0150-2.3114i}}$             &       & $2$ & ${\bf{1.6206-2.5963i}}$            &      & $2$ & ${\bf{2.1794-2.8662i}}$\\
        &      & $3$ & $3.4333-3.5877i$                    &       & $3$ & $3.7150-3.4286i$                   &      & $3$ & $4.0203-3.1637i$\\
        &      & $4$ & $4.4547-4.8654i$                    &       & $4$ & $4.6638-4.7424i$                   &      & $4$ & $4.8691-4.5548i$\\
        &      & $5$ & $5.5553-6.1016i$                    &       & $5$ & $5.7326-5.9960i$                   &      & $5$ & $5.9086-5.8446i$\\[1ex]
 \hline\hline 
 \end{tabular}
\end{table}

\begin{table}%[ht]
\centering
\caption{QNM modes for vector perturbations (spin $s = 1$) of an eleven-dimensional ($n=9$) Schwarzschild black hole with GB correction are presented in the table below for different values of $\widehat{\alpha}$ and $\ell_{9}\in\{1,2,3\}$. The corresponding results are obtained through our SM, utilising $300$ polynomials with a precision of $300$ digits. In this context, $\Omega$ is given by \eqref{ODEV} while $N$ represents the corresponding overtone. The notation 'N/A' indicates data not available. Moreover, 'SM' stands for Spectral Method.}
\label{table:D11s1L123}
\vspace*{1em}
\begin{tabular}{||c|c|c|c|c|c|c|c|c|c|c|c||}
\hline\hline
$\widehat{\alpha}$ $(D=11)$ & $\ell_{9}$ & $N$ & $\Omega$ (SM)& $\ell_{9}$ & $N$ & $\Omega$ (SM) & $\ell_{9}$ & $N$ & $\Omega$ (SM) \\ [0.5ex]
\hline\hline
$0.0$   & $1$  & $0$ & \textcolor{red}{$2.6840-1.0006i$} & $2$   & $0$ & \textcolor{red}{$3.3735-0.9709i$}  & $3$  & $0$ & \textcolor{red}{$4.0837-0.9474i$}\\
        &      & $1$ & $1.2951-2.2502i$                  &       & $1$ & $2.1475-2.5851i$                   &      & $1$ & $3.1313-2.6976i$\\
        &      & $2$ & $0.8695-4.4357i$                  &       & $2$ & $1.1123-3.6434i$                   &      & $2$ & $1.8299-3.4611i$\\
        &      & $3$ & $0.7638-8.5456i$                  &       & $3$ & $0.8011-8.1839i$                   &      & $3$ & ${\bf{0.3154-4.1251i}}$\\
        &      & $4$ & $0.7279-12.3835i$                 &       & $4$ & $0.7603-12.1336i$                  &      & $4$ & $0.7849-7.7162i$\\
$0.1$   & $1$  & $0$ & \textcolor{red}{$2.6879-0.9799i$} & $2$   & $0$ & \textcolor{red}{$3.3771-0.9530i$}  & $3$  & $0$ & \textcolor{red}{$4.0875-0.9315i$}\\
        &      & $1$ & $1.3212-2.2555i$                  &       & $1$ & $2.1982-2.5849i$                   &      & $1$ & $3.1793-2.6727i$\\
        &      & $2$ & $0.9108-4.2418i$                  &       & $2$ & $1.2226-3.5373i$                   &      & $2$ & $1.8848-3.4377i$\\
        &      & $3$ & $0.7779-8.2628i$                  &       & $3$ & $0.8356-7.8997i$                   &      & $3$ & ${\bf{0.3690-4.0960i}}$\\
        &      & $4$ & $0.7393-12.0028i$                 &       & $4$ & $0.7911-11.7635i$                  &      & $4$ & $0.8431-7.4210i$\\
        &      & $5$ & $0.7244-15.6541i$                 &       & $5$ & $0.7750-15.4779i$                  &      & $5$ & $0.8287-11.4649i$\\
$0.2$   & $1$  & $0$ & \textcolor{red}{$2.6909-0.9622i$} & $2$   & $0$ & \textcolor{red}{$3.3806-0.9376i$}  & $3$  & $0$ & \textcolor{red}{$4.0914-0.9177i$}\\
        &      & $1$ & $1.3445-2.2599i$                  &       & $1$ & $2.2445-2.5827i$                   &      & $1$ & $3.2213-2.6500i$\\
        &      & $2$ & $0.9599-4.0861i$                  &       & $2$ & $1.2982-3.4604i$                   &      & $2$ & $1.9303-3.4188i$\\
        &      & $3$ & $0.8336-8.0090i$                  &       & $3$ & $0.9151-7.6634i$                   &      & $3$ & ${\bf{0.4177-4.0902i}}$\\
        &      & $4$ & $0.8100-11.6089i$                 &       & $4$ & $0.8995-11.3955i$                  &      & $4$ & $0.9386-7.1992i$\\
        &      & $5$ & $0.7865-15.0884i$                 &       & $5$ & $0.8878-14.9395i$                  &      & $5$ & $0.9724-11.1297i$\\
$0.5$   & $1$  & $0$ & \textcolor{red}{$2.6983-0.9212i$} & $2$   & $0$ & \textcolor{red}{$3.3914-0.9014i$}  & $3$  & $0$ & \textcolor{red}{$4.1046-0.8846i$}\\
        &      & $1$ & $1.4007-2.2744i$                  &       & $1$ & $2.3661-2.5727i$                   &      & $1$ & $3.3274-2.5942i$\\
        &      & $2$ & $1.1159-3.7413i$                  &       & $2$ & $1.4457-3.3073i$                   &      & $2$ & $2.0367-3.3820i$\\
        &      & $3$ & $1.0098-7.3388i$                  &       & $3$ & $1.1987-7.0389i$                   &      & $3$ & ${\bf{0.5648-4.1040i}}$\\
        &      & $4$ & $0.9424-10.5589i$                 &       & $4$ & $1.1672-10.3895i$                  &      & $4$ & $1.3285-6.6320i$\\
        &      & $5$ & $0.8134-13.6475i$                 &       & $5$ & $1.0797-13.5553i$                  &      & $5$ & $1.3754-10.1716i$\\
$5.0$   & $1$  & $0$ & \textcolor{red}{$2.8226-0.7459i$} & $2$   & $0$ & \textcolor{red}{$3.5363-0.7501i$}  & $3$  & $0$ & \textcolor{red}{$4.2507-0.7543i$}\\
        &      & $1$ & ${\bf{1.2694-2.5125i}}$           &       & $1$ & $3.1336-2.4289i$                   &      & $1$ & $3.8981-2.3679i$\\
        &      & $2$ & $2.3426-2.5956i$                  &       & $2$ & ${\bf{1.7377-2.9593i}}$            &      & $2$ & ${\bf{2.3327-3.3752i}}$\\
        &      & $3$ & $2.6089-5.4789i$                  &       & $3$ & $2.9281-5.2483i$                   &      & $3$ & $3.2755-4.9618i$\\
$10.0$  & $1$  & $0$ & \textcolor{red}{$2.9136-0.7072i$} & $2$   & $0$ & \textcolor{red}{$3.5957-0.7106i$}  & $3$  & $0$ & \textcolor{red}{$4.2753-0.7151i$}\\
        &      & $1$ & $2.7213-2.5050i$                  &       & $1$ & $3.3744-2.3316i$                   &      & $1$ & $4.0374-2.2324i$\\
        &      & $2$ & ${\bf{1.2586-2.5093i}}$           &       & $2$ & ${\bf{1.8169-2.9694i}}$            &      & $2$ & ${\bf{2.4710-3.3648i}}$\\
        &      & $3$ & $3.4123-4.9015i$                  &       & $3$ & $3.7114-4.7043i$                   &      & $3$ & $4.0239-4.4619i$\\
        &      & $4$ & N/A                               &       & $4$ & N/A                                &      & $4$ & $4.9583-6.4851i$\\
$20.0$  & $1$  & $0$ & \textcolor{red}{$2.9966-0.6864i$} & $2$   & $0$ & \textcolor{red}{$3.6318-0.6642i$}  & $3$  & $0$ & \textcolor{red}{$4.2598-0.6469i$}\\
        &      & $1$ & $3.0168-2.3641i$                  &       & $1$ & $3.5845-2.1846i$                   &      & $1$ & $4.1732-2.0265i$\\
        &      & $2$ & ${\bf{1.3688-2.5006i}}$           &       & $2$ & ${\bf{1.9909-2.8297i}}$            &      & $2$ & ${\bf{2.5929-3.1090i}}$\\
        &      & $3$ & $3.8664-4.2264i$                  &       & $3$ & $4.1660-4.0585i$                   &      & $3$ & $4.4800-3.8258i$\\
        &      & $4$ & N/A                               &       & $4$ & $5.2354-5.6101i$                   &      & $4$ & $5.4643-5.4450i$\\
        &      & $5$ & N/A                               &       & $5$ & $6.4298-7.0442i$                   &      & $5$ & $6.6350-6.9323i$\\[1ex]
 \hline\hline 
 \end{tabular}
\end{table}

\begin{table}%[ht]
\centering
\caption{QNM modes for vector perturbations (spin $s = 1$) of a twelve-dimensional ($n=10$) Schwarzschild black hole with GB correction are presented in the table below for different values of $\widehat{\alpha}$ and $\ell_{10}\in\{1,2,3\}$. The corresponding results are obtained through our SM, utilising $300$ polynomials with a precision of $300$ digits. In this context, $\Omega$ is given by \eqref{ODEV} while $N$ represents the corresponding overtone. The notation 'N/A' indicates data not available. Moreover, 'SM' stands for Spectral Method.}
\label{table:D12s1L123}
\vspace*{1em}
\begin{tabular}{||c|c|c|c|c|c|c|c|c|c|c|c||}
\hline\hline
$\widehat{\alpha}$ $(D=12)$ & $\ell_{10}$ & $N$ & $\Omega$ (SM)& $\ell_{10}$ & $N$ & $\Omega$ (SM) & $\ell_{10}$ & $N$ & $\Omega$ (SM) \\ [0.5ex]
\hline\hline
$0.0$   & $1$  & $0$ & \textcolor{red}{$3.1113-1.0924i$} & $2$   & $0$ & \textcolor{red}{$3.8221-1.0655i$}  & $3$  & $0$ & \textcolor{red}{$4.5501-1.0430i$}\\
        &      & $1$ & $1.7045-2.4200i$                  &       & $1$ & $2.5198-2.7053i$                   &      & $1$ & $3.4554-2.8682i$\\
        &      & $2$ & $0.9494-5.0500i$                  &       & $2$ & $1.1963-3.9517i$                   &      & $2$ & $2.2107-3.7542i$\\
        &      & $3$ & $0.8496-9.7195i$                  &       & $3$ & ${\bf{0.4858-4.1436i}}$            &      & $3$ & ${\bf{0.6557-4.4422i}}$\\
        &      & $4$ & $0.8143-14.0766i$                 &       & $4$ & $0.8781-9.3669i$                   &      & $4$ & $0.8635-8.9276i$\\
$0.1$   & $1$  & $0$ & \textcolor{red}{$3.1147-1.0706i$} & $2$   & $0$ & \textcolor{red}{$3.8250-1.0464i$}  & $3$  & $0$ & \textcolor{red}{$4.5531-1.0259i$}\\
        &      & $1$ & $1.7303-2.4191i$                  &       & $1$ & $2.5598-2.7024i$                   &      & $1$ & $3.5012-2.8494i$\\
        &      & $2$ & $0.9977-4.8290i$                  &       & $2$ & $1.3998-3.8929i$                   &      & $2$ & $2.2742-3.7250i$\\
        &      & $3$ & $0.8729-9.4014i$                  &       & $3$ & ${\bf{0.3725-4.0308i}}$            &      & $3$ & ${\bf{0.7313-4.4012i}}$\\
        &      & $4$ & $0.8411-13.6444i$                 &       & $4$ & $0.9228-9.0499i$                   &      & $4$ & $0.9324-8.6053i$\\
        &      & $5$ & $0.8327-17.7867i$                 &       & $5$ & $0.8895-13.4117i$                  &      & $5$ & $0.9259-13.1288i$\\
$0.2$   & $1$  & $0$ & \textcolor{red}{$3.1172-1.0520i$} & $2$   & $0$ & \textcolor{red}{$3.8279-1.0300i$}  & $3$  & $0$ & \textcolor{red}{$4.5564-1.0111i$}\\
        &      & $1$ & $1.7529-2.4180i$                  &       & $1$ & $2.5960-2.6995i$                   &      & $1$ & $3.5419-2.8322i$\\
        &      & $2$ & $1.0582-4.6523i$                  &       & $2$ & $1.5255-3.8327i$                   &      & $2$ & $2.3263-3.6994i$\\
        &      & $3$ & $0.9457-9.1140i$                  &       & $3$ & ${\bf{0.3365-3.9723i}}$            &      & $3$ & ${\bf{0.7915-4.3828i}}$\\
        &      & $4$ & $0.9344-13.1947i$                 &       & $4$ & $1.0222-8.7816i$                   &      & $4$ & $1.0534-8.3569i$\\
        &      & $5$ & $0.9193-17.1423i$                 &       & $5$ & $1.0229-12.9875i$                  &      & $5$ & $1.0989-12.7362i$\\
$0.5$   & $1$  & $0$ & \textcolor{red}{$3.1239-1.0090i$} & $2$   & $0$ & \textcolor{red}{$3.8373-0.9915i$}  & $3$  & $0$ & \textcolor{red}{$4.5681-0.9757i$}\\
        &      & $1$ & $1.8071-2.4181i$                  &       & $1$ & $2.6907-2.6947i$                   &      & $1$ & $3.6468-2.7904i$\\
        &      & $2$ & $1.2567-4.2650i$                  &       & $2$ & $1.7589-3.6927i$                   &      & $2$ & $2.4466-3.6423i$\\
        &      & $3$ & $1.1631-8.3576i$                  &       & $3$ & ${\bf{0.3603-3.8859i}}$            &      & $3$ & ${\bf{0.9455-4.3658i}}$\\
        &      & $4$ & $1.1126-12.0179i$                 &       & $4$ & $1.3496-8.0690i$                   &      & $4$ & $1.4940-7.6994i$\\
        &      & $5$ & N/-A                               &       & $5$ & $1.3342-11.8513i$                  &      & $5$ & $1.5461-11.6435i$\\
$5.0$   & $1$  & $0$ & \textcolor{red}{$3.2471-0.8241i$} & $2$   & $0$ & \textcolor{red}{$3.9816-0.8279i$}  & $3$  & $0$ & \textcolor{red}{$4.7174-0.8317i$}\\
        &      & $1$ & ${\bf{1.7889-2.7567i}}$           &       & $1$ & $3.4538-2.6778i$                   &      & $1$ & $4.2745-2.6143i$\\
        &      & $2$ & $2.5599-2.8891i$                  &       & $2$ & ${\bf{2.2236-3.1762i}}$            &      & $2$ & ${\bf{2.8186-3.5082i}}$\\
        &      & $3$ & N/A                               &       & $3$ & $3.2399-5.9872i$                   &      & $3$ & $3.6135-5.7229i$\\
$10.0$  & $1$  & $0$ & \textcolor{red}{$3.3387-0.7810i$} & $2$   & $0$ & \textcolor{red}{$4.0460-0.7866i$}  & $3$  & $0$ & \textcolor{red}{$4.7518-0.7928i$}\\
        &      & $1$ & ${\bf{1.7308-2.7216i}}$           &       & $1$ & $3.7297-2.5958i$                   &      & $1$ & $4.4429-2.4936i$\\
        &      & $2$ & $3.0213-2.7943i$                  &       & $2$ & ${\bf{2.2824-3.1377i}}$            &      & $2$ & ${\bf{2.9092-3.4841i}}$\\
        &      & $3$ & N/A                               &       & $3$ & $4.1021-5.4035i$                   &      & $3$ & $4.4280-5.1648i$\\
$20.0$  & $1$  & $0$ & \textcolor{red}{$3.4250-0.7566i$} & $2$   & $0$ & \textcolor{red}{$4.0883-0.7408i$}  & $3$  & $0$ & \textcolor{red}{$4.7462-0.7296i$}\\
        &      & $1$ & $3.3637-2.6552i$                  &       & $1$ & $3.9718-2.4603i$                   &      & $1$ & $4.6024-2.3087i$\\
        &      & $2$ & ${\bf{1.7807-2.6870i}}$           &       & $2$ & ${\bf{2.3979-3.0305i}}$            &      & $2$ & ${\bf{3.0195-3.3140i}}$\\
        &      & $3$ & $4.3048-4.8658i$                  &       & $3$ & $4.6184-4.6887i$                   &      & $3$ & $4.9473-4.4667i$\\
        &      & $4$ & N/A                               &       & $4$ & N/A                                &      & $4$ & $6.0661-6.3339i$\\[1ex]
 \hline\hline 
 \end{tabular}
\end{table}

\begin{table}%[ht]
\centering
\caption{QNM modes for vector perturbations (spin $s = 1$) of a twenty-six-dimensional ($n = 24$) Schwarzschild black hole with GB correction are presented in the table below for different values of $\widehat{\alpha}$ and $\ell_{24}\in\{1,2,3\}$. The corresponding results are obtained through our SM, utilising $300$ polynomials with a precision of $300$ digits. In this context, $\Omega$ is given by \eqref{ODEV} while $N$ represents the corresponding overtone. The notation 'N/A' indicates data not available. Moreover, 'SM' stands for Spectral Method.}
\label{table:D26s1L123}
\vspace*{1em}
\begin{tabular}{||c|c|c|c|c|c|c|c|c|c|c|c||}
\hline\hline
$\widehat{\alpha}$ $(D=26)$ & $\ell_{24}$ & $N$ & $\Omega$ (SM)& $\ell_{24}$ & $N$ & $\Omega$ (SM) & $\ell_{24}$ & $N$ & $\Omega$ (SM) \\ [0.5ex]
\hline\hline
$0.0$   & $1$  & $0$ & \textcolor{red}{$9.4505-2.0207i$} & $2$   & $0$ & \textcolor{red}{$10.3000-2.0133i$}  & $3$  & $0$ & \textcolor{red}{$11.1521-2.0058i$}\\
        &      & $1$ & $7.6887-4.1395i$                  &       & $1$ & $8.5238-4.2521i$                    &      & $1$ & $9.3688-4.3600i$\\
        &      & $2$ & $6.0041-5.5551i$                  &       & $2$ & $6.8394-5.7839i$                    &      & $2$ & $7.6970-5.9977i$\\
        &      & $3$ & $4.3221-6.5392i$                  &       & $3$ & $5.1690-6.8927i$                    &      & $3$ & $6.0561-7.2079i$\\
        &      & $4$ & $2.6144-7.1800i$                  &       & $4$ & $3.4794-7.6760i$                    &      & $4$ & $4.4040-8.0923i$\\
        &      & $5$ & ${\bf{0.8758-7.5000i}}$           &       & $5$ & ${\bf{1.7527-8.1562i}}$                    &      & $5$ & $2.6945-8.6906i$\\
        &      & $6$ & $1.9261-13.8968i$                 &       & $6$ & $1.8192-13.4020i$                   &      & $6$ & ${\bf{0.9046-8.9937i}}$\\
        &      & $7$ & N/A                               &       & $7$ & N/A                                 &      & $7$ & $1.6715-12.9005i$\\
$0.1$   & $1$  & $0$ & \textcolor{red}{$9.4489-1.9918i$} & $2$   & $0$ & \textcolor{red}{$10.2974-1.9862i$}  & $3$  & $0$ & \textcolor{red}{$11.1488-1.9802i$}\\
        &      & $1$ & $7.7023-4.1159i$                  &       & $1$ & $8.5377-4.2294i$                    &      & $1$ & $9.3832-4.3378i$\\
        &      & $2$ & $6.0279-5.5420i$                  &       & $2$ & $6.8658-5.7714i$                    &      & $2$ & $7.7262-5.9848i$\\
        &      & $3$ & $4.3503-6.5407i$                  &       & $3$ & $5.2037-6.8960i$                    &      & $3$ & $6.0986-7.2097i$\\
        &      & $4$ & $2.6381-7.1979i$                  &       & $4$ & $3.5151-7.7009i$                    &      & $4$ & $4.4566-8.1151i$\\
        &      & $5$ & $0.8850-7.5292i$                  &       & $5$ & $1.7757-8.2029i$                    &      & $5$ & $2.7438-8.7434i$\\
$0.2$   & $1$  & $0$ & \textcolor{red}{$9.4476-1.9673i$} & $2$   & $0$ & \textcolor{red}{$10.2956-1.9631i$}  & $3$  & $0$ & \textcolor{red}{$11.1467-1.9585i$}\\
        &      & $1$ & $7.7140-4.0961i$                  &       & $1$ & $8.5499-4.2103i$                    &      & $1$ & $9.3962-4.3193i$\\
        &      & $2$ & $6.0482-5.5311i$                  &       & $2$ & $6.8886-5.7611i$                    &      & $2$ & $7.7519-5.9745i$\\
        &      & $3$ & $4.3742-6.5424i$                  &       & $3$ & $5.2336-6.8996i$                    &      & $3$ & $6.1359-7.2123i$\\
        &      & $4$ & $2.6577-7.2133i$                  &       & $4$ & $3.5452-7.7237i$                    &      & $4$ & $4.5031-8.1367i$\\
        &      & $5$ & N/A                               &       & $5$ & $1.7991-8.2509i$                    &      & $5$ & $2.7858-8.7932i$\\
$0.5$   & $1$  & $0$ & \textcolor{red}{$9.4466-1.9109i$} & $2$   & $0$ & \textcolor{red}{$10.2946-1.9095i$}  & $3$  & $0$ & \textcolor{red}{$11.1460-1.9072i$}\\
        &      & $1$ & $7.7428-4.0518i$                  &       & $1$ & $8.5815-4.1679i$                    &      & $1$ & $9.4309-4.2783i$\\
        &      & $2$ & $6.0961-5.5093i$                  &       & $2$ & $6.9441-5.7419i$                    &      & $2$ & $7.8163-5.9560i$\\
        &      & $3$ & $4.4273-6.5513i$                  &       & $3$ & $5.3030-6.9166i$                    &      & $3$ & $6.2278-7.2292i$\\
        &      & $4$ & N/A                               &       & $4$ & N/A                                 &      & $4$ & $4.6158-8.2100i$\\
$5.0$   & $1$  & $0$ & \textcolor{red}{$9.5549-1.6507i$} & $2$   & $0$ & \textcolor{red}{$10.4151-1.6545i$}  & $3$  & $0$ & \textcolor{red}{$11.2768-1.6575i$}\\
        &      & $1$ & $7.9868-3.9445i$                  &       & $1$ & $8.8582-4.0757i$                    &      & $1$ & $9.7407-4.2000i$\\
        &      & $2$ & N/A                               &       & $2$ & $7.2873-5.9293i$                    &      & $2$ & $8.2975-6.1959i$\\
$10.0$  & $1$  & $0$ & \textcolor{red}{$9.6546-1.5695i$} & $2$   & $0$ & \textcolor{red}{$10.5116-1.5802i$}  & $3$  & $0$ & \textcolor{red}{$11.3691-1.5903i$}\\
        &      & $1$ & $8.1177-3.9780i$                  &       & $1$ & $8.9961-4.1146i$                    &      & $1$ & $9.8855-4.2421i$\\
$20.0$  & $1$  & $0$ & \textcolor{red}{$9.7715-1.5076i$} & $2$   & $0$ & \textcolor{red}{$10.6078-1.5194i$}  & $3$  & $0$ & \textcolor{red}{$11.4443-1.5317i$}\\
        &      & $1$ & $8.2415-4.0542i$                  &       & $1$ & $9.1304-4.1855i$                    &      & $1$ & $10.0321-4.2994i$\\[1ex]
 \hline\hline 
 \end{tabular}
\end{table}

\section{Tensor perturbations}

\begin{figure}[ht!] 
    \includegraphics[width=0.3\textwidth]{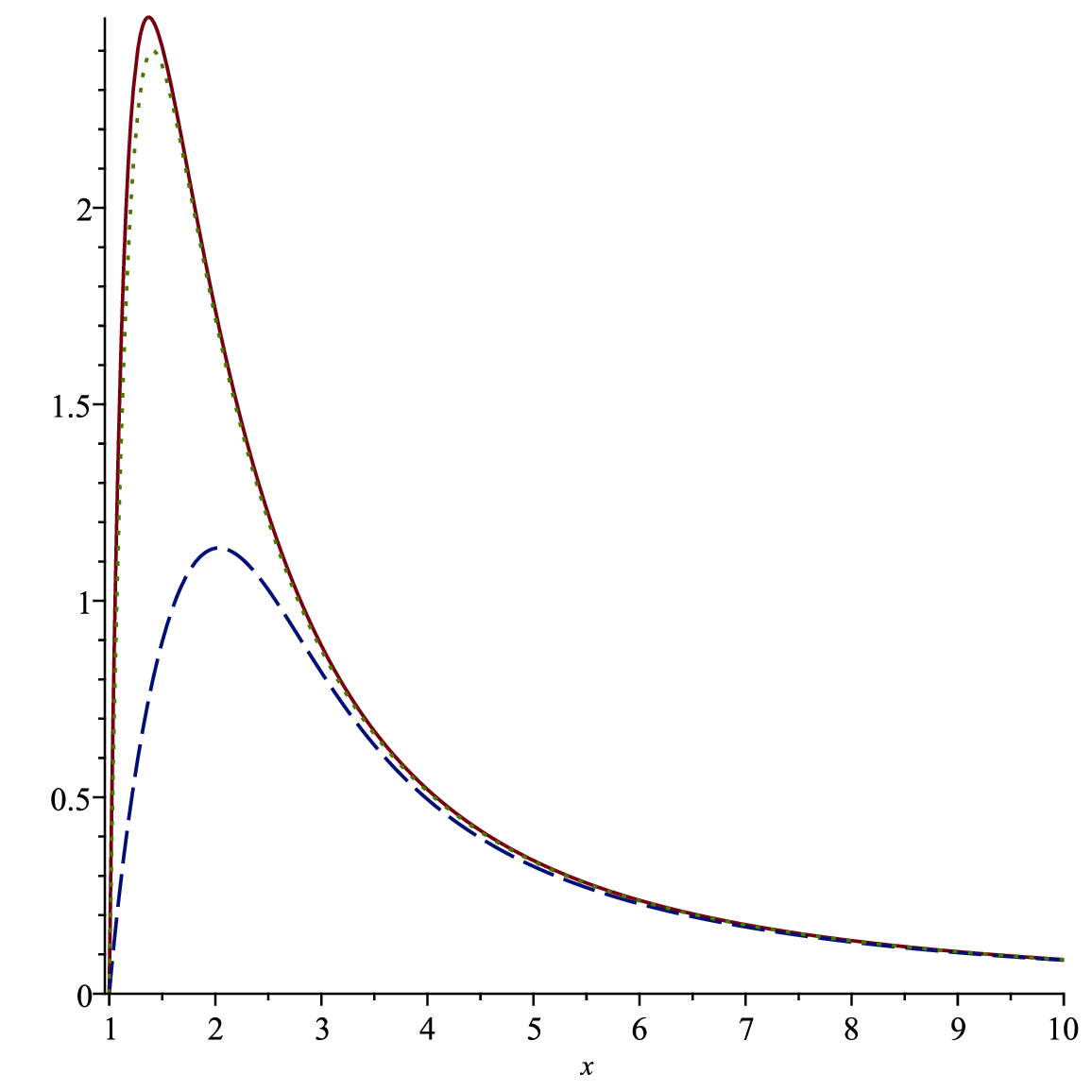}
    \includegraphics[width=0.3\textwidth]{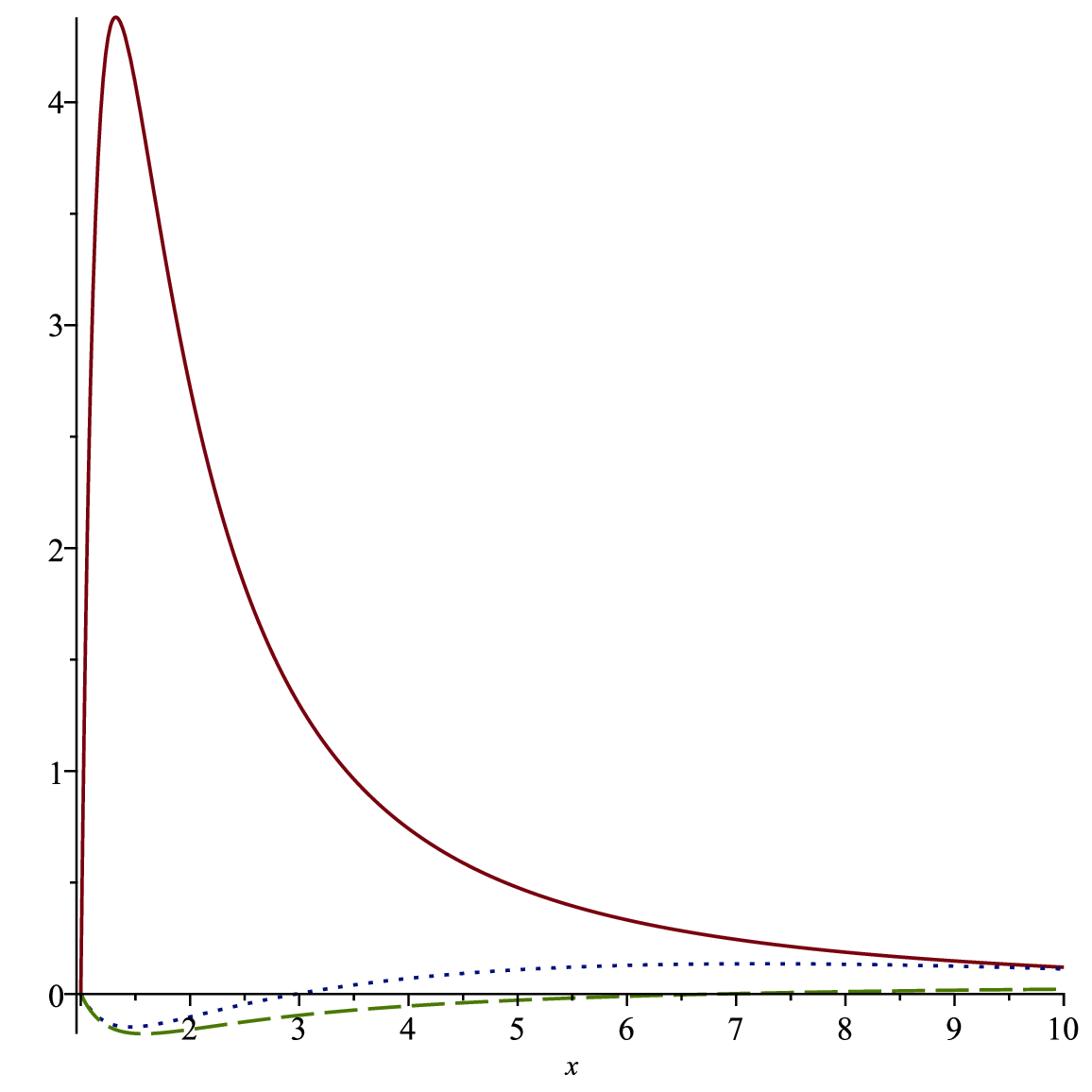}
    \includegraphics[width=0.3\textwidth]{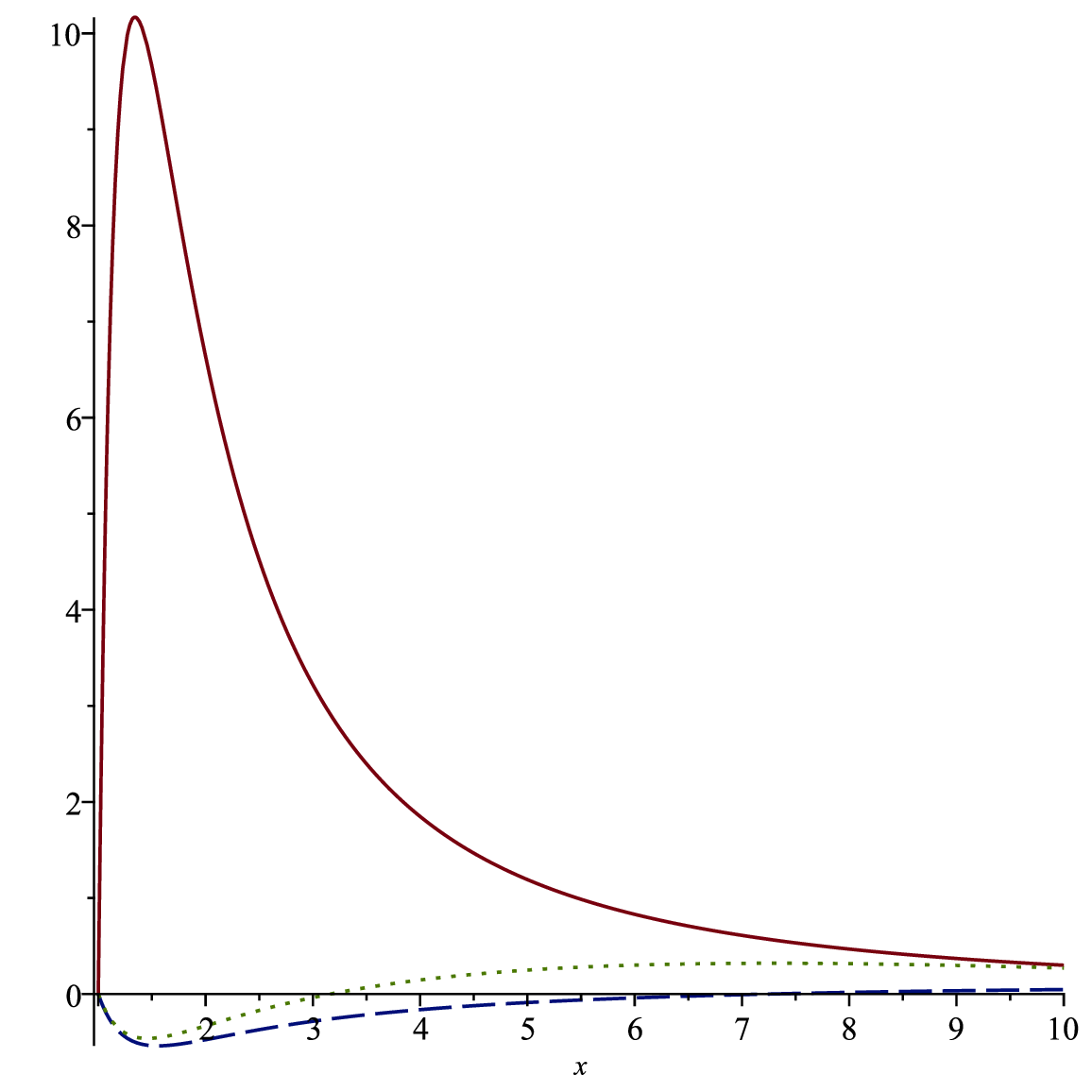}
    \caption{\label{figureUsn34}
Plots of the effective potential $\mathcal{U}_T(x)$ as a function of the rescaled variable $x$ (see equation \eqref{UT}), for different values of $\widehat{\alpha}$ and $\ell_n$. {\bf{Left panel}}: case $D=5$ and $\ell_3=2$, with $\widehat{\alpha} = 0.1$ (solid line), $\widehat{\alpha} = 0.5$  (dotted line), and $\widehat{\alpha} = 1.5$ (dashed line). {\bf{Central panel}}: $D=6$ and $\ell_4=2$, with $\widehat{\alpha}=0.1$ (solid line), $\widehat{\alpha}=10$ (dotted line), and $\widehat{\alpha}=20$ (dashed line). {\bf{Right panel}}: $D=6$ and $\ell_3=4$, with $\widehat{\alpha} = 0.1$ (solid line), $\widehat{\alpha} = 10$  (dotted line), and $\widehat{\alpha} = 20$ (dashed line).}
\end{figure}

\begin{figure}[ht!] 
    \includegraphics[width=0.3\textwidth]{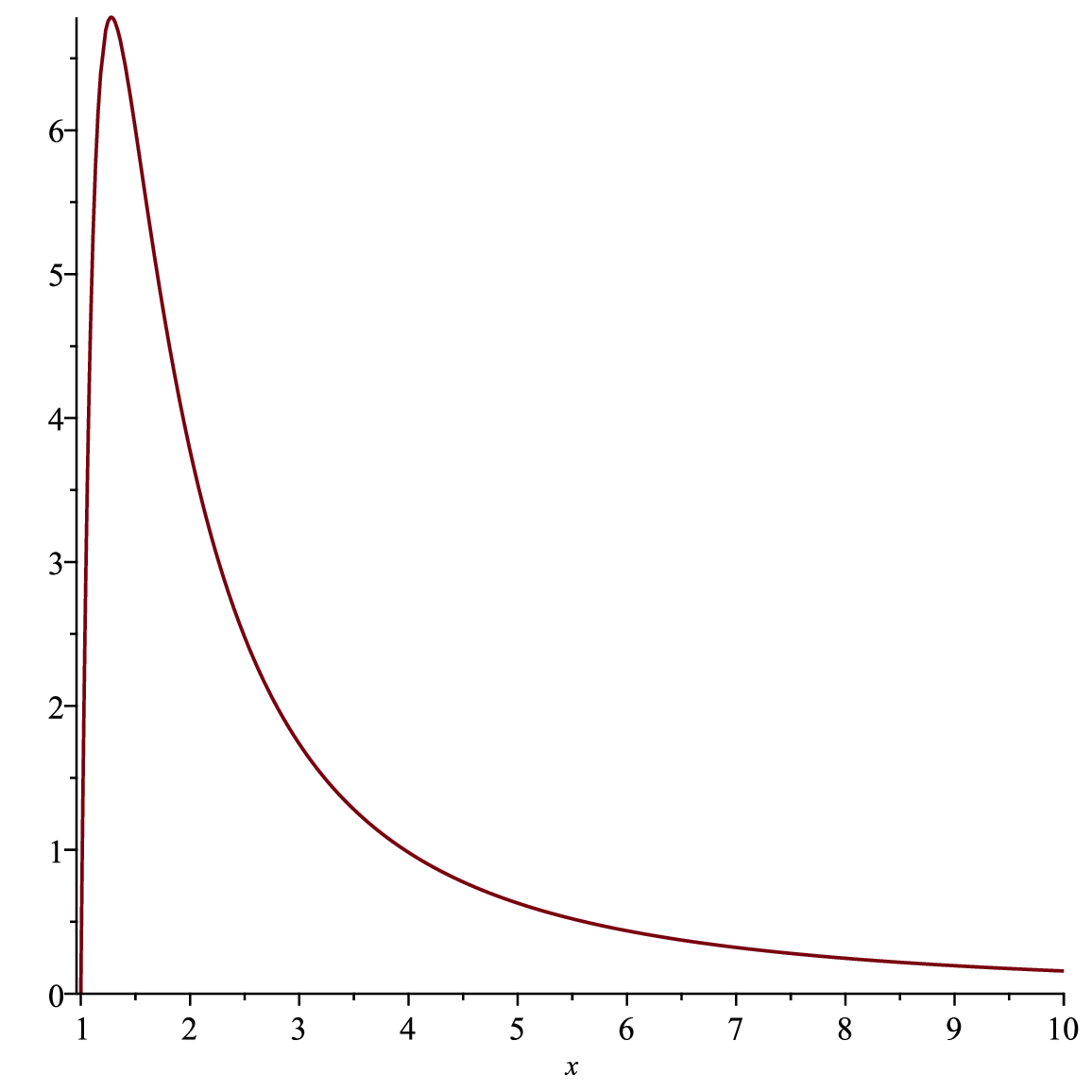}
    \includegraphics[width=0.3\textwidth]{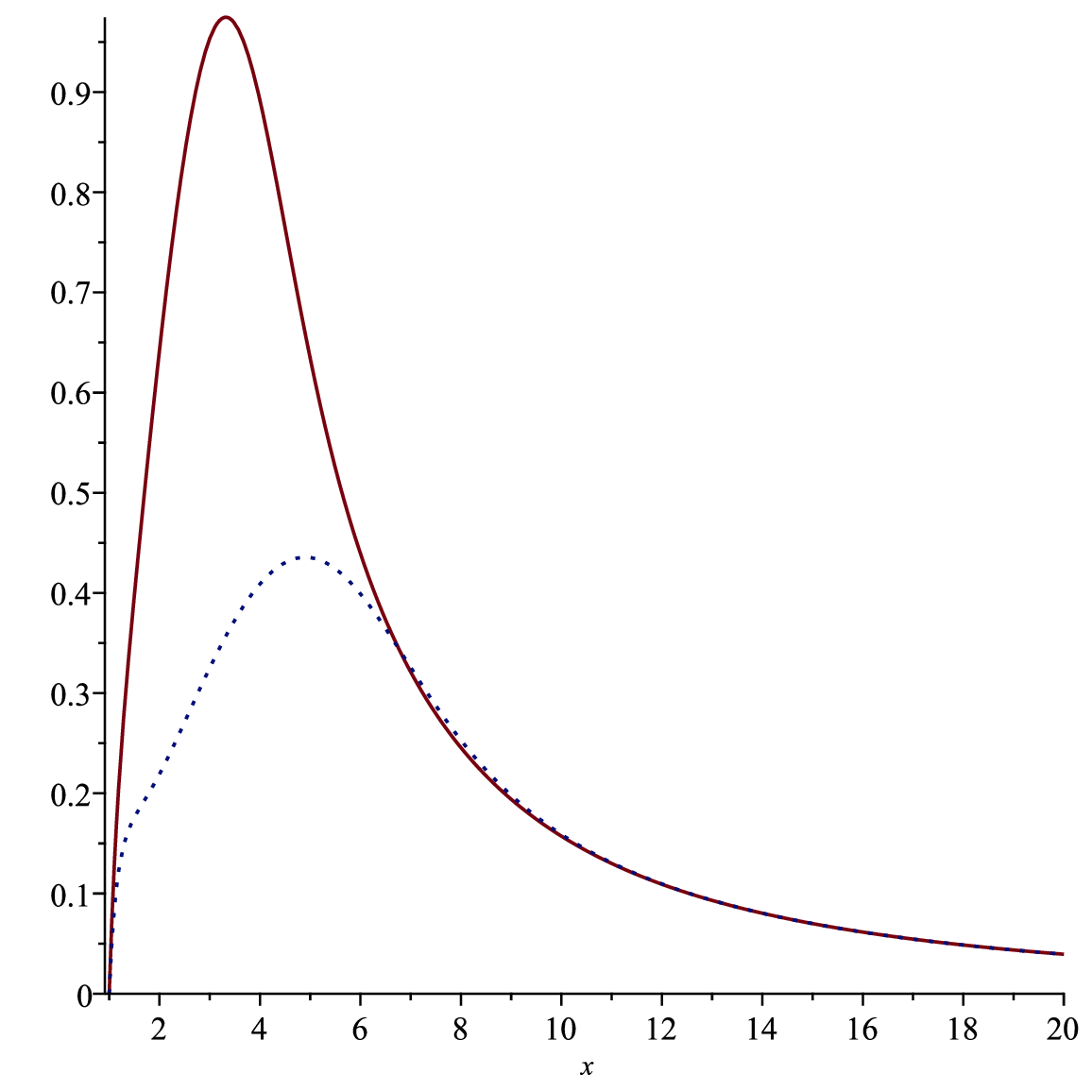}
    \includegraphics[width=0.3\textwidth]{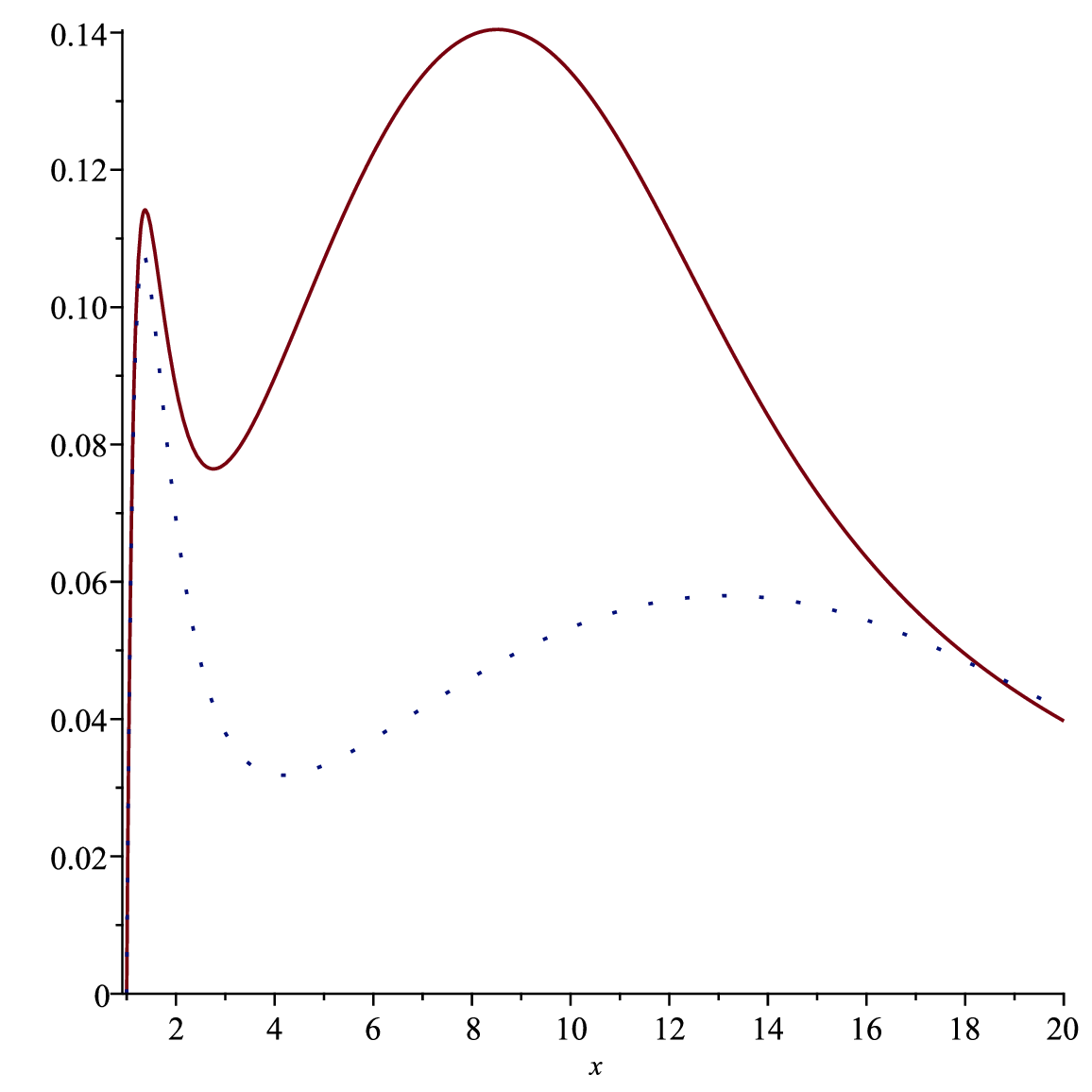}
    \caption{\label{figureUsn5}
Plots of the effective potential $\mathcal{U}_T(x)$ as a function of the rescaled variable $x$ (see equation \eqref{UT}), for different values of $\widehat{\alpha}$ and $\ell_n=2$. {\bf{Left panel}}: case $D=7$ and $\widehat{\alpha} = 0.1$. {\bf{Central panel}}: $D=7$ with $\widehat{\alpha}=10$ (solid line), and $\widehat{\alpha}=20$ (dotted line). {\bf{Right panel}}: $D=7$ with $\widehat{\alpha} = 50$ (solid line), and $\widehat{\alpha} = 100$   (spacedotted line).}
\end{figure}

\begin{figure}[ht!] 
    \includegraphics[width=0.3\textwidth]{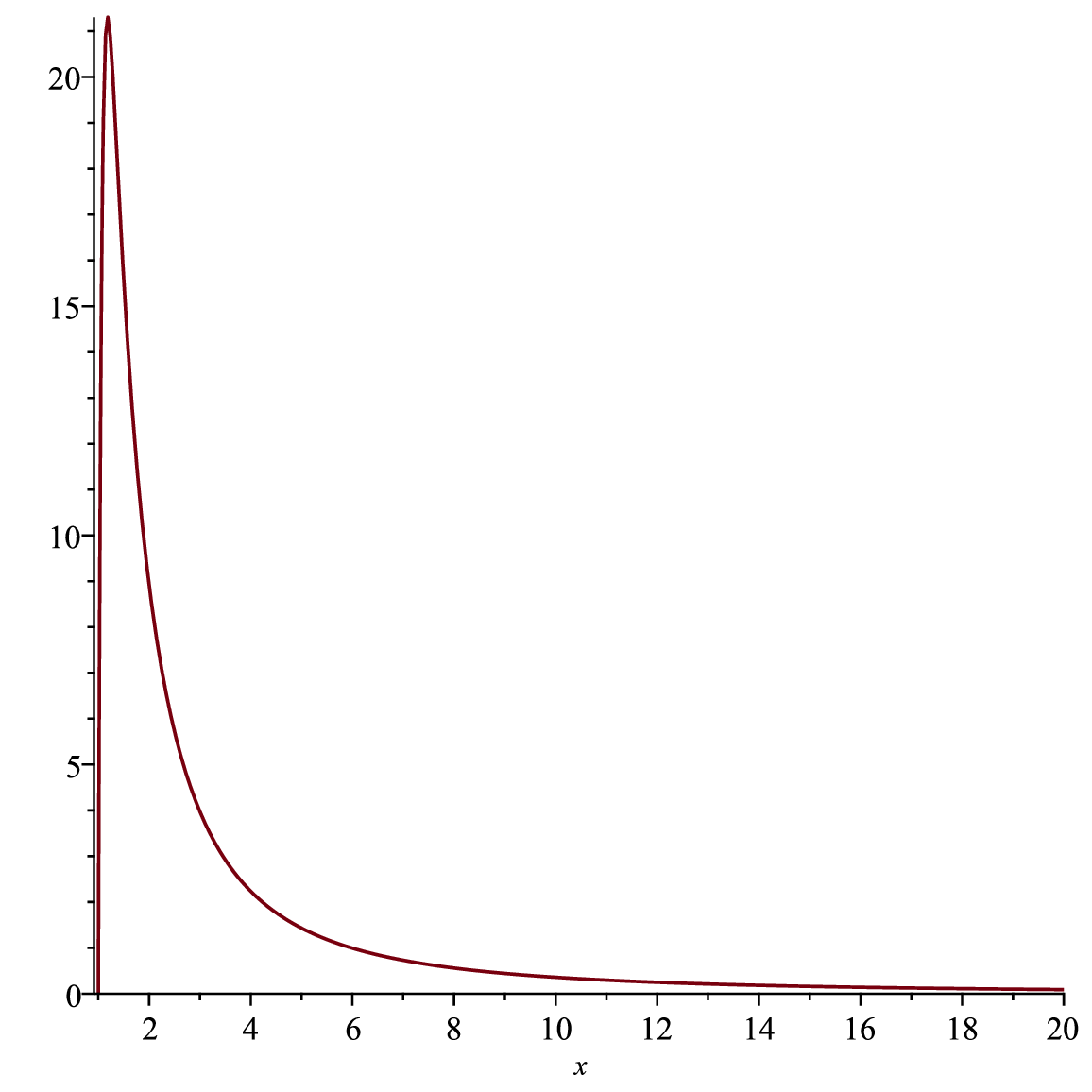}
    \includegraphics[width=0.3\textwidth]{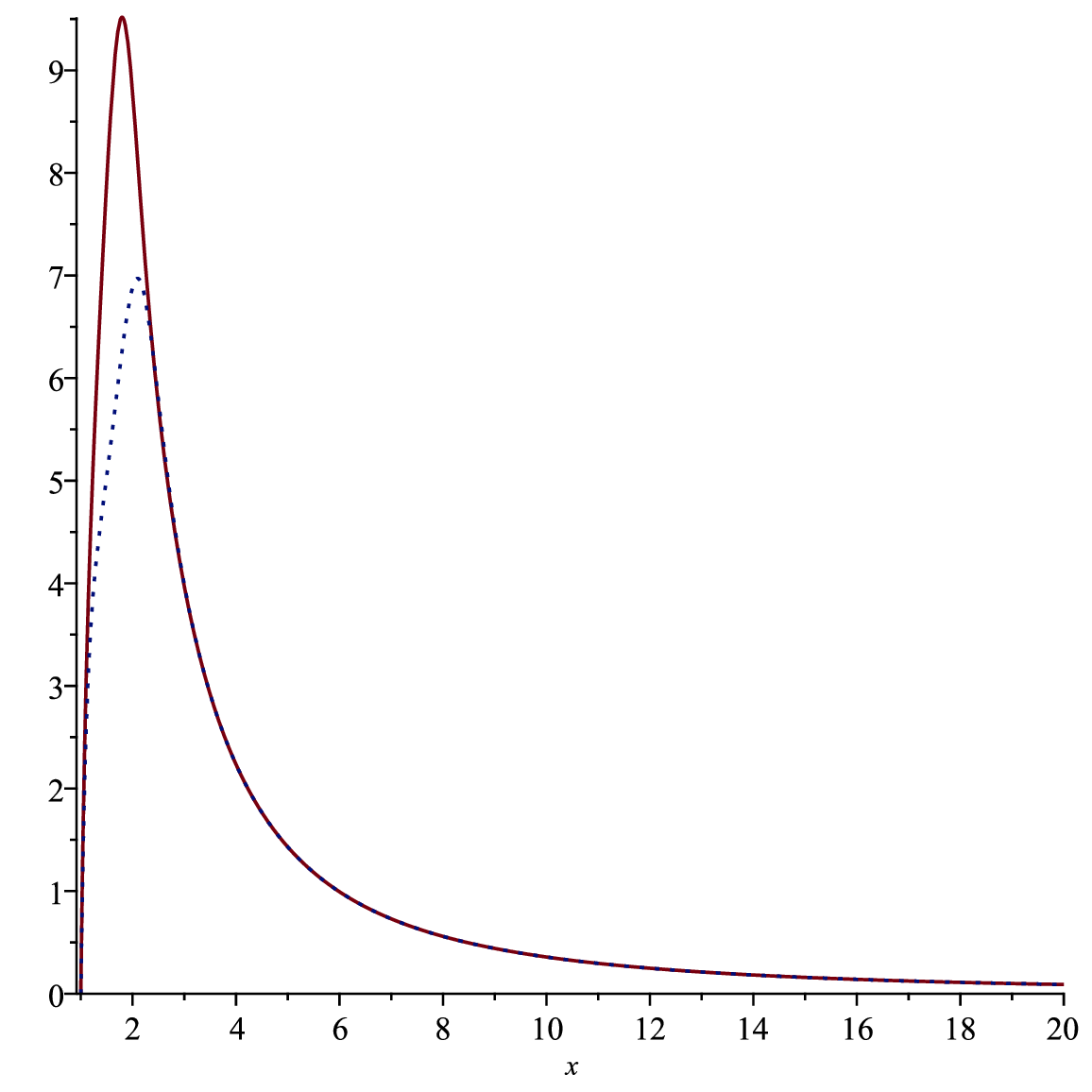}
    \includegraphics[width=0.3\textwidth]{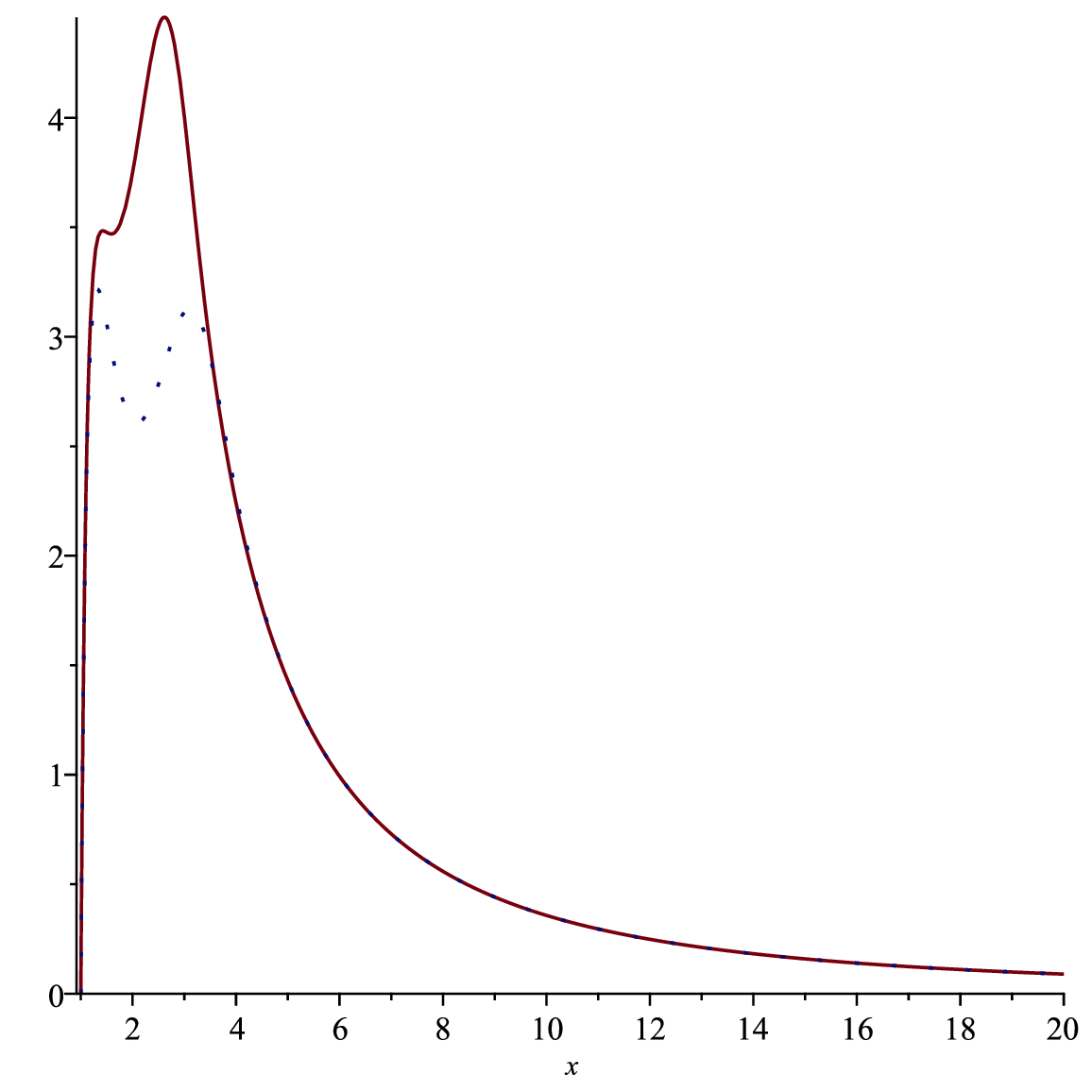}
    \caption{\label{figureUsn9}
Plots of the effective potential $\mathcal{U}_T(x)$ as a function of the rescaled variable $x$ (see equation \eqref{UT}), for different values of $\widehat{\alpha}$ and $\ell_n=2$. {\bf{Left panel}}: case $D=11$ and $\widehat{\alpha} = 0.1$. {\bf{Central panel}}: $D=11$ with $\widehat{\alpha}=10$ (solid line), and $\widehat{\alpha}=20$ (dotted line). {\bf{Right panel}}: $D=11$ with $\widehat{\alpha} = 50$ (solid line), and $\widehat{\alpha} = 100$   (spacedotted line).}
\end{figure}

Tensor perturbations $g_{\mu\nu}\to g_{\mu\nu}+h_{\mu\nu}$ of the Gauss--Bonnet-corrected Schwarzschild metric were analysed by \cite{Dotti2005CQG}. In geometrized units, and with $D=n+2$, the radial equation reads
\begin{equation}\label{ODETP}
   f(r)\frac{d}{dr}\left(f(r)\frac{dR_T}{dr}\right)+\left[\omega^2-V_T(r)\right]R_T(r)=0
\end{equation}
with
\begin{eqnarray}
V_T(r)&=&q_T(r)+\frac{f(r)}{K_T(r)}\frac{d}{dr}\left(f(r)\frac{dK_T}{dr}\right),\quad
f(r)=1+\frac{r^2}{2\alpha}\left[1-\sqrt{1+\frac{8M\alpha}{r^{n+1}}}\right],\\
K_T(r)&=&r^{\frac{n-2}{2}}\sqrt{r^2+\frac{\alpha}{n-1}\left\{(n-3)[1-f(r)]-rf^{'}(r)\right\}},\\
q_T(r)&=&\frac{\ell_n(\ell_n+n-1)f(r)}{(n-2)r^2}\cdot\frac{r^2[(n-1)(n-2)-\alpha f^{''}(r)]+\alpha(n-3)
\left\{(n-4)[1-f(r)]-2rf^{'}(r)\right\}}{(n-1)r^2+\alpha\left\{(n-3)[1-f(r)]-rf^{'}(r)\right\}},
\end{eqnarray}
where the prime denotes differentiation with respect to $r$. At this point, a clarification is necessary. The $\alpha$ that appears in the expressions for $K_T$ and $q_T$ in \cite{Dotti2005CQG} is identical to the $\alpha^{'}$ in \cite{Chakrabarti2007GRG} (see also equation (19) in \cite{Dotti2005CQG}). Furthermore, the $\alpha^{'}$ in \cite{Chakrabarti2007GRG} corresponds to the $\alpha$ in \cite{Iyer1987PRD}. Finally, the $\alpha$ appearing in \cite{Chakrabarti2007GRG} is the $\widetilde{\alpha}$ used by \cite{Iyer1989CQG}. Last but not least, the $\alpha$ in \cite{Dotti2005CQG} corresponds to the $\alpha$ used in \cite{Iyer1989CQG}. In what follows, we will adopt the notation used in \cite{Konoplya2005PRDa, Abdalla2005PRD, Chakrabarti2007GRG} to facilitate the comparison of the numerical values of the QNMs with those reported in the cited literature. If we make use of the rescalings $\rho = r/M^\frac{1}{n-1}$ and $\widehat{\alpha} = \alpha/M^\frac{2}{n-1}$, we can rewrite \eqref{ODETP} as 
\begin{equation}\label{ODET}
    f(\rho)\frac{d}{d\rho}\left(f(\rho)\frac{dR_T}{d\rho}\right)+\left[\Omega^2-U_{T}(\rho)\right]R_T(\rho) = 0,\quad \Omega=\omega M^\frac{1}{n-1}
\end{equation}
with
\begin{equation}
U_T(\rho)=q_T(\rho)+\frac{f(\rho)}{K_T(\rho)}\frac{d}{d\rho}\left(f(\rho)\frac{dK_T}{d\rho}\right),
\end{equation}
where
\begin{eqnarray}
K_T(\rho)&=&\rho^{\frac{n-2}{2}}\sqrt{\rho^2+\frac{\widehat{\alpha}}{n-1}\left\{(n-3)[1-f(\rho)]-\rho\frac{df}{d\rho}\right\}},\\
q_T(\rho)&=&\frac{\ell_n(\ell_n+n-1)f(\rho)}{(n-2)\rho^2}\cdot
\frac{\rho^2\left[(n-1)(n-2)-\widehat{\alpha}\frac{d^2 f}{d\rho^2}\right]+\widehat{\alpha}(n-3)\left\{(n-4)[1-f(\rho)]-2\rho\frac{df}{d\rho}\right\}}{(n-1)\rho^2+\widehat{\alpha}\left\{(n-3)[1-f(\rho)]-\rho\frac{df}{d\rho}\right\}}.
\end{eqnarray}
Notice that $f(\rho)$ and the equation for the event horizon are given by \eqref{frho} and \eqref{event}, respectively. Similarly to the scalar and vector cases, we introduce the rescaling $x = \rho/\rho_h$, by means of which (\ref{ODET}) becomes
\begin{equation}\label{ODET1}
  \frac{d^2 R_T}{dx^2}+P_T(x)\frac{dR_T}{dx}+Q_T(x)R_T(x)=0   
\end{equation}
with 
\begin{equation}
  P_T(x)=\frac{1}{f(x)}\frac{df}{dx},\quad 
  Q_T(x) = \frac{\rho_h^2 \Omega^2-\mathcal{U}_T(x)}{f^2(x)}
\end{equation}
and
\begin{equation}\label{UT}
  \mathcal{U}_T(x)= q_T(x)+\frac{f(x)}{K_T(x)}\frac{d}{dx}\left(f(x)\frac{dK_T}{dx}\right),
\end{equation}
where
\begin{eqnarray}
K_T(x)&=&x^{\frac{n-2}{2}}\sqrt{\rho_h^2 x^2+\frac{\widehat{\alpha}}{n-1}\left\{(n-3)[1-f(x)]-x\frac{df}{dx}\right\}},\\
q_T(x)&=&\frac{\ell_n(\ell_n+n-1)f(x)}{(n-2)x^2}\cdot
\frac{x^2\left[(n-1)(n-2)\rho_h^2-\widehat{\alpha}\frac{d^2 f}{dx^2}\right]+\widehat{\alpha}(n-3)\left\{(n-4)[1-f(x)]-2x\frac{df}{dx}\right\}}{(n-1)\rho_h^2 x^2+\widehat{\alpha}\left\{(n-3)[1-f(x)]-x\frac{df}{dx}\right\}}.
\end{eqnarray}
Here, the quantities $\kappa$ and $f(x)$ are defined as in \eqref{US}. From Figure~\ref{figureUsn34} (left panel), we observe that in $D=5$, the effective potential exhibits a qualitative behaviour similar to that of the corresponding scalar sector discussed above. However, as previously noted by \cite{Dotti2005CQG, Dotti2005PRD}, the case $D=6$ is particularly interesting, as a negative minimum in the effective potential begins to emerge for $\widehat{\alpha} \gtrsim 3.896$. This threshold corresponds to the instability condition identified in \cite{Dotti2005CQG, Dotti2005PRD}. As $\widehat{\alpha}$ increases, the minimum deepens further, signalling a stronger instability. We will not only observe but also explicitly detect these unstable QNMs using our SM. It is important to note that the WKB approximation, while often useful for analysing quasinormal spectra, is not well-suited to handle the global behaviour of such a potential, particularly in the presence of a deep negative well, which strongly affects the stability of the perturbations. Interestingly, \cite{Chakrabarti2007GRG}, while employing WKB at sixth order, hinted at the existence of such instabilities but was ultimately unable to compute them. Instead, he simply referred to the theoretical predictions of \cite{Dotti2005CQG, Dotti2005PRD}. Moreover, for $D \geqslant 7$, no negative minimum is detected. However, as $\widehat{\alpha}$ increases, a secondary local maximum emerges alongside the main peak of the effective potential (see Figures~\ref{figureUsn5} and \ref{figureUsn9}). In such cases, the WKB approximation is inadequate for accurately analysing the QNMs. In particular, for $D=11$ (see the right panel of Figure~\ref{figureUsn9}) and $\widehat{\alpha} \approx 100$, we observe the formation of two closely spaced peaks of nearly equal height. Here again, the WKB method proves unsuitable for capturing the quasinormal spectrum. Furthermore, as $D$ increases, the emergence of a secondary peak occurs at progressively higher values of $\widehat{\alpha}$, highlighting a trend in the potential's structure that further challenges the applicability of WKB-based approaches.  

Finally, the asymptotic behaviour of the solutions to equation (\ref{ODET1}) can be obtained using the method outlined in \cite{Olver1994MAA}. For this purpose, we split our analysis by examining the behaviour of the radial field in two different regions. Concerning the asymptotic behaviour as $x\to 1^+$, we observe that
\begin{equation}  
    P_T(x) = \frac{1}{x-1} + \mathcal{O}(1), \quad  
    Q_T(x) = \left(\frac{\rho_h \Omega}{f'(1)}\right)^2 \frac{1}{(x-1)^2} + \mathcal{O}((x-1)^{-1}). 
\end{equation}  
This indicated that the point $x = 1$ is a simple pole for $P_T(x)$ and a second-order pole for $Q_T(x)$. Furthermore, the exponents associated with this singularity are identical to those found in the scalar case at the event horizon. Therefore, we can follow the same approach and conclude that the QNM boundary condition at $x=1$ is given by
\begin{equation}\label{QNMBCz1T}
R_T\underset{{x\to 1^+}}{\longrightarrow} (x-1)^{-i \rho_h a\Omega},\quad
a=\frac{1}{f^{'}(1)}.
\end{equation}
Regarding the asymptotic behaviour as $x\to +\infty$, we start by observing that in the limit of $x \to +\infty$
\begin{equation}\label{pqT}
    P_T(x) = \sum_{m=0}^\infty\frac{\mathfrak{f}_m}{x^m} = \mathcal{O}\left(\frac{1}{x^n}\right), \qquad
    Q_T(x) = \sum_{m=0}^\infty\frac{\mathfrak{g}_m}{x^m}=\rho_h^2\Omega^2+\mathcal{O}\left(\frac{1}{x^2}\right).
\end{equation}
with $n\geqslant 3$. Since this result matches the scalar case, we can apply the same approach and conclude that the QNM boundary condition at positive spatial infinity is given by
\begin{equation}\label{QNMBCposinfT}
    R_T\underset{{x\to +\infty}}{\longrightarrow} e^{i\rho_h\Omega x}.
\end{equation}
At this point, we transform the radial function $R_T(x)$ into a new radial function $\Phi_T(x)$ such that the QNM boundary conditions are automatically implemented and  $\Phi_{T}(x)$ is regular at $x = 1$ and at space-like infinity. To this aim, we consider the transformation
\begin{equation}\label{AnsatzT}
    R_T(x) = x^{i\rho_h a\Omega}(x-1)^{-i\rho_h a\Omega}e^{i\rho_h\Omega(x-1)} \Phi_T(x).
\end{equation}
Multiplying \eqref{ODET1} by $f^2(x)$ and substituting \eqref{AnsatzT}, we obtain the following ordinary differential equation for the radial eigenfunctions
\begin{equation}\label{ODEznoneT}  
    P_{2T}(x) \Phi''_T(x) + P_{1T}(x) \Phi'_T(x) + P_{0T}(x) \Phi_T(x) = 0,  
\end{equation}  
where $P_{2T}(x) = P_2(x)$ and $P_{1T}(x) = P_1(x)$, with $P_1(x)$ and $P_2(x)$ given by \eqref{P1} and \eqref{P2}. The term $P_{0T}(x)$ is obtained from \eqref{P0} by replacing $\mathcal{U}_S(x)$ with $\mathcal{U}_T(x)$. Let us now introduce the transformation $x = 2/(1-y)$ mapping the point at infinity and the event horizon to $y = 1$ and $y = -1$, respectively. Furthermore, a dot denotes differentiation with respect to the new variable $y$. Then, equation \eqref{ODEznoneT} becomes
\begin{equation}\label{ODEynoneT}
    S_{2T}(y)\ddot{\Phi}_{T}(y) + S_{1T}(y)\dot{\Phi}_{T}(y) + S_{0T}(y)\Phi_{T}(y) = 0,
\end{equation}
where  $S_{2T}(y) = S_2(y)$ and $S_{1T}(y) = S_1(y)$, with $S_1(y)$ and $S_2(y)$ given by \eqref{S1onone} and \eqref{S2onone}. Also here, the term $S_{0T}(y)$ is obtained from \eqref{S0onone} by replacing $\mathcal{U}_S(y)$ with $\mathcal{U}_T(y)$ in \eqref{fv2T}. Notice that we must also require that $\Phi_{T}(y)$ is regular at $y=\pm 1$. As a result of the transformation introduced above, $f(y)$ is still given by \eqref{fv1}, while the effective potential is
\begin{eqnarray}
\mathcal{U}_T(y)&=&q_T(y)+\frac{(1-y)^3 f(y)}{4K_T(y)}\left\{(1-y)\dot{f}(y)\dot{K}_T(y)+f(y)\left[(1-y)\ddot{K}_T(y)-2\dot{K}_T(y)\right]\right\},\label{fv2T}\\
K_T(y)&=&\frac{2^{\frac{n}{2}-1}}{{(1-y)^\frac{n}{2}}}\sqrt{4\rho_h^{2}+\frac{\widehat{\alpha}}{n-1}(1-y)^2\left\{(n-3)[1-f(y)]-(1-y)\dot{f}(y)\right\}},\\
q_T(y)&=&\frac{\ell_n(\ell_n+n-1)(1-y)^2f(y)}{4(n-2)}\cdot\nonumber
\end{eqnarray}
\begin{equation}
\cdot\frac{4(n-1)(n-2)\rho_h^2-\widehat{\alpha}(1-y)^3[(1-y)\ddot{f}(y)-2\dot{f}(y)]+\widehat{\alpha}(n-3)(1-y)^2\left\{(n-4)[1-f(y)]-2(1-y)\dot{f}(y)\right\}}{4(n-1)\rho_h^2+\widehat{\alpha}(1-y)^2\left\{(n-3)[1-f(y)]-(1-y)\dot{f}(y)\right\}}.
\end{equation}

\begin{table}%[ht]
\caption{Classification of the points $y=\pm 1$ for the relevant functions defined by \eqref{S2onone}-\eqref{S0onone}, \eqref{fv1}, and \eqref{fv2T}. The abbreviations $z$ ord $n$ and $p$ ord $m$ stand for zero of order $n$ and pole of order $m$, respectively.}
\begin{center}
\begin{tabular}{ | c | c | c | c | c | c | c | c }
\hline
$y$  & $f(y)$  & $\mathcal{U}_T(y)$ & $S_{2T}(y)$ & $S_{1T}(y)$ & $S_{0T}(y)$\\ \hline
$-1$ & z \mbox{ord} 1 & z \mbox{ord} 1 & z \mbox{ord} 4& z \mbox{ord} 3 & z \mbox{ord} 2 \\ \hline
$+1$ & $+1$           & z \mbox{ord} 2 & $+4$ & p \mbox{ord} 2 & p \mbox{ord} 2\\ \hline
\end{tabular}
\label{tableEinsnoneT}
\end{center}
\end{table}

Table~\ref{tableEinsnoneT} shows that the coefficients of the differential equation \eqref{ODEynoneT} share a common zero of order $2$ at $y = -1$ while $y = 1$ is a pole of order $2$ for the coefficients $S_{1T}(y)$ and $S_{0T}(y)$. Hence, in order to apply the SM, we need to multiply (\ref{ODEynoneT}) by $(1-y)^2/(1+y)^2$. As a result, we obtain the following differential equation 
\begin{equation}\label{MT}  
    M_{2T}(y) \ddot{\Phi}_{T}(y) + M_{1T}(y) \dot{\Phi}_{T}(y) + M_{0T}(y) \Phi_{T}(y) = 0,  
\end{equation}  
where $M_{2T}(y) = M_2(y)$, $M_{1T}(y) = M_1(y)$, and $M_{0T}(y) = M_0(y)$, with $\mathcal{U}_S(y)$ replaced by $\mathcal{U}_T(y)$. We recall that $M_i(y)$, $N_0(y)$, $C_2(y)$, $C_1(y)$, and $C_0(y)$ were introduced in \eqref{S210honone} and \eqref{N0}--\eqref{C0}; in the tensor sector one replaces $\mathcal{U}_S(y)$ by $\mathcal{U}_T(y)$ in \eqref{C0}. At this point, it can be easily verified with Maple that
\begin{eqnarray}
    &&\lim_{y\to 1^{-}}M_{2T}(y)=0=\lim_{y\to -1^{+}}M_{2T}(y),\\
    &&\lim_{y\to 1^{-}}M_{1T}(y)=4i\rho_h\Omega,\quad
    \lim_{y\to -1^{+}}M_{1T}(y)=0,\\
    &&\lim_{y\to 1^{-}}M_{0T}(y)=2a\rho_h^2\Omega^2-\frac{n}{4}(n-2)-\ell_n(\ell_n+n-1),\quad
    \lim_{y\to -1^{+}}M_{0T}(y)=B_2\Omega^2,
\end{eqnarray}
where $B_2$ is defined in \eqref{B2}. It is noteworthy that the tensor and vector cases exhibit distinct limiting behaviour in the regime $y \to 1^{-}$, specifically within the functions $M_{0T}(y)$ and $M_{0V}(y)$. As a final step prior to implementing the SM, we rewrite the differential equation \eqref{MT} in the following form
\begin{equation}\label{TSCHT}
  L_{0T}\left[\Phi_{T}, \dot{\Phi}_{T}, \ddot{\Phi}_{T}\right] +  i\Omega L_{1T}\left[\Phi_{T}, \dot{\Phi}_{T}, \ddot{\Phi}_{T}\right] +  \Omega^2 L_{2T}\left[\Phi_{T}, \dot{\Phi}_{T}, \ddot{\Phi}_{T}\right] = 0
\end{equation}
with
\begin{eqnarray}
L_{0T}\left[\Phi_{T}, \dot{\Phi}_{T}, \ddot{\Phi}_{T}\right] &=& L_{00T}(y)\Phi_{T} + L_{01T}(y)\dot{\Phi}_{T} + L_{02T}(y)\ddot{\Phi}_{T},\label{L0noneT}\\
L_{1T}\left[\Phi_{T}, \dot{\Phi}_{T}, \ddot{\Phi}_{T}\right] &=&L_{10T}(y)\Phi_{T} + L_{11T}(y)\dot{\Phi}_{T}+L_{12T}(y)\ddot{\Phi}_{T}, \label{L1noneT}\\
L_{2T}\left[\Phi_{T}, \dot{\Phi}_{T}, \ddot{\Phi}_{T}\right]&=&L_{20T}(y)\Phi_{T} +L_{21T}(y)\dot{\Phi}_{T} + L_{22T}(y)\ddot{\Phi}_{T}.\label{L2noneT}
\end{eqnarray}

\begin{table}%[ht]
\caption{Behaviour of the coefficients $L_{ijT}$ at the endpoints of the interval $-1 \leq y \leq 1$ for the tensor perturbation of the Schwarzschild black hole with GB correction.}
\begin{center}
\begin{tabular}{ | c | c | c | c | c | c | c | c }
\hline
$(i,j)$ & $\displaystyle{\lim_{y\to -1^+}}L_{ijT}$  & $L_{ijT}$       & $\displaystyle{\lim_{y\to 1^-}}L_{ijT}$  \\ \hline
$(0,0)$ &  $0$                                     & $C_{0T}$          & $-\frac{n}{4}(n-2)-\ell_n(\ell_n+n-1)$\\ \hline
$(0,1)$ &  $0$                                     & $N_{0T}$          & $0$\\ \hline
$(0,2)$ &  $0$                                     & $M_{2T}$          & $0$\\ \hline 
$(1,0)$ &  $0$                                     & $C_{1T}$          & $0$\\ \hline 
$(1,1)$ &  $0$                                     & $N_{1T}$          & $4\rho_h$\\ \hline 
$(1,2)$ &  $0$                                     & $0$            & $0$\\ \hline 
$(2,0)$ &  $B_2$                                   & $C_{2T}$          & $2a\rho_h^2$\\ \hline
$(2,1)$ &  $0$                                     & $0$            & $0$\\ \hline
$(2,2)$ &  $0$                                     & $0$            & $0$\\ \hline
\end{tabular}
\label{tableDreiTensor}
\end{center}
\end{table}

Moreover, all the $L_{ijT}$ terms appearing in (\ref{L0noneT})-(\ref{L2noneT}) exhibit a regular behavior at the endpoints $y = \pm 1$ with limits given as in Table~\ref{tableDreiTensor}. The numerical implementation of the resulting quadratic eigenvalue problem is the same as in the scalar sector and is described in Sec.~III.

\section{Tensor quasinormal spectra}

In this section, we perform a detailed examination of the QNM spectra for tensor-type perturbations of Schwarzschild--Gauss--Bonnet black holes across dimensions $D \in \{5, 6, 7, 8, 10, 11, 12, 26\}$. Our analysis reveals several systematic trends as well as significant dimension-dependent differences. First, purely imaginary modes emerge prominently at moderate to strong GB couplings in lower dimensions, particularly in $D \in \{5, 6\}$, where they either form nearly equispaced towers or arise as isolated modes. As the dimension increases, however, the overdamped sector gradually fades away, with no overdamped modes detected for $D \geqslant 8$, signalling a major shift in the QNM landscape between lower and higher dimensions. Second, a characteristic non-monotonic behaviour in the real parts of higher overtones appears robustly across all dimensions, though it becomes more systematic and pronounced in higher-dimensional cases. Third, the morphology of the QNM distributions in the complex frequency plane exhibits a clear and progressive evolution with increasing coupling. At small GB couplings, typical structures such as Martini glasses, beer glasses, and 'coupe à glace' dominate, whereas at stronger couplings, bottleneck and pyramidal configurations take over, indicating a qualitative reorganization of the spectrum.

A particularly important result is the identification of tensorial instabilities in $D = 6$, consistent with earlier theoretical predictions \cite{Dotti2005CQG, Dotti2005PRD}, but confirmed here for the first time by the high-precision SM. We further derive an explicit instability threshold in the form of an inequality, $GM \leq 158.1\,\alpha^{3/2}$, relating the black hole mass to the GB coupling. This relation indicates that the instability is confined to microscopic black holes near the string scale. No analogous instabilities are found for higher-dimensional spacetimes ($D \geqslant 7$), suggesting that GB-corrected black holes in the string-theoretic regime remain dynamically stable under tensor perturbations. Additionally, our spectral analysis shows that while the lower-dimensional cases exhibit a complex and rich overtone structure including overdamped modes, spectral gaps, and morphology transitions, the higher-dimensional cases ($D\in\{10,11,12,26\}$) are much simpler, often maintaining 'coupe à glace'-type profiles across the full range of couplings. Last but not least, throughout our study, we consistently observe that the sixth-order WKB method fails to resolve key features of the spectrum, such as overdamped modes, non-monotonic behaviours, and the onset of instabilities, thus highlighting the necessity of using spectral methods for an accurate characterisation of the QNM landscape. 

In $D = 26$, for a black hole of mass $M$ and GB coupling $\widehat{\alpha} = 0.1$, the fundamental tensor QNM exhibits a dimensionless frequency of $\Omega = 11.1395 - 1.9867i$. Although a direct comparison with the four-dimensional Schwarzschild case is complicated by differences in mass scaling with spacetime dimensionality, the larger dimensionless frequency suggests an amplification phenomenon in higher dimensions. In particular, the real part is approximately thirty times larger than that of the corresponding tensor mode in $D = 4$, where $\Omega_{\text{Sch}} = 0.3737 - 0.0890i$. To assess potential observational implications, we convert the QNM frequencies into physical units (Hz) using the relation
\begin{equation}
  \nu = \frac{1}{2\pi} \left( \frac{c^3}{G M_\odot} \right)^p \frac{\Omega}{\eta^p}, \quad p = \frac{1}{n-1},
\end{equation}
where $\Omega$ is the dimensionless frequency, $M$ is the black hole mass, $M_\odot$ the solar mass, and $\eta=M/M_\odot$. For a stellar-mass black hole with $M=10 M_\odot$, the mode $\Omega=11.1395-1.9867i$ translates to a frequency $\nu=2.729-0.487i$ Hz, while for a supermassive black hole with $M=10^9 M_\odot$, it becomes $\nu=1.225-0.218i$ Hz. By comparison, in the case of superstring theory ($D=10$) with the same GB coupling, the fundamental tensor mode has $\Omega=3.5629-0.8576i$, corresponding to $\nu=2.339-0.563i$ Hz for stellar black holes and $\nu=0.168-0.041i$ Hz for supermassive black holes. Although these frequencies remain below the detection thresholds of current ground-based observatories such as LIGO and VIRGO (which have a lower sensitivity limit around 20 Hz) and lie beyond LISA’s optimal range (peaking near 1 Hz), they fall within the few-Hz to tens-of-Hz band targeted by future space-based detectors like DECIGO \cite{Kawamura2008JPCS, Kawamura2019IJMPD, Kawamura2021PTEP}. These findings thus suggest that higher-dimensional corrections predicted by string theory could leave detectable signatures in gravitational wave signals. In particular, the amplification of QNM frequencies in string-theoretically motivated dimensions ($D \in \{10, 26\}$) may render the tensorial ringdown phase of stellar and supermassive black holes observable with next-generation instruments, offering a promising avenue for empirical tests of string-inspired modifications to general relativity. Finally, we emphasise that in the string-motivated dimensions $D\in\{10,11,12,26\}$, the tensor spectrum remains distinct from both the scalar and vector spectra. No isospectral correspondence emerges in the tensor sector, whereas an exact scalar--vector match at vanishing GB coupling is established analytically in Appendix~\ref{AppendixA} and confirmed numerically in the scalar and vector sectors. This highlights a fundamental distinction in the spectral properties of gravitational perturbations compared to their lower-spin counterparts.

The results are organised dimension by dimension in the following subsections, where we highlight the main spectral features, compare the performance of different numerical methods, and provide a detailed mapping of the quasinormal mode landscape.

\subsection{$D = 5$}

\begin{figure}
    \centering
    \subfloat[$n=3$, $\ell_{3}=2$, $\widehat{\alpha}=0$ (Beer Glass)]{%
    \includegraphics[width=0.49\textwidth]{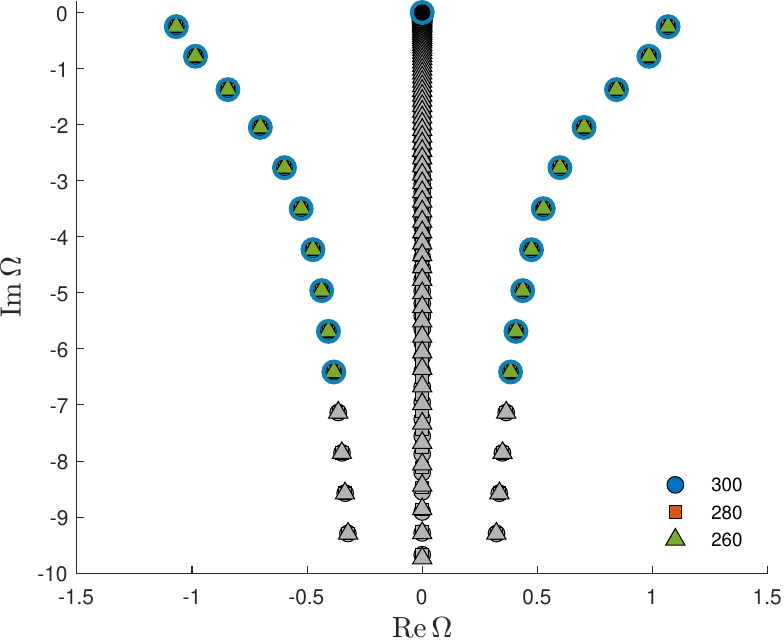}
    }
    \subfloat[$n=3$, $\ell_{3}=3$, $\widehat{\alpha}=0.5$ (Deformed Beer Glass)]{%
    \includegraphics[width=0.49\textwidth]{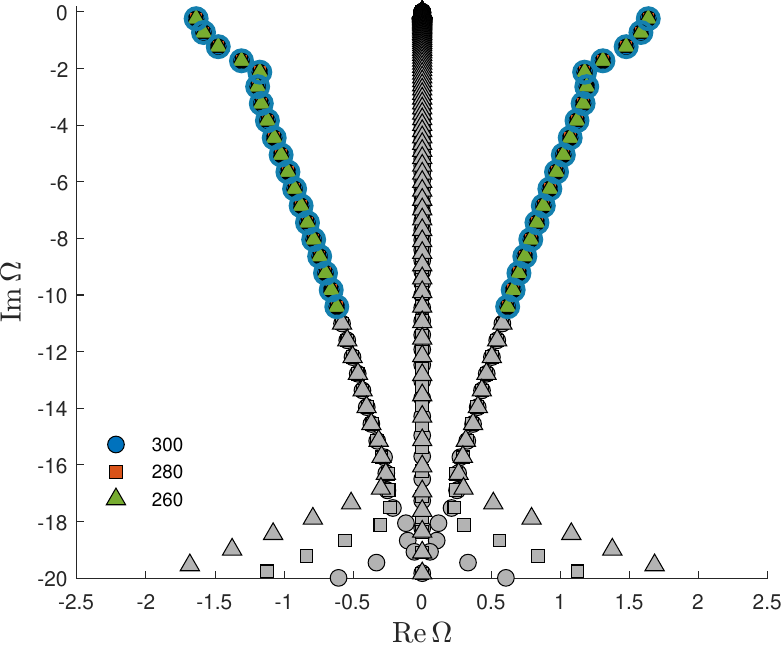}
    }
    \caption{Illustration of some QNM distribution morphologies observed in the tensorial perturbation spectrum. The left panel displays a typical beer glass configuration, while the right panel shows a deformed beer glass. Here, $n$, $\ell_n$, and $\widehat{\alpha}$ denote the number of compactified dimensions, the angular momentum, and the coupling parameter, respectively.}
    \label{atlasQNM2t}
\end{figure}

\begin{figure}
    \centering
    \subfloat[$n=3$, $\ell_{3}=3$, $\widehat{\alpha}=1.9$ (Truncated Pyramid)]{%
    \includegraphics[width=0.49\textwidth]{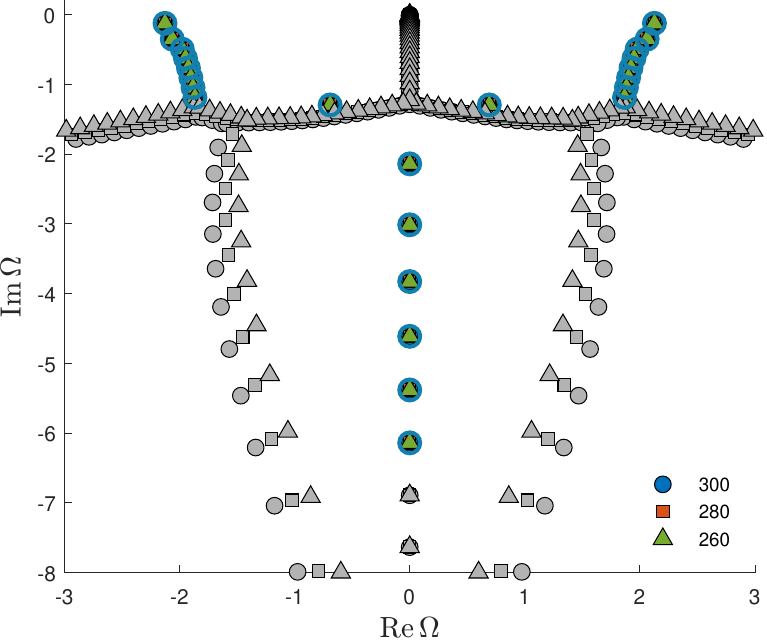}
    }
    \subfloat[$n=4$, $\ell_{4}=3$, $\widehat{\alpha}=0.5$ (Deformed Martini)]{%
    \includegraphics[width=0.49\textwidth]{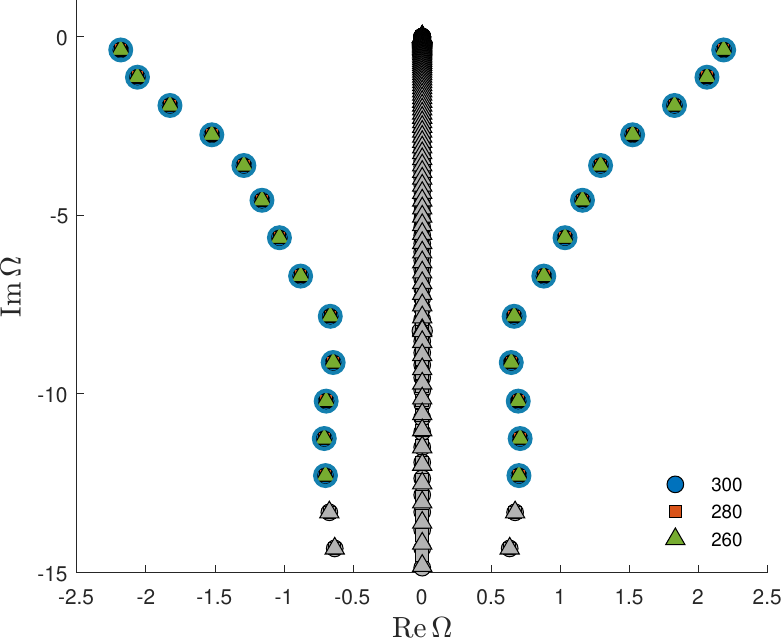}
    }
    \caption{Illustration of some typical QNM distribution morphologies observed in the tensorial perturbation spectrum. Here, $n$, $\ell_n$, and $\widehat{\alpha}$ denote the number of compactified dimensions, the angular momentum, and the coupling parameter, respectively.}
    \label{atlasQNM3t}
\end{figure}

For $\ell_3 = 2$, no overdamped modes are detected in the regime $0 \leq \widehat{\alpha} \leq 0.1$. As the coupling increases from $\widehat{\alpha}=0.2$ to $\widehat{\alpha}=1.99$, the number of purely imaginary modes steadily decreases (see Table~\ref{table:piD5L2and3}), with only two such modes remaining at $\widehat{\alpha} = 1.99$ for $\ell_3 \in \{2, 3\}$. Nevertheless, in this range, the tower of overdamped modes appears to be approximately equally spaced. For $\ell_3 = 4$, overdamped modes are again observed throughout the interval $0.2 \leq \widehat{\alpha} \leq 1.99$, and their number similarly diminishes as the coupling increases (see Table~\ref{table:piD5L2and3}). As with lower multipoles, the overdamped spectrum exhibits an approximately uniform spacing for small values of $\widehat{\alpha}$. We observe non-monotonic behaviour in the real part of the QNMs for $\ell_3 = 2$ in the range $0.5 \leq \widehat{\alpha} \leq 1.99$, with the exception of $\widehat{\alpha}=1.9$ (see bold entries in Tables~\ref{table:D5s2L2and3} and~\ref{table:D5s2L4}). For $\ell_3 = 3$, non-monotonicity is again present over the same coupling range. Although no bold entry appears in Table~\ref{table:D5s2L2and3} for $\ell_3 = 3$ and $\widehat{\alpha} = 1.9$, this is due to the relevant QNM, located at $\Omega = 0.6914 - 1.2938i$, lying beyond the fifth overtone and thus omitted for brevity. Similarly, for $\ell_3 = 4$, non-monotonic behavior sets in later for $1\leq\widehat{\alpha}\leq 1.99$. The absence of a bold entry at $\widehat{\alpha} = 1.9$ in Table~\ref{table:D5s2L4} reflects the same limitation: the relevant mode, located at $\Omega = 0.9355 - 1.4527i$, exceeds the fifth overtone and was not included in the table. Regarding the sixth-order WKB method employed by \cite{Chakrabarti2007GRG}, no results are reported for the case $\widehat{\alpha} = 0$. Across all tested angular momenta $\ell_3 \in \{2, 3, 4\}$, the WKB approximation captures only the imaginary part of the QNMs with acceptable accuracy in the low-coupling regime $0.1\leq\widehat{\alpha}\leq 0.2$, but fails to provide reliable estimates for the real part. As the coupling increases beyond $\widehat{\alpha} \geqslant 5$, the accuracy of the WKB method deteriorates significantly. Furthermore, unlike our SM, it fails to identify both the presence of overdamped modes and the non-monotonic behaviour observed in the real part of the QNM spectrum. Finally, regarding the morphology of the QNM spectrum, we observe a beer glass distribution (for a typical example, see panel (a) in Fig.~\ref{atlasQNM2t}) for all tested values of $\ell_3 \in \{2,3,4\}$ in the range $0\leqslant\widehat{\alpha}\leqslant 0.2$. This shape gradually transitions into a deformed beer glass (for a typical example, see panel (b) in Fig.~\ref{atlasQNM2t}) at $\widehat{\alpha}=0.5$ across all values of $\ell_3$. At $\widehat{\alpha}=1$, the spectrum acquires a pyramidal profile (see for example (d) in Fig.~\ref{atlasQNMdistribtens}), which then evolves into a truncated pyramid (for a typical example, see (a) in Fig.~\ref{atlasQNM3t}) at $\widehat{\alpha}=1.9$. By the time $\widehat{\alpha}=1.99$, the structure simplifies significantly: only three oscillatory QNMs remain for each $\ell_3$, accompanied by a small number of overdamped modes.

\subsection{$D = 6$}

\begin{figure}
    \centering
    \subfloat[$n=4$, $\ell_{3}=2$, $\widehat{\alpha}=10$]{%
    \includegraphics[width=0.49\textwidth]{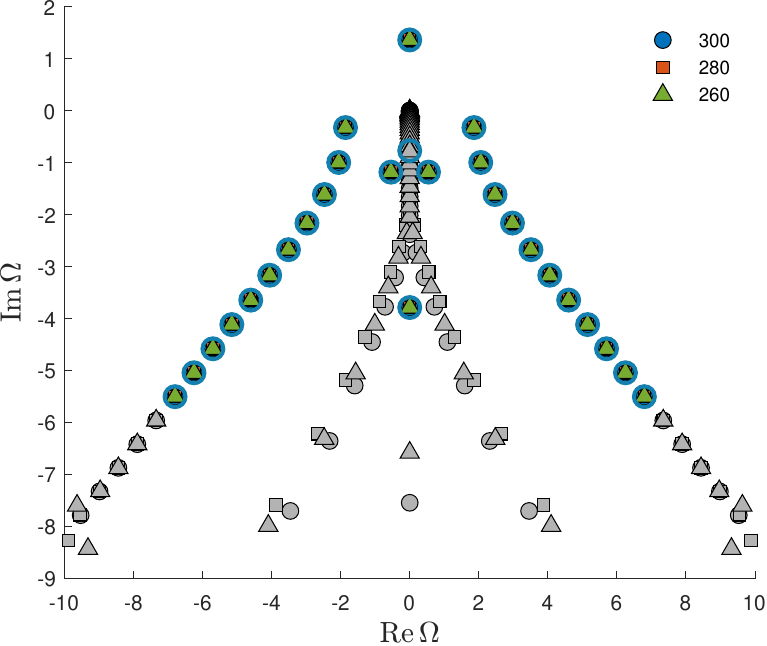}
    }
    \subfloat[$n=4$, $\ell_{3}=3$, $\widehat{\alpha}=20$]{%
    \includegraphics[width=0.49\textwidth]{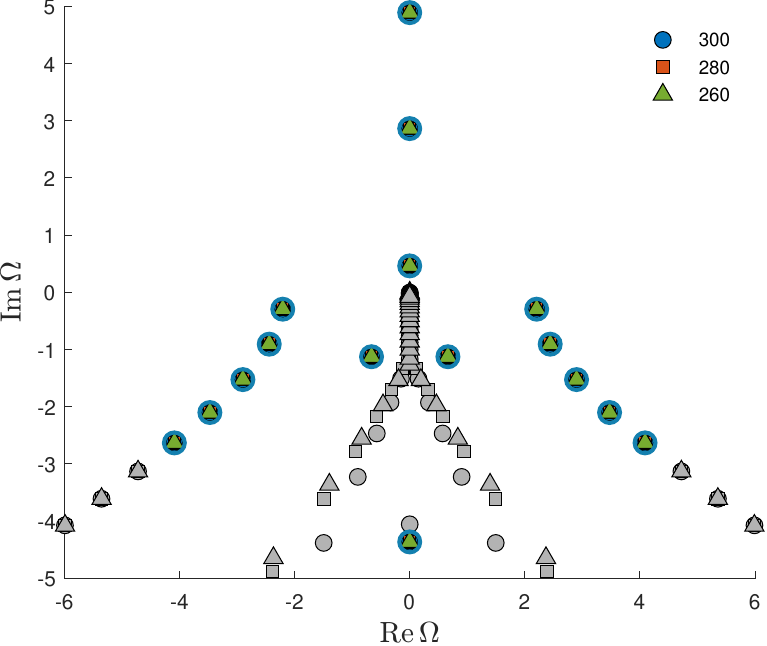}
    }
    \caption{Illustration of some QNM distribution morphologies exhibiting one (left panel) and three instabilies (right panel) observed in the tensorial perturbation spectrum. Here, $n$, $\ell_n$, and $\widehat{\alpha}$ denote the number of compactified dimensions, the angular momentum, and the coupling parameter, respectively.}
    \label{atlasQNMinst}
\end{figure}

In this scenario, we detect clear signatures of instability across all tested values of $\ell_4$ (see red-highlighted entries in Tables~\ref{table:D6s2L234} and \ref{table:D6s2L2diffalpha}). This is in agreement with the theoretical predictions by \cite{Dotti2005PRD, Dotti2005CQG}. To the best of our knowledge, this constitutes the first numerical confirmation of such instabilities. It is also worth noticing that the WKB-based analysis by \cite{Chakrabarti2007GRG} failed to capture these features. In Table~\ref{table:D6s2L2diffalpha}, we provide a refined determination of the onset of instability for $\ell_4 = 2$, pinpointing its emergence in the interval $5.2 \leqslant \widehat{\alpha} \leqslant 5.8$. The first unstable mode appears at $\Omega = 0.0688i$ for $\widehat{\alpha} = 5.4$, with the instability becoming more pronounced as the coupling increases: for example, $\Omega=0.1415i$ at $\widehat{\alpha} = 5.6$, and $\Omega = 0.2117i$ at $\widehat{\alpha} = 5.8$. Furthermore, both the magnitude and the number of unstable modes grow with increasing $\widehat{\alpha}$. At $\widehat{\alpha} = 20$, we find two unstable modes for $\ell_4 = 2$, located at $\Omega = 3.3262i$ and $\Omega = 1.2261i$, while for $\ell_4 = 4$, the number rises to three (see Table~\ref{table:D6s2L234}). Examples of the emergence of these unstable modes in the QNM spectrum are given in Figure~\ref{atlasQNMinst}. As originally argued in \cite{Dotti2005CQG}, these instabilities are rooted in the existence of a negative potential well near the event horizon, which can support bound states with purely imaginary, positive-frequency modes for a period of time long enough to lead to exponentially growing perturbations. However, the presence of such a negative minimum alone is not a sufficient condition for instability: as shown by the vector sector analysed in Sec.~VI, no instabilities arise despite the potential also going below zero in the vicinity of the event horizon. From our spectral data, we infer that for $\ell_4=2$, the instability threshold occurs precisely at $\widehat{\alpha} = 5.4$, persisting for all higher values of the GB coupling. Using the relation $\widehat{\alpha}=\widetilde{\alpha} k_n / M^{2/(n-1)}$, where $\widetilde{\alpha} = (n-1)(n-2)$ and
\begin{equation}
  k_n = \int d\Omega_n = \frac{2\pi^{(n+1)/2}}{\Gamma\left(\frac{n+1}{2}\right)}\,,
\end{equation}
we translate this instability threshold into a mass bound. For $n = 4$, the condition $\widehat{\alpha} \geqslant 5.4$ becomes equivalent to the inequality
\begin{equation}
    GM \leqslant 158.1\,\alpha^{3/2}.
\end{equation}
Note that the above inequality is dimensionally consistent. Indeed, in units where $c = G = \hbar = 1$, the GB coupling constant $\alpha$ always has dimension $L^2$, Newton's gravitational constant $G$ has dimension $L^{D-2}$, and the mass $M$ has dimension $L^{-1}$, regardless of the dimension $D$. Our inequality implies that instabilities are confined to black holes whose mass lies below a critical threshold set by the GB coupling. From a theoretical standpoint, since $\alpha$ has dimensions of length squared and is interpreted in string theory as being proportional to the square of the string length (typically assumed to be of the same order as the Planck length), that is $\alpha \sim \ell_s^2 \sim 1/M_s^2$, the threshold condition corresponds to a mass scale $M \sim \ell_s^{-3}$. This suggests that the instability regime lies in the microscopic domain, near the string scale. However, it is important to note that these instabilities appear only in the six-dimensional case. For higher spacetime dimensions such as $D \in \{10,11,12,26\}$, which are of primary interest in different string theory scenarios, our numerical analysis shows no trace of tensorial instabilities. Therefore, we conclude that while these instabilities are theoretically significant in six dimensions, they pose no threat to the stability of black holes in the dimensional frameworks relevant to string-inspired models.

\begin{figure}
    \centering
    \includegraphics[width=0.8\linewidth]{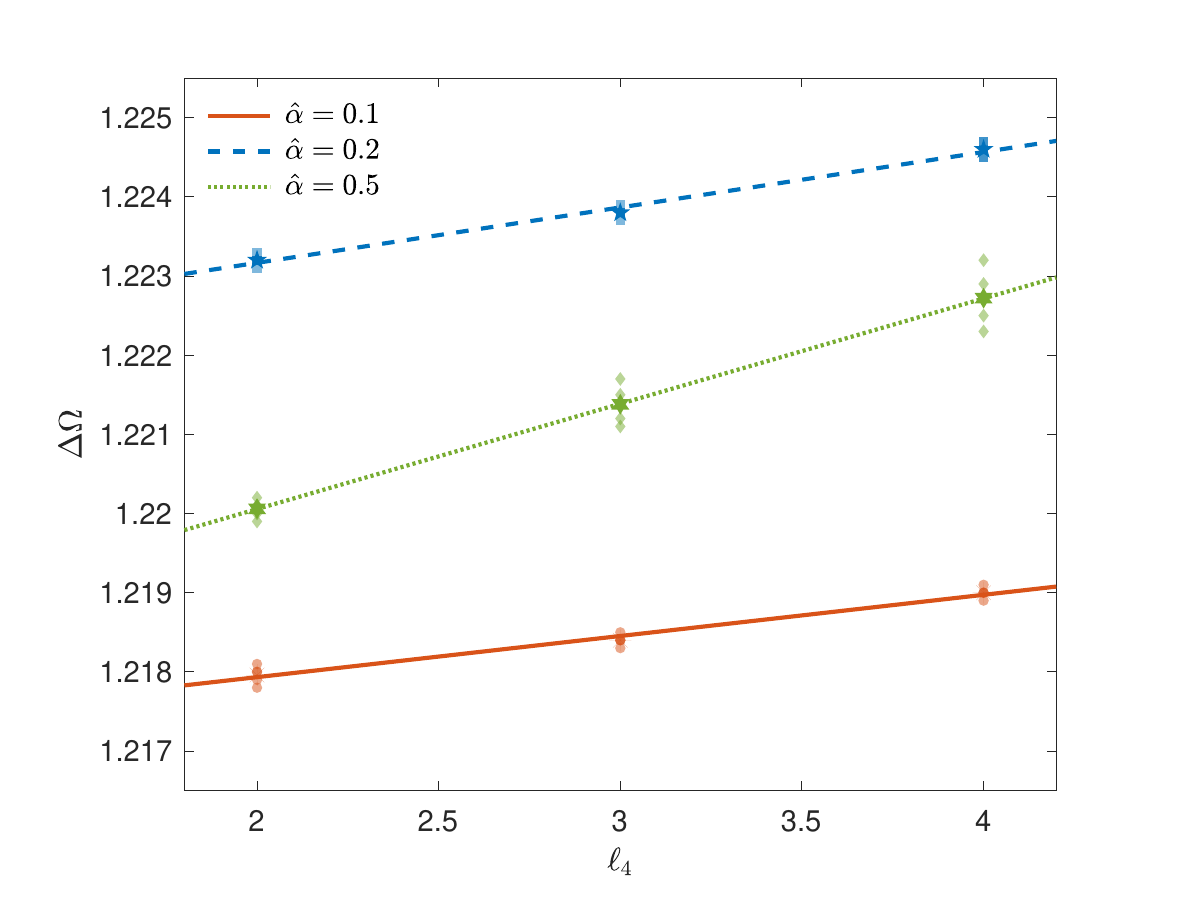}
    \caption{Linear regression analysis of the spectral gap values reported in Table~\ref{table:piD6L234}. The values of the linear regression determination coefficients for each considered value of the parameter $\widehat{\alpha}$ are as follows: $R^2 = 0.99217$ ($\widehat{\alpha}=0.1$), $R^2 = 0.99324$ ($\widehat{\alpha}=0.2$), and $R^2 = 0.99998$ ($\widehat{\alpha}=0.5$).}
    \label{fig:tens-lin01}
\end{figure}

\begin{figure}
    \centering
    \subfloat[$n=8$, $\ell_{8}=2$, $\widehat{\alpha}=0.1$ (Martini)]{%
    \includegraphics[width=0.49\textwidth]{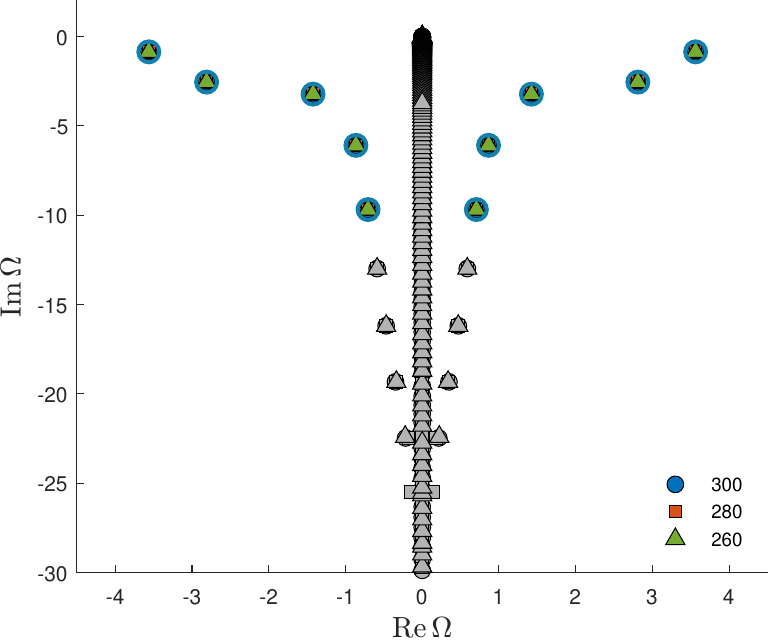}
    }
    \subfloat[$n=8$, $\ell_{8}=4$, $\widehat{\alpha}=0.1$ ('coupe à glace')]{%
    \includegraphics[width=0.49\textwidth]{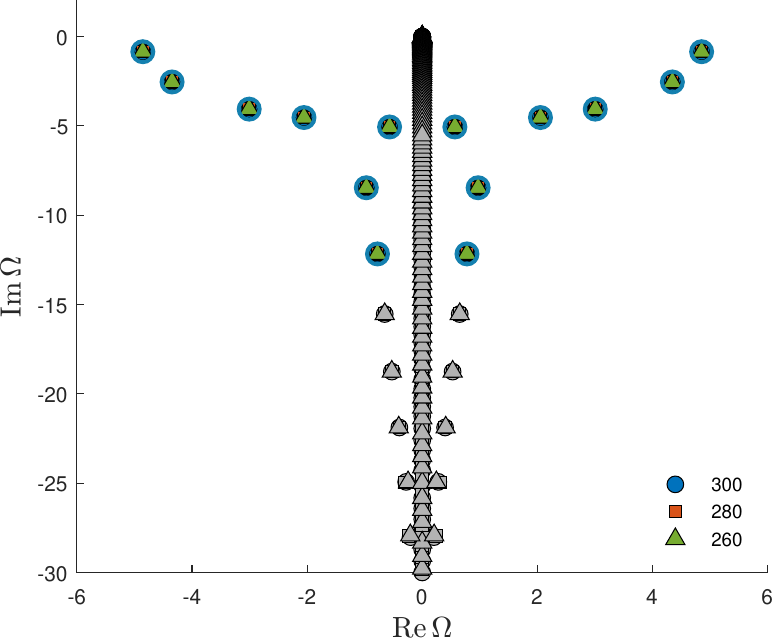}
    }\newline
    \subfloat[$n=8$, $\ell_{8}=4$, $\widehat{\alpha}=0.1$ (Bottleneck)]{%
    \includegraphics[width=0.49\textwidth]{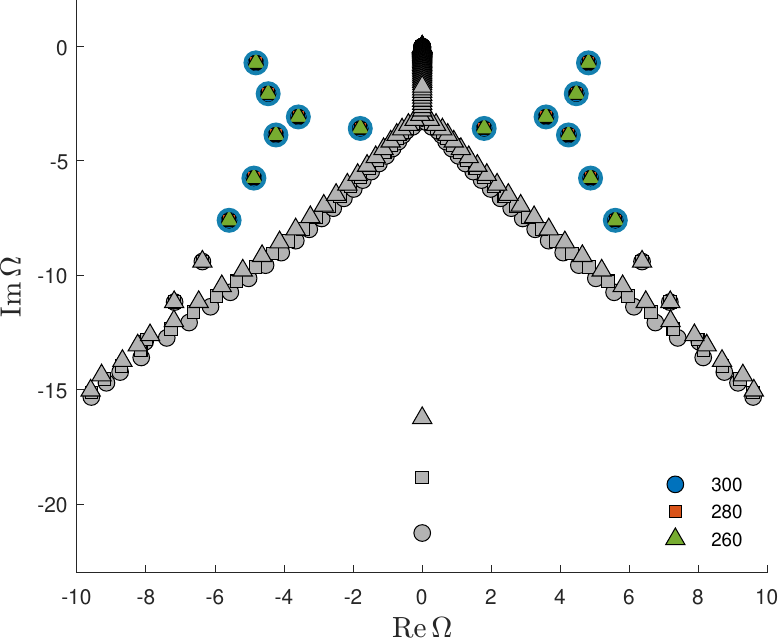}
    }
    \subfloat[$n=9$, $\ell_{9}=3$, $\widehat{\alpha}=20$ (Pyramid)]{%
    \includegraphics[width=0.49\textwidth]{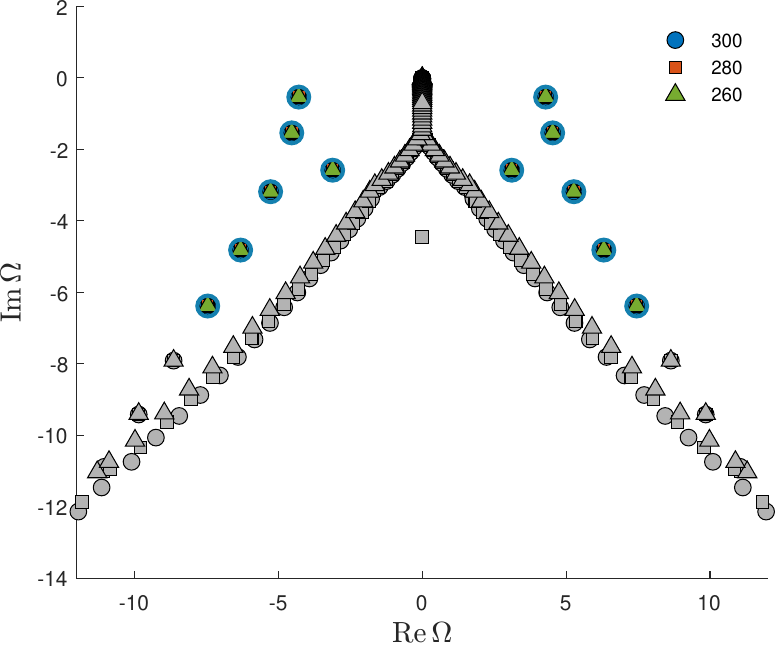}
    }
    \caption{The four main types of QNM distributions observed in the analysis of tensorial perturbations. Here, $n$, $\ell_n$, and $\widehat{\alpha}$ denote the number of compactified dimensions, the angular momentum, and the coupling parameter, respectively.}
    \label{atlasQNMdistribtens}
\end{figure}

For $D = 6$, we also observe the emergence of several equally spaced overdamped modes in the range $0.1\leqslant\widehat{\alpha}\leqslant 0.5$. In Table~\ref{table:piD6L234}, we report the measured spectral gap parameter $\Delta\Omega$ associated with these purely imaginary QNMs. To investigate the possible linear dependence of $\Delta\Omega$ on the angular momentum parameter $\ell_4$ for small values of the coupling parameter $\widehat{\alpha}$, we performed a linear regression analysis as shown in Figure~\ref{fig:tens-lin01}. The results strongly support this linear dependence, with the determination coefficient $R^2$ consistently above $0.99$ for all tested values of $\widehat{\alpha}$. We also identify specific instances of isolated overdamped modes. For example, a simulation employing $400$ Chebyshev polynomials reveals that for $\widehat{\alpha}=4$, we obtain one isolated overdamped mode at $\Omega=-0.5368i$ when $\ell_4=2$, and three isolated overdamped modes at $\Omega_1=-0.3831i$, $\Omega_2=-2.9638i$, and $\Omega_3=-8.0665i$ for $\ell_4=3$. Moreover, we systematically scanned the interval $5 \leqslant \widehat{\alpha} \leqslant 10$ to locate the onset of instability. Our analysis indicates a clear transition from stable to unstable regimes occurring between $\widehat{\alpha}=5.2$ and $\widehat{\alpha}=5.4$. Specifically, for $\widehat{\alpha}=5.2$, we observe an isolated overdamped mode at $\Omega=-0.0066i$, whereas at $\widehat{\alpha}=5.4$, this mode transitions to the positive imaginary axis, signalling an instability at $\Omega=0.0688i$. Additionally, we notice a general trend whereby the number of overdamped modes decreases as $\widehat{\alpha}$ increases (see Table~\ref{table:piD6L234}). Similar observations also apply to the cases with $\ell_4\in\{3,4\}$. Regarding the non-monotonic behaviour observed in the real part of the QNM frequency, we find that for $\ell_4\in\{2, 3\}$, this behaviour is present within the range $4 \leqslant \widehat{\alpha} \leqslant 10$. In contrast, for $\ell_4 = 4$, the non-monotonicity persists over an extended range, namely $4\leqslant\widehat{\alpha}\leqslant 20$ (see bold entries in Tables~\ref{table:D6s2L234} and~\ref{table:D6s2L2diffalpha}). Finally, regarding the morphology of the QNM spectrum, we observe a Martini-type distribution across all tested values of $\ell_4$ in the range $0\leqslant\widehat{\alpha}\leqslant 0.2$. For $\widehat{\alpha}=0.5$ and $\ell_4=2$, the Martini shape persists, though it transitions into a pyramidal structure in the range $5.2\leqslant \widehat{\alpha}\leqslant 5.8$. In contrast, for $\widehat{\alpha} = 0.5$ with $\ell_4\in\{3,4\}$, the distribution becomes a deformed Martini shape (see panel (b) in Fig.~\ref{atlasQNM3t} for a typical example). More generally, for $4\leqslant\widehat{\alpha}\leqslant 20$ and all $\ell_4\in\{2,3,4\}$, the spectrum consistently exhibits a pyramidal morphology (see, for example, (d) in Fig.~\ref{atlasQNMdistribtens}).

To assess (in)stability in the vector and tensor sectors we recall that, after separation of variables, each perturbation satisfies a one-dimensional Schrödinger equation
\begin{equation}
    H_{V,T}R_{V,T}=\left[-\frac{d^2}{du^2}+V_{V,T}(u)\right]R_{V,T}(u)=\omega^2 R_{V,T}(u),
\end{equation}
with tortoise coordinate $du/dr=f^{-1}(r)$ and purely ingoing (outgoing) boundary conditions at the horizon (infinity).  With the time dependence $e^{-i\omega t}$ adopted here and in Refs. \cite{Batic2024EPJC,Batic2024PRD,Batic2024CQG}, modes with $\omega^{2}>0$ are oscillatory, whereas a mode with $\omega^{2}<0$ (i.e. $\omega=\pm i\gamma$, $\gamma>0$) grows exponentially for the choice $\omega=+i\gamma$.  Hence, an unstable QNM mode exists if and only if $\mathcal H_{V,T}$ admits a negative-energy bound state. An upper limit on the number $N_\ell$ of such states in a given partial-wave channel is provided by the Cohn–Calogero inequality (see (29) of \cite{Brau2003JPA} and (1.4) of \cite{Brau2003JMP}; cf. \cite{Cohn1963BARB,Calogero1965CMP})
\begin{equation}\label{bound1}
N^{(V,T)}_\ell<\frac{2}{\pi}\int_{-\infty}^\infty du~\sqrt{-\mathcal{U}^{-}_{V,T}}(u),
\end{equation}
where $\mathcal{U}^{-}_{V,T}$ is the negative part of the potential $\mathcal{U}_{V,T}$ supported on the interval extending from the event horizon to the first turning point of the potential. After the rescalings $\rho=r/M^\frac{1}{n-1}$, $\widehat{\alpha}=\alpha/M^\frac{2}{n-1}$, and $x=\rho/\rho_h$, \eqref{bound1} can be rewritten as
\begin{equation}\label{bound2}
N^{(V,T)}_\ell<\frac{2}{\pi}\int_{1}^{x_t} \frac{dx}{f(x)}~\sqrt{-\mathcal{U}^{-}_{V,T}}(x).
\end{equation}
Note that the negative part of the potential, $\mathcal{U}^{-}_{V,T}$, has compact support on the interval $[1, x_t]$, vanishing identically outside it. Here, $x = 1$ marks the location of the event horizon, while $x_t > 1$ denotes the first turning point beyond which the potential becomes non-negative. Finally, since $\mathcal{U}^{-}_{V,T}(x)=\mathcal{U}_{V,T}(x)$ for all $x\in[1,x_t]$, we can rewrite \eqref{bound2} as
\begin{equation}\label{bound3}
N^{(V,T)}_\ell<\mathcal{S}(n,\widehat{\alpha},\ell_n),\quad
\mathcal{S}(n,\widehat{\alpha},\ell_n)=\frac{2}{\pi}\int_{1}^{x_t} \frac{dx}{f(x)}~\sqrt{-\mathcal{U}_{V,T}}(x),
\end{equation}
with ${\cal U}_{V}$ and ${\cal U}_{T}$ given by \eqref{UV} and \eqref{UT}, respectively. For $n = 3$ ($D = 5$), the vector potential develops a negative pocket only in the dipole ($\ell_3 = 1$) and quadrupole ($\ell_3 = 2$) channels. In both cases, the integral bound $\mathcal{S}$ remains below unity across the full range $\widehat{\alpha} \in [0,2)$, ensuring $N^{(V)}_{\ell} < 1$. Higher multipoles yield even shallower potential wells, with $\mathcal{S} \ll 1$. For $n = 4$, the potential remains strictly positive for all $\ell_4 \geq 2$, while for $n \in \{5,6,7\}$, negative pockets arise only for $\ell_n = 1,2$; for $n \in \{8,9,10\}$, only for $\ell_n = 1,2,3$; and for $n = 24$, only for $\ell_{24} < 10$. In all such cases, the maximum value of $\mathcal{S}$ remains below 1, thereby guaranteeing analytic stability via inequality \eqref{bound3}. To illustrate, for $\ell_n = 1$ (the most probable candidate for instability, as $\mathcal{S}$ decreases monotonically with increasing $\ell$), we observe the following maximal values of $\mathcal{S}$: 0.802 at $n = 4,\, \widehat{\alpha} \approx 1$; 0.799 at $n = 5,\, \widehat{\alpha} \approx 5$; 0.792 at $n = 6,\, \widehat{\alpha} \approx 1$; 0.793 at $n = 7,\, \widehat{\alpha} \approx 10$; 0.799 at $n = 8,\, \widehat{\alpha} \approx 10$; 0.803 at $n = 9,\, \widehat{\alpha} \approx 5$; 0.806 at $n = 10,\, \widehat{\alpha} \approx 10$; and 0.820 at $n = 24,\, \widehat{\alpha} \approx 10$. Because $\mathcal S<1$ throughout, inequality \eqref{bound3} guarantees $N^{(V)}_{\ell}=0$ and therefore analytic stability, in full agreement with the numerical spectrum obtained in the vector sector. In the gravitational sector, the tensor potential acquires a negative region only for $n=4$ ($D=6$). As shown in Table~\ref{CCBound}, which lists the values of $\mathcal{S}(n=4,\widehat{\alpha},\ell_4)$, the critical threshold $\mathcal{S}=1$ is crossed at $\widehat{\alpha} \simeq 5.4$ for $\ell_4 = 2$, at $\widehat{\alpha} \simeq 4.8$ for $\ell_4 = 3$, and at $\widehat{\alpha} \simeq 4.5$ for $\ell_4 = 4$. Precisely in these parameter ranges, our numerical analysis detects the onset of instabilities in the tensorial spectrum. Conversely, for all values where $\mathcal{S} < 1$, no unstable modes are observed. This remarkable agreement confirms that the Cohn–Calogero bound not only provides a rigorous upper limit on the number of bound states, but also serves as a sharp diagnostic tool.
\begin{table}%[ht]
\centering
\caption{Numerical values of the bound $\mathcal{S}$ as given by \eqref{bound3} in tensor case for $n=4$ and different values of $\ell_4$ and $\widehat{\alpha}$.}
\label{CCBound}
\vspace*{1em}
\begin{tabular}{||c|c|c|c|c|c|c|c|c|c|c|c|c|c||}
\hline\hline
$\ell_4$ & $\widehat{\alpha}$ & $\mathcal{S}$ & $\ell_4$ & $\widehat{\alpha}$ & $\mathcal{S}$ & $\ell_4$ & $\widehat{\alpha}$ & $\mathcal{S}$\\ [0.5ex]
\hline\hline
$2$   & $4$   & $0.08$ & $3$ & $4$  & $0.30$  & $4$ & $4$  & $0.49$\\
      & $5.5$ & $1.01$ &     & $5$  & $1.19$  &     & $5$  & $1.59$\\
      & $10$  & $2.44$ &     & $10$ & $3.46$  &     & $10$ & $4.43$\\
      & $20$  & $3.69$ &     & $20$ & $5.15$  &     & $20$ & $6.53$\\[1ex]
 \hline\hline 
\end{tabular}
\end{table}

\subsection{$D = 7$}
For $\ell_5\in\{2,3,4\}$, we detect equally spaced overdamped modes within the range $0.1\leqslant\widehat{\alpha}\leqslant 0.2$ (see Table~\ref{table:piD7L234}). It is worth observing that the gap between successive overdamped QNMs remains stable as $\ell_5$ varies from $2$ to $4$. A non-monotonic behaviour in the real part of the QNM frequencies is observed throughout $\ell_5\in\{2,3,4\}$ for $5\leqslant \widehat{\alpha}\leqslant 20$, becoming particularly pronounced in the interval $15\leqslant\widehat{\alpha}\leqslant 20$ (see bold entries in Tables~\ref{table:D7s2L23} and~\ref{table:D7s2L4}). We also find that the sixth-order WKB method employed by \cite{Chakrabarti2007GRG} fails to capture the non-monotonic behaviour and does not predict the presence of overdamped modes—phenomena that are clearly identified by our SM. The WKB approximation provides reasonable estimates only for the imaginary part of the QNM frequencies in the limited range $0.1\leqslant\widehat{\alpha}\leqslant 0.2$ for $\ell_5\in\{2,3,4\}$. Beyond this range, its predictions deviate significantly. Regarding the morphology of the QNM spectrum, we identify a variety of structures. Specifically, a Martini-type distribution is observed for each $\ell_5\in\{2, 3, 4\}$ within the range $0\leqslant\widehat{\alpha}\leqslant 0.2$, whereas a bottleneck-type distribution characterizes the spectrum for all $\ell_5\in\{2,3,4\}$ in the interval $5\leqslant\widehat{\alpha}\leqslant 10$. The range $0.5\leqslant\widehat{\alpha}\leqslant 1$ exhibits a transitional behaviour: a bottleneck distribution is present for $\ell_5\in\{2,3\}$, while a Martini morphology reemerges for $\ell_5=4$. Finally, in the high-coupling regime $15\leqslant\widehat{\alpha}\leqslant 20$, a pyramidal distribution arises for $\ell_5=2$, which transitions into a bottleneck morphology for $\ell_5\in\{3,4\}$.

\subsection{$D = 8$}

The case of dimension $D = 8$ is particularly distinctive, as no overdamped modes are detected for any $\ell_6\in\{2,3,4\}$ within the range $0\leqslant\widehat{\alpha}\leqslant 20$. Additionally, we observe non-monotonic behaviour in the real part of the QNM frequencies for all $\ell_6\in\{2,3,4\}$ across the entire parameter range (see bold entries in Tables~\ref{table:D8s2L23} and~\ref{table:D8s2L4}). It is worth noting that no bold entry appears in Table~\ref{table:D8s2L23} for $\ell_6=2$ and $\widehat{\alpha}=0.5$ because the non-monotonic behaviour manifests only for overtones beyond the fifth, while only the first five are reported there due to space constraints. Concerning the sixth-order WKB method employed by \cite{Chakrabarti2007GRG}, we find that it provides reasonable estimates only for the imaginary part of the fundamental mode in the narrow range $0.1\leqslant\widehat{\alpha}\leqslant 0.2$ for $\ell_6\in\{2,3\}$. For $\ell_6=4$ in the same range, the method achieves sufficient accuracy in predicting both the real and imaginary parts of the QNM. However, as in previous cases, it fails to capture the non-monotonic behaviour and does not produce higher overtones. Regarding the morphology of the QNM spectrum, we observe a consistent Martini-type distribution for all $\ell_6\in\{2,3,4\}$ throughout the range $0\leqslant\widehat{\alpha}\leqslant 20$. For the same values of $\ell_6$, the distribution transitions to a bottleneck-type morphology within the interval $0.5\leqslant\widehat{\alpha}\leqslant 20$, and further evolves into a pyramidal structure in the high-coupling regime $15\leqslant\widehat{\alpha}\leqslant 20$. Notice that the identified spectral morphologies exhibit a clear evolution with increasing $\widehat{\alpha}$, yet they appear largely insensitive to variations in $\ell_6$.

\subsection{$D = 10$ (Superstring Theory)}

In this scenario and for all tested values of $\ell_8$, no overdamped modes are detected. Non-monotonic behaviour in the real part of the QNM frequencies is observed across all values of $\ell_8$ and throughout the entire range of $\widehat{\alpha}$ (see bold entries in Table~\ref{table:D10s2L234}). It should be noted that for $\ell_8=3$ and $\widehat{\alpha}=0.2$ and $0.5$, no bold entries are shown because the non-monotonic behaviour manifests only for overtones higher than the fourth, which are not reported due to space constraints. The same applies to the case $\ell_8=4$ for $\widehat{\alpha}=0.1$, $0.2$, and $0.5$. Regarding the morphology of the QNM spectrum, we observe a 'coupe à glace'-type distribution at $\widehat{\alpha}=0$ for all $\ell_8\in\{2,3,4\}$. In the range $0.1\leqslant\widehat{\alpha}\leqslant 0.2$, a Martini-type morphology appears for $\ell_8=2$, whereas a 'coupe à glace'-type structure persists for $\ell_8\in\{3, 4\}$. For $0.5\leqslant\widehat{\alpha}\leqslant 10$, the spectrum consistently exhibits a bottleneck morphology across all tested values of $\ell_8$, while at $\widehat{\alpha}=20$, a pyramidal distribution emerges for each value of $\ell_8$.

\subsection{$D = 11$ (M-Theory)}

We observe a single overdamped mode only for $\ell_9=3$ and $\widehat{\alpha}=0.1$, located at $\Omega=-5.4825i$. Non-monotonic behaviour in the real part of the QNM frequencies is observed for all tested values of $\ell_9$ throughout the range $0\leqslant \widehat{\alpha}\leqslant 20$ (see bold entries in Table~\ref{table:D11s2L234}). It should be noted that for $\ell_9=3$ and $\widehat{\alpha}=0$, $0.1$, and $0.2$, no bold entries are reported because the non-monotonic behaviour occurs only for overtones higher than the fifth, which are omitted due to space constraints. The same applies to $\ell_9=4$ for the corresponding values of $\widehat{\alpha}$. Regarding the morphology of the QNM spectrum, we identify a 'coupe à glace'-type distribution for each $\ell_9\in\{2,3,4\}$ within the range $0\leqslant\widehat{\alpha}\leqslant 0.5$. For $5\leqslant\widehat{\alpha}\leqslant 10$, the 'coupe à glace' morphology persists for $\ell_9\in\{2, 3\}$ but transitions to a bottleneck distribution for $\ell_9=4$. Finally, at $\widehat{\alpha}=20$, a 'coupe à glace' structure remains for $\ell_9=2$, while a pyramidal morphology emerges for $\ell_9\in\{3, 4\}$.

\subsection{$D = 12$ (F-Theory)}

For all tested values of $\ell_{10}\in\{2,3,4\}$, no overdamped modes are detected. Non-monotonic behaviour in the real part of the QNM frequencies is consistently observed across all $\ell_{10}$ and throughout the range $0\leqslant \widehat{\alpha}\leqslant 20$ (see bold entries in Table~\ref{table:D12s2L234}). It is worth noting that for $\ell_{10}=3$ and $\widehat{\alpha}=0$, $0.1$, and $0.2$, no bold entries appear because the non-monotonic behaviour manifests only in overtones beyond the fifth, which are omitted for reasons of space. The same consideration applies to $\ell_{10} = 4$ for the corresponding values of $\widehat{\alpha}$. As for the morphological features of the QNM spectrum, they display a high degree of uniformity: a 'coupe à glace'-type distribution is consistently observed for all $\ell_{10}\in\{2,3,4\}$ and across the full range $0 \leqslant \widehat{\alpha} \leqslant 20$.

\subsection{$D = 26$ (Bosonic String Theory)}

For $\ell_{24} \in \{2,3,4\}$ (see Table~\ref{table:D26s2L234}), we observe that the real parts of the QNM frequencies are approximately an order of magnitude larger than those for lower-dimensional cases, such as $D = 5$. Non-monotonic behavior in the real part is detected only at $\widehat{\alpha}=0$, specifically near $\Omega_1 = 1.4839-12.3411i$ for $\ell_{24}=3$ and $\Omega_2=1.2453-11.7658i$ for $\ell_{24}=4$. These modes are not highlighted in Table~\ref{table:D26s2L234} as they correspond to overtones beyond the fourth, which are omitted due to space limitations. No overdamped modes are detected for any tested value of $\ell_{24}$ across the full range $0\leqslant\widehat{\alpha}\leqslant 20$. Similar to the case of $D=12$, the QNM spectrum exhibits a highly uniform morphology, consistently displaying a 'coupe à glace'-type distribution across all values of $\ell_{24}$ and $\widehat{\alpha}$. Finally, for $D \in \{7, 8, 10, 11, 12, 26\}$, $\ell_n = 2$, and $40\leqslant\widehat{\alpha}\leqslant 2\cdot 10^3$, Table~\ref{table:D7to26s2L2} indicates that increasing the coupling parameter $\widehat{\alpha}$ tends to suppress the emergence of overdamped modes. In this regime, the QNMs are arranged in a pyramidal configuration, and non-monotonic behaviour in the real part is still observed (see bold entries in Table~\ref{table:D7to26s2L2}), with one distinguished QNM highlighted in bold, except when $D = 26$.

\begin{table}%[ht]
\centering
\caption{QNM modes for tensor perturbations (spin $s = 2$) of a five-dimensional ($n=3$) Schwarzschild black hole with GB correction are presented in the table below for different values of $\widehat{\alpha}$ and $\ell_3\in\{2,3\}$. The corresponding results are obtained through our SM, utilising $300$ polynomials with a precision of $300$ digits. In this context, $\Omega$ and $N$ represent the dimensionless frequency and the corresponding overtone, respectively. The notation 'N/A' indicates data not available, while 'SM' stands for Spectral Method.}
\label{table:D5s2L2and3}
\vspace*{1em}
\begin{tabular}{||c|c|c|c|c|c|c|c|c|c|c|c|c|c||}
\hline\hline
$\widehat{\alpha}$ $(D=5)$ & $\ell_3$ & $N$ & $\Omega$ (WKB) \cite{Chakrabarti2007GRG} & $\Omega$ (SM) & $\ell_3$ & $N$ & $\Omega$ (WKB) \cite{Chakrabarti2007GRG} & $\Omega$ (SM) \\ [0.5ex]
\hline\hline
$0.0$              & $2$      & $0$ & N/A                 & \textcolor{red}{$1.0681-0.2528i$}  & $3$  & $0$ & N/A                 & \textcolor{red}{$1.4198-0.2516i$} \\
                   &          & $1$ & N/A                 & $0.9848-0.7810i$  &      & $1$ & N/A                 & $1.3555-0.7674i$ \\
                   &          & $2$ & N/A                 & $0.8442-1.3759i$  &      & $2$ & N/A                 & $1.2363-1.3221i$ \\
                   &          & $3$ & N/A                 & $0.7032-2.0999i$  &      & $3$ & N/A                 & $1.0875-1.9370i$ \\
                   &          & $4$ & N/A                 & $0.5982-2.7681i$  &      & $4$ & N/A                 & $0.9450-2.6114i$ \\
$0.1$              & $2$      & $0$ & \textcolor{red}{$1.07234-0.24737i$}  & \textcolor{red}{$1.1011-0.2493i$}  & $3$  & $0$ & \textcolor{red}{$1.42825-0.24648i$}  & \textcolor{red}{$1.4644-0.2483i$} \\
                   &          & $1$ & N/A                 & $1.0251-0.7675i$  &      & $1$ & N/A                 & $1.4057-0.7556i$ \\                
                   &          & $2$ & N/A                 & $0.8957-1.3413i$  &      & $2$ & N/A                 & $1.2970-1.2955i$ \\
                   &          & $3$ & N/A                 & $0.7598-1.9810i$  &      & $3$ & N/A                 & $1.1604-1.8846i$ \\               
                   &          & $4$ & N/A                 & $0.6488-2.6587i$  &      & $4$ & N/A                 & $1.0256-2.5220i$ \\               
$0.2$              & $2$      & $0$ & \textcolor{red}{$1.07949-0.24175i$}  & \textcolor{red}{$1.1336-0.2469i$}  & $3$  & $0$ & \textcolor{red}{$1.43818-0.24123i$}  & \textcolor{red}{$1.5089-0.2463i$} \\
                   &          & $1$ & N/A                 & $1.0613-0.7568i$  &      & $1$ & N/A                 & $1.4525-0.7475i$ \\ 
                   &          & $2$ & N/A                 & $0.9365-1.3116i$  &      & $2$ & N/A                 & $1.3479-1.2742i$ \\ 
                   &          & $3$ & N/A                 & $0.7981-1.9171i$  &      & $3$ & N/A                 & $1.2149-1.8374i$ \\ 
                   &          & $4$ & N/A                 & $0.6774-2.5470i$  &      & $4$ & N/A                 & $1.0827-2.4337i$ \\ 
$0.5$              & $2$      & $0$ & \textcolor{red}{$1.10393-0.22500i$}  & \textcolor{red}{$1.2260-0.2403i$}  & $3$  & $0$ & \textcolor{red}{$1.47124-0.22497i$}  & \textcolor{red}{$1.6367-0.2411i$} \\
                   &          & $1$ & N/A                 & $1.1588-0.7305i$  &      & $1$ & N/A                 & $1.5834-0.7286i$ \\
                   &          & $2$ & N/A                 & $1.0301-1.2429i$  &      & $2$ & N/A                 & $1.4766-1.2294i$ \\
                   &          & $3$ & N/A                 & $0.8455-1.7470ii$ &      & $3$ & N/A                 & $1.3085-1.7281i$ \\
                   &          & $4$ & N/A                 & ${\bf{0.7508-2.1679i}}$  &      & $4$ & N/A                 & ${\bf{1.1759-2.1303i}}$ \\
                   &          & $5$ & N/A                 & $0.7627-2.7165i$  &      & $5$ & N/A                 & $1.1913-2.6411i$ \\
$1.0$              & $2$      & $0$ & \textcolor{red}{$1.15641-0.19529i$}  & \textcolor{red}{$1.3631-0.2202i$}  & $3$  & $0$ & \textcolor{red}{$1.54074-0.19430i$}  & \textcolor{red}{$1.8243-0.2202i$} \\               
                   &          & $1$ & N/A                 & $1.3093-0.6646i$  &      & $1$ & N/A                 & $1.7849-0.6636i$ \\
                   &          & $2$ & N/A                 & $1.2043-1.1232i$  &      & $2$ & N/A                 & $1.7079-1.1175i$ \\
                   &          & $3$ & N/A                 & ${\bf{0.6290-1.4824i}}$  &      & $3$ & N/A                 & $1.6042-1.5968i$ \\
                   &          & $4$ & N/A                 & $1.0719-1.6271i$  &      & $4$ & N/A                 & ${\bf{0.9412-1.5971i}}$ \\
                   &          & $5$ & N/A                 & $1.0005-2.1374i$  &      & $5$ & N/A                 & $1.5172-2.0999i$ \\
$1.9$              & $2$      & $0$ & N/A                 & \textcolor{red}{$1.5818-0.1279i$}  & $3$  & $0$ & N/A                 & \textcolor{red}{$2.1260-0.1229i$} \\
                   &          & $1$ & N/A                 & $1.4912-0.3448i$  &      & $1$ & N/A                 & $2.0630-0.3493i$ \\
                   &          & $2$ & N/A                 & $1.4351-0.4529i$  &      & $2$ & N/A                 & $1.9758-0.4937i$ \\
                   &          & $3$ & N/A                 & $1.4090-0.5946i$  &      & $3$ & N/A                 & $1.9480-0.6251i$ \\
                   &          & $4$ & N/A                 & $1.3891-0.7341i$  &      & $4$ & N/A                 & $1.9225-0.7624i$ \\
                   &          & $5$ & N/A                 & $1.3706-0.8745i$  &      & $5$ & N/A                 & $1.9013-0.9024i$ \\
$1.99$             & $2$      & $0$ & N/A                 & \textcolor{red}{$1.6017-0.1066i$}  & $3$  & $0$ & N/A                 & \textcolor{red}{$2.1588-0.0989i$} \\
                   &          & $1$ & N/A                 & ${\bf{1.4205-0.1258i}}$  &      & $1$ & N/A                 & ${\bf{1.9455-0.1548i}}$ \\
                   &          & $2$ & N/A                 & $1.4227-0.1764i$  &      & $2$ & N/A                 & $1.9479-0.2060i$ \\
                   &          & $3$ & N/A                 & $1.4256-0.2268i$  &      & $3$ & N/A                 & $1.9498-0.2578i$ \\
                   &          & $4$ & N/A                 & N/A  &      & $4$ & N/A                 & $2.0490-0.2858i$ \\[1ex]
 \hline\hline 
 \end{tabular}
\end{table}

\begin{table}%[ht]
\centering
\caption{Overdamped QNM modes for gravitational perturbations (spin $s = 2$) of a five-dimensional ($n = 3$) Schwarzschild black hole with GB correction are presented in the table below for different values of $\widehat{\alpha}$ and $\ell_3\in\{2,3,4\}$. The corresponding results are obtained through our SM, utilising $300$ polynomials with a precision of $300$ digits. In this context, $\Omega$ and $N$ represent the dimensionless frequency and the corresponding overtone, respectively. The notation 'SM' stands for Spectral Method. Moreover, $\Delta\Omega=\Omega_{N}-\Omega_{N+1}$.}
\label{table:piD5L2and3}
\vspace*{1em}
\begin{tabular}{||c|c|c|c|c|c|c|c|c|c|c|c|c||}
\hline\hline
$\widehat{\alpha}$ $(D=5)$ & $\ell_3$ & $N$ & $\Omega$ (SM)    &$\Delta\Omega$ & $\ell_3$ & $N$ & $\Omega$ (SM)    &$\Delta\Omega$ & $\ell_3$ & $N$ & $\Omega$ (SM)    &$\Delta\Omega$\\ [0.5ex]
\hline\hline
$0.2$              & $2$      & $0$ & $0.0000-35.9891i$ & $0.7284i$ & $3$ & $0$ & $0.0000-35.9352i$ & $0.7301i$ & $4$ & $0$ & $0.0000-36.5830i$ & $0.7326i$\\
                   &          & $1$ & $0.0000-36.7175i$ & $0.7284i$ &     & $1$ & $0.0000-36.6653i$ & $0.7300i$ &     & $1$ & $0.0000-37.3156i$ & $0.7323i$\\
                   &          & $2$ & $0.0000-37.4459i$ & $0.7283i$ &     & $2$ & $0.0000-37.3953i$ & $0.7298i$ &     & $2$ & $0.0000-38.0479i$ & $0.7321i$\\
                   &          & $3$ & $0.0000-38.1742i$ & $0.7282i$ &     & $3$ & $0.0000-38.1251i$ & $0.7297i$ &     & $3$ & $0.0000-38.7800i$ & $0.7319i$\\
                   &          & $4$ & $0.0000-38.9024i$ & $0.7282i$ &     & $4$ & $0.0000-38.8548i$ & $0.7296i$ &     & $4$ & $0.0000-39.5119i$ & $0.7317i$\\
                   &          & $5$ & $0.0000-39.6306i$ & -         &     & $5$ & $0.0000-39.5844i$ & -         &     & $5$ & $0.0000-40.2436i$ & -        \\
$0.5$              & $2$      & $0$ & $0.0000-20.6519i$ & $0.7297i$ & $3$ & $0$ & $0.0000-20.5651i$ & $0.7347i$ & $4$ & $0$ & $0.0000-21.1753i$ & $0.7394i$\\
                   &          & $1$ & $0.0000-21.3816i$ & $0.7295i$ &     & $1$ & $0.0000-21.2998i$ & $0.7339i$ &     & $1$ & $0.0000-21.9147i$ & $0.7387i$\\
                   &          & $2$ & $0.0000-22.1111i$ & $0.7293i$ &     & $2$ & $0.0000-22.0337i$ & $0.7333i$ &     & $2$ & $0.0000-22.6534i$ & $0.7378i$\\
                   &          & $3$ & $0.0000-22.8404i$ & $0.7290i$ &     & $3$ & $0.0000-22.7670i$ & $0.7327i$ &     & $3$ & $0.0000-23.3912i$ & $0.7372i$\\
                   &          & $4$ & $0.0000-23.5694i$ & $0.7288i$ &     & $4$ & $0.0000-23.4997i$ & $0.7323i$ &     & $4$ & $0.0000-24.1284i$ & $0.7364i$\\
                   &          & $5$ & $0.0000-24.2982i$ & -         &     & $5$ & $0.0000-24.2320i$ & -         &     & $5$ & $0.0000-24.8648i$ & -        \\
$1.0$              & $2$      & $0$ & $0.0000-9.5153i$  & $0.7347i$ & $3$ & $0$ & $0.0000-10.0646i$ & $0.7476i$ & $4$ & $0$ & $0.0000-10.6324i$ & $0.7475i$\\
                   &          & $1$ & $0.0000-10.2500i$ & $0.7317i$ &     & $1$ & $0.0000-10.8122i$ & $0.7440i$ &     & $1$ & $0.0000-11.3799i$ & $0.7454i$\\
                   &          & $2$ & $0.0000-10.9817i$ & $0.7294i$ &     & $2$ & $0.0000-11.5562i$ & $0.7406i$ &     & $2$ & $0.0000-12.1253i$ & $0.7435i$\\
                   &          & $3$ & $0.0000-11.7111i$ & $0.7275i$ &     & $3$ & $0.0000-12.2968i$ & $0.7377i$ &     & $3$ & $0.0000-12.8688i$ & $0.7417i$\\
                   &          & $4$ & $0.0000-12.4386i$ & $0.7270i$ &     & $4$ & $0.0000-13.0345i$ & $0.7352i$ &     & $4$ & $0.0000-13.6105i$ & $0.7400i$\\
                   &          & $5$ & $0.0000-13.1656i$ & -         &     & $5$ & $0.0000-13.7697i$ & -         &     & $5$ & $0.0000-14.3505i$ & -       \\
$1.9$              & $2$      & $0$ & $0.0000-1.6433i$  & $0.8251i$ & $3$ & $0$ & $0.0000-2.1386i$  & $0.8736i$ & $4$ & $0$ & $0.0000-2.7194i$  & $0.8793i$\\
                   &          & $1$ & $0.0000-2.4684i$  & $0.8259i$ &     & $1$ & $0.0000-3.0122i$  & $0.8139i$ &     & $1$ & $0.0000-3.5987i$  & $0.8114i$\\
                   &          & $2$ & $0.0000-3.2943i$  & $0.8026i$ &     & $2$ & $0.0000-3.8261i$  & $0.7851i$ &     & $2$ & $0.0000-4.4101i$  & $0.7812i$\\
                   &          & $3$ & $0.0000-4.0969i$  & $0.7793i$ &     & $3$ & $0.0000-4.6112i$  & $0.7693i$ &     & $3$ & $0.0000-5.1913i$  & $0.7639i$\\
                   &          & $4$ & $0.0000-4.8762i$  & $0.7592i$ &     & $4$ & $0.0000-5.3805i$  & $0.7588i$ &     & $4$ & $0.0000-5.9552i$  & $0.7527i$\\
                   &          & $5$ & $0.0000-5.6354i$  & -         &     & $5$ & $0.0000-6.1393i$  & -         &     & $5$ & $0.0000-6.7079i$  & -        \\
$1.99$             & $2$      & $0$ & $0.0000-0.9261i$  & $0.7096i$ & $3$ & $0$ & $0.0000-1.1414i$  & $0.9941i$ & $4$ & $0$ & $0.0000-1.3190i$ & -\\
                   &          & $1$ & $0.0000-1.6357i$  & -         &     & $1$ & $0.0000-2.1355i$  & -         &     & $1$ & N/A              & -\\[1ex]
 \hline\hline 
 \end{tabular}
\end{table}

\begin{table}%[ht]
\centering
\caption{QNM modes for tensor perturbations (spin $s = 2$) of a five-dimensional ($n = 3$) Schwarzschild black hole with GB correction are presented in the table below for different values of $\widehat{\alpha}$ and $\ell_3=4$. The corresponding results are obtained through our SM, utilising $300$ polynomials with a precision of $300$ digits. In this context, $\Omega$ and $N$ represent the dimensionless frequency and the corresponding overtone, respectively. The notation 'N/A' indicates data not available, while 'SM' stands for Spectral Method.}
\label{table:D5s2L4}
\vspace*{1em}
\begin{tabular}{||c|c|c|c|c|c|c|c|c|c|c|c|c|c||}
\hline\hline
$\widehat{\alpha}$ $(D=5)$ & $\ell_3$ & $N$ & $\Omega$ (WKB) \cite{Chakrabarti2007GRG} & $\Omega$ (SM) \\ [0.5ex]
\hline\hline
$0.0$              & $4$      & $0$ & N/A                 & \textcolor{red}{$1.7722-0.2510i$}  \\
                   &          & $1$ & N/A                 & $1.7202-0.7611i$  \\
                   &          & $2$ & N/A                 & $1.6203-1.2960i$  \\
                   &          & $3$ & N/A                 & $1.4843-1.8722i$  \\
                   &          & $4$ & N/A                 & $1.3330-2.4999i$  \\
$0.1$              & $4$      & $0$ & \textcolor{red}{$1.78374-0.24604i$}  & \textcolor{red}{$1.8283-0.2478i$}  \\
                   &          & $1$ & N/A                 & $1.7808-0.7501i$  \\                
                   &          & $2$ & N/A                 & $1.6899-1.2733i$  \\
                   &          & $3$ & N/A                 & $1.5663-1.8306i$  \\               
                   &          & $4$ & N/A                 & $1.4282-2.4295i$  \\               
$0.2$              & $4$      & $0$ & \textcolor{red}{$1.79630-0.24094i$}  & \textcolor{red}{$1.8847-0.2460i$}  \\
                   &          & $1$ & N/A                 & $1.8387-0.7434i$  \\ 
                   &          & $2$ & N/A                 & $1.7509-1.2568i$  \\ 
                   &          & $3$ & N/A                 & $1.6317-1.7950i$  \\ 
                   &          & $4$ & N/A                 & $1.4996-2.3626i$  \\ 
$0.5$              & $4$      & $0$ & \textcolor{red}{$1.83780-0.22488i$}  & \textcolor{red}{$2.0473-0.2416i$}  \\
                   &          & $1$ & N/A                 & $2.0035-0.7282i$  \\
                   &          & $2$ & N/A                 & $1.9146-1.2242i$  \\
                   &          & $3$ & N/A                 & $1.7721-1.7254i$  \\
                   &          & $4$ & N/A                 & $1.5957-2.1484i$  \\
                   &          & $5$ & N/A                 & $1.5920-2.5869i$  \\
$1.0$              & $4$      & $0$ & \textcolor{red}{$1.92439-0.19367i$}  & \textcolor{red}{$2.2843-0.2201i$}  \\
                   &          & $1$ & N/A                 & $2.2534-0.6625i$  \\
                   &          & $2$ & N/A                 & $2.1930-1.1123i$  \\
                   &          & $3$ & N/A                 & $2.1085-1.5778i$  \\
                   &          & $4$ & N/A                 & ${\bf{1.2424-1.7583i}}$  \\
                   &          & $5$ & N/A                 & $2.0197-2.0654i$  \\
$1.9$              & $4$      & $0$ & N/A                 & \textcolor{red}{$2.6668-0.1205i$}  \\
                   &          & $1$ & N/A                 & $2.6184-0.3496i$  \\
                   &          & $2$ & N/A                 & $2.5310-0.5292i$  \\
                   &          & $3$ & N/A                 & $2.4806-0.6544i$  \\
                   &          & $4$ & N/A                 & $2.4510-0.7925i$  \\
$1.99$             & $4$      & $0$ & N/A                 & \textcolor{red}{$2.7118-0.0952i$}  \\
                   &          & $1$ & N/A                 & $2.4609-0.1845i$  \\
                   &          & $2$ & N/A                 & $2.4633-0.2354i$  \\
                   &          & $3$ & N/A                 & ${\bf{2.6278-0.2772i}}$  \\
                   &          & $4$ & N/A                 & $2.4654-0.2862i$  \\[1ex]
 \hline\hline 
 \end{tabular}
\end{table}

\begin{table}%[ht]
\centering
\caption{QNM modes for tensor perturbations (spin $s = 2$) of a six-dimensional ($n = 4$) Schwarzschild black hole with GB correction are presented in the table below for different values of $\widehat{\alpha}$ and $\ell_{4}\in\{2,3,4\}$. The corresponding results are obtained through our SM, utilising $300$ polynomials with a precision of $300$ digits. In this context, $\Omega$ is given by \eqref{ODET} while $N$ represents the corresponding overtone. The notation 'N/A' indicates data not available. Moreover, 'SM' stands for Spectral Method.}
\label{table:D6s2L234}
\vspace*{1em}
\begin{tabular}{||c|c|c|c|c|c|c|c|c|c|c|c||}
\hline\hline
$\widehat{\alpha}$ $(D=6)$ & $\ell_{4}$ & $N$ & $\Omega$ (SM)& $\ell_{4}$ & $N$ & $\Omega$ (SM) & $\ell_{4}$ & $N$ & $\Omega$ (SM) \\ [0.5ex]
\hline\hline
$0.0$   & $2$ & $0$ & \textcolor{red}{$1.5966-0.3984i$}   & $3$ & $0$ & \textcolor{red}{$2.0470-0.3960i$}       & $4$  & $0$ & \textcolor{red}{$2.4984-0.3947i$} \\ 
        &     & $1$ & $1.4176-1.2372i$                    &     & $1$ & $1.9068-1.2122i$                               &     & $1$ & $2.3834-1.2001i$ \\
        &     & $2$ & $1.0940-2.2313i$                    &     & $2$ & $1.6333-2.1147i$                               &     & $2$ & $2.1550-2.0594i$ \\
        &     & $3$ & $0.7986-3.4417i$                    &     & $3$ & $1.2880-3.1834i$                               &     & $3$ & $1.8328-3.0263i$ \\
        &     & $4$ & $0.6339-4.7224i$                    &     & $4$ & $1.0050-4.4111i$                               &     & $4$ & $1.4889-4.1462i$ \\
$0.1$   & $2$ & $0$ & \textcolor{red}{$1.6206-0.3917i$}   & $3$      & $0$ & \textcolor{red}{$2.0784-0.3897i$}  & $4$  & $0$ & \textcolor{red}{$2.5371-0.3886i$}\\
        &     & $1$ & $1.4543-1.2124i$                    &          & $1$ & $1.9476-1.1906i$                   &      & $1$ & $2.4296-1.1801i$\\
        &     & $2$ & $1.1533-2.1657i$                    &          & $2$ & $1.6943-2.0662i$                   &      & $2$ & $2.2174-2.0186i$\\
        &     & $3$ & $0.8562-3.3051i$                    &          & $3$ & $1.3704-3.0801i$                   &      & $3$ & $1.9197-2.9480i$\\
        &     & $4$ & $0.6594-4.5218i$                    &          & $4$ & $1.0837-4.2298i$                   &      & $4$ & $1.5955-4.0041i$\\
        &     & $5$ & $0.5288-5.7391i$                    &          & $5$ & $0.8785-5.4302i$                   &      & $5$ & $1.3231-5.1620i$\\
$0.2$   & $2$ & $0$ & \textcolor{red}{$1.6427-0.3866i$}   & $3$      & $0$ & \textcolor{red}{$2.1077-0.3851i$}  & $4$  & $0$ & \textcolor{red}{$2.5736-0.3844i$}\\   
        &     & $1$ & $1.4840-1.1921i$                    &          & $1$ & $1.9820-1.1738i$                   &      & $1$ & $2.4697-1.1650i$\\
        &     & $2$ & $1.1955-2.1089i$                    &          & $2$ & $1.7400-2.0247i$                   &      & $2$ & $2.2664-1.9847i$\\
        &     & $3$ & $0.8912-3.1813i$                    &          & $3$ & $1.4283-2.9872i$                   &      & $3$ & $1.9833-2.8781i$\\
        &     & $4$ & $0.6629-4.3237i$                    &          & $4$ & $1.1413-4.0582i$                   &      & $4$ & $1.6756-3.8736i$\\
        &     & $5$ & $0.5013-5.4666i$                    &          & $5$ & $0.9326-5.1701i$                   &      & $5$ & $1.4190-4.9563i$\\
$0.5$   & $2$ & $0$ & \textcolor{red}{$1.6986-0.3758i$}   & $3$      & $0$ & \textcolor{red}{$2.1826-0.3759i$}  & $4$  & $0$ & \textcolor{red}{$2.6672-0.3761i$}\\
        &     & $1$ & $1.5487-1.1471i$                    &          & $1$ & $2.0616-1.1381i$                   &      & $1$ & $2.5658-1.1347i$\\
        &     & $2$ & $1.2686-1.9765i$                    &          & $2$ & $1.8263-1.9293i$                   &      & $2$ & $2.3657-1.9097i$\\
        &     & $3$ & $0.9453-2.8616i$                    &          & $3$ & $1.5232-2.7493i$                   &      & $3$ & $2.0919-2.6997i$\\
        &     & $4$ & $0.7292-3.7772i$                    &          & $4$ & $1.2912-3.6085i$                   &      & $4$ & $1.8426-3.5190i$\\
        &     & $5$ & $0.6021-4.7901i$                    &          & $5$ & $1.1602-4.5814i$                   &      & $5$ & $1.6825-4.4530i$\\
$4.0$   & $2$ & $0$ & \textcolor{red}{$1.8861-0.3228i$}   & $3$      & $0$ & \textcolor{red}{$2.4153-0.3196i$}  & $4$  & $0$ & \textcolor{red}{$2.9448-0.3173i$}\\
        &     & $1$ & $1.8760-0.9850i$                    &          & $1$ & $2.4135-0.9707i$                   &      & $1$ & $2.9464-0.9602i$\\
        &     & $2$ & ${\bf{0.7160-1.5161i}}$             &          & $2$ & $2.4505-1.6529i$                   &      & $2$ & $2.9735-1.6262i$\\
        &     & $3$ & $1.9178-1.6823i$                    &          & $3$ & ${\bf{1.0194-1.6554i}}$            &      & $3$ & ${\bf{1.3269-1.8213i}}$\\
        &     & $4$ & $2.0061-2.3402i$                    &          & $4$ & $2.5647-2.3433i$                   &      & $4$ & $3.0527-2.3090i$\\
        &     & $5$ & $2.1764-2.8176i$                    &          & $5$ & $2.7417-2.9989i$                   &      & $5$ & $3.1769-2.9827i$\\                   
$5.0$   & $2$ & $0$ & \textcolor{red}{$1.8901-0.3212i$}   & $3$      & $0$ & \textcolor{red}{$0.0000+0.1946i$}  & $4$  & $0$ & \textcolor{red}{$0.0000+0.4355i$}\\
        &     & $1$ & $1.9184-0.9843i$                    &          & $1$ & $2.4152-0.3178i$                   &      & $1$ & $2.9409-0.3152i$\\
        &     & $2$ & ${\bf{0.6667-1.4229i}}$             &          & $2$ & $2.4469-0.9664i$                   &      & $2$ & $2.9697-0.9553i$\\
        &     & $3$ & $2.0140-1.6900i$                    &          & $3$ & ${\bf{0.9540-1.5625i}}$            &      & $3$ & $3.0473-1.6166i$\\
        &     & $4$ & ${\bf{1.9945-2.2643i}}$             &          & $4$ & $2.5460-1.6344i$                   &      & $4$ & ${\bf{1.2451-1.7252i}}$\\
        &     & $5$ & $2.3988-2.5647i$                    &          & $5$ & $2.7230-2.2753i$                   &      & $5$ & $3.1852-2.2867i$\\
        &     & $6$ & $2.8256-3.1018i$                    &          & $6$ & $2.9659-2.8514i$                   &      & $6$ & $3.3663-2.9580i$\\
        &     & $7$ & $3.2268-3.6327i$                    &          & $7$ & $3.2787-3.3899i$                   &      & $7$ & $3.6456-3.7087i$\\
        &     & $8$ & $3.6272-4.1575i$                    &          & $8$ & $3.6341-3.9270i$                   &      & $8$ & ${\bf{3.2265-3.7226i}}$\\
$10.0$  & $2$ & $0$ & \textcolor{red}{$0.0000+1.3564i$}   & $3$      & $0$ & \textcolor{red}{$0.0000+2.1145i$}  & $4$  & $0$ & \textcolor{red}{$0.0000+2.8311i$}\\
        &     & $1$ & $1.8597-0.3294i$                    &          & $1$ & $0.0000+0.5994i$                   &      & $1$ & $0.0000+1.4370i$\\
        &     & $2$ & $2.0569-1.0003i$                    &          & $2$ & $2.3465-0.3278i$                   &      & $2$ & $2.8479-0.3217i$\\
        &     & $3$ & ${\bf{0.5445-1.1865i}}$             &          & $3$ & $2.4369-1.0310i$                   &      & $3$ & $2.9341-0.9709i$\\
        &     & $4$ & $2.4699-1.6172i$                    &          & $4$ & ${\bf{0.7887-1.3202i}}$            &      & $4$ & ${\bf{1.0360-1.4696i}}$\\
        &     & $5$ & $2.9761-2.1662i$                    &          & $5$ & $2.3703-1.6054i$            &      & $5$ & $3.1286-1.5823i$\\
        &     & $6$ & $3.5112-2.6782i$                    &          & $6$ & $2.9164-1.8840i$                   &      & $6$ & $3.4815-2.1198i$\\
$20.0$  & $2$ & $0$ & \textcolor{red}{$0.0000+3.3262i$}   & $3$      & $0$ & \textcolor{red}{$0.0000+4.8889i$}  & $4$  & $0$ & \textcolor{red}{$0.0000+6.3924i$}\\
        &     & $1$ & $0.0000+1.2261i$                    &          & $1$ & $0.0000+2.8611i$                   &      & $1$ & $0.0000+4.3685i$\\
        &     & $2$ & $1.7123-0.2788i$                    &          & $2$ & $0.0000+0.4604i$                   &      & $2$ & $0.0000+2.2994i$\\
        &     & $3$ & $1.9777-0.8306i$                    &          & $3$ & $2.2063-0.2969i$                   &      & $3$ & $2.6787-0.3220i$\\
        &     & $4$ & $2.4959-1.4090i$                    &          & $4$ & $2.4409-0.9087i$                   &      & $4$ & $2.8944-0.9960i$\\
        &     & $5$ & $3.0952-1.9442i$                    &          & $5$ & ${\bf{0.6636-1.1281i}}$            &      & $5$ & ${\bf{0.8754-1.2622i}}$\\
        &     & $6$ & $3.7173-2.4443i$                    &          & $6$ & $2.8988-1.5278i$                   &      & $6$ & $1.1368-1.4555i$\\ 
        &     & $7$ & $4.3462-2.9227i$                    &          & $7$ & $3.4724-2.1037i$                   &      & $7$ & $3.3301-1.6576i$\\[1ex]
 \hline\hline 
 \end{tabular}
\end{table}

\begin{table}%[ht]
\centering
\caption{Overdamped QNM modes for gravitational perturbations (spin $s = 2$) of a six-dimensional ($n = 4$) Schwarzschild black hole with GB correction are presented in the table below for different values of $\widehat{\alpha}$ and $\ell_4 \in \{2, 3, 4\}$. The corresponding results are obtained through our SM, utilising $300$ polynomials with a precision of $300$ digits. In this context, $\Omega$ and $N$ represent the dimensionless frequency and the corresponding overtone, respectively. The notation 'SM' stands for Spectral Method. Moreover, $\Delta\Omega=\Omega_{N}-\Omega_{N+1}$.}
\label{table:piD6L234}
\vspace*{1em}
\begin{tabular}{||c|c|c|c|c|c|c|c|c|c|c|c|c||}
\hline\hline
$\widehat{\alpha}$ $(D=6)$ & $\ell_4$ & $N$ & $\Omega$ (SM)    &$\Delta\Omega$ & $\ell_4$ & $N$ & $\Omega$ (SM)    &$\Delta\Omega$ & $\ell_4$ & $N$ & $\Omega$ (SM)    &$\Delta\Omega$\\ [0.5ex]
\hline\hline
$0.1$              & $2$      & $0$ & $0.0000-88.3210i$ & $1.2178i$ & $3$ & $0$ & $0.0000-88.2938i$ & $1.2183i$ & $4$ & $0$ & $0.0000-88.2581i$ & $1.2190i$\\
                   &          & $1$ & $0.0000-89.5388i$ & $1.2181i$ &     & $1$ & $0.0000-89.5121i$ & $1.2185i$ &     & $1$ & $0.0000-89.4771i$ & $1.2191i$\\
                   &          & $2$ & $0.0000-90.7569i$ & $1.2179i$ &     & $2$ & $0.0000-90.7306i$ & $1.2184i$ &     & $2$ & $0.0000-90.6962i$ & $1.2190i$\\
                   &          & $3$ & $0.0000-91.9748i$ & $1.2180i$ &     & $3$ & $0.0000-91.9490i$ & $1.2184i$ &     & $3$ & $0.0000-91.9152i$ & $1.2189i$\\
                   &          & $4$ & $0.0000-93.1928i$ & $1.2180i$ &     & $4$ & $0.0000-93.1674i$ & $1.2184i$ &     & $4$ & $0.0000-93.1341i$ & $1.2190i$\\
                   &          & $5$ & $0.0000-94.4108i$ & -         &     & $5$ & $0.0000-94.3858i$ & -         &     & $5$ & $0.0000-94.3531i$ & -        \\
$0.2$              & $2$      & $0$ & $0.0000-70.3127i$ & $1.2233i$ & $3$ & $0$ & $0.0000-70.2835i$ & $1.2239i$ & $4$ & $0$ & $0.0000-70.2450i$ & $1.2247i$\\
                   &          & $1$ & $0.0000-71.5360i$ & $1.2232i$ &     & $1$ & $0.0000-71.5074i$ & $1.2238i$ &     & $1$ & $0.0000-71.4697i$ & $1.2247i$\\
                   &          & $2$ & $0.0000-72.7592i$ & $1.2232i$ &     & $2$ & $0.0000-72.7312i$ & $1.2238i$ &     & $2$ & $0.0000-72.6944i$ & $1.2246i$\\
                   &          & $3$ & $0.0000-73.9824i$ & $1.2232i$ &     & $3$ & $0.0000-73.9550i$ & $1.2238i$ &     & $3$ & $0.0000-73.9190i$ & $1.2245i$\\
                   &          & $4$ & $0.0000-75.2056i$ & $1.2231i$ &     & $4$ & $0.0000-75.1788i$ & $1.2237i$ &     & $4$ & $0.0000-75.1435i$ & $1.2245i$\\
                   &          & $5$ & $0.0000-76.4287i$ & -         &     & $5$ & $0.0000-76.4025i$ & -         &     & $5$ & $0.0000-76.3680i$ & -        \\
$0.5$              & $2$      & $0$ & $0.0000-43.2334i$ & $1.2201i$ & $3$ & $0$ & $0.0000-43.1919i$ & $1.2217i$ & $4$ & $0$ & $0.0000-45.5833i$ & $1.2232i$\\
                   &          & $1$ & $0.0000-44.4535i$ & $1.2202i$ &     & $1$ & $0.0000-44.4136i$ & $1.2215i$ &     & $1$ & $0.0000-46.8065i$ & $1.2229i$\\
                   &          & $2$ & $0.0000-45.6737i$ & $1.2201i$ &     & $2$ & $0.0000-45.6351i$ & $1.2214i$ &     & $2$ & $0.0000-48.0294i$ & $1.2227i$\\
                   &          & $3$ & $0.0000-46.8938i$ & $1.2199i$ &     & $3$ & $0.0000-46.8565i$ & $1.2212i$ &     & $3$ & $0.0000-49.2521i$ & $1.2225i$\\
                   &          & $4$ & $0.0000-48.1137i$ & $1.2200i$ &     & $4$ & $0.0000-48.0777i$ & $1.2211i$ &     & $4$ & $0.0000-50.4746i$ & $1.2223i$\\
                   &          & $5$ & $0.0000-49.3337i$ & -         &     & $5$ & $0.0000-49.2988i$ & -         &     & $5$ & $0.0000-51.6969i$ & -       \\
$5$                & $2$      & $0$ & -                 & -         & $3$ & $0$ & $0.0000-1.8196i$  & $8.4899i$ & $4$ & $0$ & $0.0000-1.1267i$  & -\\
                   &          & $1$ & -                 & -         &     & $1$ & $0.0000-10.3095i$ & -         &     & $1$ & -  & -\\
$10$               & $2$      & $0$ & $0.0000-0.7743i$  & $3.0166i$ & $3$ & $0$ & $0.0000-2.8414i$  & -         & $4$ & $0$ & $0.0000-0.7713i$ & $6.6065i$\\
                   &          & $1$ & $0.0000-3.7909i$  & -         &     & $1$ & -                 & -         &     & $1$ & $0.0000-7.3778i$ & -\\
$20$               & $2$      & $0$ & $0.0000-8.5685i$  & -         & $3$ & $0$ & $0.0000-4.3666i$  & $37.3544i$& $4$ & $0$ & $0.0000-15.9771i$& -\\
                   &          & $1$ & -                 & -         &     & $1$ & $0.0000-41.7210i$ & -         &     & $1$ & -              & -\\[1ex]
 \hline\hline 
 \end{tabular}
\end{table}

\begin{table}%[ht]
\centering
\caption{QNM modes for tensor perturbations (spin $s = 2$) of a six-dimensional ($n = 4$) Schwarzschild black hole with GB correction are presented in the table below for different values of $\widehat{\alpha}$ and $\ell_{4}=2$. The corresponding results are obtained through our SM, utilising $300$ polynomials with a precision of $300$ digits. In this context, $\Omega$ is given by \eqref{ODET} while $N$ represents the corresponding overtone. The notation 'N/A' indicates data not available. Moreover, 'SM' stands for Spectral Method.}
\label{table:D6s2L2diffalpha}
\vspace*{1em}
\begin{tabular}{||c|c|c|c|c|c|c|c|c|c|c|c||}
\hline\hline
$\widehat{\alpha}$ $(D=6)$ & $\ell_{4}$ & $N$ & $\Omega$ (SM)\\ [0.5ex]
\hline\hline
$5.2$              & $2$      & $0$ & \textcolor{red}{$1.8900-0.3213i$}\\
                   &          & $1$ & $1.9250-0.9860i$\\
                   &          & $2$ & ${\bf{0.6586-1.4075i}}$\\
                   &          & $3$ & $2.0324-1.7003i$\\
                   &          & $4$ & ${\bf{1.9671-2.2394i}}$\\
                   &          & $5$ & $2.4441-2.5285i$\\
$5.4$              & $2$      & $0$ & \textcolor{red}{$0.0000+0.0688i$}\\
                   &          & $1$ & $1.8896-0.3215i$\\
                   &          & $2$ & $1.9313-0.9881i$\\
                   &          & $3$ & ${\bf{0.6510-1.3930i}}$\\
                   &          & $4$ & $2.0532-1.7132i$\\
                   &          & $5$ & ${\bf{1.9319-2.2109i}}$\\
                   &          & $6$ & $2.4868-2.4969i$\\
$5.6$              & $2$      & $0$ & \textcolor{red}{$0.0000+0.1415i$}\\
                   &          & $1$ & $1.8891-0.3217i$\\
                   &          & $2$ & $1.9374-0.9908ii$\\
                   &          & $3$ & ${\bf{0.6438-1.3792i}}$\\
                   &          & $4$ & $2.0783-1.7273i$\\
                   &          & $5$ & ${\bf{1.8874-2.1811i}}$\\
                   &          & $6$ & $2.5268-2.4689i$\\
$5.8$              & $2$      & $0$ & \textcolor{red}{$0.0000+0.2117i$}\\
                   &          & $1$ & $1.8884-0.3221i$\\
                   &          & $2$ & $1.9434-0.9937i$\\
                   &          & $3$ & ${\bf{0.6370-1.3661i}}$\\
                   &          & $4$ & $2.1086-1.7396i$\\
                   &          & $5$ & ${\bf{1.8327-2.1535i}}$\\
                   &          & $6$ & $2.5640-2.4437i$\\[1ex]
 \hline\hline 
 \end{tabular}
\end{table}

\begin{table}%[ht]
\centering
\caption{QNM modes for tensor perturbations (spin $s = 2$) of a seven-dimensional ($n = 5$) Schwarzschild black hole with GB correction are presented in the table below for different values of $\widehat{\alpha}$ and $\ell_5\in\{2,3\}$. The corresponding results are obtained through our SM, utilising $300$ polynomials with a precision of $300$ digits. In this context, $\Omega$ and $N$ represent the dimensionless frequency and the corresponding overtone, respectively. The notation 'N/A' indicates data not available, while 'SM' stands for Spectral Method.}
\label{table:D7s2L23}
\vspace*{1em}
\begin{tabular}{||c|c|c|c|c|c|c|c|c|c|c|c|c|c||}
\hline\hline
$\widehat{\alpha}$ $(D=7)$ & $\ell_5$ & $N$ & $\Omega$ (WKB) \cite{Chakrabarti2007GRG} & $\Omega$ (SM) & $\ell_5$ & $N$ & $\Omega$ (WKB) \cite{Chakrabarti2007GRG} & $\Omega$ (SM) \\ [0.5ex]
\hline\hline
$0.0$              & $2$      & $0$ & N/A                 & \textcolor{red}{$2.0995-0.5313i$}  & $3$  & $0$ & N/A                 & \textcolor{red}{$2.6186-0.5278i$} \\
                   &          & $1$ & N/A                 & $1.7972-1.6491i$  &      & $1$ & N/A                 & $2.3803-1.6157i$ \\
                   &          & $2$ & N/A                 & $1.1782-3.0621i$  &      & $2$ & N/A                 & $1.8746-2.8453i$ \\
                   &          & $3$ & N/A                 & $0.7479-4.9593i$  &      & $3$ & N/A                 & $1.2235-4.4876i$ \\
                   &          & $4$ & N/A                 & $0.6020-6.8297i$  &      & $4$ & N/A                 & $0.8749-6.4024i$ \\
$0.1$              & $2$      & $0$ & \textcolor{red}{$2.09158-0.51813i$}  & \textcolor{red}{$2.1189-0.5223i$}  & $3$  & $0$ & \textcolor{red}{$2.62080-0.51715i$}  & \textcolor{red}{$2.6433-0.5193i$} \\
                   &          & $1$ & N/A                 & $1.8361-1.6167i$  &      & $1$ & N/A                 & $2.4190-1.5874i$ \\                
                   &          & $2$ & N/A                 & $1.2628-2.9598i$  &      & $2$ & N/A                 & $1.9503-2.7788i$ \\
                   &          & $3$ & N/A                 & $0.7933-4.7420i$  &      & $3$ & N/A                 & $1.3288-4.3084i$ \\               
                   &          & $4$ & N/A                 & $0.5894-6.5473i$  &      & $4$ & N/A                 & $0.9219-6.1115i$ \\               
$0.2$              & $2$      & $0$ & \textcolor{red}{$2.09568-0.50563i$}  & \textcolor{red}{$2.1360-0.5154i$}  & $3$  & $0$ & \textcolor{red}{$2.62803-0.50632i$}  & \textcolor{red}{$2.6654-0.5131i$} \\
                   &          & $1$ & N/A                 & $1.8655-1.5901i$  &      & $1$ & N/A                 & $2.4493-1.5649i$ \\ 
                   &          & $2$ & N/A                 & $1.3210-2.8730i$  &      & $2$ & N/A                 & $2.0042-2.7210i$ \\ 
                   &          & $3$ & N/A                 & $0.8168-4.5425i$  &      & $3$ & N/A                 & $1.4066-4.1505i$ \\ 
                   &          & $4$ & N/A                 & $0.5592-6.2582i$  &      & $4$ & N/A                 & $0.9648-5.8208i$ \\ 
$0.5$              & $2$      & $0$ & \textcolor{red}{$2.11523-0.47490i$}  & \textcolor{red}{$2.1761-0.5017i$}  & $3$  & $0$ & \textcolor{red}{$2.65471-0.47757i$}  & \textcolor{red}{$2.7181-0.5012i$} \\
                   &          & $1$ & N/A                 & $1.9215-1.5321i$  &      & $1$ & N/A                 & $2.5110-1.5177i$ \\
                   &          & $2$ & N/A                 & $1.4189-2.6746i$  &      & $2$ & N/A                 & $2.0961-2.5846i$ \\
                   &          & $3$ & N/A                 & $0.8963-4.0260i$  &      & $3$ & N/A                 & $1.5691-3.7693i$ \\
                   &          & $4$ & N/A                 & $0.6681-5.4681i$  &      & $4$ & N/A                 & $1.2203-5.1324i$ \\
$1.0$              & $2$      & $0$ & \textcolor{red}{$2.16399-0.44120i$}  & \textcolor{red}{$2.2179-0.4876i$}  & $3$  & $0$ & \textcolor{red}{$2.71209-0.44140i$}  & \textcolor{red}{$2.7728-0.4883i$} \\               
                   &          & $1$ & N/A                 & $1.9704-1.4743i$  &      & $1$ & N/A                 & $2.5694-1.4709i$ \\
                   &          & $2$ & N/A                 & $1.4792-2.4637i$  &      & $2$ & N/A                 & $2.1478-2.4427i$ \\
                   &          & $3$ & N/A                 & $1.0982-3.4524i$  &      & $3$ & N/A                 & $1.7320-3.3241i$ \\
                   &          & $4$ & N/A                 & $1.0174-4.7102i$  &      & $4$ & N/A                 & $1.5925-4.5003i$ \\
                   &          & $5$ & N/A                 & $1.0107-5.9656i$  &      & $5$ & N/A                 & $1.4602-5.8310i$ \\
$5.0$              & $2$      & $0$ & \textcolor{red}{$2.64723-0.33309i$}  & \textcolor{red}{$2.2412-0.4299i$}  & $3$  & $0$ & \textcolor{red}{$3.33102-0.33905i$}  & \textcolor{red}{$2.7964-0.4280i$} \\               
                   &          & $1$ & N/A                 & $2.0640-1.2925i$  &      & $1$ & N/A                 & $2.6797-1.3073i$ \\
                   &          & $2$ & N/A                 & ${\bf{1.1250-1.8744i}}$  &      & $2$ & N/A          & ${\bf{1.4818-2.0301i}}$ \\   
                   &          & $3$ & N/A                 & $2.0009-2.0220i$  &      & $3$ & N/A                 & $2.3847-2.1618i$ \\
                   &          & $4$ & N/A                 & $2.4488-2.8766i$  &      & $4$ & N/A                 & $2.6576-2.7440i$ \\
$10$               & $2$      & $0$ & \textcolor{red}{$3.31276-0.43274i$}  & \textcolor{red}{$2.1550-0.4060i$}  & $3$  & $0$ & \textcolor{red}{$4.19705-0.43708i$}  & \textcolor{red}{$2.6972-0.4019i$} \\               
                   &          & $1$ & N/A                 & $1.9895-0.9913i$  &      & $1$ & N/A                 & $2.3624-1.2417i$ \\
                   &          & $2$ & N/A                 & ${\bf{0.9667-1.5941i}}$  &      & $2$ & N/A          & $2.7106-1.4260i$ \\ 
                   &          & $3$ & N/A                 & $2.4399-1.6048i$  &      & $3$ & N/A                 & ${\bf{1.2766-1.7440i}}$ \\
                   &          & $4$ & N/A                 & $2.9687-2.4041i$  &      & $4$ & N/A                 & $3.1978-2.2809i$ \\
$15$               & $2$      & $0$ & \textcolor{red}{$3.84348-0.52104i$}  & \textcolor{red}{$2.0476-0.3966i$}  & $3$  & $0$ & \textcolor{red}{$4.87089-0.52488i$}  & \textcolor{red}{$2.6114-0.3980i$} \\               
                   &          & $1$ & N/A                 & $2.0523-0.7610i$  &      & $1$ & N/A                 & $2.2585-0.8887i$ \\
                   &          & $2$ & N/A                 & $2.5968-1.4568i$  &      & $2$ & N/A                 & $2.8776-1.3293i$ \\ 
                   &          & $3$ & N/A                 & ${\bf{0.8906-1.4630i}}$& & $3$ & N/A                 & ${\bf{1.1773-1.6074i}}$ \\ 
                   &          & $4$ & N/A                 & $3.1662-2.1904i$  &      & $4$ & N/A                 & $3.3810-2.0748i$ \\ 
$20$               & $2$      & $0$ & \textcolor{red}{$4.23657-0.59334i$}  & \textcolor{red}{$1.9387-0.3451i$}  & $3$  & $0$ & \textcolor{red}{$5.36533-0.59728i$}  & \textcolor{red}{$2.5461-0.4131i$} \\
                   &          & $1$ & N/A                 & $2.1397-0.6629i$  &      & $1$ & N/A                 & $2.2378-0.6702i$ \\
                   &          & $2$ & N/A                 & $2.6698-1.3659i$  &      & $2$ & N/A                 & $2.9384-1.2488i$ \\
                   &          & $3$ & N/A                 & ${\bf{0.8420-1.3806i}}$& & $3$ & N/A                 & ${\bf{1.1138-1.5204i}}$ \\ 
                   &          & $4$ & N/A                 & $3.2633-2.0592i$  &      & $4$ & N/A                 & $3.4686-1.9483i$ \\ [1ex]
 \hline\hline 
 \end{tabular}
\end{table}

\begin{table}%[ht]
\centering
\caption{QNM modes for tensor perturbations (spin $s = 2$) of a seven-dimensional ($n = 5$) Schwarzschild black hole with GB correction are presented in the table below for different values of $\widehat{\alpha}$ and $\ell_5=4$. The corresponding results are obtained through our SM, utilising $300$ polynomials with a precision of $300$ digits. In this context, $\Omega$ and $N$ represent the dimensionless frequency and the corresponding overtone, respectively. The notation 'N/A' indicates data not available, while 'SM' stands for Spectral Method.}
\label{table:D7s2L4}
\vspace*{1em}
\begin{tabular}{||c|c|c|c|c|c|c|c|c|c|c|c|c|c||}
\hline\hline
$\widehat{\alpha}$ $(D=7)$ & $\ell_5$ & $N$ & $\Omega$ (WKB) \cite{Chakrabarti2007GRG} & $\Omega$ (SM) \\ [0.5ex]
\hline\hline
$0.0$              & $4$      & $0$ & N/A                 & \textcolor{red}{$3.1386-0.5258i$}  \\
                   &          & $1$ & N/A                 & $2.9416-1.5991i$  \\
                   &          & $2$ & N/A                 & $2.5284-2.7557i$  \\
                   &          & $3$ & N/A                 & $1.8954-4.1459i$  \\
                   &          & $4$ & N/A                 & $1.3150-5.9409i$  \\
$0.1$              & $4$      & $0$ & \textcolor{red}{$3.14573-0.51602i$}  & \textcolor{red}{$3.1686-0.5178i$}  \\
                   &          & $1$ & N/A                 & $2.9825-1.5728i$  \\                
                   &          & $2$ & N/A                 & $2.5966-2.7017i$  \\
                   &          & $3$ & N/A                 & $2.0131-4.0248i$  \\               
                   &          & $4$ & N/A                 & $1.4363-5.6858i$  \\               
$0.2$              & $4$      & $0$ & \textcolor{red}{$3.15560-0.50603i$}  & \textcolor{red}{$3.1957-0.5120i$}  \\
                   &          & $1$ & N/A                 & $3.0157-1.5527i$  \\ 
                   &          & $2$ & N/A                 & $2.6468-2.6556i$  \\ 
                   &          & $3$ & N/A                 & $2.0990-3.9181i$  \\ 
                   &          & $4$ & N/A                 & $1.5480-5.4566i$  \\ 
$0.5$              & $4$      & $0$ & \textcolor{red}{$3.18919-0.47854i$}  & \textcolor{red}{$3.2609-0.5012i$}  \\
                   &          & $1$ & N/A                 & $3.0859-1.5118i$  \\
                   &          & $2$ & N/A                 & $2.7354-2.5485i$  \\
                   &          & $3$ & N/A                 & $2.2582-3.6481i$  \\ 
                   &          & $4$ & N/A                 & $1.8502-4.9242i$  \\ 
$1.0$              & $4$      & $0$ & \textcolor{red}{$3.25649-0.44120i$}  & \textcolor{red}{$3.3282-0.4890i$}  \\
                   &          & $1$ & N/A                 & $3.1557-1.4710i$  \\
                   &          & $2$ & N/A                 & $2.7922-2.4442i$  \\
                   &          & $3$ & N/A                 & $2.3628-3.2977i$  \\ 
                   &          & $4$ & N/A                 & $2.2007-4.3614i$  \\ 
$5.0$              & $4$      & $0$ & \textcolor{red}{$4.01271-0.34219i$}  & \textcolor{red}{$3.3507-0.4258i$}  \\
                   &          & $1$ & N/A                 & $3.2665-1.2954i$  \\
                   &          & $2$ & N/A                 & ${\bf{1.8523-2.1999i}}$  \\
                   &          & $3$ & N/A                 & $3.1090-2.2620i$  \\ 
                   &          & $4$ & N/A                 & ${\bf{0.2608-2.3905i}}$  \\ 
                   &          & $5$ & N/A                 & $2.7369-2.8492i$  \\ 
                   &          & $6$ & N/A                 & $3.3613-3.5110i$  \\ 
$10$               & $4$      & $0$ & \textcolor{red}{$5.07277-0.43935i$}  & \textcolor{red}{$3.2283-0.3955i$}  \\
                   &          & $1$ & N/A                 & $3.2148-1.2361i$  \\
                   &          & $2$ & N/A                 & $2.5936-1.6316i$  \\
                   &          & $3$ & N/A                 & ${\bf{1.5989-1.9040i}}$  \\ 
                   &          & $4$ & N/A                 & $3.4865-2.1578i$  \\ 
                   &          & $5$ & N/A                 & $3.9493-2.9472i$  \\  
$15$               &          & $0$ & \textcolor{red}{$5.88797-0.52690i$}  & \textcolor{red}{$3.1255-0.3805i$}  \\
                   &          & $1$ & N/A                 & ${\bf{2.5119-1.1383i}}$  \\
                   &          & $2$ & N/A                 & $3.2353-1.2026i$  \\ 
                   &          & $3$ & N/A                 & ${\bf{1.4756-1.7599i}}$  \\ 
                   &          & $4$ & N/A                 & $3.6467-1.9625i$  \\ 
                   &          & $5$ & N/A                 & $4.1686-2.6841i$  \\ 
$20$               &          & $0$ & \textcolor{red}{$6.48321-0.59934i$}  & \textcolor{red}{$3.0467-0.3716i$}  \\
                   &          & $1$ & N/A                 & ${\bf{2.4570-0.8694i}}$  \\
                   &          & $2$ & N/A                 & $3.2616-1.1418i$  \\ 
                   &          & $3$ & N/A                 & ${\bf{1.3965-1.6672i}}$  \\ 
                   &          & $4$ & N/A                 & $3.7216-1.8383i$  \\
                   &          & $5$ & N/A                 & $4.2735-2.5237i$  \\[1ex]
 \hline\hline 
 \end{tabular}
\end{table}

\begin{table}%[ht]
\centering
\caption{Overdamped QNM modes for gravitational perturbations (spin $s = 2$) of a seven-dimensional ($n = 5$) Schwarzschild black hole with GB correction are presented in the table below for different values of $\widehat{\alpha}$ and $\ell_5 \in \{2, 3, 4\}$. The corresponding results are obtained through our SM, utilising $300$ polynomials with a precision of $300$ digits. In this context, $\Omega$ and $N$ represent the dimensionless frequency and the corresponding overtone, respectively. The notation 'SM' stands for Spectral Method. Moreover, $\Delta\Omega=\Omega_{N}-\Omega_{N+1}$.}
\label{table:piD7L234}
\vspace*{1em}
\begin{tabular}{||c|c|c|c|c|c|c|c|c|c|c|c|c||}
\hline\hline
$\widehat{\alpha}$ $(D=7)$ & $\ell_5$ & $N$ & $\Omega$ (SM)    &$\Delta\Omega$ & $\ell_5$ & $N$ & $\Omega$ (SM)    &$\Delta\Omega$ & $\ell_5$ & $N$ & $\Omega$ (SM)    &$\Delta\Omega$\\ [0.5ex]
\hline\hline
$0.1$              & $2$      & $0$ & $0.0000-131.411i$ & $1.718i$  & $3$ & $0$ & $0.0000-131.389i$ & $1.718i$  & $4$ & $0$ & $0.0000-133.079i$ & $1.718i$\\
                   &          & $1$ & $0.0000-133.129i$ & $1.717i$  &     & $1$ & $0.0000-133.107i$ & $1.717i$  &     & $1$ & $0.0000-134.797i$ & $1.719i$\\
                   &          & $2$ & $0.0000-134.846i$ & $1.717i$  &     & $2$ & $0.0000-134.824i$ & $1.718i$  &     & $2$ & $0.0000-136.516i$ & $1.718i$\\
                   &          & $3$ & $0.0000-136.563i$ & $1.718i$  &     & $3$ & $0.0000-136.542i$ & $1.718i$  &     & $3$ & $0.0000-138.234i$ & $1.718i$\\
                   &          & $4$ & $0.0000-138.281i$ & $1.717i$  &     & $4$ & $0.0000-138.260i$ & $1.717i$  &     & $4$ & $0.0000-139.952i$ & $1.718i$\\
                   &          & $5$ & $0.0000-139.998i$ & -         &     & $5$ & $0.0000-139.977i$ & -         &     & $5$ & $0.0000-141.670i$ & -        \\
$0.2$              & $2$      & $0$ & $0.0000-109.445i$ & $1.724i$  & $3$ & $0$ & $0.0000-109.422i$ & $1.724i$  & $4$ & $0$ & $0.0000-109.393i$ & $1.724i$\\
                   &          & $1$ & $0.0000-111.169i$ & $1.723i$  &     & $1$ & $0.0000-111.146i$ & $1.724i$  &     & $1$ & $0.0000-111.117i$ & $1.725i$\\
                   &          & $2$ & $0.0000-112.892i$ & $1.724i$  &     & $2$ & $0.0000-112.870i$ & $1.724i$  &     & $2$ & $0.0000-112.842i$ & $1.725i$\\
                   &          & $3$ & $0.0000-114.616i$ & $1.724i$  &     & $3$ & $0.0000-114.594i$ & $1.724i$  &     & $3$ & $0.0000-114.567i$ & $1.724i$\\
                   &          & $4$ & $0.0000-116.340i$ & $1.724i$  &     & $4$ & $0.0000-116.318i$ & $1.724i$  &     & $4$ & $0.0000-116.291i$ & $1.725i$\\
                   &          & $5$ & $0.0000-118.064i$ & -         &     & $5$ & $0.0000-118.042i$ & -         &     & $5$ & $0.0000-118.016i$ & -        \\
$0.5$              & $2$      & $0$ & -                 & -         & $3$ & $0$ & -                 & -         & $4$ & $0$ & -                 & -\\
$5$                & $2$      & $0$ & -                 & -         & $3$ & $0$ & -                 & -         & $4$ & $0$ & -                 & -\\
$10$               & $2$      & $0$ & -                 & -         & $3$ & $0$ & -                 & -         & $4$ & $0$ & -                 & -\\
$20$               & $2$      & $0$ & -                 & -         & $3$ & $0$ & -                 & -         & $4$ & $0$ & -                 & -\\[1ex]
 \hline\hline 
 \end{tabular}
\end{table}

\begin{table}%[ht]
\centering
\caption{QNM modes for tensor perturbations (spin $s = 2$) of an eight-dimensional ($n = 6$) Schwarzschild black hole with GB correction are presented in the table below for different values of $\widehat{\alpha}$ and $\ell_6 \in \{2,3\}$. The corresponding results are obtained through our SM, utilising $300$ polynomials with a precision of $300$ digits. In this context, $\Omega$ and $N$ represent the dimensionless frequency and the corresponding overtone, respectively. The notation 'N/A' indicates data not available, while 'SM' stands for Spectral Method.}
\label{table:D8s2L23}
\vspace*{1em}
\begin{tabular}{||c|c|c|c|c|c|c|c|c|c|c|c|c|c||}
\hline\hline
$\widehat{\alpha}$ $(D=8)$ & $\ell_6$ & $N$ & $\Omega$ (WKB) \cite{Chakrabarti2007GRG} & $\Omega$ (SM) & $\ell_6$ & $N$ & $\Omega$ (WKB) \cite{Chakrabarti2007GRG} & $\Omega$ (SM) \\ [0.5ex]
\hline\hline
$0.0$              & $2$      & $0$ & N/A                 & \textcolor{red}{$2.5896-0.6534i$}  & $3$  & $0$ & N/A                 & \textcolor{red}{$3.1600-0.6488i$} \\
                   &          & $1$ & N/A                 & $2.1394-2.0137i$                   &      & $1$ & N/A                 & $2.8044-1.9784i$ \\
                   &          & $2$ & N/A                 & ${\bf{0.2130-2.6968i}}$            &      & $2$ & N/A                 & ${\bf{0.3269-3.2207i}}$ \\
                   &          & $3$ & N/A                 & $1.0364-3.9622i$                   &      & $3$ & N/A                 & $1.9422-3.4940i$ \\
                   &          & $4$ & N/A                 & $0.6912-6.6227i$                   &      & $4$ & N/A                 & $0.9843-6.0234i$ \\
$0.1$              & $2$      & $0$ & \textcolor{red}{$2.57245-0.63574i$}  & \textcolor{red}{$2.6057-0.6424i$}  & $3$  & $0$ & \textcolor{red}{$3.15722-0.63605i$}  & \textcolor{red}{$3.1802-0.6387i$} \\
                   &          & $1$ & N/A                 & $2.1832-1.9770i$                                    &      & $1$ & N/A                 & $2.8439-1.9457i$ \\                
                   &          & $2$ & N/A                 & ${\bf{0.1919-2.7059i}}$                             &      & $2$ & N/A                 & ${\bf{0.3183-3.2109i}}$ \\
                   &          & $3$ & N/A                 & $1.1588-3.7822i$                                    &      & $3$ & N/A                 & $2.0533-3.4150i$ \\               
                   &          & $4$ & N/A                 & $0.7070-6.3386i$                                    &      & $4$ & N/A                 & $1.0718-5.7395i$ \\               
$0.2$              & $2$      & $0$ & \textcolor{red}{$2.57391-0.61982i$}  & \textcolor{red}{$2.6195-0.6342i$}  & $3$  & $0$ & \textcolor{red}{$3.16206-0.62266i$}  & \textcolor{red}{$3.1977-0.6313i$} \\
                   &          & $1$ & N/A                 & $2.2150-1.9466i$                                    &      & $1$ & N/A                 & $2.8734-1.9194i$ \\ 
                   &          & $2$ & N/A                 & ${\bf{0.1676-2.7068i}}$                             &      & $2$ & N/A                 & ${\bf{0.3110-3.1975i}}$ \\ 
                   &          & $3$ & N/A                 & $1.2499-3.6364i$                                    &      & $3$ & N/A                 & $2.1316-3.3444i$ \\ 
                   &          & $4$ & N/A                 & $0.7004-6.0666i$                                    &      & $4$ & N/A                 & $1.1452-5.4725i$ \\ 
$0.5$              & $2$      & $0$ & \textcolor{red}{$2.58863-0.58251i$}  & \textcolor{red}{$2.6501-0.6183i$}  & $3$  & $0$ & \textcolor{red}{$3.18328-0.58829i$}  & \textcolor{red}{$3.2374-0.6172i$} \\
                   &          & $1$ & N/A                 & $2.2715-1.8795i$                                    &      & $1$ & N/A                 & $2.9280-1.8635i$ \\
                   &          & $2$ & N/A                 & $1.4294-3.3275i$                                    &      & $2$ & N/A                 & ${\bf{0.3003-3.1372i}}$ \\
                   &          & $3$ & N/A                 & $0.7923-5.3014i$                                    &      & $3$ & N/A                 & $2.2676-3.1718i$ \\
                   &          & $4$ & N/A                 & $0.6898-7.1821i$                                    &      & $4$ & N/A                 & $1.4224-4.8373i$ \\
                   &          & $5$ & N/A                 & $0.6690-9.0545i$                                    &      & $5$ & N/A                 & $1.1070-6.8146i$ \\
$1.0$              & $2$      & $0$ & \textcolor{red}{$2.63419-0.54656i$}  & \textcolor{red}{$2.6785-0.6025i$}  & $3$  & $0$ & \textcolor{red}{$3.23448-0.54912i$}  & \textcolor{red}{$3.2742-0.6028i$} \\               
                   &          & $1$ & N/A                 & $2.3132-1.8105i$                                    &      & $1$ & N/A                 & $2.9707-1.8069i$ \\
                   &          & $2$ & N/A                 & $1.5844-3.0181i$                                    &      & $2$ & N/A                 & $2.3556-2.9690i$ \\
                   &          & $3$ & N/A                 & ${\bf{1.1820-4.5058i}}$                                    &      & $3$ & N/A                 & ${\bf{0.3115-3.0164i}}$ \\
                   &          & $4$ & N/A                 & $1.2498-6.1876i$                                    &      & $4$ & N/A                 & $1.8181-4.2371i$ \\
                   &          & $5$ & N/A                 & $1.4311-7.8179i$                                    &      & $5$ & N/A                 & $1.6914-5.9525i$ \\
$5.0$              & $2$      & $0$ & \textcolor{red}{$3.02722-0.42703i$}  & \textcolor{red}{$2.6557-0.5253i$}  & $3$  & $0$ & \textcolor{red}{$3.70504-0.42783i$}  & \textcolor{red}{$3.2426-0.5261i$} \\               
                   &          & $1$ & N/A                 & $2.4215-1.5174i$                                    &      & $1$ & N/A                 & $3.0198-1.5380i$ \\
                   &          & $2$ & N/A                 & ${\bf{1.5882-2.2219i}}$                             &      & $2$ & N/A                 & $2.0216-2.3986i$ \\     
                   &          & $3$ & N/A                 & $2.5250-2.7153i$                                    &      & $3$ & N/A                 & ${\bf{0.3602-2.5118i}}$ \\
                   &          & $4$ & N/A                 & $3.0737-3.9606i$                                    &      & $4$ & N/A                 & $2.9425-2.5672i$ \\
                   &          & $5$ & N/A                 & $3.4668-5.1551i$                                    &      & $5$ & N/A                 & $3.3823-3.7969i$ \\             
$10$               & $2$      & $0$ & \textcolor{red}{$3.43874-0.47634i$}  & \textcolor{red}{$2.5500-0.4500i$}  & $3$  & $0$ & \textcolor{red}{$4.24267-0.48196i$}  & \textcolor{red}{$3.1096-0.4637i$} \\               
                   &          & $1$ & N/A                 & $2.5561-1.1832i$                                    &      & $1$ & N/A                 & $3.0216-1.1585i$ \\
                   &          & $2$ & N/A                 & ${\bf{1.4192-1.9229i}}$                             &      & $2$ & N/A                 & $3.4118-2.0541i$ \\
                   &          & $3$ & N/A                 & $3.0173-2.2087i$                                    &      & $3$ & N/A                 & ${\bf{1.7904-2.0907i}}$ \\
                   &          & $4$ & N/A                 & $3.6484-3.2806i$                                    &      & $4$ & N/A                 & $3.9542-3.1233i$ \\
$15$               & $2$      & $0$ & \textcolor{red}{$3.78370-0.53669i$}  & \textcolor{red}{$2.4692-0.3705i$}  & $3$  & $0$ & \textcolor{red}{$4.67515-0.54131i$}  & \textcolor{red}{$2.9868-0.3934i$} \\               
                   &          & $1$ & N/A                 & $2.6682-1.0090i$                                    &      & $1$ & N/A                 & $3.1119-0.9331i$ \\
                   &          & $2$ & N/A                 & ${\bf{1.3296-1.7818i}}$                             &      & $2$ & N/A                 & $3.5892-1.8261i$ \\    
                   &          & $3$ & N/A                 & $3.2039-1.9731i$                                    &      & $3$ & N/A                 & ${\bf{1.6756-1.9438i}}$ \\
                   &          & $4$ & N/A                 & $3.8772-2.9489i$                                    &      & $4$ & N/A                 & $4.1814-2.7964i$ \\
$20$               & $2$      & $0$ & \textcolor{red}{$4.05205-0.58548i$}  & \textcolor{red}{$2.4250-0.3022i$}  & $3$  & $0$ & \textcolor{red}{$5.00776-0.58973i$}  & \textcolor{red}{$2.9107-0.3120i$} \\
                   &          & $1$ & N/A                 & $2.7339-0.9137i$                                    &      & $1$ & N/A                 & $3.1787-0.8257i$ \\
                   &          & $2$ & N/A                 & ${\bf{1.2707-1.6931i}}$                             &      & $2$ & N/A                 & $3.6777-1.6809i$ \\
                   &          & $3$ & N/A                 & $3.2988-1.8234i$                                    &      & $3$ & N/A                 & ${\bf{1.6008-1.8504i}}$ \\
                   &          & $4$ & N/A                 & $3.9953-2.7377i$                                    &      & $4$ & N/A                 & $4.2987-2.5887i$ \\[1ex]
 \hline\hline 
 \end{tabular}
\end{table}

\begin{table}%[ht]
\centering
\caption{QNM modes for tensor perturbations (spin $s = 2$) of an eight-dimensional ($n = 6$) Schwarzschild black hole with GB correction are presented in the table below for different values of $\widehat{\alpha}$ and $\ell_6=4$. The corresponding results are obtained through our SM, utilising $300$ polynomials with a precision of $300$ digits. In this context, $\Omega$ and $N$ represent the dimensionless frequency and the corresponding overtone, respectively. The notation 'N/A' indicates data not available, while 'SM' stands for Spectral Method.}
\label{table:D8s2L4}
\vspace*{1em}
\begin{tabular}{||c|c|c|c|c|c|c|c|c|c|c|c|c|c||}
\hline\hline
$\widehat{\alpha}$ $(D=8)$ & $\ell_6$ & $N$ & $\Omega$ (WKB) \cite{Chakrabarti2007GRG} & $\Omega$ (SM) \\ [0.5ex]
\hline\hline
$0.0$              & $4$      & $0$ & N/A                 & \textcolor{red}{$3.7310-0.6462i$}  \\
                   &          & $1$ & N/A                 & $3.4357-1.9604i$  \\
                   &          & $2$ & N/A                 & $2.7678-3.3706i$  \\
                   &          & $3$ & N/A                 & ${\bf{0.5436-3.7438i}}$  \\
                   &          & $4$ & N/A                 & $1.5926-5.3564i$  \\
$0.1$              & $4$      & $0$ & \textcolor{red}{$3.73477-0.63489i$}  & \textcolor{red}{$3.7553-0.6366i$}  \\
                   &          & $1$ & N/A                 & $3.4748-1.9298i$  \\                
                   &          & $2$ & N/A                 & $2.8515-3.3095i$  \\
                   &          & $3$ & N/A                 & ${\bf{0.5212-3.7472i}}$  \\               
                   &          & $4$ & N/A                 & $1.7668-5.1384i$  \\               
$0.2$              & $4$      & $0$ & \textcolor{red}{$3.74238-0.62289i$}  & \textcolor{red}{$3.7767-0.6297i$}  \\
                   &          & $1$ & N/A                 & $3.5049-1.9060i$  \\ 
                   &          & $2$ & N/A                 & $2.9119-3.2556i$  \\ 
                   &          & $3$ & N/A                 & ${\bf{0.5144-3.7400i}}$  \\ 
                   &          & $4$ & N/A                 & $1.9118-4.9596i$  \\ 
$0.5$              & $4$      & $0$ & \textcolor{red}{$3.77012-0.59069i$}  & \textcolor{red}{$3.8254-0.6169i$}  \\
                   &          & $1$ & N/A                 & $3.5626-1.8568i$  \\
                   &          & $2$ & N/A                 & $3.0169-3.1229i$  \\
                   &          & $3$ & N/A                 & ${\bf{0.5644-3.6769i}}$  \\
                   &          & $4$ & N/A                 & $2.2297-4.5631i$  \\
$1.0$              & $4$      & $0$ & \textcolor{red}{$3.82896-0.55030i$}  & \textcolor{red}{$3.8706-0.6035i$}  \\               
                   &          & $1$ & N/A                 & $3.6100-1.8078i$  \\
                   &          & $2$ & N/A                 & $3.0762-2.9681i$  \\
                   &          & $3$ & N/A                 & ${\bf{0.6641-3.5044i}}$  \\
                   &          & $4$ & N/A                 & $2.5299-4.1131i$  \\
$5.0$              & $4$      & $0$ & \textcolor{red}{$4.38417-0.42912i$}  & \textcolor{red}{$3.8302-0.5259i$}  \\
                   &          & $1$ & N/A                 & $3.6333-1.5602i$  \\
                   &          & $2$ & N/A                 & $3.4340-2.5056i$  \\
                   &          & $3$ & N/A                 & $2.4549-2.5794i$  \\
                   &          & $4$ & N/A                 & ${\bf{0.7328-2.8373i}}$  \\
                   &          & $5$ & N/A                 & $3.7567-3.6478i$  \\
                   &          & $6$ & N/A                 & $4.2086-4.8578i$  \\
$10$               & $4$      & $0$ & \textcolor{red}{$5.04128-0.48505i$}  & \textcolor{red}{$3.6771-0.4711i$}  \\
                   &          & $1$ & N/A                 & $3.4992-1.1799i$  \\
                   &          & $2$ & N/A                 & $3.8485-1.9325i$  \\
                   &          & $3$ & N/A                 & $2.1740-2.2582i$  \\
                   &          & $4$ & N/A                 & ${\bf{0.6984-2.5634i}}$  \\
                   &          & $5$ & N/A                 & $4.3181-2.9736i$  \\
                   &          & $6$ & N/A                 & $4.8879-4.0163i$  \\
$15$               &          & $0$ & \textcolor{red}{$5.55942-0.54389i$}  & \textcolor{red}{$3.5238-0.4241i$}  \\
                   &          & $1$ & N/A                 & $3.5517-0.8837i$  \\
                   &          & $2$ & N/A                 & $4.0161-1.7060i$  \\
                   &          & $3$ & N/A                 & ${\bf{2.0340-2.1045i}}$  \\
                   &          & $4$ & N/A                 & $4.5382-2.6501i$  \\
$20$               &          & $0$ & \textcolor{red}{$5.95548-0.59212i$}  & \textcolor{red}{$3.4032-0.3354i$}  \\
                   &          & $1$ & N/A                 & $3.6321-0.7545i$  \\
                   &          & $2$ & N/A                 & $4.0958-1.5623i$  \\
                   &          & $3$ & N/A                 & ${\bf{1.9428-2.0061i}}$  \\
                   &          & $4$ & N/A                 & $4.6516-2.4451i$  \\[1ex]
 \hline\hline 
 \end{tabular}
\end{table}

\begin{table}%[ht]
\centering
\caption{QNM modes for tensor perturbations (spin $s = 2$) of a ten-dimensional ($n = 8$) Schwarzschild black hole with GB correction are presented in the table below for different values of $\widehat{\alpha}$ and $\ell_{8} \in \{2, 3, 4\}$. The corresponding results are obtained through our SM, utilising $300$ polynomials with a precision of $300$ digits. In this context, $\Omega$ is given by \eqref{ODET}, while $N$ represents the corresponding overtone. The notation 'N/A' indicates data not available. Moreover, 'SM' stands for Spectral Method.}
\label{table:D10s2L234}
\vspace*{1em}
\begin{tabular}{||c|c|c|c|c|c|c|c|c|c|c|c||}
\hline\hline
$\widehat{\alpha}$ $(D=10)$ & $\ell_{8}$ & $N$ & $\Omega$ (SM)& $\ell_{8}$ & $N$ & $\Omega$ (SM) & $\ell_{8}$ & $N$ & $\Omega$ (SM) \\ [0.5ex]
\hline\hline
$0.0$              & $2$      & $0$ & \textcolor{red}{$3.5513-0.8714i$}   & $3$      & $0$ & \textcolor{red}{$4.1945-0.8654i$}  & $4$  & $0$ & \textcolor{red}{$4.8379-0.8618i$} \\
                   &          & $1$ & $2.7543-2.5826i$                    &          & $1$ & $3.5603-2.5820i$                   &      & $1$ & $4.3063-2.5774i$ \\
                   &          & $2$ & $1.3872-3.2355i$                    &          & $2$ & $2.0124-3.6009i$                   &      & $2$ & $2.8894-4.0520i$ \\
                   &          & $3$ & $0.8077-6.3972i$                    &          & $3$ & $\bf{0.4212-4.6162i}$              &      & $3$ & $1.9924-4.6181i$ \\
                   &          & $4$ & $0.7353-10.0697i$                   &          & $4$ & $1.0032-5.2943i$                   &      & $4$ & $\bf{0.5318-5.0588i}$ \\
$0.1$              & $2$      & $0$ & \textcolor{red}{$3.5629-0.8576i$}   & $3$      & $0$ & \textcolor{red}{$4.2087-0.8527i$}  & $4$  & $0$ & \textcolor{red}{$4.8548-0.8499i$}\\
                   &          & $1$ & $2.8101-2.5509i$                    &          & $1$ & $3.6051-2.5474i$                   &      & $1$ & $4.3465-2.5430i$\\
                   &          & $2$ & $1.4230-3.2243i$                    &          & $2$ & $2.0607-3.6240i$                   &      & $2$ & $3.0065-4.0630i$\\
                   &          & $3$ & $0.8642-6.0887i$                    &          & $3$ & ${\bf{0.3300-4.6914i}}$            &      & $3$ & $2.0545-4.5337i$\\
                   &          & $4$ & $0.7058-9.6829i$                    &          & $4$ & $1.2615-4.9951i$                   &      & $4$ & $0.5675-5.0611i$\\
$0.2$              & $2$      & $0$ & \textcolor{red}{$3.5723-0.8474i$}   & $3$      & $0$ & \textcolor{red}{$4.2205-0.8435i$}  & $4$  & $0$ & \textcolor{red}{$4.8691-0.8413i$}\\
                   &          & $1$ & $2.8501-2.5221i$                    &          & $1$ & $3.6372-2.5188i$                   &      & $1$ & $4.3756-2.5156i$\\
                   &          & $2$ & $1.4489-3.2177i$                    &          & $2$ & $2.1027-3.6483i$                   &      & $2$ & $3.1064-4.0544i$\\
                   &          & $3$ & $0.8910-5.8140i$                    &          & $3$ & $1.4454-4.7899i$                   &      & $3$ & $2.0871-4.4677i$\\
                   &          & $4$ & $0.6226-9.2870i$                    &          & $4$ & $0.7944-8.7687i$                   &      & $4$ & $0.5846-5.0464i$\\
$0.5$              & $2$      & $0$ & \textcolor{red}{$3.5911-0.8282i$}   & $3$      & $0$ & \textcolor{red}{$4.2448-0.8263i$}  & $4$  & $0$ & \textcolor{red}{$4.8989-0.8256i$}\\
                   &          & $1$ & $2.9197-2.4521i$                    &          & $1$ & $3.6921-2.4540i$                   &      & $1$ & $4.4252-2.4560i$\\
                   &          & $2$ & $1.4968-3.2172i$                    &          & $2$ & $2.2363-3.7392i$                   &      & $2$ & $3.3207-3.9746i$\\
                   &          & $3$ & $0.9882-5.0711i$                    &          & $3$ & $1.6987-4.3312i$                   &      & $3$ & $2.1150-4.3304i$\\
                   &          & $4$ & $0.6930-7.9285i$                    &          & $4$ & $1.0824-7.4840i$                   &      & $4$ & $0.5469-4.9044i$\\
$5.0$              & $2$      & $0$ & \textcolor{red}{$3.5303-0.7117i$}   & $3$      & $0$ & \textcolor{red}{$4.1725-0.7127i$}  & $4$  & $0$ & \textcolor{red}{$4.8157-0.7139i$}\\
                   &          & $1$ & $3.1003-2.0595i$                    &          & $1$ & $3.7890-2.0536i$                   &      & $1$ & $4.4612-2.0630i$\\
                   &          & $2$ & $2.4312-2.8169i$                    &          & $2$ & $2.9925-2.9273i$            &      & $2$ & $3.5871-3.0776i$\\
                   &          & $3$ & ${\bf{0.9899-2.9708i}}$             &          & $3$ & ${\bf{1.3689-3.2855i}}$            &      & $3$ & ${\bf{1.7938-3.5814i}}$\\
                   &          & $4$ & $3.3807-4.2816i$                    &          & $4$ & $3.7713-4.0668i$                   &      & $4$ & $4.2359-3.8667i$\\
$10.0$             & $2$      & $0$ & \textcolor{red}{$3.4137-0.6128i$}   & $3$      & $0$ & \textcolor{red}{$4.0266-0.6176i$}  & $4$  & $0$ & \textcolor{red}{$4.6436-0.6238i$}\\
                   &          & $1$ & $3.3396-1.7548i$                    &          & $1$ & $3.9247-1.6813i$                   &      & $1$ & $4.5032-1.6466i$\\
                   &          & $2$ & ${\bf{2.3390-2.4417i}}$             &          & $2$ & ${\bf{2.8216-2.6187i}}$            &      & $2$ & ${\bf{3.3125-2.7957i}}$\\
                   &          & $3$ & $3.9731-3.4872i$                    &          & $3$ & $4.3853-3.2745i$                   &      & $3$ & $4.8649-3.0794i$\\
                   &          & $4$ & $4.8979-5.0987i$                    &          & $4$ & $5.2148-4.9190i$                   &      & $4$ & $5.5854-4.7285i$\\
$20.0$             & $2$      & $0$ & \textcolor{red}{$3.2756-0.4559i$}   & $3$      & $0$ & \textcolor{red}{$3.8410-0.4516i$}  & $4$  & $0$ & \textcolor{red}{$4.4079-0.4545i$}\\
                   &          & $1$ & $3.5749-1.4126i$                    &          & $1$ & $4.1140-1.3040i$                   &      & $1$ & $4.6559-1.2221i$\\
                   &          & $2$ & ${\bf{2.1706-2.1925i}}$             &          & $2$ & ${\bf{2.6078-2.3698i}}$            &      & $2$ & $5.2308-2.4962i$\\
                   &          & $3$ & $4.3376-2.8517i$                    &          & $3$ & $4.7596-2.6638i$                   &      & $3$ & ${\bf{3.0499-2.5373i}}$\\
                   &          & $4$ & $5.3174-4.2312i$                    &          & $4$ & $5.6475-4.0568i$                   &      & $4$ & $6.0251-3.8773i$\\[1ex]
 \hline\hline 
 \end{tabular}
\end{table}

\begin{table}%[ht]
\centering
\caption{QNM modes for tensor perturbations (spin $s = 2$) of an eleven-dimensional ($n = 9$) Schwarzschild black hole with GB correction are presented in the table below for different values of $\widehat{\alpha}$ and $\ell_{9} \in \{2,3,4\}$. The corresponding results are obtained through our SM, utilising $300$ polynomials with a precision of $300$ digits. In this context, $\Omega$ is given by \eqref{ODET} while $N$ represents the corresponding overtone. The notation 'N/A' indicates data not available. Moreover, 'SM' stands for Spectral Method.}
\label{table:D11s2L234}
\vspace*{1em}
\begin{tabular}{||c|c|c|c|c|c|c|c|c|c|c|c||}
\hline\hline
$\widehat{\alpha}$ $(D=11)$ & $\ell_{9}$ & $N$ & $\Omega$ (SM)& $\ell_{9}$ & $N$ & $\Omega$ (SM) & $\ell_{9}$ & $N$ & $\Omega$ (SM) \\ [0.5ex]
\hline\hline
$0.0$              & $2$      & $0$ & \textcolor{red}{$4.0275-0.9699i$}   & $3$      & $0$ & \textcolor{red}{$4.6979-0.9635i$}  & $4$  & $0$ & \textcolor{red}{$5.3681-0.9595i$}\\ 
                   &          & $1$ & $3.0588-2.7797i$                    &          & $1$ & $3.9128-2.8212i$                   &      & $1$ & $4.7047-2.8345i$\\
                   &          & $2$ & $1.8259-3.5297i$                    &          & $2$ & $2.4776-3.7548i$                   &      & $2$ & $3.2090-4.0737i$\\
                   &          & $3$ & $\bf{0.2592-4.1602i}$               &          & $3$ & $1.0005-4.6683i$                   &      & $3$ & $2.1014-5.1217i$\\
                   &          & $4$ & $0.8367-7.6450i$                    &          & $4$ & $0.8232-5.8206i$                   &      & $4$ & $0.9904-5.7814i$\\
$0.1$              & $2$      & $0$ & \textcolor{red}{$4.0375-0.9550i$}   & $3$      & $0$ & \textcolor{red}{$4.7099-0.9498i$}  & $4$  & $0$ & \textcolor{red}{$5.3824-0.9466i$}\\
                   &          & $1$ & $3.1157-2.7566i$   &                           & $1$ & $3.9600-2.7894i$                   &      & $1$ & $4.7463-2.8006i$\\
                   &          & $2$ & $1.8719-3.5007i$   &                           & $2$ & $2.5225-3.7502i$                   &      & $2$ & $3.2711-4.0824i$\\
                   &          & $3$ & ${\bf{0.2987-4.1579i}}$   &                           & $3$ & $1.0729-4.7210i$                   &      & $3$ & $2.2578-5.0920i$\\
                   &          & $4$ & $0.8756-7.3052i$   &                           & $4$ & $0.9372-6.4615i$                   &      & $4$ & $1.0639-5.6778i$\\
                   &          & $5$ & $0.7503-11.3582i$   &                          & $5$ & $0.8226-10.9092i$                  &      & $5$ & $0.8944-10.3130i$\\
$0.2$              & $2$      & $0$ & \textcolor{red}{$4.0453-0.9441i$}   & $3$      & $0$ & \textcolor{red}{$4.7198-0.9399i$}  & $4$  & $0$ & \textcolor{red}{$5.3944-0.9375i$}\\   
                   &          & $1$ & $3.1577-2.7336i$   &                           & $1$ & $3.9939-2.7622i$                   &      & $1$ & $4.7761-2.7731i$\\
                   &          & $2$ & $1.9032-3.4783i$   &                           & $2$ & $2.5574-3.7474i$                   &      & $2$ & $3.3243-4.0890i$\\
                   &          & $3$ & ${\bf{0.3280-4.1668i}}$   &                           & $3$ & $1.1404-4.7844i$                   &      & $3$ & $2.3787-5.0381i$\\
                   &          & $4$ & $0.8764-7.0014i$   &                           & $4$ & $0.9917-6.1299i$                   &      & $4$ & $1.0994-5.5883i$\\
                   &          & $5$ & $0.6249-10.9137i$   &                          & $5$ & $0.7623-10.4331i$                  &      & $5$ & $0.9050-9.8170i$\\
$0.5$              & $2$      & $0$ & \textcolor{red}{$4.0599-0.9236i$}   & $3$      & $0$ & \textcolor{red}{$4.7389-0.9215i$}  & $4$  & $0$ & \textcolor{red}{$5.4180-0.9207i$}\\
                   &          & $1$ & $3.2332-2.6718i$   &                           & $1$ & $4.0519-2.6980i$                   &      & $1$ & $4.8258-2.7115i$\\
                   &          & $2$ & $1.9564-3.4379i$   &                           & $2$ & $2.6331-3.7503i$                   &      & $2$ & $3.4642-4.0991i$\\
                   &          & $3$ & ${\bf{0.3707-4.2186i}}$   &                    & $3$ & $1.6033-4.9909i$                   &      & $3$ & $2.5864-4.8313i$\\
                   &          & $4$ & $0.8649-6.1177i$   &                           & $4$ & ${\bf{0.7711-5.2164i}}$                   &      & $4$ & ${\bf{1.0997-5.3644i}}$\\
                   &          & $5$ & $0.6476-9.2229i$   &                           & $5$ & $1.0477-8.8218i$                   &      & $5$ & $1.4498-8.2491i$\\
$5.0$              & $2$      & $0$ & \textcolor{red}{$3.9751-0.8002i$}   & $3$      & $0$ & \textcolor{red}{$4.6406-0.8014i$}  & $4$  & $0$ & \textcolor{red}{$5.3067-0.8030i$}\\
                   &          & $1$ & $3.4104-2.2842i$   &                           & $1$ & $4.1463-2.2907i$                   &      & $1$ & $4.8552-2.3029i$\\
                   &          & $2$ & $2.8293-3.1360i$   &                           & $2$ & $3.4144-3.1942i$                   &      & $2$ & $4.0392-3.2956i$\\
                   &          & $3$ & ${\bf{1.4457-3.2971i}}$   &                    & $3$ & ${\bf{1.8756-3.6032i}}$            &      & $3$ & ${\bf{2.3370-3.8909i}}$\\
                   &          & $4$ & $3.8075-5.0660i$   &                           & $4$ & $4.1897-4.8453i$                   &      & $4$ & $4.6488-4.6296i$\\
                   &          & $5$ & $4.7750-7.2149i$   &                           & $5$ & $5.0719-7.0344i$                   &      & $5$ & $5.4204-6.8299i$\\
$10.0$             & $2$      & $0$ & \textcolor{red}{$3.8498-0.6980i$}   & $3$      & $0$ & \textcolor{red}{$4.4879-0.7017i$}  & $4$  & $0$ & \textcolor{red}{$5.1287-0.7071i$}\\
                   &          & $1$ & $3.6775-2.0248i$   &                           & $1$ & $4.3143-1.9400i$                   &      & $1$ & $4.9341-1.8969i$\\
                   &          & $2$ & $2.8033-2.6720i$   &                           & $2$ & $3.3192-2.8352i$                   &      & $2$ & ${\bf{3.8582-3.0019i}}$\\
                   &          & $3$ & ${\bf{1.3301-3.0510i}}$   &                    & $3$ & ${\bf{1.7420-3.3487i}}$            &      & $3$ & $5.3181-3.6992i$\\
                   &          & $4$ & $4.4460-4.1420i$   &                           & $4$ & $4.8492-3.9192i$                   &      & $4$ & $6.2018-5.6345i$\\
                   &          & $5$ & $5.5243-6.0133i$   &                           & $5$ & $5.8389-5.8320i$                   &      & $5$ & $7.2041-7.4828i$\\
$20.0$             & $2$      & $0$ & \textcolor{red}{$3.6952-0.5388i$}   & $3$      & $0$ & \textcolor{red}{$4.2897-0.5353i$}  & $4$  & $0$ & \textcolor{red}{$4.8848-0.5385i$}\\
                   &          & $1$ & $3.9593-1.6562i$   &                           & $1$ & $4.5331-1.5365i$                   &      & $1$ & $5.1062-1.4474i$\\
                   &          & $2$ & ${\bf{2.6299-2.4018i}}$   &                    & $2$ & ${\bf{3.1112-2.5808i}}$            &      & $2$ & ${\bf{3.5966-2.7499i}}$\\
                   &          & $3$ & $4.8442-3.3853i$   &                           & $3$ & $5.2687-3.1821i$                   &      & $3$ & $5.7464-2.9919i$\\
                   &          & $4$ & $5.9793-4.9939i$   &                           & $4$ & $6.3107-4.8154i$                   &      & $4$ & $6.6866-4.6271i$\\
                   &          & $5$ & $7.1738-6.5348i$   &                           & $5$ & $7.4534-6.3809i$                   &      & $5$ & $7.7682-6.2114i$\\[1ex]
 \hline\hline 
 \end{tabular}
\end{table}

\begin{table}%[ht]
\centering
\caption{QNM modes for tensor perturbations (spin $s = 2$) of a twelve-dimensional ($n = 10$) Schwarzschild black hole with GB correction are presented in the table below for different values of $\widehat{\alpha}$ and $\ell_{10} \in \{2, 3, 4\}$. The corresponding results are obtained through our SM, utilising $300$ polynomials with a precision of $300$ digits. In this context, $\Omega$ is given by \eqref{ODET}, while $N$ represents the corresponding overtone. The notation 'N/A' indicates data not available. Moreover, 'SM' stands for Spectral Method.}
\label{table:D12s2L234}
\vspace*{1em}
\begin{tabular}{||c|c|c|c|c|c|c|c|c|c|c|c||}
\hline\hline
$\widehat{\alpha}$ $(D=12)$ & $\ell_{10}$ & $N$ & $\Omega$ (SM)& $\ell_{10}$ & $N$ & $\Omega$ (SM) & $\ell_{10}$ & $N$ & $\Omega$ (SM) \\ [0.5ex]
\hline\hline
$0.0$   & $2$ & $0$ & \textcolor{red}{$4.5022-1.0627i$}   & $3$      & $0$ & \textcolor{red}{$5.1956-1.0560i$}  & $4$  & $0$ & \textcolor{red}{$5.8887-1.0517i$}\\ 
        &     & $1$ & $3.3882-2.9298i$                    &          & $1$ & $4.2627-3.0193i$                   &      & $1$ & $5.0898-3.0577i$\\
        &     & $2$ & $2.2130-3.8302i$                    &          & $2$ & $2.9205-3.9677i$                   &      & $2$ & $3.6362-4.2006i$\\
        &     & $3$ & $\bf{0.6403-4.4577i}$               &          & $3$ & $1.3969-4.8525i$                   &      & $3$ & $2.2931-5.2691i$\\
        &     & $4$ & $0.8946-8.8735i$                    &          & $4$ & $0.8461-8.1890i$                   &      & $4$ & $0.8015-6.9259i$\\
$0.1$   & $2$ & $0$ & \textcolor{red}{$4.5107-1.0467i$}   & $3$      & $0$ & \textcolor{red}{$5.2059-1.0414i$}  & $4$  & $0$ & \textcolor{red}{$5.9008-1.0380i$}\\
        &     & $1$ & $3.4407-2.9146i$                    &          & $1$ & $4.3107-2.9921i$                   &      & $1$ & $5.1326-3.0259i$\\
        &     & $2$ & $2.2730-3.7930i$                    &          & $2$ & $2.9662-3.9494i$                   &      & $2$ & $3.6858-4.1933i$\\
        &     & $3$ & ${\bf{0.6952-4.4346i}}$                    &          & $3$ & $1.4668-4.8698i$                   &      & $3$ & $2.3986-5.2875i$\\
        &     & $4$ & $0.9235-8.5020i$                    &          & $4$ & $0.9455-7.7970i$                   &      & $4$ & $1.0969-6.6406i$\\
        &     & $5$ & $0.8074-13.0298i$                   &          & $5$ & $0.8633-12.6188i$                  &      & $5$ & $0.9090-12.1014i$\\
$0.2$   &     & $0$ & \textcolor{red}{$4.5172-1.0352i$}   & $3$      & $0$ & \textcolor{red}{$5.2140-1.0309i$}  & $4$  & $0$ & \textcolor{red}{$5.9108-1.0283i$}\\   
        &     & $1$ & $3.4806-2.8980i$                    &          & $1$ & $4.3455-2.9679i$                   &      & $1$ & $5.1631-2.9995i$\\
        &     & $2$ & $2.3127-3.7611i$                    &          & $2$ & $2.9996-3.9345i$                   &      & $2$ & $3.7247-4.1871i$\\
        &     & $3$ & ${\bf{0.7325-4.4246i}}$                    &          & $3$ & $1.5238-4.8878i$                   &      & $3$ & $2.4955-5.2968i$\\
        &     & $4$ & $0.9050-8.1706i$                    &          & $4$ & $0.9904-7.4427i$                   &      & $4$ & $1.2890-6.3987i$\\
        &     & $5$ & $0.6432-12.5440i$                   &          & $5$ & $0.7531-12.0884i$                  &      & $5$ & $0.8738-11.5389i$\\
$0.5$   & $2$ & $0$ & \textcolor{red}{$4.5284-1.0136i$}   & $3$      & $0$ & \textcolor{red}{$5.2289-1.0115i$}  & $4$  & $0$ & \textcolor{red}{$5.9294-1.0105i$}\\
        &     & $1$ & $3.5559-2.8481i$                    &          & $1$ & $4.4060-2.9075i$                   &      & $1$ & $5.2140-2.9381i$\\
        &     & $2$ & $2.3758-3.6948i$                    &          & $2$ & $3.0637-3.9058i$                   &      & $2$ & $3.8109-4.1763i$\\
        &     & $3$ & ${\bf{0.7933-4.4292i}}$                    &          & $3$ & $1.6639-4.9722i$                   &      & $3$ & $2.7686-5.2574i$\\
        &     & $4$ & $0.7956-7.1657i$                    &          & $4$ & $1.0826-6.4051i$                   &      & $4$ & $1.5159-5.8860i$\\
        &     & $5$ & $0.5933-10.5095i$                   &          & $5$ & $1.0215-10.1226i$                  &      & $5$ & $1.3924-9.6497i$\\
$5.0$   & $2$ & $0$ & \textcolor{red}{$4.4227-0.8852i$}   & $3$      & $0$ & \textcolor{red}{$5.1088-0.8867i$}  & $4$  & $0$ & \textcolor{red}{$5.7952-0.8888i$}\\
        &     & $1$ & $3.7248-2.4675i$                    &          & $1$ & $4.4942-2.4980i$                   &      & $1$ & $5.2356-2.5211i$\\
        &     & $2$ & $3.2025-3.4785i$                    &          & $2$ & $3.8166-3.4801i$                   &      & $2$ & $4.4647-3.5421i$\\
        &     & $3$ & ${\bf{1.9012-3.6048i}}$             &          & $3$ & ${\bf{2.3728-3.8966i}}$            &      & $3$ & ${\bf{2.8721-4.1704i}}$\\
        &     & $4$ & $4.2340-5.8494i$                    &          & $4$ & $4.6076-5.6261i$                   &      & $4$ & $5.0555-5.3938i$\\
        &     & $5$ & N/A                                 &          & $5$ & $5.6311-8.1176i$                   &      & $5$ & $5.9709-7.9145i$\\
$10.0$  & $2$ & $0$ & \textcolor{red}{$4.2884-0.7812i$}   & $3$      & $0$ & \textcolor{red}{$4.9489-0.7848i$}  & $4$  & $0$ & \textcolor{red}{$5.6112-0.7900i$}\\
        &     & $1$ & $3.9842-2.2709i$                    &          & $1$ & $4.6780-2.1828i$                   &      & $1$ & $5.3384-2.1365i$\\
        &     & $2$ & ${\bf{3.2789-2.9072i}}$             &          & $2$ & $3.8046-3.0465i$                   &      & $2$ & $4.3762-3.1894i$\\
        &     & $3$ & $4.9215-4.7999i$                    &          & $3$ & ${\bf{2.2138-3.6357i}}$            &      & $3$ & ${\bf{2.6921-3.9038i}}$\\
        &     & $4$ & $6.1541-6.9287i$                    &          & $4$ & $5.3171-4.5740i$                   &      & $4$ & $5.7729-4.3431i$\\
        &     & $5$ & N/A                                 &          & $5$ & $6.4655-6.7483i$                   &      & $5$ & $6.8204-6.4586i$\\
$20.0$  & $2$ & $0$ & \textcolor{red}{$4.1171-0.6212i$}   & $3$      & $0$ & \textcolor{red}{$4.7367-0.6196i$}  & $4$  & $0$ & \textcolor{red}{$5.3566-0.6241i$}\\
        &     & $1$ & $4.3256-1.8937i$                    &          & $1$ & $4.9324-1.7626i$                   &      & $1$ & $5.5343-1.6668i$\\
        &     & $2$ & ${\bf{3.0946-2.5947i}}$             & $3$      & $2$ & ${\bf{3.6106-2.7710i}}$            &      & $2$ & ${\bf{4.1343-2.9375i}}$\\
        &     & $3$ & $5.3503-3.9260i$                    &          & $3$ & $5.7724-3.7129i$                   &      & $3$ & $6.2485-3.5047i$\\
        &     & $4$ & $6.6442-5.7609i$                    &          & $4$ & $6.9751-5.5809i$                   &      & $4$ & $7.3470-5.3880i$\\
        &     & $5$ & $7.9935-7.5207i$                    &          & $5$ & $8.2741-7.3659i$                   &      & $5$ & $8.5860-7.1959i$\\[1ex]
 \hline\hline 
 \end{tabular}
\end{table}

\begin{table}%[ht]
\centering
\caption{QNM modes for tensor perturbations (spin $s = 2$) of a twenty-six-dimensional ($n = 24$) Schwarzschild black hole with GB correction are presented in the table below for different values of $\widehat{\alpha}$ and $\ell_{24} \in \{2,3,4\}$. The corresponding results are obtained through our SM, utilising $300$ polynomials with a precision of $300$ digits. In this context, $\Omega$ is given by \eqref{ODET}, while $N$ represents the corresponding overtone. The notation 'N/A' indicates data not available. Moreover, 'SM' stands for Spectral Method.}
\label{table:D26s2L234}
\vspace*{1em}
\begin{tabular}{||c|c|c|c|c|c|c|c|c|c|c|c||}
\hline\hline
$\widehat{\alpha}$ $(D=26)$ & $\ell_{24}$ & $N$ & $\Omega$ (SM)& $\ell_{24}$ & $N$ & $\Omega$ (SM) & $\ell_{24}$ & $N$ & $\Omega$ (SM) \\ [0.5ex]
\hline\hline
$0.0$   & $2$ & $0$ & \textcolor{red}{$11.1408-2.0097i$}   & $3$ & $0$ & \textcolor{red}{$11.9828-2.0060i$}      & $4$  & $0$ & \textcolor{red}{$12.8234-2.0031i$} \\ 
        &     & $1$ & $9.3566-4.3662i$                     &     & $1$ & $10.1973-4.4757i$                              &     & $1$ & $11.0429-4.5818i$ \\
        &     & $2$ & $7.6840-6.0071i$                     &     & $2$ & $8.5467-6.2131i$                               &     & $2$ & $9.4213-6.4011i$ \\
        &     & $3$ & $6.0435-7.2221i$                     &     & $3$ & $6.9424-7.5043i$                               &     & $3$ & $7.8434-7.7426i$ \\
        &     & $4$ & $4.3945-8.1116i$                     &     & $4$ & $5.3225-8.4607i$                               &     & $4$ & $6.2294-8.7552i$ \\
$0.1$   & $2$ & $0$ & \textcolor{red}{$11.1395-1.9867i$}   & $3$ & $0$ & \textcolor{red}{$11.9815-1.9844i$}      & $4$  & $0$ & \textcolor{red}{$12.8222-1.9828i$}\\
        &     & $1$ & $9.3715-4.3460i$                     &          & $1$ & $10.2129-4.4559i$                  &      & $1$ & $11.0594-4.5621i$\\
        &     & $2$ & $7.7131-5.9955i$                     &          & $2$ & $8.5786-6.2001i$                   &      & $2$ & $9.4558-6.3858i$\\
        &     & $3$ & $6.0861-7.2242i$                     &          & $3$ & $6.9912-7.5007i$                   &      & $3$ & $7.8949-7.7317i$\\
        &     & $4$ & $4.4487-8.1342i$                     &          & $4$ & $5.3870-8.4704i$                   &      & $4$ & $6.2951-8.7526i$\\
        &     & $5$ & $2.7430-8.7666i$                     &          & $5$ & $3.7162-9.2087i$                   &      & $5$ & $4.6641-9.5684i$\\
$0.2$   & $2$ & $0$ & \textcolor{red}{$11.1372-1.9706i$}   & $3$      & $0$ & \textcolor{red}{$11.9793-1.9695i$} & $4$  & $0$ & \textcolor{red}{$12.8204-1.9688i$}\\   
        &     & $1$ & $9.3813-4.3302i$                     &          & $1$ & $10.2234-4.4403i$                  &      & $1$ & $11.0706-4.5467i$\\
        &     & $2$ & $7.7349-5.9846i$                     &          & $2$ & $8.6025-6.1879i$                   &      & $2$ & $9.4813-6.3716i$\\
        &     & $3$ & $6.1206-7.2228i$                     &          & $3$ & $7.0299-7.4937i$                   &      & $3$ & $7.9342-7.7185i$\\
        &     & $4$ & $4.4964-8.1482i$                     &          & $4$ & $5.4403-8.4711i$                   &      & $4$ & $6.3464-8.7437i$\\
        &     & $5$ & N/A                                  &          & $5$ & $3.7878-9.2293i$                   &      & $5$ & $4.7338-9.5677i$\\
$0.5$   & $2$ & $0$ & \textcolor{red}{$11.1255-1.9418i$}   & $3$      & $0$ & \textcolor{red}{$11.9682-1.9429i$} & $4$  & $0$ & \textcolor{red}{$12.8100-1.9444i$}\\
        &     & $1$ & $9.3958-4.2947i$                     &          & $1$ & $10.2391-4.4050i$                  &      & $1$ & $11.0874-4.5111i$\\
        &     & $2$ & $7.7780-5.9542i$                     &          & $2$ & $8.6491-6.1533i$                   &      & $2$ & $9.5297-6.3316i$\\
        &     & $3$ & $6.1968-7.2068i$                     &          & $3$ & $7.1104-7.4625i$                   &      & $3$ & $8.0111-7.6744i$\\
        &     & $4$ & N/A                                  &          & $4$ & $5.5542-8.4461i$                   &      & $4$ & $6.4469-8.7033i$\\
$5.0$   & $2$ & $0$ & \textcolor{red}{$10.8642-1.7951i$}   &  $3$     & $0$ & \textcolor{red}{$11.6949-1.8039i$} & $4$  & $0$ & \textcolor{red}{$12.5248-1.8123i$}\\
        &     & $1$ & $9.3326-3.9943i$                     &          & $1$ & $10.1621-4.0875i$                  &      & $1$ & $10.9933-4.1759i$\\
        &     & $2$ & $7.9187-5.6297i$                     &          & $2$ & $8.7735-5.7900i$                   &      & $2$ & $9.6280-5.9355i$\\
$10.0$  & $2$ & $0$ & \textcolor{red}{$10.6491-1.7041i$}   & $3$      & $0$ & \textcolor{red}{$11.4676-1.7159i$} & $4$  & $0$ & \textcolor{red}{$12.2854-1.7273i$}\\
        &     & $1$ & $9.2578-3.7736i$                     &          & $1$ & $10.0714-3.8525i$                  &      & $1$ & $10.8845-3.9273i$\\
        &     & $2$ & N/A                                  &          & $2$ & $8.8449-5.6436i$                   &      & $2$ & $9.7389-5.7854i$\\
$20.0$  & $2$ & $0$ & \textcolor{red}{$10.3491-1.6039i$}   & $3$      & $0$ & \textcolor{red}{$11.1515-1.6217i$} & $4$  & $0$ & \textcolor{red}{$11.9540-1.6394i$}\\
        &     & $1$ & $9.2114-3.4654i$                     &          & $1$ & $10.0119-3.5119i$                  &      & $1$ & $10.8060-3.5517i$\\[1ex]
 \hline\hline 
 \end{tabular}
\end{table}

\begin{table}%[ht]
\centering
\caption{QNMs for tensor perturbations (spin $s = 2$) with $\ell_n = 2$ in higher-dimensional Schwarzschild black holes with a GB correction, focusing on large values of the parameter $\widehat{\alpha}$. Results have been computed using our SM, employing $300$ polynomials and a numerical precision of $300$ digits. The frequency parameter \(\Omega\) is defined by equation~\eqref{ODET}, and \(N\) denotes the overtone number.}
\label{table:D7to26s2L2}
\vspace*{1em}
\begin{tabular}{||c|c|c|c|c|c|c|c|c|c|c|c||}
\hline\hline
$D$  & $n$ & $\widehat{\alpha}$ & $\ell_{n}$ & $N$ & $\Omega$ (SM) \\ [0.5ex]
\hline\hline
$7$  & $5$ & $40$               & $2$        & $0$ & \textcolor{red}{$1.8074-0.1750i$}\\
     &     &                    &            & $1$ & $2.2144-0.5638i$\\
     &     &                    &            & $2$ & $2.7487-1.1844i$\\
     &     &                    &            & $3$ & ${\bf{0.7396-1.2091i}}$\\
     &     &                    &            & $4$ & $3.3732-1.7964i$\\
$8$  & $6$ & $40$               & $2$        & $0$ & \textcolor{red}{$2.3680-0.1762i$}\\
     &     &                    &            & $1$ & $2.8234-0.7347i$\\
     &     &                    &            & $2$ & ${\bf{1.1424-1.5083i}}$\\
     &     &                    &            & $3$ & $3.4274-1.5122i$\\
     &     &                    &            & $4$ & $4.1535-2.2980i$\\
     &     & $200$              & $2$        & $0$ & \textcolor{red}{$2.1277-0.1762i$}\\
     &     &                    &            & $1$ & $2.8257-0.6556i$\\
     &     &                    &            & $2$ & $3.5000-1.0094i$\\
     &     &                    &            & $3$ & ${\bf{0.8997-1.1785i}}$\\
     &     &                    &            & $4$ & $4.0876-1.5389i$\\
$10$ & $8$ & $60$               & $2$        & $0$ & \textcolor{red}{$3.1385-0.2316i$}\\
     &     &                    &            & $1$ & $3.7285-0.9914i$\\
     &     &                    &            & $2$ & ${\bf{1.9182-1.8998i}}$\\
     &     &                    &            & $3$ & $4.5542-2.0715i$\\
     &     &                    &            & $4$ & $5.5531-3.1437i$\\
$11$ & $9$ & $80$               & $2$        & $0$ & \textcolor{red}{$3.4855-0.2398i$}\\
     &     &                    &            & $1$ & $4.1573-1.0632i$\\
     &     &                    &            & $2$ & ${\bf{2.2906-2.0410i}}$\\
     &     &                    &            & $3$ & $5.0978-2.2542i$\\
     &     &                    &            & $4$ & $6.2357-3.4281i$\\
$12$ & $10$& $100$              & $2$        & $0$ & \textcolor{red}{$3.8294-0.2502i$}\\
     &     &                    &            & $1$ & $4.5744-1.1381i$\\
     &     &                    &            & $2$ & ${\bf{2.6770-2.1816i}}$\\
     &     &                    &            & $3$ & $5.6279-2.4458i$\\
     &     &                    &            & $4$ & $6.9048-3.7247i$\\
$26$ & $24$& $2000$             & $2$        & $0$ & \textcolor{red}{$8.0646-0.2038i$}\\
     &     &                    &            & $1$ & $9.7992-1.1301i$\\
     &     &                    &            & $2$ & $11.9071-3.1972i$\\[1ex]
 \hline\hline 
 \end{tabular}
\end{table}

\section{Discussion and conclusions}

We have presented a single, unified analysis of scalar, vector, and tensor QNMs of Schwarzschild--Gauss--Bonnet black holes in higher dimensions. Taken together, the three sectors reveal a broad and internally consistent spectral landscape that is considerably richer than suggested by earlier WKB-based studies. Across all sectors, higher-curvature effects generate overdamped modes, overtone reordering, and pronounced deformations of the QNM distributions in the complex-frequency plane. The spectral method resolves these structures with high accuracy even in regimes where the effective potential develops negative wells or multiple barriers.

Two results stand out. First, the scalar and vector sectors exhibit an exact isospectrality at vanishing GB coupling: the scalar monopole and vector dipole share the same QNM spectrum for all $D\geq 5$. Appendix~\ref{AppendixA} shows that this degeneracy is rooted in a Darboux pairing of the corresponding Schr\"odinger operators, and the numerical data confirm that it extends from the fundamental mode to the overtone towers. The degeneracy is lifted immediately when the GB term is switched on, and no analogous pairing appears in the tensor sector.

Second, the tensor sector displays a genuine dynamical instability in six dimensions. Our spectral computations provide the first direct numerical confirmation of the instability anticipated in \cite{Dotti2005CQG, Dotti2005PRD}. The onset of the unstable modes is sharply correlated with the Cohn--Calogero bound, and the corresponding threshold translates into the explicit mass inequality $GM\leq 158.1\,\alpha^{3/2}$. This indicates that the instability is confined to microscopic black holes near the string scale. By contrast, no tensor instability was detected for $D\geq 7$, and the vector sector remains stable throughout the parameter range examined despite the presence of shallow negative pockets in several effective potentials.

A further overarching feature is the strong amplification of the dimensionless QNM frequencies in higher dimensions, especially in the string-motivated cases $D\in\{10,11,12,26\}$. When translated into physical units, these amplified frequencies fall in a range that is inaccessible to current ground-based detectors but may overlap with the design band of future space-based interferometers such as DECIGO. From a phenomenological perspective, this raises the possibility that higher-dimensional and higher-curvature signatures might eventually be constrained through precision ringdown spectroscopy.

Overall, the merged analysis provides a systematic benchmark for QNMs in Einstein--Gauss--Bonnet gravity, clarifies the rôle of stability bounds and isospectral structures, and shows that strong-coupling and high-overtone effects require tools beyond the standard WKB approximation. Natural extensions of the present work include rotating backgrounds, asymptotically anti-de Sitter geometries, and related higher-curvature theories in which similar spectral mechanisms may arise.

\section*{Code availability}

All analytical computations reported in this manuscript have been rechecked in the computer algebra system \textsc{Maple}. Sector-specific \textsc{Maple} worksheets for the scalar, vector, and tensor cases, together with the \textsc{Matlab} routines used to discretise the differential operators \eqref{L0none}--\eqref{L2none}, \eqref{L0noneV}--\eqref{L2noneV}, and \eqref{L0noneT}--\eqref{L2noneT}, solve the resulting quadratic eigenvalue problems, and perform the linear-regression analyses of the spectral gaps, are freely available in the repository below:

\begin{itemize}
    \item \url{https://github.com/dutykh/GaussBonnet/}
\end{itemize}

\clearpage
\appendix

\section{Analytic Proof of Scalar--Vector Isospectrality}
\label{AppendixA}

Let us first recall that the scalar master equation \eqref{ODE0}, in the limit $\alpha\to 0$ and for $\ell_n=0$, reduces to
\begin{equation}\label{odes1}
 f_0(r)\frac{d}{dr}\left(f_0(r)\frac{dR_S}{dr}\right)+\left[\omega^2-\widetilde{V}_S(r)\right]R_S(r)=0,
\end{equation}
where
\begin{equation}\label{A}
 f_0(r)=\lim_{\alpha\to 0}f(r)=1-2Mr^{1-n},\quad
 \widetilde{V}_S(r)=\frac{n(n-2)}{4r^2}f_0^2(r)+\frac{n}{4r}\frac{d f_0^2}{dr}.
\end{equation}
In contrast, the vector master equation \eqref{ODEVP}, in the same limit $\alpha\to 0$ but with $\ell_n=1$, becomes
\begin{equation}\label{odev1}
 f_0(r)\frac{d}{dr}\left(f_0(r)\frac{dR_V}{dr}\right)+\left[\omega^2-\widetilde{V}_V(r)\right]R_V(r)=0,
\end{equation}
with effective potential
\begin{equation}\label{B}
 \widetilde{V}_V(r)=\frac{n(n+2)}{4r^2}f_0^2(r)-\frac{n}{4r}\frac{d f_0^2}{dr}.
\end{equation}
As expected, the potentials displayed in \eqref{A} and \eqref{B} are not identical. Isospectrality does not require pointwise equality, but rather that the two potentials be related through a Darboux transformation. Introducing the tortoise coordinate $u$ by $du/dr=1/f_0$, the radial equations \eqref{odes1} and \eqref{odev1} take the Schr\"odinger form
\begin{eqnarray}
-\frac{d^2 R_S}{du^2}+\widetilde{V}_S(u)R_S(u)=\omega^2 R_S(u),\quad
-\frac{d^2 R_V}{du^2}+\widetilde{V}_V(u)R_V(u)=\omega^2 R_V(u),
\end{eqnarray}
where $\widetilde{V}_{S,V}(u)=\widetilde{V}_{S,V}(r(u))$, with $\widetilde{V}_S$ and $\widetilde{V}_V$ given by \eqref{A} and \eqref{B}, respectively. Define the auxiliary function
\begin{equation}
    W(r)=\frac{n}{2r}f_0(r)
\end{equation}
with $r=r(u)$. This choice is motivated by the fact that both potentials \eqref{A} and \eqref{B} contain the same two basic building blocks, namely $f_0^2/r^2$ and $r^{-1} d f_0^2/dr$. Using the definition of the tortoise coordinate, it is straightforward to verify that
\begin{equation}
    \frac{dW}{du}=-\frac{n}{2r^2}f_0^2(r)+\frac{n}{4r}\frac{d f_0^2}{dr}.
\end{equation}
It then follows that
\begin{equation}
    W^2\pm\frac{dW}{du}=\frac{n(n\mp 2)}{4r^2}f_0^2(r)\pm\frac{n}{4r}\frac{d f_0^2}{dr}.
\end{equation}
This identity shows that the scalar and vector effective potentials can be written compactly as
\begin{equation}\label{C}
    W^2+\frac{dW}{du}=\widetilde{V}_S(u),\quad
    W^2-\frac{dW}{du}=\widetilde{V}_V(u),
\end{equation}
where $\widetilde{V}_S(u)$ and $\widetilde{V}_V(u)$ are defined as in \eqref{A} and \eqref{B}, respectively. Introduce now the first-order differential operators
\begin{equation}
    A=\frac{d}{du}+W(u),\quad
    A^\dagger=-\frac{d}{du}+W(u),
\end{equation}
following the standard notation of supersymmetric quantum mechanics \cite{Cooper2001}. A direct computation yields the factorised form of the scalar and vector Hamiltonians,
\begin{equation}
\mathcal{H}_S=AA^\dagger=-\frac{d^2}{du^2}+ \widetilde{V}_S(u),\quad
\mathcal{H}_V=A^\dagger A=-\frac{d^2}{du^2}+ \widetilde{V}_V(u).
\end{equation}
In the framework of supersymmetric quantum mechanics, the factorised forms above imply that $\mathcal{H}_S$ and $\mathcal{H}_V$ constitute a Darboux pair. As such, they possess identical spectra, leading to the isospectrality condition
\begin{equation}
\omega^{(S)}_{N,\ell=0}(\alpha=0) = \omega^{(V)}_{N,\ell=1}(\alpha=0), \qquad \forall~N \geq 0,
\end{equation}
independently of the spacetime dimension $D$. The only constraint is $D \geq 5$, since the vector potential is defined only for $n=D-2\geq 3$; see \eqref{n3}. This analytical result is fully corroborated by the numerical data reported in the scalar and vector sections of the present work. We emphasise that this isospectrality is a purely Einsteinian feature emerging in higher-dimensional general relativity and is unrelated to higher-curvature corrections such as the Gauss--Bonnet term.

%\bibliography{QNMS}

\end{document}